\documentclass[11pt]{article}
\usepackage{mathrsfs}
\usepackage{amssymb}
\usepackage{fancybox}
\newenvironment{bxd}%
  {\bigskip\noindent\begin{Sbox}\begin{minipage}{0.9\textwidth} }%
  {\end{minipage}\end{Sbox}\fbox{\TheSbox}\bigskip}%
\newenvironment{lilbox}%
  {\bigskip\noindent\begin{Sbox}\begin{minipage}{0.65\textwidth} }%
  {\end{minipage}\end{Sbox}\fbox{\TheSbox}\bigskip}%
\usepackage{amsthm}
\usepackage{amsfonts}
\usepackage{amsmath}

\newcommand{\alg}[1]{\mathfrak{#1}}
\newcommand{\norm}[1]{\| #1\|}
\newcommand{\bh}{\mathfrak{B}(\mathcal{H})}

\newcommand{\ip}[2]{\langle #1,#2\rangle}

\newcommand{\abs}[1]{|#1|}
\theoremstyle{definition}
\newtheorem{assumption}{Assumption}
\newtheorem{proposal}{Proposal}
\swapnumbers
\newtheorem{thm}{Theorem}[section]
\newtheorem*{thm*}{Theorem}
\newtheorem{fact}[thm]{Fact}
\newtheorem*{conj}{Conjecture}
\newtheorem{prop}[thm]{Proposition}
\newtheorem{cor}[thm]{Corollary}
\newtheorem{lemma}[thm]{Lemma}
\newtheorem{defprop}[thm]{Definition/Proposition}

\newtheorem*{emb}{The Embedding Theorem}
\newtheorem*{DR}{Doplicher-Roberts Reconstruction Theorem}

\theoremstyle{definition}
\newtheorem{defn}[thm]{Definition}
\newtheorem{example}[thm]{Example}
\newtheorem{rema}[thm]{Remark}
\newtheorem{note}[thm]{Remark}

\newcommand{\mb}[1]{#1}

\theoremstyle{remark}

\newenvironment{eqn}{\begin{equation}}{\end{equation}}

  {\bigskip\noindent\begin{Sbox}\begin{minipage}{0.9\textwidth} \textbf{Question for MM:}}%
  {\end{minipage}\end{Sbox}\ovalbox{\TheSbox}\bigskip}%

  {\bigskip\noindent\begin{Sbox}\begin{minipage}{0.9\textwidth} }%
  {\end{minipage}\end{Sbox}\ovalbox{\TheSbox}\bigskip}%

\usepackage{amsfonts,amssymb,amsmath,diagrams,bbm}

\newcommand{\Vect}{\mathrm{Vect}}
\newcommand{\SVect}{\mathrm{SVect}}

\newcommand{\restr}{\upharpoonright}
\newcommand{\oli}{{\overline{\imath}}}

\newcommand{\DS}{\displaystyle}
\newcommand{\gp}{\,\cdot \,}

\newcommand{\Obj}{\mathrm{Obj}}

\newcommand{\obj}{\mathrm{Obj}}
\newcommand{\op}{{\mbox{\scriptsize op}}}
\newcommand{\impl}{\Rightarrow}

\newcommand{\ve}{\varepsilon}

\def\wt#1{{\tilde #1}}
\def\2#1{{\mathcal #1}}
\def\3#1{{\mathfrak #1}}
\def\7#1{{\mathbb #1}}
\def\8#1{{\mathbbm #1}}
\def\1#1{{\bf #1}}
\def\ol#1{{\overline #1}}
\def\4#1{{\mathscr #1}}
\def\al#1{{\mathfrak #1}}

\def\a{\alpha}  \def\g{\gamma} \def\d{\delta}
\def\e{\varepsilon} \def\f{\varphi}  
    
\def\om{\omega} \def\Om{\Omega} \def\dd{\partial} \def\D{\Delta}
\def\G{\Gamma}

\newcommand{\Hom}{\mathrm{Hom}}
\newcommand{\Mor}{\mathrm{Mor}}
\newcommand{\End}{\mathrm{End}}
\newcommand{\Aut}{\mathrm{Aut}}
\newcommand{\Inn}{\mathrm{Inn}}
\newcommand{\Rep}{\mathrm{Rep}}
\newcommand{\mcirc}{\,\circ\,}
\newcommand{\tr}{\mathrm{Tr}}
\newcommand{\rarr}{\rightarrow}
\def\id{\mathrm{id}}

\newcommand{\Nat}{\mathrm{Nat}}

\usepackage{wasysym}
\usepackage{tan-gle}
\usepackage{named} 
\usepackage[mathscr]{euscript}

\def\@greek#1{%
 \ifcase#1\or$\alpha$\or$\beta$\or$\gamma$\or$\delta$\or$\varepsilon$%
  \or$\zeta$\or$\eta$\or$\vartheta$\or$\iota$\or$\kappa$\or$\lambda$\or$%
  \mu$\or$\nu$\or$\xi$\or$ o$\or$\varpi$\or$\varrho$\or$\varsigma$\or$\tau$%
  \or$\upsilon$\or$\varphi$\or$\chi$\or$\psi$\or$\omega$\else\@ctrerr\fi}
\def\greek#1{\@greek{\csname c@#1\endcsname}}

\newcommand{\lra}{\:\Longrightarrow \:}

\author{Hans Halvorson\thanks{Department of Philosophy, Princeton
    University, Princeton NJ, US. {\tt hhalvors@princeton.edu}} \\ (with an appendix by Michael
  M{\"u}ger\thanks{Department of Mathematics, Radboud University,
    Nijmegen, NL.  {\tt mueger@math.ru.nl}} )} \title{Algebraic Quantum Field
  Theory\footnote{Forthcoming in \textit{Handbook of the Philosophy of
      Physics}, edited by Jeremy Butterfield and John Earman.  HH
    wishes to thank: Michael M{\"u}ger for teaching him about the
    Doplicher-Roberts Theorem; the editors for their helpful feedback
    and patience; and David Baker, Tracy Lupher, and David Malament
    for corrections.  MM wishes to thank Julien Bichon for a critical reading of the appendix and useful
    comments.}}

\newcommand{\et}{\textit{et al.} }

\usepackage{graphicx}
\newcommand{\fields}{(\pi ,\2H ,\alg{F},(G,k))}

\date{February 14, 2006}
\begin{document}
\maketitle

\begin{abstract}
  Algebraic quantum field theory provides a general, mathematically
  precise description of the structure of quantum field theories, and
  then draws out consequences of this structure by means of various
  mathematical tools --- the theory of operator algebras, category
  theory, etc..  Given the rigor and generality of AQFT, it is a
  particularly apt tool for studying the foundations of QFT.  This
  paper is a survey of AQFT, with an orientation towards foundational
  topics.  In addition to covering the basics of the theory, we
  discuss issues related to nonlocality, the particle concept, the
  field concept, and inequivalent representations.  We also provide a
  detailed account of the analysis of superselection rules by
  S. Doplicher, R. Haag, and J. E. Roberts (DHR); and we give an
  alternative proof of Doplicher and Roberts' reconstruction of fields
  and gauge group from the category of physical representations of the
  observable algebra.  The latter is based on unpublished ideas due to
  Roberts and the abstract duality theorem for symmetric tensor
  $*$-categories, a self-contained proof of which is given in the
  appendix.
\end{abstract}

{\small \tableofcontents }

\vspace{3em} 

\noindent {\LARGE \textbf{Introduction}}

\bigskip

\noindent From the title of this Chapter, one might
suspect that the subject is some \emph{idiosyncratic}
approach to quantum field theory (QFT).  The approach
is indeed idiosyncratic in the sense of demographics:
only a small proportion of those who work on QFT work
on \emph{algebraic} QFT (AQFT).  However, there are
particular reasons why philosophers, and others
interested in foundational issues, will want to study
the ``algebraic'' approach.

In philosophy of science in the analytic tradition, studying the
foundations of a theory $T$ has been thought to presuppose some
minimal level of clarity about the referent of $T$.  (Moreover, to
distinguish philosophy from sociology and history, $T$ is not taken to
refer to the activities of some group of people.)  In the early
twentieth century, it was thought that the referent of $T$ must be a
set of axioms of some formal, preferably first-order, language.  It
was quickly realized that not many interesting physical theories can
be formalized in this way.  But in any case, we are no longer in the
grip of axiomania, as Feyerabend called it.  So, the standards were
loosened somewhat --- but only to the extent that the standards were
simultaneously loosened within the community of professional
mathematicians.  There remains an implicit working assumption among
many philosophers that studying the foundations of a theory requires
that the theory has a mathematical description.  (The philosopher's
working assumption is certainly satisfied in the case of statistical
mechanics, special and general relativity, and nonrelativistic quantum
mechanics.)  In any case, whether or not having a mathematical
description is mandatory, having such a description greatly
facilitates our ability to draw inferences securely and efficiently.

So, philosophers of physics have taken their object of study to be
theories, where theories correspond to mathematical objects (perhaps
sets of models).  But it is not so clear where ``quantum field
theory'' can be located in the mathematical universe.  In the absence
of some sort of mathematically intelligible description of QFT, the
philosopher of physics has two options: either find a new way to
understand the task of interpretation, or remain silent about the
interpretation of quantum field theory.\footnote{For the first option,
  see \cite{wallace}.}

It is for this reason that AQFT is of particular interest for the
foundations of quantum field theory.  In short, AQFT is our best story
about where QFT lives in the mathematical universe, and so is a
natural starting point for foundational inquiries.

\section{Algebraic Prolegomena}

This first section provides a minimal overview of the mathematical
prerequisites of the remainder of the Chapter.

\subsection{von Neumann algebras}

The standard definition of a von Neumann algebra involves reference to a topology,
and it is then shown (by von Neumann's double commutant theorem) that this
topological condition coincides with an algebraic condition (condition 2 in the
Definition \ref{vNA}).  But for present purposes, it will suffice to take the
algebraic condition as basic.

\begin{defn} Let $\2H$ be a Hilbert space.  Let $\bh$ be the set of
  bounded linear operators on $\2H$ in the sense that for each $A\in
  \bh$ there is a smallest nonnegative number $\norm{A}$ such that
  $\langle Ax,Ax\rangle ^{1/2}\leq \norm{A}$ for all unit vectors
  $x\in \2H$.  [Subsequently we use $\norm{\cdot }$ ambiguously for
  the norm on $\2H$ and the norm on $\bh$.]  We use juxtaposition $AB$
  to denote the composition of two elements $A,B$ of $\bh$.  For each
  $A\in \bh$ we let $A^*$ denote the unique element of $\bh$ such that
  $\langle A^*x,y\rangle =\langle x,Ay\rangle$, for all $x,y\in
  \alg{R}$.  \end{defn}

\begin{defn} Let $\alg{R}$ be a $*$-subalgebra of $\bh$, the bounded operators on the
  Hilbert space $\2H$.  Then $\alg{R}$ is a von Neumann algebra if
  \begin{enumerate}
\item $I\in \alg{R}$, 
\item $\alg(\alg{R}')'=\alg{R}$,
\end{enumerate}
where $\alg{R}'=\{ B\in \bh :[B,A]=0 ,\forall A\in \alg{R} \} $. \label{vNA}
\end{defn}

\begin{defn} We will need four standard topologies on the set $\bh$ of bounded linear
  operators on $\2H$.  Each of these topologies is defined in terms of a family of
  seminorms --- see \cite[Chaps.\ 1,5]{kr} for more details.
\begin{itemize}
\item The uniform topology on $\bh$ is defined in terms of a single norm:
$$ \norm{A} =\sup \{ \norm{Av}:v\in \2H ,\norm{v}\leq 1 \} ,$$
where the norm on the right is the given vector norm on $\2H$.  Hence, an operator
$A$ is a limit point of the sequence $(A_i)_{i\in \7N}$ iff $(\norm{A_i-A})_{i\in
  \7N}$ converges to $0$.
\item The weak topology on $\bh$ is defined in terms of the family $\{ p_{u,v}:u,v\in
  \2H \}$ of seminorms where
$$ p_{u,v}(A)=\langle u,Av\rangle .$$
The resulting topology is not generally first countable, and so the closure of a
subset $S$ of $\bh$ is generally larger than the set of all limit points of sequences
in $S$.  Rather, the closure of $S$ is the set of limit points of generalized
sequences (nets) in $S$ --- see \cite[Chap.\ 1]{kr} for more details.  A net
$(A_i)_{i\in \2I}$ in $\bh$ converges weakly to $A$ just in case
$(p_{u,v}(A_i))_{i\in \2I}$ converges to $p_{u,v}(A)$ for all $u,v\in \2H$.
\item The strong topology on $\bh$ is defined in terms of the family $\{ p_v :v\in
  \2H \}$ of seminorms where
$$ p_v(A) =\norm{Av} .$$
Thus, a net $(A_i)_{i\in \2I}$ converges strongly to $A$ iff $(p_v(A_i))_{i\in \2I}$
converges to $p_v(A)$, for all $v\in \2H$.
\item The ultraweak topology on $\bh$ is defined in terms of the family $\{ p_{\rho}
  :\rho \in \2T (\2H ) \}$ where $\2T (\2H )$ is the set of positive, trace 1
  operators (``density operators'') on $\2H$ and 
$$ p_\rho (A)=\mathrm{Tr} (\rho A) .$$
Thus a net $(A_i)_{i\in \2I}$ converges ultraweakly to $A$ just in case
$(\mathrm{Tr}(\rho A_i))_{i\in \2I}$ converges to $\mathrm{Tr}(\rho A)$, for all
$\rho \in \2T (\2H )$.
\end{itemize}
\end{defn}

\begin{fact}
  The topologies are ordered as follows:

  \vspace{1em} \begin{tabular}{lllll}
                &                              & norm &                             &        \\ 
                & \rotatebox{45}{$\subseteq$}  &      & \rotatebox{-45}{$\supseteq$} &        \\
ultraweak       &                              &      &                             & strong \\
                & \rotatebox{-45}{$\supseteq$}  &      & \rotatebox{45}{$\subseteq$} & \\
                &                              & weak &                             & \end{tabular}\vspace{1em}

Since closed sets are just the complements of open sets, this means that a weakly
closed set is ultraweakly closed, and an ultraweakly closed subset is norm closed.
Furthermore, the four topologies on $\bh$ coincide iff $\2H$ is finite dimensional.
\end{fact}

\begin{fact} If $S$ is a bounded, convex subset of $\bh$, then the weak, ultraweak,
  and norm closures of $S$ are the same. \end{fact}

\begin{fact} For a $*$-algebra $\alg{R}$ on $\2H$ that contains $I$, the following
  are equivalent: (i) $\alg{R}$ is weakly closed; (ii) $\alg{R}''=\alg{R}$.  This is
  von Neumann's double commutant theorem. \end{fact}

\begin{defn} Let $\alg{R}$ be a subset of $\bh$.  A vector $x\in \2H$ is said to be
  \emph{cyclic} for $\alg{R}$ just in case $[\alg{R}x]=\2H$, where $\alg{R}x =\{
  Ax:A\in \alg{R} \}$, and $[\alg{R}x]$ is the closed linear span of $\alg{R}x$.  A
  vector $x\in \2H$ is said to be \emph{separating} for $\alg{R}$ just in case $Ax
  =0$ and $A\in \alg{R}$ entails $A=0$.  \end{defn}

\begin{fact} Let $\alg{R}$ be a von Neumann algebra on $\2H$, and let $x\in \2H$.
  Then $x$ is cyclic for $\alg{R}$ iff $x$ is separating for $\alg{R}'$.
\end{fact}

\begin{defn} A \emph{normal} state of a von Neumann algebra $\alg{R}$
  is an ultraweakly continuous state.  We let $\al R_*$ denote the
  normal state space of $\al R$.  \end{defn}

\subsection{$C^*$-algebras and their representations}

\begin{defn} A $C^*$-algebra is a pair consisting of a $*$-algebra
  $\al A$ and a norm $\norm{\cdot}:\al A\to \7C$ such that
  \begin{eqnarray*}
    \norm{AB}\leq \norm{A}\cdot \norm{B} ,\qquad \norm{A^*A}=\norm{A}^2
    ,\end{eqnarray*} 
  for all $A,B\in \alg{A}$.  We usually use $\al A$ to denote the
  algebra and its norm.    
\end{defn}

In this Chapter, we will only use $C^*$-algebras that contain a
multiplicative identity $I$.

\begin{defn} A state $\om$ on $\alg{A}$ is a linear functional such
  that $\om (A^*A)\geq 0$ for all $A\in \alg{A}$, and $\om (I)=1$.
\end{defn}

\begin{defn} A state $\om$ of $\alg{A}$ is said to be \emph{mixed} if
  $\om =\frac{1}{2}(\om _1+\om _2)$ with $\om _1\neq \om _2$.
  Otherwise $\om$ is said to be \emph{pure}.  \end{defn}

\begin{defn} Let $\alg{A}$ be a $C^*$-algebra.  A
  \emph{representation} of $\alg{A}$ is a pair $(\2H ,\pi )$, where
  $\2H$ is a Hilbert space and $\pi$ is a $*$-homomorphism of
  $\alg{A}$ into $\bh$.  A representation $(\2H ,\pi )$ is said to be
  \emph{irreducible} if $\pi (\alg{A})$ is weakly dense in $\bh$.  A
  representation $(\2H ,\pi )$ is said to be \emph{faithful} if $\pi$
  is an isomorphism.  \end{defn}

\begin{defn} Let $(\2H ,\pi )$ and $(\2K ,\phi )$ be representations
  of a $C^*$-algebra $\alg{A}$.  Then $(\2H ,\pi )$ and $(\2K ,\phi )$
  are said to be:
  \begin{enumerate}
  \item \emph{unitarily equivalent} if there is a unitary $U:\2H \to
    \2K$ such that $U\pi (A)=\phi (A)U$ for all $A\in \alg{A}$.
  \item \emph{quasiequivalent} if the von Neumann algebras $\pi
    (\alg{A})''$ and $\phi (\alg{A})''$ are $*$-isomorphic.
  \item \emph{disjoint} if they are not quasiequivalent.
  \end{enumerate} \label{reps}
\end{defn}

\begin{defn} A representation $(\2K ,\phi )$ is said to be a
  \emph{subrepresentation} of $(\2H ,\pi )$ just in case there is an
  isometry $V:\2K \to \2H$ such that $\pi (A)V=V\phi (A)$ for all
  $A\in \al A$.
\end{defn}

\begin{fact} Two representations are quasiequivalent iff they have
  unitarily equivalent subrepresentations.  \end{fact}

The famous Gelfand-Naimark-Segal (GNS) theorem shows that every
$C^*$-algebraic state can be represented by a vector in a Hilbert
space.

\begin{thm}[GNS] Let $\om$ be a state of $\alg{A}$.  Then there is a
  representation $(\2H ,\pi )$ of $\alg{A}$, and a unit vector $\Om
  \in \2H$ such that:
  \begin{enumerate}
  \item $\om (A)=\langle \Om ,\pi (A)\Om \rangle$, for all $A\in
    \alg{A}$;
  \item $\pi (\alg{A})\Om$ is dense in $\2H$.
  \end{enumerate}
  Furthermore, the representation $(\2H ,\pi )$ is the unique one (up
  to unitarily equivalence) satisfying the two conditions.  \label{gns}
\end{thm}

Since we will later need to invoke the details of the GNS
construction, we sketch the outlines of its proof here.

\begin{proof}[Sketch of proof] We construct the Hilbert space $\2H$
  from equivalence classes of elements in $\al A$, and the
  representation $\pi$ is given by the action of left multiplication.
  In particular, define a bounded sesquilinear form on $\al A$ by
  setting
$$ \langle A,B\rangle _\om = \om (A^*B ),\qquad A,B\in \al A .$$
Let $\2H _0$ be the quotient of $\al A$ induced by the norm
$\norm{A}_{\om}=\langle A,A\rangle _{\om}^{1/2}$.  Let $\2H$ be the
unique completion of the pre-Hilbert space $\2H _0$.  Thus there is an
inclusion mapping $j:\al A\to \2H$ with $j(\al A)$ dense in $\2H$.
Define the operator $\pi (A)$ on $\2H$ by setting
$$ \pi (A)j(B) = j(AB) ,\qquad B\in \al A .$$
One must verify that $\pi (A)$ is well-defined, and extends uniquely
to a bounded linear operator on $\2H$.  One must also then verify that
$\pi$ is a $*$-homomorphism.  Finally, if we let $\Om =j(I)$, then
$\Om$ is obviously cyclic for $\pi (\alg{A})$.
\end{proof}

\begin{prop} Let $\om$ be a state of $\alg{A}$. The GNS representation
  $(\2H ,\pi )$ of $\alg{A}$ induced by $\om$ is irreducible iff $\om$
  is pure.  \label{pure-irred}
\end{prop}

\bigskip \noindent {\small \textit{Notes:} Standard references on
  $C^*$-algebras include \cite{kr} and \cite{tak1}.}

\subsection{Type classification of von Neumann algebras}

\begin{defn} Two projections $E,F$ in a von Neumann algebra $\alg{R}$ are said to be
  \emph{equivalent}, written $E\sim F$ just in case there is a $V\in \alg{R}$ such
  that $V^*V=E$ and $VV^*=F$.  \end{defn}

\begin{note} If we were being really careful, we would replace ``equivalent'' in the
  previous definition with ``equivalence modulo $\alg{R}$'', and similarly ``$\sim$''
  with ``$\sim _{\alg{R}}$.''  But we will not run into trouble by omitting the
  reference to $\alg{R}$.  The operator $V$ in the previous definition is called a
  \emph{partial isometry} with \emph{initial projection} $E$ and \emph{final
    projection} $F$. \end{note}

\begin{defn} For von Neumann algebras $\al R_1$ and
  $\al R_2$, we let $\al R_1\wedge \al R_2=\al R_1\cap
  \al R_2$.  We let $\al R_1\vee \al R_2$ denote the
  von Neumann algebra generated by $\al R_1$ and $\al
  R_2$, i.e.\ the intersection of all von Neumann
  algebras containing $\al R_1$ and $\al
  R_2$.  \end{defn}

\begin{defn} $Z(\al R)=\al R\wedge \al R'$ is called
  the \emph{center} of the von Neumann algebra $\al R$.
  A von Neumann algebra $\al R$ is called a
  \emph{factor} just in case $Z(\al R)=\7C I$,
  equivalently, $\al R\vee \al R'=\bh$.  A projection
  $E\in Z(\al R)$ is called a \emph{central projection}
  in $\al R$.
\end{defn}

\begin{defn} Let $E\in \alg{R}$ be a projection, and
  let $E\alg{R}E=\{ EAE:A\in \alg{R}\}$.  Then clearly,
  $E\alg{R}E$ is a linear subspace of $\alg{R}$.
  Furthermore, since for $A,B\in \alg{R}$, $AEB\in
  \alg{R}$ and $A^*\in \alg{R}$, it follows that
  $E\alg{R}E$ is closed under products, as well as
  under $*$.  It is also not difficult to see that
  $E\alg{R}E$ is weakly closed, and hence is a von
  Neumann algebra on $E\2H$.
\end{defn}

\begin{defn} Let $\alg{R}$ be a von Neumann algebra.  A
  projection $E\in \alg{R}$ is said to be:
  \begin{enumerate}
  \item \emph{minimal} just in case $\alg{R}$ contains
    no proper subprojection of $E$.
  \item \emph{abelian} just in case the algebra $E\alg{R}E$ is abelian.
  \item \emph{infinite} just in case there is a
    projection $E_0\in \alg{R}$ such that $E_0<E$ and
    $E\sim E_0$.
  \item \emph{finite} just in case it is not infinite.
  \item \emph{properly infinite} just in case $E$ is infinite and for
    each central projection $P$ of $\al R$, either $PE=0$ or $PE$ is
    infinite.
  \end{enumerate}
\end{defn}

\begin{fact} We have the following relations for projections:
\begin{quote} \begin{center} minimal $\lra$ abelian $\lra$ finite  \\
    properly infinite $\lra$ infinite $\:\Longleftrightarrow \: \neg\,$finite
  \end{center} \end{quote} For factors, the first arrows on both lines can be
reversed.  \end{fact}

We now give the Murray-von Neumann type classification of factors (for
more on this, see \cite[Chap.\ 7]{kr} or \cite[Chap.\ 1]{sunder}).

\begin{defn} A von Neumann factor $\alg{R}$ is said to be: 
  \begin{enumerate}
  \item type I if it contains an abelian projection;
  \item type II if it contains a finite projection, but no abelian projection;
  \item type III if it is neither type I nor type II.
  \end{enumerate}
\end{defn}

The type I factors were already completely classified by Murray and
von Neumann: for each cardinal number $\kappa$ there is a unique (up
to isomorphism) type I$_\kappa$ factor, namely $\bh$ where $\2H$ is a
Hilbert space of dimension $\kappa$.  The type II factors can be
further subdivided according to whether or not the identity projection
$I$ is finite (type II$_1$) or infinite (type II$_{\infty}$).  The
type III factors can be subdivided into types III$_{\lambda}$ with
$\lambda \in [0,1]$, although the basis for this subclassification
depends on Tomita-Takesaki modular theory (see Section \ref{tt}).

For general von Neumann algebras, the type classification must be just
a bit more sophisticated: a type I algebra is defined as an algebra
that has an abelian projection $E$ such that no nontrivial projection
in $Z(\alg{R})$ majorizes $E$.  Similarly, a type II algebra is
defined as an algebra having a finite projection $E$ such that no
nontrivial projection in $Z(\alg{R})$ majorizes $E$.  Thus we have:

\begin{prop} Let $\alg{R}$ be a von Neumann algebra.  Then
  $\alg{R}=\alg{R}_{\mathrm{I}}\oplus \alg{R}_{\mathrm{II}}\oplus
  \alg{R}_{\mathrm{III}}$, where $\alg{R}_{\mathrm{X}}$ is type X for X=I,II,III.
  \label{threeway}
\end{prop}

\begin{proof} See \cite[Thm.\ 6.5.2]{kr}. \end{proof}

We will soon see that the local algebras in QFT are ``typically'' type III, and this
has many interesting implications.  The fact that type III algebras do not have
abelian projections is connected to questions of locality in Section \ref{bell}.  The
fact that the state space of type III$_1$ factors is homogeneous is also connected to
questions of locality in Section \ref{bell}.  The fact that type III algebras do not
contain representatives of their states (i.e.\ density operators) is connected to the
modal interpretation of QFT in Section \ref{modal}.

The following classification of von Neumann algebras is also natural,
but it cuts across the Murray-von Neumann classification.

\begin{defn} A von Neumann algebra $\alg{R}$ is said to be:
\begin{itemize}
\item of \emph{infinite type} if $I$ is infinite in $\alg{R}$;
\item \emph{properly infinite} if $I$ is properly infinite in $\alg{R}$.
\item \emph{semi-finite} if the central projection $E_{\mathrm{III}}$ in $\alg{R}$
  (defined in Prop.\ \ref{threeway}) is zero.
\end{itemize}\end{defn}

The finite factors include the type I$_n$ and type II$_1$ factors.
The infinite factors include the type I$_{\infty}$ factors as well as
the type II$_{\infty}$ and III factors.  The distinction between
finite and infinite factors coincides with the existence of a tracial
state.

\begin{defn} A faithful normalized trace on a von Neumann algebra $\alg{R}$ is a
  state $\rho$ on $\alg{R}$ such that:
\begin{enumerate}
\item $\rho$ is tracial; i.e.\ $\rho (AB)=\rho (BA)$, for all $A,B\in \alg{R}$,
\item $\rho$ is faithful; i.e.\ $\rho (A^*A)=0$ only if $A=0$.
\end{enumerate}
\end{defn}

\begin{fact} A von Neumann factor $\alg{R}$ is finite iff there is a
  faithful normal tracial state $\rho$ on $\alg{R}$.  A von Neumann
  factor $\alg{R}$ is semifinite iff there is a ``faithful normal
  semifinite trace'' on $\alg{R}$; but we do not pause here to define
  this notion.  \end{fact}

\subsection{Modular theory} \label{tt}

We state here without proof some of the basic facts about
Tomita-Takesaki modular theory.  These facts are necessary in order to
understand the classification of type III von Neumann algebras, which
in turn is essential to understanding the mathematical structure of
AQFT.  

\begin{defn} Let $\al R$ be a von Neumann algebra acting on a Hilbert
  space $\2H$, and suppose that $\Om \in \2H$ is cyclic and separating
  for $\al R$.  In such a case, we say that $(\al R,\Om )$ is \emph{in
    standard form}.  Define an operator $S_0$ on $\2H$ by setting
  \[ S_0A\Om = A^*\Om ,\qquad A\in \al R .\] Then $S_0$
  extends to a closed anti-linear operator $S$ on
  $\2H$.  Let $S=J\D ^{1/2}$ be the polar decomposition
  of $S$, so that $\D$ is positive (but generally
  unbounded), and $J$ is anti-unitary.  (Recall that a
  positive operator has spectrum in $\7R^+$.)  We call
  $\D$ the \emph{modular operator} and $J$ the
  \emph{modular conjugation} associated with the pair
  $(\al R,\Om )$.  \end{defn}

\begin{thm} Let $\al R$ be a von Neumann algebra with
  cyclic and separating vector $\Om$.  Then $J\Om =\Om
  =\D \Om$ and
  \begin{eqnarray*} \D ^{it}\al R\D ^{-it} &=&
    \al R ,\qquad \forall t\in \7R ,\\
    J\al R J&=&\al R' .\end{eqnarray*} \end{thm}

\begin{proof} See \cite[Thm.\ 9.2.9]{kr}, or \cite[Thm.\
  2.3.3]{sunder}.
\end{proof}

\begin{defn} Let $(\al R,\Om )$ be in standard form, and let $\om$ be
  the state of $\al R$ induced by $\Om$.  For each $t\in \7R$, define
  the \emph{modular automorphism} $\sigma ^{\om}_t$ of $\al R$ by
$$\sigma ^{\om}_t(A)=\D ^{it}A\D ^{-it} ,\qquad A \in  \al R ,$$
for all $A\in \al R$.  Define a $*$ anti-isomorphism
$\g :\al R\to \al R'$ by setting $\g (A)=JA^*J$, for
all $A\in \al R$.  \end{defn}

\begin{defn} If $\al A$ is a $C^*$-algebra, we let $\Inn \al A$ denote
  the group of inner automorphisms of $\al A$; i.e.\ $\a \in \Inn \al
  A$ just in case there is a unitary $U\in \al A$ such that $\a
  (A)=UAU^*$ for all $A\in \al A$.  \end{defn}

The spectrum of the modular operator $\D$ gives a rough
measure of the periodicity of the modular automorphism
group $( \sigma ^{\om}_t)_{t\in \7R}$; i.e.\ the
smaller the spectrum of $\D$, the closer the
automorphism $\sigma ^{\om}_t$ is to the identity
$\iota :\al R\to \al R$.  In the extreme case, if
$\mathrm{sp}\D=\{ 1\}$, then $\sigma ^{\om}_t=\iota$
for all $t\in \7R$.  Conversely, as $\D$ goes up to
$\7R ^+$, the group $(\sigma _t^{\om})_{t\in \7R}$
tends toward being ergodic (i.e.\ having no fixed
points).

\begin{defn} Define the \emph{modular spectrum} $S(\al
  R)$ of $\al R$ by \begin{eqnarray*} S(\al R) &=&
    \bigcap _{\om }\mathrm{sp} (\D _{\om }
    ),\end{eqnarray*} where $\om$ runs over the family
  of faithful normal states of $\al R$, and $\D _\om$
  are the corresponding modular operators.
\end{defn}

\begin{prop} Let $\al R$ be a von Neumann factor with cyclic and
  separating vector $\Om $.  Then the following are equivalent:
  \begin{enumerate}
  \item $\al R$ is semifinite.
\item For all $t\in \7R$, the modular automorphism $\sigma ^{\om}_t$
  is inner; i.e.\ there is a unitary $U\in \al R$ such that $\sigma
  ^{\om}_t(A)=UAU^*$ for all $A\in \al R$.
\item $S(\al R)=\{ 1\}$. \end{enumerate}
\end{prop}

\begin{proof} See \cite[p.\ 122]{tak2} and \cite[p.\ 111]{sunder}.
\end{proof}

We now proceed to Connes' subclassification of the type III factors.
This subclassification uses the notion of the ``period of the flow of
weights'' (where a weight is a generalization of the notion of a
state).  However, in order to bypass some background material, we use
the following (provably equivalent) definition.

  \begin{defn} A factor $\al R$ of type III is said to be:
\begin{enumerate}
\item Type III$_0$ if $S(\al R)=\{ 0,1\}$.
\item Type III$_\lambda$, $\lambda \in (0,1)$, if $S(\al R) = \{
  \lambda ^n :n\in \7Z \} \cup \{ 0\}$.
\item Type III$_1$ if $S(\al R)=\7R ^+$.
\end{enumerate} \label{connes}
\end{defn} 

The conditions in Defn.\ \ref{connes} do not bear their physical
interpretation on their sleeve.  That is, it is not immediately clear
how the physics of type III$_{\lambda}$ algebras differs (if at all)
from that of type III$_\mu$ algebras, for $\lambda \neq \mu$.
However, a result of Connes and St{\o}rmer [\citeyear{consto}] cashes
out some of the significance of the distinctions between different
types of factors.

\begin{defn} Let $\al R$ be a von Neumann algebra, and let $\al R_*$
  be its normal state space.  We define the \emph{diameter of the
    state orbit space} $d(\al R)$ by 
$$ d (\al R) = \sup \Bigl\{ \inf \bigl\{ \norm{ (\om _1\circ \a )-\om _2 } :
\a \in \Inn \al R \bigr\} : \om _1,\om _2 \in \al R_{*}
\Bigr\} .$$ Alternatively, let $[\om ]$ denote the norm
closure of $\{ \om \circ \a :\a \in \Inn \al R\}$ (the
orbit of the state under inner automorphisms), and let
$K$ denote the quotient of the normal state space $\al
R_*$.  Then $d(\al R)$ is the diameter of $K$ relative
to the induced metric
\[ \ol d([\om _1],[\om _2]) = \inf \{ \norm{\om _1'-\om _2'} :\om '_i\in
[\om _i] \} .\] 
\end{defn}

Clearly $d(\al R )\in [0,2]$, with $d(\al R)=0$ iff the
orbit of every state is dense in the normal state
space.  If $\al R$ is not a factor, then there are
states $\om _1,\om _2$ such that $\norm{\om _1\circ \a
  -\om _2}=2$ for all $\a \in \Inn \al R$, and so
$d(\al R)=2$.  For type I$_n$ factors, the distance
between normal states is the same as the trace norm
distance of the corresponding density operators.  In
this case, we have
$$d(\al R)= 2\Bigl( 1-\frac{1}{n} \Bigr) =\norm{\tau -\om },$$
where $\tau$ is the trace and $\om$ is any pure state.  We also have
$d(\al R)=2$ for factors of type I$_\infty$ and of type II \cite[p.\
430]{tak2}.

If $d(\al R)$ gives some sort of measure of ``how noncommutative'' the
algebra $\al R$ is, then type III$_1$ factors are the most
noncommutative.

\begin{defn} A von Neumann algebra $\alg{R}$ is said to be
  \emph{countably decomposable} just in case any family of mutually
  orthogonal projection operators in $\alg{R}$ is countable.
\end{defn}

\begin{prop} If $\al R$ is a countably decomposable
  factor of type III$_\lambda$, then
$$ d(\al R) = 2\frac{1-\lambda ^{1/2}}{1+\lambda
  ^{1/2}}  .$$ \end{prop}

\begin{proof} See \cite{consto} and \cite[p.\ 427]{tak2}. \end{proof}

The function $f(\lambda )=2(1-\lambda ^{1/2})/(1+\lambda ^{1/2})$ is
monotonically decreasing on $[0,1]$.  In particular, $f(1)=0$ so that,
for type III$_1$ factors, the orbit of any normal state $\om$ is norm
dense in the state space.  According to Connes [\citeyear[p.\
473]{con}] this means that ``one cannot distinguish between two states
of a factor of type III$_1$ by means of a property that is closed and
invariant under inner automorphisms.''  In particular, since two
unitarily equivalent states must be considered to be ``equally
mixed,'' there are no distinctions to be drawn in terms of the
mixedness of states of a type III$_1$ factor.

\bigskip \noindent {\small \textit{Notes:} For an overview of modular
  theory, see \cite{modular} or \cite{con}.  For a full treatment, see
  \cite{tak2}.  For a detailed exposition of applications of modular
  theory in QFT, see \cite{brch}.}

\section{Structure of the Net of Observable Algebras}

\label{basics}

\subsection{Nets of algebras, basic properties} 

AQFT proceeds by isolating some structural assumptions that hold in
most known QFT models.  It formalizes these structural assumptions,
and then uses ``abstract but efficient nonsense'' to derive
consequences of these assumptions.

The basic formalism of AQFT is a ``net of local observable algebras''
over spacetime.  Although this formalism can be applied to a very wide
class of spacetimes, we restrict attention in this Chapter mostly to
Minkowski spacetime.

An open \emph{double cone} in Minkowski spacetime is the intersection
of the causal future of a point $x$ with the causal past of a point
$y$ to the future of $x$.  Let $\2K$ be the set of open double cones
in Minkowski spacetime, and let $O\mapsto \alg{A}(O)$ be a mapping
from $\2K$ to $C^*$-algebras.  We assume that all our $C^*$-algebras
are unital, i.e.\ have a multiplicative identity.  We assume that the
set $\{ \alg{A}(O):O\in \2K \}$ of $C^*$-algebras (called a \emph{net
  of observable algebras over Minkowski spacetime}) is an inductive
system in the sense that:

\begin{quote} If $O_1\subseteq O_2$, then there is an embedding (i.e.\
  an isometric $*$-homomorphism) ${\a _{12}:\alg{A}(O_1)\to
  \alg{A}(O_2)}$.  \end{quote}

\begin{assumption}[Isotony] The mapping $O\mapsto \al A (O)$ is an
  inductive system.  \end{assumption}

The isotony assumption is sometimes motivated by the idea that an
observable measurable in a region $O_1$ is \emph{a fortiori}
measurable in any region $O_2$ containing $O_1$.  But the isotony
axiom is also justified by its utility: for, if $\{ \alg{A}(O):O\in
\2K \}$ is an inductive system, then there is an inductive limit
$C^{*}$-algebra $\alg{A}$ generated by all the local algebras.  We
call $\alg{A}$ the \emph{quasilocal} algebra, because it contains
observables that can be uniformly approximated by local observables.

\begin{note} In some spacetimes, the set of double cones is not
  directed.  In many such cases, it is still possible to define the
  quasilocal algebra by means of more sophisticated techniques
  \cite{glob}.
\end{note}

Now we turn to the main \emph{relativistic} assumption of AQFT.
\begin{assumption}[Microcausality] A net $\alg{A}$ of $C^*$-algebras
  is said to satisfy \emph{microcausality} just in case if $O_1,O_2$
  are spacelike separated double cones, then
  $[\alg{A}(O_1),\alg{A}(O_2)]=\{ 0\}$.
  \label{micro}
\end{assumption}
This assumption is thought to reflect the constraints on spacetime structure imposed
by the theory of relativity.  

\begin{note} It is \emph{not} a tenet of AQFT that quantities that are associated to
  spacelike separated regions must be represented by commuting operators.  In fact,
  Fermi field operators assigned to spacelike separated regions will anticommute.
  So, AQFT has need of a distinction between observable (represented by elements of
  $\alg{A}(O)$) and unobservable quantities (represented by ``field operators'').
  For more on this distinction, see Sections \ref{proto-DHR} and following on DHR
  superselection theory.
\end{note}

In this Chapter, we will not attempt to justify or to dispute the microcausality
assumption.  However, we will briefly discuss its connection to issues of locality in
Section \ref{summers}.

\subsection{Existence/uniqueness of vacuum states/representations}

\label{uniq}

\subsubsection{The existence of translation-invariant states}

In this section, we inquire concerning the existence and uniqueness of
vacuum states and representation.  For this, recall that an
\emph{affine space} (e.g.\ Minkowski spacetime) is a triple consisting
of a set $S$, a vector space $V$, and a map $+:S\times V\to S$
satisfying certain properties.  In this case, $V$ is called the
\emph{translation group}.

\begin{assumption}[Translation Covariance] If $\alg{A}$ is a net of
  operator algebras on an affine space, then we assume that there is a
  faithful, continuous representation $x\mapsto \a _x$ of the
  translation group in the group $\Aut \alg{A}$ of automorphisms of
  $\alg{A}$, and $$ \a _x(\alg{A}(O))=\alg{A}(O+x) ,$$ for any double
  cone $O$, and translation $x$.
\end{assumption}

\begin{note} For the case of Minkowski spacetime, the translation
  group is a subgroup of the Poincar{\'e} group.  In many cases of
  physical interest, $x\to \a _x$ extends to a representation of the
  full Poincar{\'e} group in the group $\Aut \al A$ of automorphisms
  of $\al A$.  But we will only need that fact for one result (Prop.\
  \ref{kami}).
\end{note}

Translation invariance has traditionally been thought to be a necessary condition on
a vacuum state.

\begin{fact} If there is an action $\a$ of the translation group on
  $\al A$, then translation-invariant states of $\al A$ exist.
  Indeed, since the translation group is abelian, it has an invariant
  mean $\mu$ --- i.e.\ a translation invariant, positive linear
  functional on the algebra $L^{\infty}(G)$ of essentially bounded
  measurable (with respect to the Haar measure) functions on the group
  $G$.  Given a state $\om$ of $\al A$, we can then define an averaged
  state $\rho$ by
$$ \rho (A):=\int \om (\a _xA)d\mu (x) .$$ 
The state $\rho$ is translation invariant.  (See Emch,
this volume, Section 3.5.)
\end{fact}

\begin{note} The preceding argument cannot be used to show the
  existence of Lorentz invariant states.  The Lorentz group is not
  \emph{amenable}, and so does not admit an invariant mean.  Hence, we
  cannot use these general methods to prove the existence of Lorentz
  invariant states.  Of course, in concrete models (e.g.\ free Bose
  and Fermi fields) there are other way to establish the existence of
  such states.
\end{note}

Let $G$ be a group acting by automorphisms on $\alg{A}$.  A
generalization of the GNS theorem shows that a $G$-invariant state
$\om$ of $\alg{A}$ gives rise to a GNS Hilbert space $\2H$ that
carries a unitary representation $U$ of $G$, and the GNS vector $\Om$
is invariant under the $G$-action on $\2H$.

\begin{fact} Let $\a$ be a strongly continuous action of the group $G$
  by automorphisms of $\alg{A}$.  If $\om$ is a $G$-invariant state of
  $\alg{A}$, then the GNS representation $(\2H ,\pi )$ of $\alg{A}$
  induced by $\om$ is $G$-covariant in the sense that there is a
  strongly continuous representation $U$ of $G$ in the unitary group
  of $\bh$ such that
  \begin{enumerate}
  \item $U(g)\pi (A)U(g)^* =\pi (\a _g(A))$, for all $A\in \alg{A}$,
  \item $U(g)\Om =\Om$ for all $g\in G$.
\end{enumerate} \label{unitary}
\end{fact}

\subsubsection{Only one vacuum per Hilbert space} \label{vacuum}

\begin{note} When considering the group $\Aut \alg{A}$ of
  automorphisms of a $C^*$-algebra, we take as our standard topology
  the strong topology on the set $L(\alg{A})$ of bounded linear
  mappings on $\alg{A}$ (considered as a Banach space).  That is, $\a
  _i$ converges to $\a$ just in case for each $A\in \alg{A}$, $\a
  _i(A)$ converges to $\a (A)$ in the norm on $\alg{A}$.  \end{note}

\begin{defn} We use the GNS representation theorem (Thm.\ \ref{gns})
  to transfer terminology about representations (Defn.\ \ref{reps}) to
  terminology about states.  So, e.g., we say that two states are
  \emph{disjoint} if their GNS representations are disjoint.
\end{defn}

A vacuum state should be at least translation invariant.  Furthermore,
the microcausality assumption on the net $\alg{A}$ entails that any
two observables commute ``in the limit'' where one is translated out
to spacelike infinity.  That is, for any $A,B\in \alg{A}$, and for any
spacelike vector $x$, $$ \lim _{t\rightarrow \infty}\norm{ [\a
  _{tx}(A),B] } =0 .$$ This in turn entails that $G$ acts on $\alg{A}$
as a \emph{large group of automorphisms} in the following sense:
\begin{quote} If $\om$ is a $G$-invariant state and $(\2H ,\pi )$ is
  the GNS representation of $\alg{A}$ induced by $\om$, then for any
  $A\in \alg{A}$, $$ \overline{\mathrm{conv}}\bigl\{ \pi (\a
  _g(A)):g\in G \bigr\} , $$ has nonempty intersection with $\pi
  (\alg{A})'$.  \end{quote} Here we use $\overline{\mathrm{conv}}S$ to
denote the weakly closed convex hull of $S$.  (See \cite{storm} for
the relevant proofs.)  Note however that we would also expect the same
to be true in a non-relativistic setting, because we would expect
observables associated with disjoint regions of space to commute.  (We
have not invoked the fact that any vector in Minkowski spacetime is
the sum of two spacelike vectors.)

Thanks to extensive research on ``$C^*$-dynamical systems,'' much is
known about $G$-invariant states when $G$ acts as a large group of
automorphisms of $\alg{A}$.  In particular, the set of $G$-invariant
states is convex and closed (in the weak* topology), hence the set has
extreme points, called \emph{extremal invariant states}.  (Obviously
if a pure state of $\alg{A}$ is $G$-invariant, then it is extremal
invariant.)  Furthermore, we also have the following result concerning
the disjointness of $G$-invariant states.

\begin{prop} Let $\om$ be a $G$-invariant state of $\alg{A}$, let $\2H$ be its GNS
  Hilbert space, and let $\Om$ be the GNS vector.  Then the following are equivalent:
\begin{enumerate}
\item $\om$ is clustering in the sense that
$$ \lim _{t\to \infty }\om (\a _{tx}(A)B)=\om (A)\om (B) .$$
\item $\om$ is extremal invariant.
\item If a $G$-invariant state $\rho$ is quasiequivalent to $\om$,
  then $\rho =\om$.  In other words, no other $G$-invariant state is
  quasiequivalent to $\om$.
\item The ray spanned by $\Om$ is the unique (up to scalar multiples) $G$-invariant
  subspace of $\2H$.
\end{enumerate} \label{G-inv}
\end{prop}

\begin{proof} See \cite{storm}.  For related details,
  see also \cite[pp.\ 183, 287]{emch} and Emch, this
  volume, Section 3.
\end{proof}

So, if a (vacuum) state is clustering, then no other translation
invariant state is in its folium (i.e.\ the set of states that are
quasiequivalent to that state).  Similarly, if a state is
\emph{extremal} invariant (\emph{a fortiori} if it is pure) then it is
the unique translation invariant state in its folium.

\begin{note} The existence of disjoint vacua is related to spontaneous
  symmetry breaking.  See Section \ref{ssb}.  \end{note}

\begin{note} Prop.\ \ref{G-inv} plays a central role in the proof of
  ``Haag's theorem'' given in \cite[p.\ 248]{emch}.  In particular,
  the uniqueness of extremal $G$-invariant states is equated with the
  nonexistence of ``vacuum polarization.''
\end{note}

\subsection{The Reeh-Schlieder Theorem} \label{rst}

We have assumed that a vacuum state is translation invariant.  But we
expect a vacuum state to obey a stronger constraint that reflects the
relativistic nature of the theory.  In particular, the unitary
representation defined in Fact \ref{unitary} is generated
infinitesimally by the four momentum operator $\mathbf{P}$.  (The idea
of a four momentum operator can be made precise in the ``SNAG
[Stone-Naimark-Ambrose-Gelfand] Theorem,'' which generalizes Stone's
theorem on the existence of self-adjoint operators generating
one-parameter unitary groups.)  We require that the energy is positive
in every Lorentz frame, equivalently, that the spectrum of
$\mathbf{P}$ lies in the forward light cone.

We now generalize this requirement by abstracting away
from the details of the forward lightcone.  The forward
lightcone $G_+$ has the following property: $G_+\cap
(-G^+)=\{ 0\}$ where $-G_+=\{ -g:g\in G_+\}$.  So, the
spectrum condition only requires that the unitary
representation of the translation group has spectrum in
a set that is asymmetric under taking additive
inverses.

\begin{assumption}[Spectrum Condition] Let $G$ be the translation
  group, and let $\om$ be a $G$-invariant state of $\al A$.  We say
  that the pair $(\al A,\om )$ satisfies the \emph{spectrum condition}
  just in case: there is a subset $G_+$ of $G$ such that $G_+\cap
  (-G_+)=\{ 0\}$, and in the GNS representation $(\2H ,\pi )$ of $\al
  A$ induced by $\om$, the spectrum $\mathrm{sp}(U)$ of the induced
  unitary representation of $G$, is contained in $G_+$.
  \label{spectrum}
\end{assumption}

The Reeh-Schlieder Theorem shows that the spectrum
condition entails that the vacuum vector $\Om$ is
cyclic for \emph{every} local algebra.  For this
theorem, we suppose that a translation invariant vacuum
state $\om$ on $\alg{A}$ has been chosen, and that
$(\2H ,\pi )$ is the GNS representation of $\alg{A}$
induced by $\om$.  We then define a corresponding net
$\alg{R}$ of von Neumann algebras on $\2H$ by
$$O\mapsto \alg{R}(O)\equiv \pi
(\alg{A}(O))'' .$$ If the net $\alg{A}$ satisfies microcausality, then so will
$\alg{R}$.  Since $\Om$ is cyclic for $\pi (\alg{A})$, the set $\{ \alg{R}(O)\Om
:O\in \2K \}$ is dense in $\2H$.

To prove the theorem, we need one additional assumption.
\begin{assumption} The net $O\mapsto \alg{R}(O)$ is
  said to satisfy \emph{additivity} just in case for
  any double cone $O$, the set $\{ \al R(O+x):x\in G\}$
  generates $\al R$ as a $C^*$-algebra. (Here again,
  $G$ denotes the translation group.)  \label{add}
\end{assumption}
The additivity assumption is sometimes justified on the grounds that there should be
no smallest length scale in the theory --- i.e.\ any observable is generated by
taking products, sums, etc.\ of observables from arbitrarily small regions.

\begin{thm}[Reeh-Schlieder] Suppose that the net $O\mapsto \alg{R}(O)$ satisfies the
  spectrum condition and additivity.  Then for all double cones $O$, $\Om$ is cyclic
  for $\alg{R}(O)$.  If the net $\alg{R}$ also satisfies microcausality, then $\Om$
  is separating for every local algebra.
\end{thm}

The Reeh-Schlieder (RS) Theorem has been one of the
more intensely studied issues in the foundations of
relativistic QFT.  In a pair of articles
\cite{red1,red2}, Redhead shows that the RS Theorem entails
that the vacuum state displays nonlocal
correlations. (See also \cite{gbell}).  Redhead also
points out since the vacuum is separating for each
local algebra, every local event has a nonzero
probability of occurring in the vacuum state; in
particular, there can be no local number operators
(since they would have the vacuum state as an
eigenvector).  Finally, \cite{flem} argues that RS Theorem
entails a pernicious sort of nonlocality, worse than
the nonlocality in non-relativistic QM, and so
indicates a need to revise the standard formulation of
AQFT.  (For one possible reply, see \cite{me}.)

Due to the use of the spectrum condition, it would seem
that RS Theorem is a ``purely relativistic result,''
without analogue in non-relativistic QM or QFT (see
\cite{saund}).  Furthermore, we might expect that many
other results of relativistic QFT that are derived from
RS Theorem would fail for non-relativistic theories.
Indeed, non-relativistic QFT \emph{does} admit local
number operators.  However, a version of the spectrum
condition, and consequently a version of RS Theorem has
been shown to hold for non-relativistic theories
\cite{req}.

\bigskip \noindent {\small \textit{Notes:} The original Reeh-Schlieder
  Theorem was formulated in the axiomatic approach to QFT, and can be
  found in \cite{reeh}.  More up-to-date presentations of the theorem
  can be found in \cite{horuzhy,antoni,russ}, and \cite{araki}.}

\subsection{The funnel property}

\begin{defn} Let $\al R_1,\al R_2$ be von Neumann
  algebras on $\2H$ such that $\al R_1\subseteq \al
  R_2$.  If there is a vector $\Om \in \2H$ that is
  cyclic and separating for $\al R_1,\al R_2$, and $\al
  R_1'\cap \al R_2$, then the pair $(\al R_1,\al R_2)$
  is said to be a \emph{standard inclusion} of von
  Neumann algebras.  \end{defn}

\begin{note} Let $O\mapsto \al R(O)$ be a net of von
  Neumann algebras on Minkowski spacetime.  Suppose
  that the Reeh-Schlieder property holds for $\Om$,
  i.e.\ for each double cone $O$, $\Om$ is cyclic and
  separating for $\al R(O)$.  Then if $O_1,O_2$ are
  double cones such that the closure $\ol{O}_1$ of
  $O_1$ is contained in $O_2$, then the pair
  $(\alg{R}(O_1),\alg{R}(O_2))$ is a standard inclusion
  of von Neumann algebras.  \end{note}

\begin{defn} Let $\al R_1,\al R_2$ be von Neumann
  algebras on $\2H$ such that $\al R_1\subseteq \al
  R_2$.  The pair $(\al R_1,\al R_2)$ is said to be a
  \emph{split inclusion} if there is a type I factor
  $\al N$ such that $\al R_1\subseteq \al N\subseteq
  \al R_2$.  \label{banana} \end{defn}

\begin{assumption}[Funnel Property] The net $O\mapsto
  \al R(O)$ of von Neumann algebras is said to satisfy
  the \emph{funnel property} if for any double cones
  $O_1,O_2$ with $\ol O_1$ contained in $O_2$, the pair
  $(\al R(O_1),\al R(O_2))$ is a split inclusion.
\end{assumption}

\begin{note} A type I factor $\alg{N}$ is countably decomposable iff
  $\alg{N}$ is isomorphic to $\bh$ with $\2H$ separable iff $\alg{N}$
  is separable in the ultraweak topology (see \cite[Exercise
  5.7.7]{kr}).
\end{note}

In our discussion of superselection theory (Sections
\ref{proto-DHR}--\ref{last-DHR}), at one crucial juncture (Prop.\
\ref{uto}, p.\ \pageref{uto}) we will have to invoke the assumption
that the vacuum Hilbert space is separable.  This will be the only
place in the Chapter where we need to assume that a Hilbert space is
separable.  In particular, the separability assumption is needed to
establish the correspondence between two notions of superselection
sectors, one of which is physically motivated, and one of which is
mathematically useful.  The following result is the only attempt we
will make to connect the separability assumption to something with
(perhaps) more clear physical significance.  (In general, we are
highly suspicious of the physical warrant for the separability
assumption; compare with Section \ref{pointy}, and with \cite{hans}.)

\begin{prop} Let $\alg{R}$ be a net of von Neumann algebras on $\2H$,
  and suppose that $\Om \in \2H$ be cyclic and separating for all
  local algebras.  If the net satisfies the funnel property, then
  $\2H$ is separable.  \label{separable}
\end{prop}

\begin{proof} (Compare with Prop.\ 1.6 of \cite{split}.) Let $O_1,O_2$
  be double cones with $\ol O_1\subseteq O_2$.  Let $\alg{N}$ be a
  type I factor such that $\alg{R}(O_1)\subseteq \alg{N}\subseteq
  \alg{R}(O_2)$, and let $\om$ be the state of $\alg{N}$ induced by
  $\Om$.  Recall that $\alg{N}$ is isomorphic to $\alg{B}(\2K )$ for
  some Hilbert space $\2K$.  Since $\alg{N}\subseteq \alg{R}(O_2)$ and
  $\Om$ is separating for $\alg{R}(O_2)$, $\om$ is faithful and
  normal.  Hence $\2K$ is separable, and there is a countable set
  $\alg{N}_0$ that is ultraweakly dense in $\alg{N}$.  Since
  $\alg{R}(O_1)\subseteq \alg{N}$, and $\Om$ is cyclic for
  $\alg{R}(O_1)$ it follows that $[\alg{N}_0\Om ]=[\alg{N}\Om ]=\2H$.
  Hence $\2H$ is separable.  \end{proof}

If one wanted to justify an assumption that the vacuum
Hilbert space is separable, Prop.\ \ref{separable}
shows that it is enough to justify the funnel property.
There are concrete models where the funnel property
demonstrably does \emph{not} hold \cite[p.\
23]{horuzhy}.  But the physical significance of these
models is not clear, and there are a couple of other
considerations that might favor the funnel property:
(i): In Section \ref{bell}, we show that connection of
the funnel property with issues about nonlocality.
(ii): Buchholz and Wichmann [\citeyear{bucwic}] argue
that the funnel property is a \emph{sufficient}
condition for a particle interpretation of QFT.  Of
course, the interpreter of QFT will want to critically
examine Buchholz and Wichmann's notion of a ``particle
interpretation.'' (Compare with Section \ref{scat},
where particle interpretations are discussed further.
Compare also with Section \ref{aleph}, which hints at
connections between nonseparable Hilbert space and
field interpretations of QFT.)

\begin{note} The funnel property for free fields is shown in
  \cite{pstates}.
\end{note}

\subsection{Type of local algebras} \label{type-loc}

We now collect the currently known information on the type of local algebras in
physically relevant representations of the net of local observable algebras.

\begin{defn} Let $\alg{R}_1$ and $\alg{R}_2$ be nets of von Neumann algebras on a
  Hilbert space $\2H$.  We say that $\alg{R}_1$ and $\alg{R}_2$ are \emph{locally
    quasiequivalent} just in case for each double cone $O$ there is an isomorphism
  $\f _O:\alg{R}_1(O)\to \alg{R}_2(O)$.  \end{defn}

\begin{note} Although it is not an ``axiom'' of AQFT, there are good
  reasons to believe that representations of physical interest (in
  particular for elementary particle physics) are locally
  quasiequivalent to some vacuum representation, where a vacuum
  representation is the GNS representation of some privileged (e.g.\
  perhaps translation invariant) state.  For example local
  quasi-equivalence holds between any two physical representations
  according to the selection criterion of Doplicher-Haag-Roberts (see
  Section \ref{proto-DHR} and following), and according to the more
  liberal selection criterion of \cite{buch-fred}.  Thus, any
  conclusion we draw concerning the structure of \emph{local algebras}
  in a vacuum representation can be inferred to hold as well for these
  other representations.
\end{note}

\subsubsection{Local algebras are properly infinite}

Some relatively simple results narrow down the possible options for
the type of local algebras.  For this, we define the important
``property B,'' because it is a consequence of plausible assumptions
(viz.\ additivity and the spectrum condition), because it also makes
sense in situations where there is no translation group (unlike the
spectrum condition), and because it is all we need to infer various
results, in particular that local algebras are properly infinite.

\begin{defn} Let $O\to \alg{R}(O)$ be a net of von Neumann algebras on
  some Hilbert space $\2H$.  We say that the net $\alg{R}$ satisfies
  \emph{property B} just in case for any two double cones $O_1$ and
  $O_2$ such that $\ol {O}_1\subseteq O_2$, if $E\in \alg{R}(O_1)$ is
  a nonzero projection, then $E$ is equivalent in $\alg{R}(O_2)$ to
  the identity projection $I$; i.e.\ there is an isometry $V\in
  \alg{R}(O_2)$ such that $VV^*=E$.
\end{defn}

\begin{note} If for each $O$, the algebra $\alg{R}(O)$ is type III, then the net
  $\alg{R}$ satisfies property B.  \end{note}

We expect property B to hold for a net of observable algebras because
it follows from the physically motivated postulates of weak additivity
and the spectrum condition.

\begin{prop} Let $O\mapsto \alg{R}(O)$ be a net of von Neumann
  algebras satisfying microcausality, the spectrum condition, and weak
  additivity.  Then the net $O\mapsto \alg{R}(O)$ satisfies property
  B. \label{prop-B}
\end{prop}

\begin{proof} For the original proof, see \cite{borch}.  For a recent exposition, see
  \cite{antoni}. \end{proof}

\begin{assumption}[Nontriviality] A net $O\mapsto \alg{A}(O)$ of
  $C^*$-algebras is said to satisfy \emph{non-triviality} just in case
  for each double cone $O$, $\alg{A}(O)\neq \7C I$.
\end{assumption}

\begin{prop} Let $O\to \alg{R}(O)$ be a net of von Neumann algebras that satisfies
  microcausality, property B, and non-triviality.  Then for every double cone $O$, the
  von Neumann algebras $\alg{R}(O)$ and $\alg{R}(O')'$ are properly infinite.
  \label{infinite}
\end{prop}

\begin{proof} We first show that $\alg{R}(O)$ is
  properly infinite; that is, that every central
  projection in $\alg{R}(O)$ is infinite.  Let $C$ be a
  central projection in $\alg{R}(O)$.  Choose a
  nontrivial double cone $O_1$ whose closure is
  contained in $O$.  Then by property B, for each
  nonzero projection $E\in \alg{R}(O_1)$, $E$ is
  equivalent to $I$ modulo $\alg{R}(O)$.  Since
  $\alg{R}(O_1)\neq \7C I$, there is a projection $E\in
  \alg{R}(O_1)$ such that $E\sim (I-E)\sim I$ modulo
  $\alg{R}(O)$.  It then follows that $EC\sim
  (I-E)C\sim C$ modulo $\alg{R}(O)$.  It is clear that
  $EC=CEC\leq C$.  If $EC=C$ then $(I-E)C=0$, a
  contradiction.  Therefore $EC<C$ and $EC\sim C$
  modulo $\alg{R}(O)$.  That is, $C$ is an infinite
  projection in $\alg{R}(O)$, and $\alg{R}(O)$ is
  properly infinite.  By microcausality,
  $\alg{R}(O_1)\subseteq \alg{R}(O')'$; thus the
  preceding argument also shows that $\alg{R}(O')'$ is
  properly infinite.
\end{proof}

In particular, the preceding proposition rules out the cases of type I$_n$ and type
II$_1$ von Neumann algebras.  Already this result has implications for questions
about nonlocality; see Prop.\ \ref{genbell} in Section \ref{bell}.  However, the
previous proposition leaves open the possibility that local algebras might be type
I$_\infty$ factors, and it also leaves open the case that local algebras might be
direct sums of heterogeneous types of von Neumann algebras.

\subsubsection{Local algebras are hyperfinite}

We will shortly see that the best results we have point toward the fact that local
algebras are type III, which were originally thought to be unruly anomalies with no
relevance for physics.  However, we first show that under some physically plausible
conditions, local algebras are approximated by finite-dimensional algebras (i.e.\
they are ``hyperfinite''), which shows that after all they are not so unruly.

\begin{defn} Let $\alg{R}$ be a von Neumann algebra.  Then $\alg{R}$ is said to be
  \emph{hyperfinite} just in case there is a family $(\alg{R}_a)_{a\in \7A}$ of
  finite dimensional von Neumann algebras in $\alg{R}$ such that $\alg{R}=(\cup
  _{a\in \7A }\alg{R}_a)''$.  \end{defn}

Hyperfiniteness turns out to be an extremely useful condition for mathematical
purposes.  Indeed, hyperfiniteness is intimately linked to the existence of normal
conditional expectations (see \cite[Chap.\ 8]{kr}), and there is a unique type II$_1$
hyperfinite factor, and a unique type III$_1$ hyperfinite factor.  From a
physical/foundational point of view, one might also think that a failure of
hyperfiniteness for $\alg{R}$ might make it difficult to find a correspondence between
elements of the algebra $\alg{R}$ and real-life laboratory procedures which can only
involve a finite number of tasks.

\begin{fact} Every type I von Neumann algebra is hyperfinite.  See \cite[Exercise
  8.7.26]{kr}.
\end{fact}

\begin{assumption}[Inner/Outer Continuity] A net $O\mapsto \al R(O)$
  of von Neumann algebras is said to be \emph{inner continuous} if for
  any monotonically increasing net $(O_{a})_{a\in \7A}$ with least
  upper bound $O$, we have
$$ \bigvee _{a\in \7A } \al R(O_a) = \al R(O) ,$$ 
where $\al R_1\vee \al R_2$ denotes the von Neumann algebra generated
by $\al R_1$ and $\al R_2$.  Outer continuity is defined by taking a
decreasing net of regions, and the intersection of the corresponding
von Neumann algebras.  \end{assumption}

\begin{note} The condition that the net $\al R$ be continuous from the
  inside is satisfied whenever $\al R$ is the ``minimal'' net
  constructed in the standard way from underlying Wightman fields.
  See \cite{typeIII}.  Similarly, the maximal net satisfies outer
  continuity.
\end{note}

\begin{prop} Suppose that the net $O\mapsto \al R(O)$
  satisfies the funnel property and either inner or outer continuity.
  Then for each double cone $O$, $\al R(O)$ is hyperfinite.
\end{prop}

\begin{proof}[Sketch of proof] (Compare \cite[p.\ 134]{typeIII}.) We
  just look at the case where the net is inner continuous.  By the
  funnel property there is a type I factor $\alg{N}_i$ interpolating
  between $\alg{R}(O_i)$ and $\alg{R}(O)$.  It then follows that the
  union of the ascending sequence $\alg{N}_i$ of hyperfinite factors
  is dense in $\alg{R}(O)$, hence $\alg{R}(O)$ is hyperfinite.
\end{proof}

\subsubsection{Local algebras are type III$_1$ factors}

A series of results, accumulated over a period of more than thirty
years, indicates that the local algebras of relativistic QFT are type
III von Neumann algebras, and more specifically, hyperfinite type
III$_1$ factors.  We gather some of these results in this section.
The first result, due to Longo [\citeyear{lon}], improved on some
earlier results by Driessler.

\begin{prop} Let $\al R$ be a von Neumann algebra acting on $\2H$,
  $\Om \in \2H$ a separating unit vector for $\al R$, $G$ a locally
  compact abelian group with dual $\G$, and $U$ a continuous unitary
  representation of $G$ on $\2H$ such that $U\Om =\Om$ and the ray
  $\7C\, \Om$ is the unique $U(G)$ invariant subspace of $\2H$.
  Suppose that there exist subsets $G_+\subseteq G$ and $\G
  _+\subseteq \G$ such that
\begin{enumerate}
\item $G_+\cup (-G_+)=G$ and $U(g)\al RU(g)^*\subseteq \al R$, for all
  $g\in G_+$.
\item $\G _+\cap (-\G _+)=\{ 0\}$ and $\mathrm{sp}(U)\subseteq \G _+$.
\end{enumerate}
Then either $\al R=\7C I$ or $\al R$ is a type III$_1$ factor. \label{half-sided}
\end{prop}

\begin{proof}[Sketch of proof] (See \cite[p.\ 203]{lon} for details.)
  Let $\om$ be the state of $\al R$ given by $\om (A)=\langle \Om
  ,A\Om \rangle $.  The proof of this result proceeds by showing that
  $\al R_\om =\7C I$, where $\al R_\om$ is the centralizer of the
  state $\om$.  In particular, let $E$ be a projection in $\al R_\om$,
  and define the function $f:G\to \7C$ by
  $$ f(g)=\langle \Om ,EU(g)E\Om \rangle =\langle \Om ,EU(g)EU(-g)\Om
  \rangle .$$ Using the constraint on $\mathrm{sp}(U)$, it can be
  shown that $f$ is constant, and hence $U(g)E\Om =E\Om$ for all $g\in
  G$.  Since $\7C \,\Om$ is the unique invariant subspace under
  $U(G)$, it follows that $E\Om =\Om$, and since $\Om$ is separating
  for $\al R$, $E=0$ or $E=I$.
\end{proof}

The preceding proposition applies to algebras of the
form $\pi (\al A(W))''$, where $W$ is a wedge region,
and $\pi$ is a vacuum representation of the quasilocal
algebra $\al A$.  Indeed, we can take $G_+$ to be a
one-parameter semi-group of lightlike translations with
origin at the apex of $W$, in which case $\7R =G_+\cup
(-G_+)$.  Let $\om$ be a translation invariant state on
$\al A$ such that $(\al A,\om )$ satisfies that
spectrum condition (Assumption \ref{spectrum}).  We
then have that the dual group $\Gamma$ of $G$ in $\7R
^4$ is also a lightlike line, and hence the spectrum
condition entails that there is a subset $\Gamma _+$ of
$\Gamma$, namely those vectors that point toward the
future, such that $\Gamma _+\cap (-\Gamma _+)=\{ 0\}$.
Finally, we saw in Section \ref{vacuum} that when $\om$
is extremal invariant, the ray $\7C \,\Om$ is the
unique $U(G)$ invariant subspace of $\2H$.

For results relevant to \emph{local} algebras, we must impose one
further condition on the net $\al R$.  The first result
(\cite{typeIII}) requires reference to axiomatic QFT with unbounded
operators smeared by test-functions (see \cite{SW}).  That is, we must
assume that the net $\al R$ arises from an underlying Wightman field
theory that satisfies a certain condition --- asymptotic scale
invariance.

Recall that in the axiomatic approach, fields are essentially
self-adjoint operators of the form $\Phi (f)$, where $f$ is a
test-function on spacetime.  The presence of these test-functions allows
the definition of a notion of asymptotic scale invariance.

\begin{defn} Let $N:\7R ^+\to \7R^+$ be a monotone function.  Then a
  scaling transformation of the test-functions is given by $f\mapsto
  f_\lambda$, where $f_\lambda (x)=N(\lambda )f(\lambda ^{-1}x)$.  Let
  $\Phi _\a$ be a set of Wightman fields generating the net $O\mapsto
  \alg{R}(O)$.  We say that the fields satisfy \emph{asymptotic scale
    invariance} just in case there is some field $\Phi$ with vanishing
  vacuum expectation values:
$$ \bigl\langle \Om ,\Phi (f)\Om \rangle =0 ,$$
and for a suitable choice of $N(\lambda )$, the scaled field operators $\Phi
(f_\lambda )$ have the following properties:
\begin{enumerate}
\item The expectation values $\langle \Om ,\Phi
  (f_\lambda )^*\Phi (f_\lambda )\Om \rangle$ converge
  for all test-functions in the limit $\lambda \to 0$,
  and are nonzero for some $f$;
\item The norms $\norm{\Phi (f_\lambda )^*\Phi (f_\lambda )\Om }$ and $\norm{\Phi
    (f_\lambda )\Phi (f_\lambda )^*\Om }$ stay bounded in this limit.
\end{enumerate}
\end{defn}

When a net of von Neumann algebras arises from a Wightman theory with
asymptotic scale invariance, it follows that local algebras are
hyperfinite type III$_1$ factors.

\begin{prop}[\cite{typeIII}] Let $\al R$ be a net of von Neumann
  algebras that satisfies microcausality, the spectrum condition, and
  the funnel property.  Suppose also that $\al R$ can be constructed
  from an underlying Wightman theory that satisfies asymptotic scale
  invariance.  Then for each double cone $O$, $\al R (O)=\al
  M\,\overline{\otimes}\,\al Z$, where $\al M$ is the unique type
  III$_1$ hyperfinite factor and $\al Z$ is the center of $\al R(O)$.
\end{prop}

\begin{note} In \cite{typeIII}, the funnel property is derived from a
  more basic postulate called ``nuclearity,'' which imposes bounds on
  the number of local degrees of freedom.  \end{note}

Of course, one wishes for a result that is more intrinsic to AQFT.
Such a result is provided in \cite{scaling}, using the method of
scaling algebras that allows the computation of the short distance
(scaling) limit of a net $\alg{A}$ of local observables.  (For a short
exposition of scaling algebras, we refer the reader to \cite{bucky}.)
In summary, besides the basic assumptions on the net, the only
additional assumption needed to derive the type III$_1$ property is
that the net has a nontrivial scaling limit.

\begin{note} In some concrete models, it can be shown directly that
  local algebras are the unique type III$_1$ hyperfinite factor.  For
  example, for the free Bose field of mass $m=0$ (in the Minkowski
  vacuum representation), local algebras are isomorphic to algebras
  for wedge regions.  Thus Prop.\ \ref{half-sided} shows that local
  algebras are type III$_1$ factors.  Furthermore, the free Bose field
  of mass $m>0$ is locally quasiequivalent to the case of $m=0$, and
  so its local algebras are also type III$_1$ hyperfinite factors.
  See \cite[p.\ 254]{horuzhy}.
\end{note}

The derivation of the type III$_1$ property is one of the most
surprising and interesting results of contemporary mathematical
physics.  But what is the foundational significance of the result?
How would the world be different if local algebras were, say, type
III$_{1/2}$, or even more radically different, if they were type
II$_\infty$?  For one, there is a crucial difference between the
structure of states on familiar type I algebras, and the structure of
states on type III algebras: since type III algebras have no atomic
projections, and the support projection of a pure normal state is
atomic, it follows that type III algebras have no pure normal states.
(But of course the same is true for type II algebras.)  As pointed out
in \cite{clif} and \cite{rut}, this absence of pure states is a
further obstacle to an ignorance interpretation of quantum
probabilities.  (See also Section \ref{open}.)  

Yngvason [\citeyear{yng}] makes several interesting claims about the
conceptual importance of type III algebras, especially in relation to
questions of nonlocality.  First, according to Yngvason, ``type I
intuitions'' can lead to paradoxes, such as that encountered in
Fermi's famous two-atom system.  However, claims Yngvason, these
paradoxes disappear if we model these situation appropriately with
type III algebras.  Second, Yngvason claims that the homogeneity of
the state space of a type III$_1$ factor $\al R$ can be interpreted as
saying that for any two states $\om _1,\om _2$ on $\al R$, $\om _2$
can be prepared from $\om _1$ (within arbitrarily good accuracy) via a
unitary operation.  Such an operation is, of course, nonselective, and
so does not change the statistics of measurements of observables in
$\al R'$.  So, in one sense, an observer with a type III algebra has
more control over his state space than an observer with a type I
algebra.

\section{Nonlocality and Open Systems in AQFT}  \label{summers}

\begin{note} For this section, we use the following notational
  conventions: uppercase roman letters for algebras, lowercase roman
  letters for operators, and $\81$ for the multiplicative identity in
  an algebra. \end{note}

It is a basic assumption of AQFT that the observable algebras $A(O_1)$
and $A(O_2)$ are mutually commuting when $O_1$ and $O_2$ are spacelike
separated.  This requirement --- which we have called
``microcausality'' --- is sometimes also called ``Einstein
causality,'' because of a suggested connection between the
commutativity of the algebras $A(O_1),A(O_2)$ and the relativistic
prohibition on ``superluminal signaling.''  Implicit in this
connection is a claim that if $[a,b]\neq 0$ for $a\in A(O_1)$ and
$b\in A(O_2)$, then a measurement of $a$ could change the statistics
of a measurement of $b$.

Despite the fact that nonrelativistic QM makes no reference to
spacetime, it has a footprint of the relativistic prohibition of
superluminal signalling.  In particular, the state space of two
distinct objects is a tensor product $H_1\otimes H_2$, and their joint
algebra of observables is $B(H_1)\otimes B(H_2)$.  In this tensor
product construction we represent observables for system $A$ as simple
tensors $a\otimes \81$ and observables of system $B$ as $\81 \otimes
b$.  Thus, we have a version of microcausality.  But we also have
stronger independence properties.  For example, for every state $\f
_1$ of system $A$ and state $\f _2$ of system $B$, there is a state
$\f$ of $A\otimes B$ such that $\f |_{A}=\f _1$ and $\f |_B = \f _2$.

In this section, we investigate the extent to which two local algebras
$A(O_1),A(O_2)$ can be thought to represent distinct, independent
parts of reality.  In Sections \ref{liberty} and \ref{lib2}, we
discuss the relations between microcausality and other independence
assumptions for the algebras $A(O_1),A(O_2)$.  In Section \ref{bell},
we summarize some results concerning violation of Bell's inequality in
AQFT.  Finally, in Section \ref{open} we ask whether a local algebra
$A(O)$ can be isolated from the influences of its environment.

\subsection{Independence of $C^*$ and von Neumann algebras}
\label{liberty}

We first consider notions of independence between a general pair of
von Neumann or $C^*$-algebras.

\begin{defn} If $e,f$ are projection operators on a Hilbert space
  $\2H$, then we let $e\wedge f$ denote the projection onto the closed
  subspace $e(\2H )\cap f(\2H )$.  \end{defn}

\begin{fact} Let $R$ be a von Neumann algebra acting on $\2H$.  If
  $e,f\in R$ then $e\wedge f\in R$.  \end{fact}

\begin{defn}[Schlieder Property] Let $R_1,R_2$ be von Neumann algebras
  acting on the Hilbert space $\2H$.  We say that the pair $(R_1,R_2)$
  satisfies the \emph{Schlieder property} just in case if $e\in R_1$
  and $f\in R_2$ are nonzero projections, then $e\wedge f\neq 0$.
\end{defn}

\noindent The Schlieder property entails that for $e\in R_1 ,f\in
R_2$, if $e,f\neq 0$ and $e,f\neq \81$ then:
$$ e\wedge f\neq 0,\quad \neg e\wedge \neg f \neq 0,\quad e\wedge \neg
f\neq 0, \quad \neg e \wedge f\neq 0 ,$$ where $\neg x=\81 -x$ is the
projection onto the orthogonal complement of $x(\2H )$.  Hence if
``$\wedge$'' is the analogue of conjunction in classical logic, then
the Schlieder property is the analogue of logical independence.

\begin{defn} If $A,B$ are $C^*$-subalgebras of some $C^*$-algebra $C$,
  we let $A\vee B$ denote the $C^*$-algebra generated by $A\cup B$.
\end{defn}

\begin{defn}[$C^*$-Independence] Let $A,B$ be $C^*$-algebras.  We say
  that the pair $(A,B)$ is $C^*$-\emph{independent} just in case for
  any state $\om _1$ of $A$ and any state $\om _2$ of $B$, there is a
  state $\om$ of $A\vee B$ such that $\om |_A=\om _1$ and $\om
  |_{B}=\om _2$.  In other words, each state of $A$ is compatible with
  each state of $B$.
\end{defn}

The $C^*$-independence assumption has an obvious operationalist
motivation: if Alice is an observer at $O_1$ and Bob is an observer at
$O_2$, then $C^*$-independence amounts to the claim that Alice's
choice to prepare a state cannot in any way obstruct Bob's ability to
prepare a state.  Indeed, \cite{sumbuc} claim that a failure of
$C^*$-independence could be detected by local observers.  On the other
hand, $C^*$-independence could also be regarded as an explication of
the notion of the independence of objects:
\begin{quote} \textit{Two objects $A,B$ are truly independent just in
    case any state of $A$ is compatible with any state of $B$; i.e.\
    there are no logical relations between predications of states to $A$ and
    $B$.}  \end{quote} 

Unfortunately, $C^*$-independence does not imply microcausality.

\begin{example} We show that $C^*$-independence does not entail
  microcausality.  (Compare with \cite{napi}.) Consider the finite
  dimensional $*$-algebra $C(\7Z _4)\oplus M_2$, where $C(\7Z _4)$ is
  the abelian $*$-algebra of dimension 4, and $M_2$ is the $2\times 2$
  matrices over $\7C$.  The projection lattice of $C(\7Z _4)$ is the
  Boolean algebra with two atoms; hence it contains logically
  independent elements $e_1,e_2$.  Now choose two projections
  $f_1,f_2\in M_2$ such that $[f_1,f_2]\neq 0$, and let $R_i$ be the
  abelian $*$-subalgebra of $C(\7Z _4)\oplus M_2$ generated by the
  projection $e_i\oplus f_i$.

  To see that $(R_1,R_2)$ is $C^*$-independent, let $\om _i$ be states
  on the $R_i$, and let $\lambda _i=\om _i(e_i\oplus f_i)$.  By the
  logical independence of $e_1,e_2$, there is a state $\rho$ of $C(\7Z
  _4)$ such that $\rho (e_i)=\lambda _i$.  Then the state $\rho \oplus
  0$ on $C(\7Z _4)\oplus M_2$ is a common extension of the $\om _i$
  since
$$ (\rho \oplus 0)(e_i+f_i)=\rho (e_i)=\lambda  _i ,$$
and a state's value on $e_i\oplus f_i$ determines its value on $R_i$.
Therefore, $(R_1,R_2)$ is $C^*$-independent.  On the other hand,
$[e_1+f_1,e_2+f_2]=[f_1,f_2]\neq 0$, whence $(R_1,R_2)$ does not
satisfy microcausality.
\end{example}
 
In the previous example, the algebras $R_1$ and $R_2$ share a common
superselection sector: each commutes with the projection $p=\81 \oplus
0$.  However, the reduced algebras $pR_ip$ are not $C^*$-independent.
In fact, the diagnosis of this example can be generalized into the
following result.

\begin{prop} Let $R_1$ and $R_2$ be von Neumann algebras acting on a
  Hilbert space $\2H$.  If for every projection $e\in Z(R_1\vee R_2
  )$, the pair $(eR_1e,eR_2e)$ is $C^*$-independent, then
  $[R_1,R_2]=\{ 0\}$.
\end{prop}

\begin{proof} See \cite{sumbuc}. \end{proof}

\begin{defn}[Split Property] Let $R_ 1$ and $R_2$ be von Neumann
  algebras on $\2H$ such that $R_1\subseteq R_2'$.  Then the pair
  $(R_1,R_2)$ is said to satisfy the \emph{split property} just in
  case there is a type I factor $M$ such that $R_1\subseteq M\subseteq
  R_2'$.  \end{defn}

\begin{rema} (i): It is clear that the previous
  definition is equivalent to saying that $(R_1,R_2')$
  is a `split inclusion' as per Definition
  \ref{banana}.

  (ii): If $(R_1,R_2)$ satisfies the split property,
  then under some fairly standard conditions (e.g.\
  $R_1$ or $R_2$ is type III), there is a natural
  $*$-isomorphism $\a$ between $\overline{R_1\vee R_2}$
  and the von Neumann algebra tensor product
  $R_1\overline{\otimes}R_2$; by saying that $\a$ is
  `natural', we mean that it extends the map $AB\mapsto
  A\otimes B$.  Furthermore, the $*$-isomorphism $\a$
  is spatial, i.e.\ there is a unitary operator $u$
  such that $\a (x)=uxu^*$.  See \cite[p.\ 212]{sum}.

  (iii): On the other hand, suppose that $R$ is a factor, so that
  $R\cup R'$ generates $B(\2H )$ as a von Neumann algebra, i.e.\
  $\overline{R\vee R'}=B(\2H )$.  Then $R'$ is of the same type (I,
  II, or III) as $R$ \cite[Thm.\ 9.1.3]{kr}, and so the von Neumann
  algebra tensor product $R\ol \otimes R'$ is of the same type as $R$
  \cite[p.\ 830]{kr}.  So if $R$ is type II or III, then
  $\overline{R\vee R'}$ is strictly larger than, and not isomorphic to
  $R\ol \otimes R'$.
\end{rema}

\begin{defn}[$W^*$-Independence] Let $R_1$ and $R_2$ be von Neumann
  algebras acting on $\2H$.  The pair $(R_1,R_2)$ is said to be
  $W^*$-\emph{independent} just in case for every normal state $\f _1$
  of $R_1$ and for every normal state $\f _2$ of $R_2$, there is a
  normal state $\f$ of $R_1\vee R_2$ such that $\f |_{R_i}=\f _i$.
\end{defn}

With the assumption of the mutual commutativity of
$R_1$ and $R_2$ (i.e.\ microcausality), we have the
following implications (see \cite[p.\ 222]{sum}):

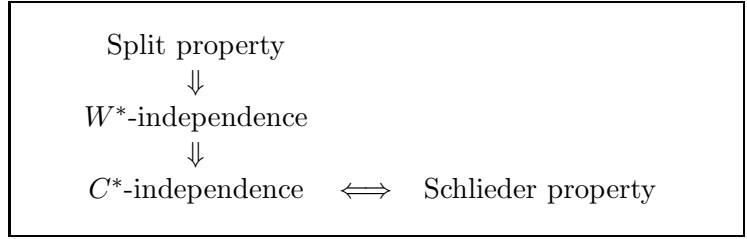
\begin{lilbox} \begin{tabular}{cccc}
    & Split property \\
    & $\Downarrow$    \\
    &$W^*$-independence \\
    & $\Downarrow $   \\
    & $C^*$-independence & $\Longleftrightarrow$ &
    Schlieder property
\end{tabular} \end{lilbox}

\subsection{Independence of local algebras} \label{lib2}

We now consider which independence properties hold
between pairs of algebras associated with spacelike
separated regions.  In general, not much can be said
about the independence of such algebras.  In order to
get such results off the ground, we need a stronger
notion of spacelike separation.

\begin{defn} Two double cones $O_1,O_2$ are said to be \emph{strictly
    spacelike separated} just in case there is a neighborhood $N$ of
  zero such that $O_1+x$ is spacelike separated from $O_2$ for all
  $x\in N$.  \end{defn}

\begin{prop} Suppose that the net $O\mapsto R(O)$ satisfies
  microcausality, weak additivity, and the spectrum condition.  If
  $O_1$ and $O_2$ are strictly spacelike separated, then
  $(R(O_1),R(O_2))$ satisfies the Schlieder property.
  \label{schlieder} \end{prop}

\begin{proof} See \cite{schlied}. \end{proof}

In terms of logical strength, the following concept lies between
spacelike separation and strict spacelike separation; furthermore,
this concept makes sense for spacetimes without a translation group.

\begin{defn} Two double cones $O_1$ and $O_2$ are said to be
  \emph{strongly spacelike separated} just in case there are double
  cones $\wt O_i$ such that $\ol O_i\subseteq \wt O_i$, and $\wt
  O_1,\wt O_2$ are spacelike.  \end{defn}

\begin{fact} If $O_1$ and $O_2$ are strictly spacelike separated, then
  they are strongly spacelike separated.  \end{fact}

Of course, the assumptions of Proposition \ref{schlieder}
(microcausality, additivity, spectrum) are precisely what is used to
derive property B for the net (Proposition \ref{prop-B}).  So, it is
perhaps illustrative to give a simple derivation of the Schlieder
property from property B.  (Such a result also applies in contexts ---
e.g.\ QFT on curved spacetime --- where the spectrum condition does
not make sense.)

\begin{prop} Suppose that the net $O\mapsto R(O)$ of von Neumann
  algebras satisfies microcausality and property B.  If $O_1$ and
  $O_2$ are strongly spacelike separated, then $(R(O_1),R(O_2))$
  satisfies the Schlieder property.
  \label{frees}
\end{prop}

\begin{proof} Let $O_1$ and $O_2$ be strictly spacelike separated, and
  let $e_i\in \al R (O_i)$ be projections.  Then there are regions
  $\wt O_i$ such that $\ol O_i\subseteq \wt O_i$, and $\wt O_1$ is
  spacelike to $\wt O_2$.  By property B, there are isometries $v_i\in
  R(\wt O_i)$ such that $v_iv_i^*=e_i$.  Furthermore, $[v_1,v_2]=0$
  and hence $e_1e_2=v_1v_2(v_1v_2)^*$.  But $v_1v_2$ is an isometry,
  and so $v_1v_2(v_1v_2)^*\neq 0$.
\end{proof}

The split property clearly does not hold for
$(R(W),R(W'))$ where $W$ is a wedge region and $W'$ is
its causal complement.  Indeed, since $R(W)$ and
$R(W')$ are type III$_1$ factors, there can be no
$*$-isomorphism between $R(W)\overline{\otimes}R(W')$
and $\overline{R(W)\vee R(W)'}=B(H)$.  However, if the
funnel property holds for the net $O\mapsto R(O)$, then
$(R(O_1),R(O_2))$ satisfies the split property when
$O_1$ and $O_2$ are strictly spacelike separated double
cones.

\subsection{Bell correlation between von Neumann algebras}
\label{bell}

We first define a generalized notion of Bell type measurements for a
pair of von Neumann algebras.

\begin{defn} Let $A$ and $B$ be mutually commuting $C^*$-subalgebras
  of some $C^*$-algebra $C$.  Then we set \begin{eqnarray*} \7B (A,B)
    &\equiv & \Bigl\{ (1/2) [a_1(b_1+b_2)+a_2(b_1-b_2)] : a_i=a_i^*\in
    A, b_i=b_i^{*}\in B , \\
&& \ \ -\81 \leq a_{i},b_{i} \leq \81 \Bigr\}
    .\end{eqnarray*} Elements of $\7B (A,B)$ are called \emph{Bell
    operators} for $(A,B)$.
\end{defn}

Let $r$ be a Bell operator for $(A,B)$. It can be shown that $\abs{\f
  (r)}\leq \sqrt{2}$ for each state $\f$ on $C$ \cite{early-sum}.  It
is also straightforward to check that if $\f$ is a separable state
(i.e.\ a mixture of product states) then $\abs{\f (r)}\leq 1$.
Indeed, the Bell measurement correlations in the state $\f$ can be
reproduced by a local hidden variable model iff $\abs{\f (r)}\leq 1$
\cite{early-sum,baebel}.

\begin{defn} Define the Bell correlation coefficient of a state $\f$
  of $A\vee B$ by 
$$ \beta (\f ,A,B) = \sup \{ \, \abs{\f (r) }:r\in \7B (A,B) \,\} . $$
If $\abs{\beta (\f ,A,B)}>1$, then $\f$ is said to \emph{violate a
  Bell inequality}, or to be \emph{Bell correlated}.
\end{defn}

It is a straightforward exercise to show that if $R_1$ is an abelian
von Neumann algebra and $R_1\subseteq R_2'$, then for any state $\f$,
$\beta (\f ,R_1,R_2)\leq 1$.  For a sort of converse, Landau
[\citeyear{lan}] shows that if $R_1$ and $R_2$ are nonabelian von
Neumann algebras such that $R_1\subseteq R_2'$, and if $(R_1,R_2)$
satisfies the Schlieder property, then there is \emph{some} state $\f$
that violates Bell's inequality maximally relative to $(R_1,R_2)$.
Similarly, Bacciagaluppi [\citeyear{bacc}] shows that if $A$ and $B$
are $C^*$-algebras, then some state violates a Bell inequality for
$A\otimes B$ iff both $A$ and $B$ are nonabelian.

When $A$ and $B$ have further properties, we can derive even stronger
results.  For present purposes, we will simply apply a couple of the
known results to the case of AQFT.  (See \cite{sum} for many more
details.)

\begin{prop} Let $R$ be a type III$_1$ factor acting on a separable
  Hilbert space $\2H$.  Then every normal state $\f$ of $B(\2H )$ is
  maximally Bell correlated across $(R,R')$, that is $\beta (\f
  ,R,R')=\sqrt{2}$. \label{max}
\end{prop}

\begin{proof} See \cite{sumwe2,sumwe}. \end{proof}

\begin{note} Prop.\ \ref{half-sided} tells us that under quite generic
  conditions, the wedge algebra $R(W)$ is a type III$_1$ factor.
  In this case, Prop.\ \ref{max} tells us that the vacuum is maximally
  Bell correlated across $(R(W),R(W)')$.  \end{note}

\begin{prop} Suppose that $R_1$ and $R_2$ are von Neumann algebras on
  $\2H$ such that $R_1\subseteq R_2'$, and $(R_1,R_2)$ satisfies the
  Schlieder property.  If $R_1$ and $R_2$ are properly infinite, then
  there is a dense set of vectors in $\2H$ that induce Bell correlated
  states across $(R_1,R_2 )$.  \label{genbell} \label{car} \end{prop}

\begin{proof} See \cite{gbell}. \end{proof}

\begin{note} If a net $O\mapsto R(O)$ of von Neumann algebras on $\2H$
  satisfies property B and nontriviality, then the hypotheses of
  Prop.\ \ref{car} apply to algebras $R(O_1)$ and $R(O_2)$ when $O_1$
  and $O_2$ are strongly spacelike separated.  \end{note}

\bigskip \noindent {\small \textit{Notes:} For a comprehensive review
  of pre-1990 results on independence of local algebras in AQFT, see
  \cite{sum}.  For some more recent results, see
  \cite{bellst,flo,redei,gbell,sumbuc}.}

\subsection{Intrinsically entangled states} \label{open}

According to Clifton and Halvorson [\citeyear{clif}], the type III
property of local algebras in AQFT shows that it is impossible to
disentangle local systems from their environment.  To see the
argument, recall that it is a standard (perhaps somewhat justified)
assumption that the general form of a dynamical evolution $T$ of
observables, represented by self-adjoint elements of a $C^*$-algebra
$A$ is given by a completely positive (CP) linear mapping $T$ of $A$
such that $T(\81 )=\81 $.  (Such an assumption is certainly
commonplace in, say, quantum information theory.)  Here we recall the
pertinent definition.

\begin{defn} Let $A$ be a $C^*$-algebra.  A linear map
  $T$ of $A$ is said to be \emph{positive} if
  $T(a^*a)\geq 0$ for each $a\in A$.  $T$ is said to be
  \emph{completely positive} if for each $n\in \7N$,
  the map $T\otimes \id _n:A\otimes M_n\to A\otimes
  M_n$ defined on elementary tensors by
$$ (T\otimes \id _n)(a\otimes b)=T(a)\otimes b ,$$
is positive.  Here $M_n$ is the $C^*$-algebra of
$n\times n$ matrices over $\7C$.  \end{defn}

\begin{note} If $T:\al A\to \al A$ is positive and
  $T(I)=I$, then for each state $\om$ of $\al A$, we
  define $T^*(\om )$ by $T^*(\om )(A)=\om (T(A))$.  It
  follows that $T^*$ is an affine mapping of the state
  space into itself.  \end{note}  

%

For type I factors, Kraus' theorem \cite{kraus} shows
that CP maps are ``inner.''
\begin{thm}[Kraus Representation] If $R$ is a type I$_n$ factor then
  the following are equivalent for a linear map $T:R\to R$.
\begin{enumerate}
\item $T$ is completely positive and $T(\81 )=\81 $.
\item $T$ is the restriction of an automorphism $x\mapsto uxu^*$ on an
  algebra of the form $R\otimes B(H)$.
\item There are positive operators $a_1,\dots ,a_n\in R$ such that
  $\sum _{i=1}^{n}a_i=\81$ and
  \begin{equation} T(x)=\sum _{i=1}a_i^{1/2}xa_i^{1/2} .\label{kraus} \end{equation}
\end{enumerate}
\end{thm}

\noindent One special case of Eqn.\ (\ref{kraus}) is the L{\"u}ders
rule with projection operators $e$ and $\81 -e$:
$$ T_e(x)=exe+(\81 -e)x(\81 -e) .$$
Furthermore, if the algebra $R$ is type I, we can
choose $e\in R$ to be an abelian projection.  We have
the following result:
\begin{quote} If the local algebra $R$ is a type I factor, then there
  is a universal disentangling operation $T_e$.  That is, no matter
  what the initial state, the outcome of applying $T_e$ is that the
  final state is separable.
\end{quote} However, suppose that $R$ has no abelian projections
(e.g.\ $R$ is type III).  Then for each nonzero projection $e\in R$,
the algebras $eRe$ and $eR'e$ are nonabelian, and hence there is some
entangled state $\f$ for the pair $(eRe,eR'e)$.  This entangled state
is the image under the operation $(T_e)^*$ of some state on
$\overline{R\vee R'}$.  Hence, the operation $T_e$ does not
disentangle all states.

This heuristic argument can be tightened up into a ``proof'' that no
operation on $R$ can disentangle the states of $\overline{R\vee R'}$.
See \cite{clif} for details.

\begin{note} (i): The Kraus representation theorem is not valid as it
  stands for type III algebras.  Indeed, the Kraus representation
  theorem is a special case of the Stinespring decomposition theorem
  \cite{sti}.

  (ii): A CP operation on a von Neumann algebra is typically also
  assumed to be ultraweakly continuous.  The continuity of $T$ might
  be justified on the grounds that it is necessary if $T^*$ is to map
  normal states to normal states.  For objections to the continuity
  requirement, see \cite{sri}.
\end{note}

\section{Prospects for Particles} \label{parts}

The main application of relativistic QFT is to fundamental
\emph{particle} physics.  But it is not completely clear that
fundamental particle physics is really about particles.  Indeed,
despite initial signs that QFT permits a particle interpretation (via
Fock space), there are many negative signs concerning the possibility
of particle ontology of relativistic QFT.  This section is devoted to
assessing the status of particles from the point of view of AQFT.

\subsection{Particles from Fock space} \label{focked}

We begin our investigation of particles with the ``story from mother's
knee'' about how to give QFT a particle interpretation.  (See
\cite{tell} for one philosopher's interpretation of this story.)  The
story begins with a special Hilbert space, called \emph{Fock space}.
Now Fock space is just another separable infinite dimensional Hilbert
space (and so isomorphic to all its separable infinite dimensional
brothers).  But the key is writing it down in a fashion that
\emph{suggests} a particle interpretation.  In particular, suppose
that $H$ is the one-particle Hilbert space, i.e.\ the state space for
a single particle.  Now depending on whether our particle is a Boson
or a Fermion, the state space of a pair of these particles is either
$E_s(H \otimes H)$ or $E_a(H \otimes H)$, where $E_s$ is the
projection onto the vectors invariant under the permutation $\Sigma
_{H,H}$ on $H\otimes H$, and $E_a$ is the projection onto vectors that
change signs under $\Sigma _{H,H}$.  For present purposes, we ignore
these differences, and simply use $H\otimes H$ to denote one
possibility or the other.  Now, proceeding down the line, for $n$
particles, we have the Hilbert space $H^n\equiv H\otimes \dots \otimes
H$, etc..

A state in $H^n$ is definitely a state of $n$ particles.  To get
disjunctive states, we make use of the direct sum operation
``$\oplus$'' on Hilbert spaces.  So we define the Fock space $\2F (H)$
over $H$ as the infinite direct sum:
$$ \2F (H)= \7C \oplus H \oplus (H\otimes H)\oplus (H\otimes H\otimes H)\oplus \cdots
.$$ So, the state vectors in Fock space include a state where the are
no particles (the vector lies in the first summand), a state where
there is one particle, a state where there are two particles, etc..
Furthermore, there are states that are superpositions of different
numbers of particles.

One can spend time worrying about what it means to say that particle
numbers can be superposed.  But that is the ``half empty cup'' point
of view.  From the ``half full cup'' point of view, it makes sense to
count particles.  Indeed, the positive (unbounded) operator
$$ N=0\oplus 1\oplus 2\oplus 3 \oplus 4 \oplus \cdots ,$$ 
is the formal element of our model that permits us to talk about the
number of particles.

\begin{note} In the category of Hilbert spaces, all separable Hilbert
  spaces are isomorphic --- there is no difference between Fock space
  and the single particle space.  If we are not careful, we could
  become confused about the bearer of the name ``Fock space.''

  The confusion goes away when we move to the appropriate category.
  According to Wigner's analysis \cite{wig}, a particle corresponds to
  an irreducible unitary representation of the identity component
  $\2P$ of the Poincar{\'e} group.  Then the single particle space and
  Fock space are distinct objects in the category of representations
  of $\2P$.  The underlying Hilbert spaces of the two representations
  are both separable (and hence isomorphic as Hilbert spaces); but the
  two representations are most certainly not equivalent (one is
  irreducible, the other reducible).
\end{note}

\subsection{Fock space from the algebra of observables}

The Fock space story is not completely abandoned within the algebraic
approach to QFT.  In fact, when conditions are good, Fock space
emerges as the GNS Hilbert space for some privileged vacuum state of
the algebra of observables.  We briefly describe how this emergence
occurs before proceeding to raise some problems for the naive Fock
space story.  (We look here only at the symmetric --- Bosonic ---
case.  A similar treatment applies to the antisymmetric --- Fermionic
--- case.)

The algebraic reconstruction of Fock space arises from
the algebraic version of canonical quantization.
Suppose that $S$ is a real vector space (equipped with
some suitable topology), and that $\sigma$ is a
symplectic form on $S$.  So, $S$ represents a classical
phase space (see Butterfield, this volume).  The
\emph{Weyl algebra} $\alg{A}[S,\sigma ]$ is a specific
$C^*$-algebra generated by elements of the form $W(f)$,
with $f\in S$ and satisfying the canonical commutation
relations in the Weyl-Segal form:
$$ W(f)W(g)=e^{-i\sigma (f,g)/2}W(f+g) .$$ 
Suppose that there is also some notion of spacetime localization for
elements of $S$, i.e.\ a mapping $O\mapsto S(O)$ from double cones in
Minkowski spacetime to subspaces of $S$.  Then, if certain constraints
are satisfied, the pair of mappings
$$ O\mapsto S(O)\mapsto \alg{A}(O)\equiv C^*\{ W(f):f\in S(O) \} ,$$ 
can be composed to give a net of $C^*$-algebras over Minkowski
spacetime. (Here $C^*X$ is the $C^*$-algebra generated by the set
$X$.)

Now if we are given some dynamics on $S$, then we can --- again, if
certain criteria are satisfied --- define a corresponding dynamical
automorphism group $\a _t$ on $\alg{A}[S,\sigma ]$.  There is then a
unique dynamically stable pure state $\om _0$ of $\alg{A}[S,\sigma ]$,
and we consider the GNS representation $(\2H ,\pi )$ of
$\alg{A}[S,\sigma ]$ induced by $\om _0$.  To our delight, we find
that the infinitesimal generators $\Phi (f)$ of the one-parameter
groups $\{ \pi (W(f)) \}_{t\in \7R}$ behave just like the field
operators in the old-fashioned Fock space approach.  Furthermore (now
speaking non-rigorously), if we define operators
\begin{eqnarray*} 
  a(f) &=& 2^{-1/2}\left( \Phi (f)+i\Phi (Jf) \right) , \\
  a^*(f) &=& 2^{-1/2}\left( \Phi (f)-i\Phi (Jf) \right) ,
\end{eqnarray*} 
we find that they behave like creation and annihilation operators of
particles.  (Here $J$ is the unique ``complex structure'' on $S$ that
is compatible with the dynamics.)  In particular, by applying them to
the vacuum state $\Om$, we get the entire GNS Hilbert space $\2H$.
Finally, if we take an orthonormal basis $\{ f_i\}$ of $S$, then the
sum
\begin{eqnarray*} \sum _{i=1}^{\infty} a ^*(f_i)a(f_ i) ,\end{eqnarray*}
is the number operator $N$.  Thus, the traditional Fock space
formalism emerges as one special case of the GNS representation of a
state of the Weyl algebra.

\begin{note} The Minkowski vacuum representation $(\2H _0,\pi _0 )$ of
  $\al A$ is Poincar{\'e} covariant, i.e.\ the action $\a _{(a,\Lambda
    )}$ of the Poincar{\'e} group by automorphisms on $\al A$ is
  implemented by unitary operators $U(a,\Lambda )$ on $\2H$.  When we
  say that $\2H$ is isomorphic to Fock space $\2F (H)$, we do not mean
  the trivial fact that $\2H$ and $\2F (H)$ have the same dimension.
  Rather, we mean that the unitary representation $(\2H ,U)$ of the
  Poincar{\'e} group is a Fock representation.  \end{note}

\noindent {\small \textit{Notes:} See \cite[Section 5.2]{brat2} for a
  detailed account of the reconstruction of Fock space from the Weyl
  algebra.  See also \cite{rindler} and \cite{me} for shorter
  expositions.}

\subsection{Nonuniqueness of particle interpretations}

If we have a representation $(\2H ,\pi )$ of the quasilocal algebra
$\alg{A}$ such that $\2H$ is isomorphic to Fock space, then we can
make sense of talk about particles.  Furthermore, such representations
exist, e.g., the GNS representation of the Minkowski vacuum state $\om
_0$ of the free Bose field.  So, in the most simple cases (e.g.\ free
fields on flat spacetime), there is no problem concerning the
existence of particle interpretations of the theory.

But there is a problem about \emph{uniqueness}: there are unitarily
inequivalent representations of $\al A$, each of which is isomorphic
to Fock space.  Furthermore, a result from \cite{cha2,cha1} shows that
two inequivalent Fock representations correspond to two number
operators that cannot be thought of as notational variants of the same
description of reality.  Indeed, there are no states of $\al A$ that
assign sharp values to both number operators.  Hence, the particle
interpretations provided by the two Fock representations are mutually
exclusive.

The issue of inequivalent Fock representations is treated in depth in
\cite{rindler}.  For present purposes, we simply note that this worry
about nonuniqueness is tied in to a more general worry about
inequivalent representations of the quasilocal $C^*$-algebra
$\alg{A}$.  But this more general issue cannot be resolved without
reference to recent developments in the theory of superselection
sectors (see Sections \ref{proto-DHR} and following).  We return to
this question in Section \ref{inter}.

\subsection{Problems for localized particles}

Suppose that we have settled the uniqueness problem that is raised in
the previous subsection --- e.g.\ we have found a good reason for
preferring a particular Fock representation $(\2H, \pi )$ of $\al A$,
and so we have a preferred global number operator $N$ on $\2H$.  The
next question is whether relativistic QFT is consistent with an
ontology of \emph{localized} particles --- that is, whether it makes
sense to talk about the number of particles in a bounded region $O$ of
space.

As pointed out in Section \ref{rst}, the Reeh-Schlieder
(RS) Theorem entails that the local algebras of AQFT do
not contain operators that annihilate the vacuum.
Hence if a number operator has the vacuum as an
eigenstate, then there are no local number operators.
That is perhaps enough to convince most readers that
localized particles are not possible in relativistic
QFT.  Nonetheless, there have been attempts to bypass
the RS Theorem, most notably the proposal of Newton and
Wigner (recently resurrected in \cite{flem}).  It has
been argued that such attempts are not promising
\cite{me}.  Furthermore, it can be shown independently
of the full framework of AQFT, and without the RS
Theorem, that a positive energy condition combined with
microcausality rules out local number operators
\cite{noplace}.

Despite the various No Go results for localized
particles in relativistic QFT, the interpretation of
experiments in high energy physics seems to require a
notion of something causing clicks in detectors, and
that a ``detector'' is fairly well localized in some
bounded region of spacetime.  A detector corresponds to
a positive operator $C$ in $\al A$, and is ``completely
reliable'' only if it registers $0$ identically in the
vacuum state, i.e.\ $C\Om =0$.  Hence the
Reeh-Schlieder Theorem entails that $C$ is not
contained in any local algebra.  Nonetheless, a notion
of approximate localization of $C$ can be salvaged:
choose some $A\in \al A (O)$ with $0\leq A\leq I$, and
set
$$ C =\int f(x)\a _x(A) dx ,$$
where $f$ is a smooth function whose Fourier transform has support in
the complement of the forward light cone.  (The function $f$
automatically has unbounded support.)  Then $C\Om =0$, and the
function $f$ can also be chosen so that $C$ is ``close'' in the norm
topology to an operator in $\al A(O)$.

The notion of approximately localized detectors is employed
extensively in Haag-Ruelle scattering theory and recent developments
thereof, to which we now turn.

\subsection{Particle interpretations generalized: Scattering theory
  and beyond} \label{scat}

It is not true that a representation $(\2K ,\pi )$ of $\al A$ must be
a Fock representation in order for states in the Hilbert space $\2K$
to have an interpretation as particle states.  Indeed, one of the
central tasks of ``scattering theory,'' is to provide criteria --- in
the absence of full Fock space structure --- for defining particle
states.  These criteria are needed in order to describe scattering
experiments which cannot be described in a Fock representation, but
which need particle states to describe the input and output states.

Haag and Swieca [\citeyear{haags}] propose to pick out the
$n$-particle states by means of localized detectors; we call this the
\emph{detector criterion}:
\begin{quote} A state with at least $n$-particles is a state that
  would trigger $n$ detectors that are far separated in space.
\end{quote} 
Philosophers might worry that the detector criterion is too
operationalist.  Indeed, some might claim that detectors themselves
are made out of particles, and so \emph{defining} a particle in terms
of a detector would be viciously circular.

If we were trying to give an analysis of the concept of a particle,
then we would need to address such worries.  However, scattering
theory does not end with the detector criterion.  Indeed, the goal is
to tie the detector criterion back to some other more intrinsic
definition of particle states.  The traditional intrinsic definition
of particle states is in terms of Wigner's symmetry criterion:
\begin{quote} A state of $n$ particles (of spins $s_i$
  and masses $m_i$) is a state in the tensor product of
  the corresponding representations of the Poincar{\'e}
  group.  \end{quote} Thus, scattering theory --- as
originally conceived --- needs to show that the states
satisfying the detector criterion correspond to an
appropriate representation of the Poincar{\'e} group.
In particular, the goal is to show that there are
isometries $\Om ^{\mathrm{in}},\Om ^{\mathrm{out}}$
that embed Fock space $\2F (H)$ into $\2K$, and that
intertwine the given representations of the
Poincar{\'e} group on $\2F (H)$ and $\2K$.

Based on these ideas, detailed models have been worked
out for the case where there is a mass gap.
Unfortunately, as of yet, there is no model in which
$\2H ^{\mathrm{in}}=\2H ^{\mathrm{out}}$, which is a
necessary condition for the theory to have an S-matrix,
and to define transition probabilities between incoming
and outgoing states.  (Here $\2H ^{\mathrm{in}}$ is the
image of Fock space in $\2K$ under the isometry $\Om
^{\mathrm{in}}$, and similarly for $\2H
^{\mathrm{out}}$.)

Recently, Buchholz and collaborators have claimed that Wigner's
symmetry criterion is too stringent --- i.e.\ there is a more general
definition of particle states.  They claim that it is only by means of
this more general criterion that we can solve the ``infraparticles''
problem, where massive particles carry a cloud of photons (see
\cite{buc-por}).

\bigskip \noindent {\small \textit{Note:} For a review of progress in
  scattering theory in AQFT, see \cite[Chapter 6]{haag} and
  \cite{buc-sum}.}

\section{The Problem of Value-Definiteness in AQFT} \label{modal}

The ``measurement problem'' of nonrelativistic QM shows that the
standard approach to the theory is impaled on the horns of a dilemma:
either (i) one must make ad hoc adjustments to the dynamics
(``collapse'') when needed to explain the results of measurements, or
(ii) measurements do not, contrary to appearances, have outcomes (see
Dickson, this volume, Section 5).

There are two main responses to the dilemma: On the one hand, some
suggest that we abandon the unitary dynamics of QM in favor of
stochastic dynamics that accurately predicts our experience of
measurement outcomes.  On the other hand, some suggest that we
maintain the unitary dynamics of the quantum state, but that certain
quantities (e.g.\ position of particles) have values even though these
values are not specified by the quantum state.  (See Dickson, this
volume, Section 5.5 for a more nuanced discussion of the possible
responses.)

Both approaches --- the approach that alters the dynamics, and the
approach with additional values --- are completely successful as
responses to the measurement problem in nonrelativistic QM.  But both
approaches run into obstacles when it comes to synthesizing quantum
mechanics with relativity.  In particular, the additional values
approach (e.g.\ the de Broglie--Bohm pilot-wave theory) appears to
require a preferred frame of reference to define the dynamics of the
additional values (see \cite[pp.\ 188--191, 196--198]{cush},
\cite{holland}, and \cite[Chaps.\ 11 \& 12]{hiley}), and in this case
it would fail the test of Lorentz invariance.

The ``modal'' interpretation of quantum mechanics is
similar in spirit to the de Broglie--Bohm theory, but
begins from a more abstract perspective on the question
of assigning definite values to some observables.
(Following \cite{jbell}, we might call these the
``beables'' of the theory.)  Rather than making an
intuitively physically motivated choice of the
determinate values (e.g.\ particle positions), the
modal interpretation makes the mathematically motivated
choice of the spectral decomposition of the quantum
state (i.e.\ the density operator) as determinate.
(See \cite{ver,piet} for reviews of the modal
interpretation; see \cite{rob-kd} for motivation.)

Unlike the de Broglie--Bohm theory, it is not obvious
that the modal interpretation must violate the spirit
or letter of relativistic constraints, e.g.\ Lorentz
invariance \cite[p.\ 9]{dic}.  So, it seems that there
should be some hope of developing a modal
interpretation within the framework of AQFT.  This is
the starting point for Dieks' [\citeyear{diek}]
proposal for a modal interpretation of AQFT.  Rather
than expound Dieks' original proposal, we move directly
to the criticism in \cite{cli9}, to which we also refer
the reader for further elaboration.

\subsection{Clifton-Kitajima classification of modal algebras}

Clifton's critique of the modal interpretation of AQFT is based on a
remarkable theorem which classifies all possible ``modal subalgebras''
of a local von Neumann algebra $\alg{R}(O)$ relative to a state
$\rho$.  According to Clifton --- and the modal interpreters seem to
agree on this point --- the algebra $\alg{D}$, $\alg{D}\subseteq
\alg{R}(O)$ of definite local observables should satisfy the following
constraints relative to a given state $\rho$ of $\alg{R}(O)$:

\begin{defn} Let $\alg{R}$ be a von Neumann algebra, and let $\rho$ be
  a state of $\alg{R}$.  Then a von Neumann subalgebra $\alg{D}$ of
  $\alg{R}$ is said to be a \emph{modal algebra} for $(\alg{R},\rho )$
  just in case:
  \begin{enumerate}
  \item (\emph{Value definiteness}) The restricted state $\rho
    |_{\alg{D}}$ is a mixture of dispersion-free states.  (Definition:
    A state is \emph{dispersion free} iff it assigns each projection
    operator either $0$ or $1$.)
  \item (\emph{Definability}) $\alg{D}$ is left invariant under all
    symmetries of $\alg{R}$ that leave the state $\rho$ invariant.
  \item (\emph{Maximality}) $\alg{D}$ is maximal, subject to the first
    two conditions.
  \end{enumerate} \label{modal-alg}
\end{defn}

The last requirement is imposed simply to rule out trivial
counterexamples to uniqueness --- e.g.\ one could always pick the
algebra $\7C I$ of scalar multiples of the identity.  The second
requirement is supposed to explicate the idea that $\alg{D}$ is
``picked out by'' (i.e.\ is definable in terms of) the state $\rho$.
We have left the notion of a ``symmetry'' vague (and we will return to
this question in the next subsection), but Clifton takes the
symmetries to coincide with the $*$-automorphisms of $\alg{R}$, and
this is needed for the main result (Theorem \ref{kitty}).

To state this result, we need to define the notion of the centralizer
of a state.  The following proposition establishes the equivalence of
two possible definitions of the centralizer.

\begin{prop} Let $\alg{R}$ be a von Neumann algebra, let $\om$ be a
  faithful normal state of $\alg{R}$, and let $\sigma _t^\om$ be the
  modular automorphism group of $\alg{R}$.  Then the following two
  sets are coextensive:
  \begin{enumerate}
  \item $\{ A\in \alg{R}:\sigma _t^{\om}(A)=A, \forall t\in \7R \}$
  \item $\{ A\in \alg{R}:\om (AB)=\om (BA), \forall B\in \alg{R} \}$
  \end{enumerate} \label{cent}
\end{prop}

The proof of Prop.\ \ref{cent} depends on the full apparatus of
modular theory.  We refer the reader to \cite[Chap. 8]{tak2} for
details.

\begin{defn} It is clear that the set defined in the previous
  proposition is in fact a von Neumann subalgebra of $\alg{R}$.  We
  call this subalgebra the \emph{centralizer} of $\om$ in $\alg{R}$,
  and we denote it by $\al R_\om$.
\end{defn}

\begin{example} Let $\al R=\bh$, and let $\om$ be a faithful normal
  state of $\al R$.  Then $\om$ has the form
$$ \om (A)=\tr
(DA),\qquad A\in \al R ,$$ for some density operator $D\in \al R$.
Then $\al R_\om = \{ D\}'$, and $Z(\al R_\om )$ is the abelian von
Neumann algebra $\{ D\}''$.  In particular, if $\om$ is the maximally
mixed state of a type I$_n$ factor, then $\al R_\om =\bh$, and $Z(\al
R_\om )=\7C I$.
\end{example}

The Clifton-Kitajima Theorem shows that there is a unique modal
algebra for $(\alg{R},\om )$, and in the case that the state $\om$ is
faithful, it is $Z(\al R_\om )$, the center of the centralizer of
$\om$.

\begin{thm}[Clifton-Kitajima] Let $\alg{R}$ be a von Neumann algebra
  acting on a Hilbert space $\2H$, and let $\om$ be a normal state of
  $\alg{R}$.
  \begin{enumerate} \item If $\om$ is faithful then $Z(\al R_\om )$ is
    the unique modal algebra for $(\alg{R} ,\om )$.
  \item Generally, the unique modal algebra for $(\alg{R},\om )$ is
    $\alg{N}\oplus Z(\al R_\om )E$, where $E$ is the smallest
    projection in $\al R$ such that $\om (E)=1$, and $\al N$ is the
    algebra of all bounded operators on $(I-E)(\2H)$.
  \end{enumerate}
  \label{kitty}
\end{thm}

\noindent The result is proven for the case where $\om$ is faithful in
\cite{cli9}, and for the general case in \cite{kit}.

As pointed out by Clifton [\citeyear{cli9}], Thm. \ref{kitty} spells
trouble for a modal interpretation of AQFT, because there are many
cases where the algebra $Z(\al R_\om )$ is trivial.  (See
\cite{rue-ear} for further development of this point.)

\begin{enumerate}
\item Let $W$ be a wedge region in Minkowski spacetime, and let $\Om$
  be the vacuum state.  Then there are no fixed points in $\al R(W)$
  of the modular automorphism group $\sigma ^\om _t$ (see the proof of
  Proposition \ref{half-sided}, and also \cite{wulf}). Hence, $\al
  R_\om =\7C I$, and $Z (\al R_\om )=\7C I$.
\item In relativistic QFT, local algebras are the type III$_1$
  hyperfinite factor $\al R$ (see Section \ref{type-loc}).  But $\al
  R$ has a dense set of ergodic states --- states with trivial
  centralizer.  For all these states, $Z(\al R_\om)=\7C I$.
\end{enumerate}

Thus, it makes an enormous difference --- at least for the feasibility
of the modal interpretation --- that local algebras are type III$_1$.
For if local algebras were either type I$_{\infty}$ or III$_0$, then
there would be good news for the modal interpretation.

\begin{prop} Let $\al R$ be a type I$_\infty$ factor.  Then for every
  normal state $\om$ of $\al R$, the unique modal algebra $\al D_\om$
  is nontrivial.  \end{prop}

\begin{proof} We have $\D _\om = Z(\al R_\om )=\{ D\}''$, where $D$ is
  the density operator, i.e.\ the positive operator in $\al R$ that
  implements the state $\om$ via the trace formula.  Furthermore, when
  $\al R$ is type I$_\infty$, $D$ cannot be a multiple of the
  identity.
\end{proof}

\begin{prop} Let $\al R$ be a type III$_0$ factor.  Then for every
  faithful normal state $\om$ of $\al R$, the unique modal algebra
  $\al D_\om$ is nontrivial.
\end{prop}

\begin{proof} Prop.\ 3.15 in \cite[p.\ 402]{tak2} entails that $\al
  D_\om$ has no atomic projections, and hence is infinite dimensional.
\end{proof}

\subsection{What is a symmetry in AQFT?}

We note here just one problem with application of the Clifton-Kitajima
theorem to AQFT: the notion of symmetry invoked might be too liberal
for the setting where we have a \emph{net of algebras over spacetime},
as opposed to a single von Neumann algebra.  Clifton's application of
the theorem assumes that any automorphism of $\al R$ is a symmetry.
However, if $\al R=\al R(O)$ is just one algebra of an entire net
$O\mapsto \al R(O)$, then it is not clear that every automorphism of
$\al R$ is a symmetry of the relevant system.  What we need is a
notion of a symmetry of the net $O\mapsto \al R(O)$.

\begin{note} A partially ordered set $\2K$ can be regarded as a
  category where for $x,y\in \2K$, $\Hom (x,y)=\{ (x,y) \}$ if $x\leq
  y$, and otherwise $\Hom (x,y)=\emptyset$.  Let $\mathbf{C^*}$ be the
  category with $C^*$-algebras as objects and $*$-homomorphisms as
  arrows.  On this conception, a net of $C^*$-algebras on Minkowski
  spacetime is a functor $\al A:\2K \to \mathbf{C^*}$ where $\2K$ is
  the category of double cones in Minkowski spacetime, ordered by
  inclusion, and such that $\alg{A}(\Hom (O_1,O_2))$ is an isometry
  when $\Hom (O_1,O_2)$ is not empty.  (For definitions of functors
  and natural transformations, see p.\ \pageref{nat-trans}.)
\end{note}

\begin{defn} Let $\2K$ be a partially ordered set (e.g.\ regions in
  some manifold ordered by inclusion).  Let $O\mapsto \al A(O)$ and
  $O\mapsto \al B(O)$ be nets of $C^*$-algebras over $\2K$.  A
  \emph{net morphism} $\a :\al A\to \al B$ is a natural transformation
  between the functors.  That is, $\a$ consists of a collection of
  morphisms
$$ \bigl\{ \a _O:\alg{A}(O)\to \alg{B}(O) : O\in \2K \bigr\},$$ that is natural in
$O$.  In other words, for each $f\in \Hom (O_1,O_2)$, $\a _{O_2}\circ
\al A(f)=\al B(f)\circ \a _{O_1}$, which just means that the following diagram
commutes
\begin{diagram}
  \alg{A}(O_1) & \rTo^{\a _{O_1}}  &  \alg{B}(O_1) \\
  \dTo^{\al A(f)}  & & \dTo_{\al B(f)} \\
  \alg{A}(O_2) & \rTo_{\a _{O_2}} & \alg{B}(O_2) \end{diagram}
\label{net-morphism} 
\end{defn}

\begin{fact} Net automorphisms correspond to automorphisms of the
  quasilocal algebra that leave each local subalgebra globally
  invariant.  To state this precisely, let $\alg{A}$ denote the
  functor from $\2K$ into $\mathbf{C^*}$, and let $\alg{B}$ denote the
  inductive limit of $\alg{A}$.  We identify $\alg{A}(O)$ with its
  image in $\alg{B}$.  Then $\a$ is a net automorphism of $\alg{A}$
  iff there is an automorphism $\beta$ of $\alg{B}$ such that
$$ \beta |_{\alg{A}(O)} = \a _O .$$
\end{fact}

Now, given a net $\alg{A}$ with inductive limit $\alg{B}$, what should we consider as
a symmetry of $\alg{B}$?

\begin{proposal} A symmetry of the net $\alg{A}$ corresponds to a net
  automorphism $\a$; i.e.\ a natural transformation of $\alg{A}$.
  That is, a symmetry of $\alg{A}$ corresponds to an automorphism of
  the quasilocal algebra that leaves each local subalgebra globally
  invariant.
\end{proposal}

This first proposal is surely too strict, because it excludes the case
of symmetries induced by underlying symmetries of the spacetime.  But
if $\2K$ consists of an appropriate set of regions of spacetime $M$
(i.e.\ a set that is closed under symmetries of the spacetime), then a
symmetry of $M$ will induce an order-preserving bijection $F$ on
$\2K$.  Note that since $F$ is a functor, $\alg{A}\circ F$ is also a
functor.  Thus, we consider the following liberalized definition.

\begin{proposal} A symmetry of the net $\alg{A}$ consists of a pair $(F,\a )$ where
  $F$ is an order-preserving bijection of $\2K$, and $\a$ is a net morphism (natural
  transformation) from $\alg{A}$ to $\alg{A}\circ F$.
\end{proposal}

If we accept this proposal, then we must replace Clifton's definability condition
with the following modified condition:

\begin{quote} \textit{Definability-2:} Given $O\in \2K$, let $\2K _0$
  be the full subcategory of $\2K$ with objects $\{ O_0:O_0\leq O \}$,
  and let $\alg{R}_{O}$ denote the restriction of the von Neumann
  algebra valued functor $\alg{R}$ to $\2K _0$.  Then the algebra
  $\alg{D}$ must be left invariant by all symmetries of $\alg{R}_O$
  that preserve the state $\rho$ on $\alg{R}(O)$.
\end{quote} Since not all automorphisms of $\alg{R}(O)$ are symmetries of the net
$\alg{R}_O$, the new definability condition is weaker than the old one: there will
typically be more candidates for the role of $\alg{D}$.   

The Clifton-Kitajima Theorem does not apply under the revised
definition of symmetries of $\al R(O)$.  On the other hand, we are not
aware of a positive result showing the existence and uniqueness of
subalgebras of $\al R(O)$ that are definite in the state $\om$ and
invariant under all net automorphisms that preserve $\om$.  There are
suggestive hints such as the result in \cite{split}:

\begin{prop} Let $(\al R_1\subseteq \al R_2 ,\om )$ be
  a standard split inclusion of von Neumann algebras.
  Then there is a \emph{unique} type I factor $\al N$
  such that: (i) $\al R_1\subseteq \al N\subseteq \al
  R_2$, and (ii) $\al N$ is invariant under all
  automorphisms of $\al R_2$ that preserve both $\al
  R_1$ and the state $\om$. \label{midd} \end{prop}

Of course, the algebra $\al N$ itself does not have dispersion-free
states, and so cannot be the algebra of definite observables.
However, the state $\om |_{\al N}$ is normal, and since $\al N$ is a
type I factor, there is a density operator $D\in \al N$ that induces
the state in the sense that $\om (A)=\tr (DA)$, for all $A\in \al N$.
Then assuming that $\al R_1$ must for some reason be left invariant
under symmetries of $\al R_2$, the algebra $\al D =\{ D\}''$ looks
like a good candidate for the modal interpreter's set of
definite-value observables in $\al R_2$ in the state $\om$.

To apply Prop.\ \ref{midd} to AQFT with $\al R_i=\al R(O_i)$, and
$O_1\subseteq O_2$, we would have to assume that the split property
holds.  Although the split property does not hold in every model,
failure of the split property implies a sort of pathology, and it
might not be too surprising if there were certain physically
pathological cases where the modal interpretation yields a trivial set
of definite quantities.

\bigskip \noindent {\small \textit{Notes:} For recent discussions of
  adapting the modal interpretation to a relativistic setting, see
  \cite{myr,rue-ear}.}

\section{Quantum Fields and Spacetime Points} \label{pointy}

In standard/heuristic presentations of QFT, the fundamental physical
quantities (observables, or more generally quantum fields) are
operators indexed by spacetime points: $\Phi (x)$ (see t'Hooft, this
volume).  Based on this fact, at least one philosopher (\cite{tell})
describes the ontology of QFT in terms of the idea a \emph{field of
  operators and their expectation values}.  On the other hand, the
mathematical approach to QFT (e.g.\ the Wightman approach) eschews the
use of operators at points in favor of operators smeared over
space(time) by test-functions: $\Phi (f)$.  According to Arnteznius
[\citeyear{franky}], this fact supports the view that spacetime has no
pointlike events, and \emph{a fortiori} that there are no field values
at spacetime points.

As QFT became more mathematically rigorous, an intuition developed
that it is not only difficult to define the value of a field at a
point, but that it is impossible to do so --- such quantities simply
do not exist.  (Compare von Neumann's critique of Dirac's delta
functions and the notion of pointlike localized particles.)  This
intuition has sometimes been buttressed by heuristic and
operationalist arguments --- e.g.\ Bohr and Petersen's
[\citeyear{bohr}] argument that it is impossible to measure field
strengths at a point.  For example, Haag [\citeyear[p.\ 58]{haag}]
claims that, ``a quantum field $\Phi (x)$ at a point cannot be a
proper observable.''  Even philosophers can be found claiming that,
``field operators need to be `smeared' in space'' \cite[p.\ 631, fn.\
8]{nhug}.

But the arguments against field operators at a point often confuse
questions of measurability with questions of existence, and rarely
rise to a level of rigor that is acceptable for drawing metaphysical
conclusions.  In this section, we review some of the rigorous
arguments that exist for and against field quantities at points.  We
will see that these results do not decisively rule out field
quantities at points, but they clarify the interpretive tradeoffs that
must be made.

\subsection{No Go theorems}

In the following three subsections, we review No Go theorems for field
operators at spacetime points.

\subsubsection{Translation covariance rules out operators at a point}

The first no go theorem shows that if there is a continuous unitary representation of
the translation group, then for any fixed time $t$, the field configuration operators
$\phi (\mb{x},t)$ commute with the field momentum operators $\pi (\mb{x}',t)$,
\emph{even when these operators are associated with the same point}.  This result is
a serious problem, because $\phi (\mb{x},t)$ and $\pi (\mb{x},t)$ are supposed to be
canonically conjugate (see \cite[p.\ 131]{ryder}, \cite{huggett}):
\begin{equation} [\phi (\mb{x},t),\pi (\mb{x}',t)]=i\delta
  (\mb{x}-\mb{x}') .\end{equation} Moreover, this bad outcome cannot
be blamed on any sort of ``conflict'' between quantum mechanics and
relativity, because the bad outcome also holds for non-relativistic
theories.

\begin{thm} Let $\phi (x,t)$ and $\pi (y,t)$ be fields of operators, either bounded
  or unbounded and self-adjoint, such that
  \[ {[}\phi (x,t),\pi (y,t){]}=0 ,\] when $\mb{x}\neq y$.  (In the
  unbounded case, we mean that $\phi (x,t)$ and $\pi (y,t)$ are defined on a common
  dense set $\2D$, and they commute on this set.)  If $y\mapsto U(y)$ is a continuous
  representation of the translation group such that $U(y)\pi (x,t)U(y)^*=\pi
  (x+y,t)$, for all $x,y\in \mathbb{R}^{3}$, then
  \[ {[}\phi (x,t),\pi (x,t){]}=0 ,\] for all
  $\mb{x}\in \mathbb{R}^{3}$. \end{thm}

\begin{proof} Since this proof only uses field operators on the same time slice, we
  will suppress reference to $t$.  Suppose first that $\phi (x)$ and $\pi (y)$ are
  bounded operators.  In this case, the mapping:
  \begin{equation} f(y):=[\phi (x),\pi (x+y)]=[\phi (x),U(y)\pi (x)U(y)^*]
    ,\end{equation} is a weak-operator continuous function from $\mathbb{R}^{3}$ into
  the bounded operators on $\2H$.  Choose a sequence $(y_n)_{n\in \7N}$ of nonzero
  vectors that converges to $0$.  Since $f$ is continuous, and $f(y _n )=0$ for all
  $n\in \mathbb{N}$,
  \begin{equation} [\phi (x),\pi (x)]=f(0)=\lim _{n\rightarrow \infty}f(y_n )=0
    .\end{equation}

  Now suppose that $\phi (x)$ and $\pi (y)$ are unbounded but self-adjoint.  Then
  replace $\pi (x)$ with one of its spectral projections $E_S(x)$, where $S$ is a
  Borel subset of $\7R$, and replace $\pi (x)$ with one of its spectral projections
  $F_{S'}(y)$, where $S'$ is a Borel subset of $\7R$.  By the preceding argument,
  $E_S(x)$ and $F_{S'}(y)$ commute.  Since this is true for all such pairs of
  spectral projections, it follows that the spectral projections of $\phi (x)$
  commute pairwise with the spectral projections of $\pi (x)$.  Hence $\phi (x)$ and
  $\pi (x)$ are defined on a common dense set $\2D$ in $\2H$, and they commute on
  this dense set.
\end{proof}

\subsubsection{Poincar{\'e} covariance rules out operators at a point}

For our next two no go theorems, we will need to gather a couple of classic results.

\begin{defn} A function $f:\7R ^n\to \7C$ is said to be \emph{of
    positive type} just in case for each $c_1,\dots ,c_n\in \7C$, and
  each $x_1,\dots ,x_n\in \7R ^n$, we have
$$ \sum _{i=1}^n \sum _{j=1}^n\ol c_j c_i f(x_i-x_j) \geq 0 .$$
\end{defn}

\begin{thm}[Bochner] Let $f:\7R ^n\to \7C$ be a continuous function of positive type.
  Then the Fourier transform of $f$ is a bounded measure on $\7R ^n$.
\end{thm}

\begin{proof} For a proof of Bochner's theorem, see \cite[p.\ 303]{rudin2} and
  \cite[p.\ 95]{folland}.
\end{proof}

\begin{note} Only the group structure of $\7R ^n$ is really needed for Bochner's
  theorem.  So, we are not making any mistake by thinking of Minkowski spacetime as
  $\7R ^4$ in this context.  \end{note}

We will need the following key lemma in all of our subsequent results.

\begin{lemma} Let $f$ be a continuous positive definite function on
  $\7R ^n$.  Then $f$ is the constant $1$ function iff the Fourier
  transform of $f$ is the probability measure with support $\{ 0\}$.
  \label{pmass}
\end{lemma}

The proof of the above lemma is trivial: the Fourier transform of the
measure $\mu$ with support $\{ 0\}$ is the function $f$ defined by
$$ f(x)=\int _{\7R ^n}e^{i(x\cdot p)} d\mu (p) = e^{i(x\cdot 0)}= 1 .$$
But the Fourier transformation is a bijection between complex Radon measures on $\7R
^n$ and bounded continuous functions on $\7R ^n$.

\begin{defn} We say that a measure $\mu$ on Minkowski spacetime is
  \emph{Lorentz invariant} just in case $\mu (\Lambda (S))=\mu (S)$
  for each Borel subset $S$ of $M$, and each homogeneous Lorentz
  transformation $\Lambda$, where $\Lambda (S)=\{ \Lambda (x):x\in
  S\}$.  \end{defn}

Clearly, the only Lorentz invariant probability measure on Minkowski
spacetime is the measure supported on $\{ 0\}$ (the unique fixed point
of the homogeneous Lorentz group).  The following result is the
``Fourier transformed'' version of that fact.

\begin{lemma} Let $M$ be Minkowski spacetime.  If $f:M\to \7C$ is a continuous
  function of positive type such that $f(\Lambda x)=f(x)$ for each Lorentz
  transformation $\Lambda$, then $f$ is constant. \label{constant}
\end{lemma}

\begin{proof}[Sketch of proof.] By Bochner's theorem, if $f:M\to \7C$ is a continuous
  function of positive type, then $f$ is the Fourier transform of a bounded measure
  $\mu$ on $M$.  It is straightforward to verify that if $f$ is Lorentz invariant
  then so is $\mu$.  But a bounded, Lorentz invariant measure is supported on $\{
  0\}$.  By Lemma \ref{pmass}, the Fourier transform of $\mu$ is a constant function.
  Therefore, $f=1$ is constant.
\end{proof}

\begin{fact} Let $U$ be a unitary representation of the translation group on a
  Hilbert space $\2H$.  Then the following are equivalent:
\begin{enumerate}
\item The spectrum of the representation $U$ is $\D$;
\item For every $u,v\in \2H$, the function $f:\7R ^n\to \7C$ given by
$$ f(x)=\langle u,U(x)v\rangle ,\qquad x\in \7R ^n ,$$
has Fourier transform with support in $\D$.
\end{enumerate}
\end{fact}

Finally, the following is our core lemma for the next two results.

\begin{lemma} Let $A:M\to \bh$ be an operator valued function, and let $U$ be a
  unitary representation of the translation group on $\2H$ such that
  $U(x)A(0)U(x)^*=A(-x)$ for all $x\in M$.  Define a function $f:M\to \7C$ by
$$ f(x)=\langle \Om ,A(x)A(0)\Om \rangle =\langle \Om ,U(x)^*A(0)U(x)A(0)\Om \rangle .$$
If $f$ is constant, then there is a $c\in \7C$ such that $A(x)\Om =c\,\Om$ for each
$x\in M$.  \label{two-point} \end{lemma}

\begin{proof} Let $\psi =A(0)\Om$.  Then $f(x)=f(0)$ is expressed as
$$\langle \psi ,U(x)\psi \rangle = \langle \psi ,\psi \rangle =\norm{\psi}^2 . $$
But we also have $\norm{\psi }=\norm{U(x)\psi}$ since $U(x)$ is unitary.  Hence
$$ \langle \psi ,U(x)\psi \rangle =\norm{\psi}\cdot \norm{U(x)\psi } ,$$
and the Cauchy-Schwartz inequality entails that $U(x)\psi =\psi$ for all $x$.  That
is, $U(x)A(0)\Om =A(0)\Om$.  Note in addition that $U(x)A(y)\Om = U(x+y)A(0)\Om
=A(0)\Om$.  Hence all vectors $A(x)\Om$ are invariant under the translation group.
\end{proof}

Now, the second no go theorem (due to \cite{wizi}) shows that there is no nontrivial
Poincar{\'e} covariant field of bounded operators on Minkowski spacetime.  

\begin{thm} Suppose that $A:M\to \bh$ is an operator-valued function, and $U$ is a
  continuous unitary representation of the Poincar{\'e} group on $\2H$ such that:
\begin{enumerate}
\item $U(y,\Lambda )A(x)U(y,\Lambda )^*=A((\Lambda x)-y)$, for all $(y,\Lambda )\in
  \2P$ and $x\in M$;
\item There is a unique (up to scalar multiples) translation-invariant vector $\Om
  \in \2H$.
\end{enumerate} Then there is a $c\in \7C$ such that $A(x)\Om =c\,\Om$ for all $x\in
M$.
\label{kami}
\end{thm}

\begin{note} (i): The assumption of the uniqueness of
  $\Om$ might seem unwarranted.  But under some fairly
  standard conditions, this assumption can be derived.
  See Section \ref{uniq}.  (ii): This theorem makes no
  assumption about commutation relations between
  operators $A(x)$ and $A(y)$.  \end{note}

\begin{proof}[Proof of Theorem \ref{kami}.] Define a function $f:M\to \7C$ by
  $$ f(x)= \bigl\langle \Omega ,A(x)^*A(0)\Omega \big\rangle  ,\qquad x\in M .$$
  By condition 2 we have $U(x)\Om =\Om$.  Hence by condition 1 we have
  $A(x)^*=U(x)A(0)^*U(x)^*$, and hence
$$ f(x) =\bigl\langle A(0)\Om ,U(x)^*A(0)\Om \bigr\rangle  , $$
which is obviously positive definite.  Furthermore, since $x\mapsto U(x)^*$ is weakly
continuous, $f$ is continuous.

Now we establish that $f(\Lambda (x))=f(x)$ for all $x\in M$ and all Lorentz
transformations $\Lambda$.  We have 
\begin{eqnarray*} 
  f(\Lambda x) &=& \bigl\langle \Om ,A(\Lambda x)^*A(0)\Om \bigr\rangle  \\
  &=& \bigl\langle \Om ,U(0,\Lambda )A^*(x)U(0,\Lambda )^{-1}A(0)\Om \bigr\rangle \\
  &=& \bigl\langle U(0,\Lambda )^{-1}\Om ,A(x)^*U(0,\Lambda )^{-1}A(0)U(0,\Lambda )\Om
  \bigr\rangle \\
  &=& \bigl\langle \Om ,A(x)^*A(\Lambda (0))\Om \bigr\rangle \\
  &=& \bigl\langle \Om ,A(x)^*A(0)\Om \bigr\rangle \\
  &=& f(x) .
\end{eqnarray*}
Thus, Lemma \ref{constant} entails that $f$ is constant, and Lemma \ref{two-point}
entails that there is a $c\in \7C$ such that $A(x)\Om =c\,\Om$ for all $x\in M$. 
\end{proof}

\subsubsection{Microcausality and Spectrum Condition rule out
  operators at a point}

The final no go theorem, originally by Wightman [\citeyear{wightman}]
invokes both microcausality and the spectrum condition.  (See
\cite[p.\ 46]{horuzhy} and \cite[p.\ 115]{russ} for alternative proofs.)

\begin{thm} Suppose that $A:M\to \bh$ is an operator valued function, and $U$ is a
  continuous unitary representation of the translation group on $\2H$ such that:
\begin{enumerate}
\item $[A(x),A(y)]=0$ when $x$ and $y$ are spacelike separated;
\item $U(x)A(y)U(x)^* = A(y-x)$, for all $x,y\in M$;
\item $U$ satisfies the spectrum condition.
\item There is a unique translation invariant vector $\Om\in \2H$.
\end{enumerate}
Then there is a $c\in \7C$ such that $A(x)\Om =c\,\Om$ for all $x\in M$.
\label{wight}
\end{thm}

\begin{proof} As above, define $f:M\to \7C$ by $$ f(x)= \bigl\langle \Omega
  ,A(x)A(0)\Omega \big\rangle ,\qquad x\in M .$$ Fix a nonzero spacelike vector $x$.
  Then by condition 1, $$ U(x)^*A(0)U(x)A(0)=A(x)A(0)=A(0)A(x)=A(0)U(x)^*A(0)U(x) .$$
  Therefore, \begin{eqnarray*} \lefteqn{ f(x)=\langle \Om ,U(x)^*A(0)U(x)A(0)\Om
      \rangle = \langle \Om A(0)U(x)^*A(0)\Om \rangle } \\ &=& \langle \Om
    A(0)U(-x)A(0)\Om \rangle =f(-x) . \end{eqnarray*} Now consider the function
  $F:\7R \to \7C$ given by $F(t)=f(tx)$, so that $F(t)=F(-t)$.  By condition 3, the
  Fourier transform of $f$ is supported in the forward light cone.  Hence, the
  Fourier transform of $F$ is supported in $[0,+\infty )$.  But since $F(t)=F(-t)$,
  the Fourier transform of $F$ is also supported in $(-\infty ,0]$.  Therefore, the
  Fourier transform of $F$ is the point mass at $\{ 0\}$.  By Lemma \ref{constant},
  $F$ is constant.  Finally, since any two points in $M$ can be connected by two
  spacelike vectors, we can apply the previous procedure twice to show that $f$ is
  constant.  Therefore, by Lemma \ref{two-point}, there is a $c\in \7C$ such that
  $A(x)\Om =c\, \Om$ for all $x\in M$. \end{proof}

\begin{cor} Let $O\mapsto \al R(O)$ be a net of von Neumann algebras acting
  irreducibly on a Hilbert space $\2H$, and let $U$ be a strongly continuous unitary
  representation that implements the action of the translation group on the net $\al
  R$.  Suppose that the net satisfies microcausality (assumption \ref{micro}).
  Suppose that $U$ satisfies the spectrum condition, and that there is a translation
  invariant vector $\Om\in \2H$.  Then for each point $x\in M$,
$$ \bigcap _{\{ O\in  \2K\,:\,x\in O \}}\al R(O) =\7C I .$$
\end{cor}

\begin{proof} Fix $x\in M$, and fix a double cone $x\in O$.  Choose an arbitrary
  operator, denoted by $A(x)$, in
$$ \bigcap _{\{ O\in  \2K\,:\,x\in O \}}\al R(O) .$$  Now for general
$y\in M$, define
$$ A(y) =U(x-y)A(x)U(x-y)^* ,$$
so that the mapping $A:M\to \bh$ automatically satisfies condition 2
of Theorem \ref{wight}.  Furthermore, since the net $\al R$ satisfies
microcausality, and the unitary group $U$ implements the translations
on $\al R$, the mapping $A$ satisfies condition 1 of Theorem
\ref{wight}.  It then follows that there is a $c\in \7C$ such that
$A(x)=cI$.  Since $x$ was an arbitrary element of $M$, the result is
proven.
\end{proof}

\subsection{Go theorems}

Why should we care if $\Phi (x)$ cannot be taken to denote any non-trivial operator
on Hilbert space?  Does this have any implications for the interpretation of QFT?
After all, for any neighborhood $O$ of $x$, we can find a test-function $f$ that is
supported in $O$, and we can replace the non-denoting term ``$\Phi (x)$'' with the
denoting term ``$\Phi (f)$''.  In fact, couldn't we think of ``$\Phi (x)$'' as a name
for the sequence $\{ \Phi (f_n ) \} _{n=1}^{\infty}$, where $\{ f_n
\}_{n=1}^{\infty}$ is a sequence of test-functions that converges to the
delta-function at $x$?  More precisely, it seems that we could even attempt to define
an expectation value for the pseudo-operator $\Phi (x)$ as follows: If $\rho$ is a
state of the quantum field, define:
\begin{equation} \rho (\Phi (x)):= \lim _{n\rightarrow \infty}\rho (\Phi (f_{n}))
  .\end{equation}

In this section, we make this idea precise in two Go Theorems for
field quantities at points.  The first result we report from the work
of Rehberg and Wollenberg \cite{wollenberg86a,wollenberg86b} (see also
\cite{fred81}, \cite{bostelmann,bostelmann-b}).  This result shows
that within the Wightman framework, a quantum field at a point can be
represented by a sesquilinear form.  The second result shows that if
we drop the requirement of continuity on our representation of the
translation group, then quantum fields at points can be represented by
self-adjoint operators.

\subsubsection{Quantum fields as sesquilinear forms}

\begin{defn} Let $\2H$ be a Hilbert space.  A \emph{sesquilinear form}
  on $\2H$ is a linear subspace $D(t)$ of $\2H$ and a mapping $t:D(t)
  \times D(t) \to \7C$ that is antilinear in the first argument, and
  linear in the second argument.  The form $t$ is said to be
  \emph{densely defined} just in case $D(t)$ is dense in $\2H$.  The
  form $t$ said to be \emph{symmetric} just in case $t(\f ,\psi
  )=\overline{t(\psi ,\f )}$ for all $\f ,\psi \in D(t)$.  The form
  $t$ is said to be \emph{positive} just in case $t(\psi ,\psi )\geq
  0$ for all $\psi \in D(t)$.
\end{defn}

\begin{defn} If $t$ is a sesquilinear form on $\2H$ then we define the
  associated \emph{quadratic form} by $t(\psi )=t(\psi ,\psi )$ for
  all $\psi \in D(t)$.  A positive quadratic form $t$ is said to be
  \emph{closed} just in case for any sequences $(\psi _n)_{n\in \7N }$
  in $D(t)$ if $\psi _n\to \psi$ and $t(\psi _n-\psi _m)\to 0$, then
  $\psi \in D(t)$ and $t(\psi _n-\psi )\to 0$.  \end{defn}

\begin{note} A densely defined, symmetric sesquilinear form is a
  \emph{prima facie} candidate to represent a physical quantity or an
  observable.  Since $t$ is symmetric, the corresponding quadratic
  form is real-valued.  Hence, for each unit vector $\psi \in D(t)$,
  we might say that the ``expectation value'' of $t$ in state $\psi$
  is $t(\psi )$.  Indeed, at first glance, the expectation value
  mapping $t\mapsto t(\psi )$ seems to have all the same properties as
  the corresponding expectation mapping for operators.
\end{note}

\begin{thm} Let $\Phi (\cdot )$ be a Wightman field on the Hilbert
  space $\2H$.  That is, $\Phi $ maps elements of a test-function
  space $\2S (\7R ^4)$ to unbounded operators on $\2H$ with some
  common dense domain $\2D$.  Let $(\delta _n)_{n\in \7N}$ be a
  sequence of test-functions whose support shrinks to the point $x$.
  Then for each $u,v\in \2D$, the sequence
  $$ \langle u,\Phi (\delta _1)v\rangle , \, \langle u,\Phi (\delta _2)v\rangle ,\,\langle u,\Phi (\delta _3)v\rangle , \dots ,$$
  converges to a finite real number, which we denote by $\langle
  u,\Phi (x)v\rangle $.  The map $u,v\mapsto \langle u,\Phi
  (x)v\rangle$ is a sesquilinear form with domain $\2D$, which we
  denote by $\Phi (x)$.  \label{russ}
\end{thm}

\begin{proof} See \cite[p.\ 332]{russ} and
  \cite{wollenberg86a,wollenberg86b}.  \end{proof}

\begin{note} One naturally wishes to have a version of this theorem in
  a more purely algebraic setting.  Such a result might be available
  in the context of the scaling algebras of \cite{scaling,bucky}.
\end{note}

The result is surprising for a couple of reasons.  We might have
thought that the reason $\Phi (x)$ is not an operator is because the
expectation values $\langle u,\Phi (\d _n)u\rangle $ grow without
bound as the test-functions $\d _n$ shrink to a point --- i.e.\ there
is some sort of divergence.  But Theorem \ref{russ} shows that
conjecture to be false.  The obstruction to $\Phi (x)$'s becoming an
operator must lie somewhere else.

So, we have several No Go Theorems against quantum fields as operators
(even unbounded operators), and one Go Theorem for quantum fields as
sesquilinear forms.  What should we conclude from these apparently
conflicting results?  Should we say that there is a field quantity at
the point $x$, or not?

To answer this question we need to think harder about the relation
between operators on a Hilbert space and physical quantities.  Why was
it that we thought that physical quantities correspond to operators?
If we suppose that an unbounded operator can represent a quantity,
then must that operator be self-adjoint (i.e.\ must $A$ coincide with
$A^*$ on a common dense domain), or does it suffice to satisfy some
weaker condition?  Does any symmetric sesquilinear form have all the
features necessary to represent a physical quantity?  In order to get
clear on these questions, it might help to get clear on the
mathematical details of the relationship between sesquilinear forms
and operators.  Fortunately, there are quite a few results in this
direction.

Clearly, every linear (possibly unbounded) operator $T$ on $\2H$
defines a sesquilinear form with domain $D(T)$ via the equation
\begin{equation} t(\psi ,\f )=\langle \psi ,T\f \rangle .\label{sesqu}
\end{equation} On the other hand, it is less clear when an arbitrary
form $t$ corresponds an operator via Eqn.\ (\ref{sesqu}).  

\begin{defn} A sesquilinear form $t$ on $\2H$ is said to be
  \emph{bounded} just in case there is a $n \in \7N$ such that
  $\abs{t(\f ,\psi )}\leq n$ whenever $\f ,\psi \in D(t)$ with
  $\norm{\f },\norm{\psi }\leq 1$. \end{defn}

\begin{prop} There is a one-to-one correspondence between densely
  defined, bounded sesquilinear forms on $\2H$ and elements of $\bh$.
  In particular, if $t$ is bounded sesquilinear form on $\2H$ then
  there is a unique operator $T\in \bh$ such that $t(\f ,\psi)=\langle
  \f,T\psi \rangle$ for all $\f ,\psi \in \2H$.  Furthermore, $t$ is
  symmetric iff $T$ is self-adjoint.
\end{prop}

\begin{proof} See \cite[Theorem 2.4.1]{kr}.  \end{proof}

\begin{prop} If $t$ is a densely defined, positive, closed, quadratic
  form, then there exists a unique positive operator $T$ on $\2H$ such
  that the domain of $T^{1/2}$ is $D(t)$ and
$$ t(\f ,\psi )=\langle T^{1/2}\f ,T^{1/2}\psi \rangle ,$$
for all $\f ,\psi \in D(t)$.  In particular, $t(\f ,\psi )=\langle \f
,T\psi \rangle$ for all $\f ,\psi \in D(t)$.  \end{prop}

\begin{note} The previous Proposition is useful in showing when a
  number operator $N$ can be defined in a representation of the Weyl
  algebra $\alg{A}[S,\sigma ]$.  For details, see \cite{rindler} and
  \cite[p.\ 27]{brat2}. \end{note}

The previous two propositions do not apply to the sesquilinear form
$\Phi (x)$ because it is neither bounded nor positive.  Furthermore,
there is no known (to the author) characterization of when a symmetric
sesquilinear form admits a representation as an operator --- although
there are some partial results in this direction (see \cite{mcin}).
It is clear that $\Phi (x)$ is not an operator; it is unclear what
features operators have that $\Phi (x)$ lacks, and whether these
features are necessary for a mathematical object to represent a
quantity.  Accordingly, it is unclear whether or not $\Phi (x)$
represents an element of reality.

\subsubsection{Quantum fields as operators on non-separable Hilbert space}
\label{aleph}

Our second Go result for quantum field operators at a point is really just a sketch
of an example.  We take a nonseparable Hilbert space $H$ that can represent states of
particles with point positions (compare with \cite{hans}).  We then apply the
standard second quantization procedure --- which does not depend on the one-particle
space being separable --- to obtain a Fock space $\2F (H)$, and self-adjoint field
operators $\phi (x),\pi (x)$ indexed by points in $\7R$.

Let $H=l_2(\mathbb{R})$ be the Hilbert space of square-summable sequences over $\7R$;
i.e.\ an element $f$ of $l_2(\7R )$ is a mapping from $\7R$ into $\7C$ such that $f$
vanishes at all but countably many points, and $\sum _{x\in
  \mathbb{R}}\abs{f(x)}^{2}<\infty $.  The inner product on $l_2(\mathbb{R})$ is
given by
\begin{equation}
  \langle f,g\rangle = \sum _{x\in \mathbb{R}}\overline{f(x)}g(x)
  .\end{equation} 
Let $\2F (H)$ be the Fock space over $H$.  For each $x\in
\mathbb{R}$, we let $\delta _x \in l_2(\7R )$ denote the
characteristic function of $\{ x\}$; the set $\{ \delta _x:x\in
\mathbb{R}\}$ is an (uncountably infinite) orthonormal basis for
$l_2(\mathbb{R})$.  For any $x\in \7R$, we define the creation
operator $a(x)$ by:
\begin{equation}
  a(x)(f_1\otimes \dots \otimes f_n) := \delta _x \otimes f_1 \otimes \cdots \otimes
  f_n .\end{equation}
As in the standard case, we verify that $a^-(x)+ia^+(x)$ and
$a^{+}(x)-ia^-(x)$ are preclosed, i.e.\
that the closure of the graphs of these operators are
graphs of linear operators (see \cite[p.\ 155]{kr}), which we denote by 
\begin{eqnarray}
  \phi (x) &= & \overline{a^-(x)+ia^+(x)} , \\
  \pi (x) &= & \overline{a^+(x)-ia^-(x)} .\end{eqnarray}
It then follows that $\phi (x)$ and $\pi (x)$ are self-adjoint, and on a dense domain
$\2D$ in $\2F (H)$, we have 
\begin{equation} [\pi (x) ,\phi (x')]= i \,\langle\delta _x,\delta
  _x'\rangle = i\,\delta _0(x-x') ,\end{equation} where now $\d _0$ is
a completely legitimate mathematical object --- viz.\ the probability
measure supported on $\{ 0\}$.

Consider the (discontinuous) representation $x\mapsto
V(x)$ of the translation group on $l_2(\mathbb{R})$
defined on the basis elements $\{ \delta _y:y\in
\mathbb{R}\}$ by
\begin{equation}
  V(x)\delta _{y} =\delta _{y-x} .\end{equation}
Let $\Gamma$ be the `Fock functor'; i.e.\ $\Gamma$ maps
a unitary operator $V$ on the single particle
space $H$ to the corresponding operator 
\[ I\oplus V\oplus (V\otimes V)\oplus \cdots , \] on
$\2F (H)$.  Then $x\mapsto U(x):=\Gamma (V(x))$ is a
\emph{discontinuous} representation of the translation
group on $\2F (H)$, and
\begin{equation}
  U(x)^*\phi (y)U(x)=\Phi (y-x) .\end{equation}
Thus, $(\phi (\cdot ),\pi (\cdot ),\2F (H),U)$ is a
field system over $l_2(\2R )$,
where $x\mapsto U(x)$ is a discontinuous unitary representation.  We could then use
the field system to define a net $O\mapsto \al R(O)$ of von Neumann algebras on $\2F
(H)$.  But this net of course fails the spectrum condition, because the
representation of the translation group is not continuous.    

The model just described is probably too unwieldy to be
of much use in describing real physical situations.
Furthermore, there is no good reason to think that the
procedure we followed generalizes to the case of
interacting theories, where pointlike localized
operators are needed to make sense of products of field
operators.  However, we hoped to show that it is
conceivable that we can get by without some of the
technical assumptions of the No Go Theorems.  So, we
should think very carefully before we try to use these
theorems to draw conclusions about how QFT must be
interpreted.

\subsection{Field interpretations of QFT}

In Section \ref{parts}, we saw that there are severe obstacles to a
particle interpretation of QFT.  One might argue then, by a process of
elimination, that we should adopt a ``field interpretation'' of QFT
(see e.g.\ \cite{tell,nhug}).  But if we examine the field
interpretation on its own merits, its not clear that it is better off
than the particle interpretation.  

In constructing canonical free theories (e.g.\ the free Bose and Fermi
fields), one begins with a Hilbert space $H$ which can be interpreted
either as a ``single particle space'' (i.e.\ space of wavefunctions of
a single quantum mechanical particle) or as a space of configurations
of a classical field.  Corresponding to these two interpretations,
there are two ways to construct the Hilbert space of the quantum field
theory:

\begin{enumerate}
\item Second quantization: the Hilbert space of the quantum field is
  the Fock space $\2F (H)$ over $H$.  (See Section \ref{focked}).
\item Field quantization: the Hilbert space of the quantum field is is
  space $L_2(H ,d)$ of `square integrable' functions from $H$ into
  $\7C$ relative to the isonormal distribution $d$ on $H$.

\end{enumerate}
(In a rigorous treatment, elements of $L_2(H,d )$ are not really
functions.  See \cite[Section 1.3]{baez} for details.)  The free field
theories constructed by these two methods are known to be unitarily
equivalent.  However, the field quantization approach lends itself
more naturally to a field interpretation.  Indeed, in a recent survey
of the foundations of QFT (\cite{nhug}), one finds the suggestive
notation:
\[ \Psi (\phi ),\qquad \phi \in L_2(\7R ^{3n}) ,\] for a function on
the space $H:=L_2(\7R ^{3n})$ of wavefunctions. Thus, it seems that a
quantum field state can be interpreted as a superposition of classical
field configurations in the same sense that a wavefunction of $n$
particles can be interpreted as a superpositions of classical
configurations of $n$ particles.

However, there are difficulties with this approach.  First, the field
operators $\Phi (x)$ on $L_2(H ,d)$ are the precise analogues of the
position operators $Q_i$ for a system of particles.  That is, there is
a natural interpretation of a function $\Psi \in L_2(H, d)$ as a
probability distribution over the family $\{ \Phi (x):x\in M\}$ of
mutually commuting operators.  But the No Go theorems for the
operators $\Phi (x)$ militate against interpreting $\Psi$ as a
probability distribution over classical field configurations.  More
directly, since $d$ assigns zero measure to points in $H$ (i.e.\ to
individual field configurations), characteristic functions of
singleton subsets of $H$ --- i.e.\ determinate field configurations
--- are identified with the zero vector in $L_2(H ,d)$.  That is,
there is no state of the quantum field in which it is in a definition
configuration.

It follows from the preceeding considerations that the No Go theorems
for fields operators at a point undermine the field interpretation of
QFT in the same way that No Go theorems for number operators undermine
the particle interpretation.  Thus, we should be wary of arguments for
field interpretations based on problems with particle interpretations.    

\subsection{Points of time?}

The preceding results were aimed at answering the question of whether
there can be field operators at a spacetime point.  Suppose that we
concede that there cannot be, and we proceed with the standard
mathematically rigorous approach to, say, the free Bose field, where
field operators are smeared over \emph{space} by test functions (see
e.g.\ \cite{arak-bose}).  In this case, quantities are not tied to
spacetime points, but they are tied to pointlike times.  However, some
claim that in general, the quantities will also have to be fuzzy in
time.  For example, according to Haag,
\begin{quote}
  Renormalization theory suggests that it is essential to smear out
  $\Phi$ both in space and time, in contrast to the case of free
  fields, where an averaging over $3$-dimensional space at a fixed
  time is sufficient.  Due to the stronger singularities, one cannot
  assume well-defined commutation relations of fields at equal time.
  \cite[p.\ 59]{haag} \end{quote} But such claims are speculative ---
we know of no theorems that prove that interacting fields must be
smeared out in time.  So, at the present time we have no particularly
good reason to conclude that time is pointless.

\section{The Problem of Inequivalent Representations} \label{inter}

The philosophy of local quantum physics (in Haag's terminology) is
that the theoretical parts of QFT (e.g.\ unobservable fields, gauge
group) should not count as part of the given data.  Instead, the
abstract net $\al A$ of observable algebras should be taken as the
primitive.  Following the terminology in \cite{ruetsche}, we define
`Algebraic Imperialism' as the position that:
\begin{quote}
  The physical content of a quantum field theory is encoded in the net
  $O\mapsto \al A(O)$, the subgroup of $\Aut (\al A)$ corresponding to
  physical symmetries (including dynamics), and the states on the
  quasilocal algebra $\al A$.  A representation $(\2H ,\pi )$ of $\al
  A$ may be an aid to calculation, but has no ontological
  significance.
\end{quote}
Such an attitude might seem incomprehensible to those steeped in the
traditional Hilbert space formalism of QM. Indeed, where is the
Hamiltonian, where are the transition probabilities, and how do we
describe measurements?  The very abstractness and generality of the
algebraic formalism seems to empty it of a great deal of the content
we expect in a physical theory.

However, some of these worries about lack of content of the abstract
algebraic formalism are ill founded.  Indeed, the GNS theorem (Thm.\
\ref{gns}) shows that all the Hilbert spaces we will ever need are
hidden inside the algebra itself.  Furthermore, much of the vocabulary
one learns to speak in elementary QM can be defined within this purely
abstract setting.  For example, for a definition of transition
probabilities between states, see \cite{robrop}; and for a definition
of measurement probabilities, see \cite{wald}.

But it is not true --- at least on the face of it --- that \emph{all}
pieces the traditional vocabulary of QFT can be reproduced in the
algebraic setting.  For example, the quasilocal algebra does not
contain a number operator, and probability distributions over the
spectrum of the number operator cannot be defined in terms of
expectation values on $\al A$ (see \cite{rindler}).  What is perhaps
even worse is that by beginning with a net $O\mapsto \al A(O)$ of
\emph{observable} algebras, we have effectively closed our eyes to the
existence of unobservable fields, which do not generally commute with
each other at spacelike separation.  Thus, we seem to have no way to
account for deep theoretical facts of QFT such as the connection
between spin (commutation relations of field operators) and
statistics.

Worries such as these might push us towards the second main position
on the issue of representations, which Ruetsche [\citeyear{ruetsche}]
calls Hilbert Space Conservatism:
\begin{quote}
  The theory is not the net $O\mapsto \al A (O)$, but the net plus one
  specific representation $(\2H ,\pi )$.
\end{quote}
In fact, Hilbert Space Conservatism might be thought of as the default
view of most workers in mainstream (Lagrangian) QFT, since the
abstract algebra (and its representations) do not play a central role
there.

But as with many realist views, the Conservative view faces
epistemological difficulties: How do we decide which is the correct
representation?  In this case, the difficulty is particularly severe,
because it can be proven mathematically that the predictions of states
within any one representation can be reproduced to arbitrarily high
accuracy by the states in \emph{any other}
representation.\footnote{This way of stating the problem is biased,
  and depends on taking ``predictions of a representation'' to mean
  expectation values of observables in the abstract algebra.  If we
  also include expectation values of observables in the weak closure
  $\pi (\al A)^{-}$, and expectation values of unbounded operators on
  $\2H$, then the story becomes more complicated.  Compare with
  \cite{clif}.}  (This is due to the fact that since $\al A$ is
simple, Fell's theorem implies that the states in any folium are weak*
dense in the state space.)

Nonetheless, it is tempting to think that the algebraic formalism is
creating an interpretive problem.  That is, it is tempting to think
that if we stick to the old-fashioned way of doing QFT, the issue of
inequivalent representations does not arise, and so neither does this
interpretive dilemma.  So, are inequivalent representations telling us
something of foundational importance, or are they just mathematical
playthings?  

The motivating arguments for the algebraic approach have included the
existence of inequivalent representations of the canonical commutation
relations, as well as physical effects associated with Rindler-Fulling
quanta.  Nonetheless, these arguments have been resisted for various
reasons, e.g.\ there is a suspicion that the Rindler vacuum
representation is inferior, as a description of reality, to the
Minkowski vacuum representation.  So, in the following sections, we
discuss another motivating argument for the algebraic approach ---
viz.\ superselection rules.  It is in the analysis of superselection
rules that the algebraic approach most clearly displays its beauty,
utility, and foundational importance.  

\label{sec-DHR}

 \subsection{Superselection rules}

In a now famous paper, Wick, Wightman, and Wigner
[\citeyear{www}] argue that there is a physical system
with state space $H$, and state vectors $\psi _1,\psi
_2 \in H$ such that the linear combinations
\begin{equation} 2^{-1/2}(\psi _1+e^{i\theta }\psi _2 ), \qquad \qquad
  \theta \in [0,2\pi ) , \label{phase} \end{equation} give rise to
`empirically indistinguishable' states.  When this occurs, Wick \et
say that there is a `superselection rule' between $\psi _1$ and $\psi
_2$; alternatively, $\psi _1$ and $\psi _2$ lie in different
`superselection sectors.'  We put `empirically indistinguishable' in
scare quotes, because the literature has been anything but clear about
the nature of the relation between the states in Eqn.\ (\ref{phase}).
Are the states in Eqn.\ (\ref{phase}) merely \emph{empirically}
indistinguishable, or is there a stronger sense in which these two
states are equivalent?  If the indistinguishability is empirical, how
strong is the modal force?  Do these states predict the same empirical
phenomena in all physically possible worlds, or is their
indistinguishability due to some special features (e.g.\ initial
conditions) in our world?  In this paper, we will not attempt to
resolve these important questions about the nature of superselection
rules.\footnote{Superselection rules are also of foundational interest
  because they have been thought to help with the measurement problem
  --- see e.g.\ \cite[p.\ 74]{belt}, \cite{klaas}, \cite[pp.\
  264--272]{bas} --- and more generally because of their connection
  with the emergence of a classical realm \cite{giul}.  However, we do
  not take up those specific issues in this Chapter.}  Rather, we will
content ourselves with explaining the proposal of Doplicher, Haag, and
Roberts (DHR) for making Wick \textit{et al.}'s notion precise within
the context of AQFT.

The first approaches to superselection rules involved
an \textit{ad hoc} butchery of the state space and of
the algebra of observables, with a resulting confusion
about what should count as the states and observables
(or quantities) of the resulting theory.  One begins
with a Hilbert space $H$ with unit vectors giving pure
states, and with self-adjoint elements of $B(H)$ as
observables (or quantities).  One then gives a
heuristic argument for the claim that a superselection
rule holds between the state vectors in some subspace
$H _1$ and the state vectors in the complementary
subspace $H_2:=(H _1)^{\perp}$.  On the basis of this
argument, the state space $H$ is reduced to the union
of $H_1$ and $H_2$; that is, a linear combination of a
vector in $H _1$ and $H _2$ is no longer thought to be
a possible (pure state) of the theory; the pure state
vectors lie in \emph{either} $H_1$ \emph{or} $H _2$.
Equivalently, the algebra of observables $B(H)$ is
reduced to $B(H _1)\oplus B(H _2)$.  The operators in
$B(H_1\oplus H_2)$ that could distinguish between the
states in Eqn.\ (\ref{phase}) are demoted to the status
of ``unobservable.''  Thus, the algebra of observables
is actually $B(H _1)\oplus B(H _2)$, and not
$B(H_1\oplus H_2)$.

Now, the algebraic approach provides two inversely related approaches
to superselection rules.  

\begin{enumerate} \item First, we can follow the original ``state
  space butchery'' approach in a slightly more principled fashion:
  suppose that we are given some fields acting as operators on some
  big Hilbert space $\2H$.  Let $\al F$ denote the algebra of field
  operators.  [Here $\al F$ is the analogue of the algebra $B(H
  _1\oplus H_2)$, and $\2H$ is the analogue of $H_1\oplus H_2$.  In
  this case, however, we are not given an \emph{a priori}
  decomposition of $\2H$ into a direct sum.]  Suppose that we are also
  given a gauge group $G$ that acts on the fields.  We then
  \emph{define} the observables as the gauge invariant fields.  Let
  $\al A$ denote the algebra of observables.  We also define the
  physical pure states as those vectors in $\2H$ that transform
  irreducibly under the gauge group.  A series of mathematical results
  (partially described in Section \ref{sec-localfield}) shows then
  that $\2H$ decomposes into a direct sum $\bigoplus \2H _{\xi}$ of
  subspaces that transform irreducibly under the gauge group; and each
  subspace $\2H _\xi$ corresponds to an irreducible representation of
  the algebra of observables $\al A$.  We sketch this ``top down''
  approach to superselection rules in Section \ref{sec-localfield}.

\item Instead of beginning with the field algebra $\al F$ and deriving
  the superselection structure (i.e.\ the set of physically
  interesting representations of the algebra $\al A$ of observables),
  we can begin with $\al A$ and consider its set of physical
  representations.  What is a `physical' representation of $\al A$?
  According to the criterion proposed by Doplicher, Haag, and Roberts
  (the DHR selection criterion), the physical representations are
  those that differ observably from the vacuum representation only in
  local regions.  In this case, we still have the notion of
  superselection sectors, but we do not yet have a notion of fields or
  of a gauge group.  It is not immediately clear that we have enough
  structure to explain the phenomena.

  However, it is at this point that the deep mathematical analysis
  begins.  First, one proves that the category of DHR representations
  corresponds precisely to the set $\D$ of localized transportable
  endomorphisms of the observable algebra $\al A$ (see Section
  \ref{ss-dhr}).  Second, one proves that the set $\D$ naturally has
  the structure of a symmetric tensor $*$-category (see Section
  \ref{tensor}).  Finally, the Doplicher-Roberts Reconstruction
  Theorem shows that the unobservable fields $\al F$ and gauge group
  $G$ can be uniquely reconstructed from the category
  $\D$. \end{enumerate}

The following sections outline some of the most important insights
that have been gained in the study of superselection rules, and how
this analysis bears on the foundational questions about the role of
inequivalent representations.  In short, our conclusion is that
inequivalent representations are \emph{not} irrelevant, and nor are
they a problem.  Rather, it is the \emph{structure of the category of
  representations} that provides the really interesting theoretical
content of QFT.

\subsection{Minimal assumptions about the algebra of observables} \label{proto-DHR}

For our discussion of superselection theory we need only a
considerably pared down set of assumptions about the net of observable
algebras.  So, we now effectively cancel all assumptions we made about
the net in Section \ref{basics}.  We begin with a \emph{tabula rasa},
and add only those assumptions that we will need in the following
sections.

\setcounter{assumption}{0} 

By calling $\al A$ a ``net,'' we are assuming that if $O_1\subseteq
O_2$ then $\al A(O_1)\subseteq \al A(O_2)$.  But we do not promote
this to the status of an Assumption.

\begin{bxd}
\begin{assumption}[Microcauality] If $O_1$ and $O_2$ are spacelike
  separated then $[\alg{A}(O_1),\alg{A}(O_2)]=\{ 0\}$.
\end{assumption}
\begin{assumption}[Property B] The net $O\to \alg{R}_0(O)\equiv \pi
  _0(\alg{A}(O))''$ of von Neumann algebras satisfies property B,
  where $(\2H _0,\pi _0)$ is the GNS representation of $\alg{A}$
  induced by $\om _0$.
  \label{borchers}
\end{assumption}
\begin{assumption}[Duality] The pair $(\alg{A},\om _0)$ satisfies
  \emph{Haag duality}, i.e.\
  $$ \pi _0(\alg{A}(O'))'=\pi _0(\alg{A}(O))'' ,$$ for each double cone $O$, where
  $(\2H _0 ,\pi _0)$ is the GNS representation of $\alg{A}$ induced by
  $\omega _0$.
\end{assumption}
\begin{assumption}[Separability] The vacuum Hilbert space $\2H _0$ is
  separable.  \end{assumption}
\begin{assumption}[Nontriviality] For each double cone $O$, $\pi
  _0(\alg{A}(O))$ contains an operator that is not a multiple of the
  identity; i.e.\ $\pi _0(\alg{A}(O))\neq \7C I$.
  \label{nontrivial}
\end{assumption} \end{bxd}

A few remarks on these assumptions: (i) The first assumption is about
the net $\al A$, but the remaining assumptions apply to a pair $(\al
A,\om _0)$, where $\al A$ is the quasilocal algebra and $\om _0$ is
some fixed state.  (ii) The duality assumption states that not only
are the observables in $\alg{R}_0(O')$ compatible with the observables
in $\alg{R}_0(O)$, but that $\alg{R}_0(O')$ contains \emph{all}
observables that are compatible with the collection of observables in
$\alg{R}_0(O)$.  We will assume in the following two sections (on DHR
superselection theory) that the net $\alg{A}$ satisfies duality
relative to some privileged vacuum state $\om _0$.  But, it does not
follow from this that the net satisfies duality relative to every
physical representation.  In fact, a representation satisfies duality
iff that sector has normal (Bose/Fermi) statistics; and every
representation satisfies duality iff the gauge group is abelian.
(iii) Duality in the vacuum sector is equivalent to the non-existence
of spontaneously broken gauge symmetries.  For the case of broken
symmetries, we would impose a weaker requirement: essential duality.
cf.\ Section \ref{ssb}.  (iv) The separability assumption will only be
invoked once --- to show all superselection sectors induced by local
fields are strongly locally equivalent (Prop.\ \ref{uto}).

To be clear, note that we are now making no assumptions about the
following: (i) No assumptions about the action of spacetime symmetries
(e.g.\ translation symmetries, Lorentz symmetries) on the algebra $\al
A$; (ii) No assumptions to the effect that the vacuum state $\om _0$
is translation invariant; (iii) No assumptions about the action of
spacetime symmetries on the vacuum Hilbert space; (iv) No assumptions
about the spectrum condition.

\section{The Category $\D$ of Localized Transportable Endomorphisms}

In this Section we study the category $\D (\al A)$ of localized
transportable endomorphisms of the observable algebra $\al A$.  Since
the physical motivation for this study might not be initially clear,
we pause to note the relation between endomorphisms and
representations.

Suppose that $\pi _0$ is a fixed representation of $\al A$ of some
physical significance --- e.g.\ the vacuum representation.  Then for
any endomorphism $\rho$ of $\al A$, the composition $\pi _0\circ \rho$
is also a representation of $\al A$.  Thus, endomorphisms of $\alg{A}$
correspond naturally to representations of $\alg{A}$, and we can hope
to gain insight into the structure of the representations of $\alg{A}$
by studying the endomorphisms of $\al A$.  However, the set $\End \al
A$ of endomorphisms of $\al A$ has more intrinsic structure than the
set $\Rep \alg{A}$ of representations of $\alg{A}$ --- e.g., there is
a product (viz.\ composition) operation on $\End \al A$, and some
endomorphisms have inverses.  Thus, besides the traditional notions of
equivalence and disjointness of representations, there are additional
relations of physical importance on the set of representations of the
form $\pi _0\circ \rho$ with $\rho \in
\End \alg{A}$.

If the problem of Lagrangian QFT is that there is only one Hilbert
space, the problem of AQFT is that there are too many Hilbert spaces!
Surely, not all of the representations of $\alg{A}$ are physical.  In
Section \ref{sec-localfield}, we look at the problem from a more
traditional point of view.  In particular, we begin with a field
algebra $\alg{F}$ of operators acting on a Hilbert space $\2H$, and a
gauge group $G$ of unitary operators on $\2H$.  (We may suppose that
$G$ is the image of some representation of a fundamental symmetry
group, e.g.\ $SU(2)$.)  We also suppose that $\2H$ contains a vacuum
state $\Om$.  We then \emph{define} the observable algebra $\alg{A}$
as the \emph{gauge invariant fields}.  But then we are again in the
domain of AQFT: we have a reducible representation $\pi$ of $\al A$ on
$\2H$, and the irreducible subrepresentations of $\pi$ are the
\emph{superselection sectors that can be reached from the vacuum
  sector by the action of local (unobservable) fields}.  Not all
representations of $\alg{A}$ appear in the decomposition of $\pi$ ---
those that do not are surplus structure.  However, all representations
that appear in the decomposition of $\pi$ are of the form $\pi _0\circ
\rho$, with $\rho$ an endomorphism from the category $\D (\al A)$!
So, the motivation for studying these endomorphisms is that they
correspond to representations that arise in this traditional,
physically motivated way by acting on the vacuum representation with
(unobservable) fields.\footnote{The DHR representations do \emph{not}
  include those that can be reached from the vacuum by \emph{nonlocal}
  fields, and so the domain of DHR superselection theory does not
  include theories with long range forces.  But the case of local
  fields is already complicated enough, and is good training for the
  more general case.}

There is yet another motivation for studying the DHR category: we want
to understand the nature of gauge symmetries, and the DR
Reconstruction Theorem provides crucial insight.  In particular, the
Theorem shows that DHR categories are in duality (in a mathematically
precise sense) with compact groups.  So, wherever there is a compact
group, there is a DHR category, and vice versa.  The study of DHR
categories and the study of compact gauge groups are one and the same;
or, to take a more controversial stance, the structure of the category
of physical representations of $\al A$ explains why there is a compact
gauge group (see \cite{rob7}).

We now define the category $\D = \D (\al A)$ and uncover some of its
natural structures.  As stated above, the objects of our category $\D$
will be a subset of the set $\End \al A$ of $*$-endomorphisms of $\al
A$.

\begin{defn} Let $\rho$ be a $*$-endomorphism of $\al
  A$, i.e. $\rho :\al A\to \al A$ is a $*$-homomorphism
  (not necessarily surjective).  Let $O$ be a double
  cone in (Minkowski) spacetime.  Then $\rho$ is said
  to be \emph{localized} in $O$ just in case $\rho
  (A)=A$, for all $A\in \al A(O')$, where $O'$ is the
  spacelike complement of $O$.  We say that $\rho$ is
  \emph{localized} just in case there is a double cone
  $O$ in which it is localized.
\end{defn}

\begin{note} By definition, a localized endomorphism satisfies $\rho (I)=I$, where
  $I$ is the identity in $\alg{A}$. \end{note}

\begin{defn} If $\rho$ is localized in $O$ then $\rho$ is said to be
  \emph{transportable} just in case for any other double cone $O_1$, there is a
  morphism $\rho _1$ localized in $O_1$ and a unitary operator $U\in \alg{A}$ such
  that $U\rho (A)=\rho _1(A)U$ for all $A\in \alg{A}$.
\end{defn}

\begin{defn} For each double cone $O\in \2K$, we let $\D (O)$ denote the set of
  transportable morphisms that are localized in $O$, and we let $\D = \bigcup _{O\in
    \2K}\D (O)$.  Elements of $\D$ are the \emph{objects} of the DHR category.
\end{defn}

We must now define arrows between the objects.

\begin{defn} Let $\rho ,\rho '\in \D$.  We define the set $\Hom (\rho ,\rho ')$ of
  arrows between $\rho$ and $\rho '$ as follows:
$$ \Hom (\rho ,\rho '):= \{ T\in \alg{A}:T\rho (A)=\rho '(A)T ,\: \forall A\in
\alg{A} \} .$$ If $T\in \Hom (\rho ,\rho ')$ and $S\in \Hom (\rho ',\sigma )$ then we
define $S\circ T=ST$, which is obviously in $\Hom (\rho ,\sigma )$.
\end{defn}

Obviously, the identity $I\in \alg{A}$ functions as the identity arrow for all
objects; i.e. $I=\id _{\rho}\in \End (\rho )$ for all $\rho \in \Obj (\D )$.
Occasionally, we will write $I_\rho$ to indicate that we are considering $I$ as the
identity of $\End (\rho )$.

  \begin{lemma} Suppose that $\rho _i\in \D (O_i)$ for $i=1,2$, and that $T\in \Hom
    (\rho _1,\rho _2)$.  Then for any double cone $O$ containing $O_1\cup O_2$, we
    have $T\in \alg{A}(O)$.  \label{twine} \end{lemma}

\begin{proof} Let $B\in \alg{A}(O')$.  Then 
  \begin{eqnarray*} TB = T\rho _1(B)=\rho _2(B)T =BT .\end{eqnarray*} Hence $T\in
  \alg{A}(O')'$.  By duality in the vacuum sector, $T\in \alg{A}(O)$.  \end{proof}

\begin{prop} With the definition of hom-sets given above, $\D$ is a category.
\end{prop}

\begin{proof} Completely straightforward.  \end{proof}

So, we have shown that $\D$ is a category.  In the remainder of this
Section, we uncover more structure on $\D$.  We first show that $\D$
is a $C^*$-category; this involves showing that $\D$ has direct sums
(an $\oplus$ operation), subobjects, and the hom-sets of $\D$ are
vector spaces with a $*$-operation and norm $\norm{\cdot}$ that obey
the appropriate analogue of the $C^*$-algebra norm property.  We then
drop reference to the norms on the hom-sets, and show that there is a
product operation $\otimes$ on $\D$ such that $(\D ,\otimes,\iota )$
is a tensor $*$-category.

\label{DHR-star}

\begin{defn} A category $\2C$ is said to be a \emph{linear category} over the complex
  field $\mathbb{C}$, or a $\7C$-\emph{linear category}, just in case for all $X,Y\in
  \Obj (\2C )$, $\Hom (X,Y)$ is a complex vector space, and the composition $\circ$
  of morphisms is bilinear.  When speaking of $\7C$-linear categories, we stipulate
  that all functors should be $\mathbb{C}$-linear.
\end{defn}

\begin{defn} A $*$-\emph{operation} on a $\7C$-linear category $\2C$ is a map which
  assigns to an arrow $s\in\Hom(X,Y)$ another arrow $s^*\in\Hom(Y,X)$.  This map has
  to be antilinear, involutive ($s^{**}=s$), and contravariant ($(s\,\circ\,
  t)^*=t^*\mcirc s^*$).  A $*$-operation is \emph{positive} iff $s^*\mcirc s=0$
  implies $s=0$. A \emph{$*$-category} is a $\7C$-linear category with a positive
  $*$-operation. \end{defn}

\begin{note} If $\2C$ is a $*$-category, then for each $X\in \Obj (\2C )$, $\End (X)$
  is a $*$-algebra.  \end{note}

\begin{defn} A $*$-category is called a $C^*$-\emph{category} if for all $X,Y\in \Obj
  (\2C )$, there is a norm $\norm{\cdot }_{X,Y}$ on $\Hom (X,Y)$ such that $\langle
  \Hom (X,Y),\norm{\cdot }_{X,Y}\rangle$ is a Banach space and \[ \begin{array}{lll}
    \norm{s\circ t}_{X,Z} \leq \norm{s}_{Y,Z}\cdot \norm{t}_{X,Y}, & & \forall s\in
    \Hom (Y,Z) ,\forall t\in
    \Hom (X,Y) \\
    \norm{s^*\circ s}_{X,X}=\norm{s}_{X,Y}^2 ,& & \forall s\in \Hom (X,Y)
    .\end{array} \]
\end{defn} 

We borrow some definitions from the theory of $*$-algebras.

\begin{defn} Let $\2C$ be a $*$-category.  An arrow
  $f\in \Hom (X,Y)$ is said to be an \emph{isometry}
  just in case $f^*\circ f=\id _X$.  An arrow $f\in
  \Hom (X,Y)$ is said to be \emph{unitary} just in case
  $f$ and $f^*$ are isometries.  An arrow $p\in
  \End (Y)=\Hom (Y,Y)$ is said to be a \emph{projection} if $p=p^*$ and $p\circ p=p$.
\end{defn}

\begin{note} If $s \in \Hom (Y,X)$ is an isometry then
  the arrow $p\equiv s\circ s^*\in \End (X)$ is a
  projection. \end{note}

\begin{defn} Let $\2C$ be a $*$-category.  If $X,Y\in
  \Obj (\2C )$, then $X$ is said to be a
  \emph{subobject} of $Y$ just in case there is an
  isometry $f\in \Hom (X,Y)$.  (Roughly speaking, there
  is an isometric embedding of $X$ into $Y$.)  The
  $*$-category $\2C$ is said to \emph{have subobjects}
  just in case for each $Y\in \Obj (\2C )$ and
  projection $g\in \End (Y)$, there is an $X\in \Obj
  (\2C )$ and an isometry $f\in \Hom (X,Y)$ such that
  $f\circ f^*=g$. The $*$-category $\2C$ is said to
  \emph{have direct sums} just in case for any two
  objects $X,Y$ in $\2C$, there is an object $Z$ in
  $\2C$ and isometries $f\in \Hom (X,Z)$, $g\in \Hom
  (Y,Z)$ such that $f\circ f^*+g\circ g^*=\id _Z$.
\end{defn}

We begin by verifying that the DHR category $\D$ is a $*$-category, i.e.\ the hom
sets are vector spaces over $\7C$, and there is a positive $*$-operation.

\begin{lemma} The DHR category $\D$ is a $*$-category.  That is, if $\rho ,\sigma \in
  \Obj (\D )$, then $\Hom (\rho ,\sigma )$ is a vector space over $\7C$ with the
  operations inherited from $\alg{A}$ (which is a vector space over $\7C$), and the
  composition of arrows is bilinear.  Furthermore, the $*$-operation inherited from
  $\alg{A}$ is antilinear, involutive, contravariant, and positive.
\end{lemma}

\begin{proof} Completely straightforward.  \end{proof}

\begin{prop} The DHR category $\D$ has direct sums.  \end{prop}

\begin{proof} Let $\rho _1\in \D (O_1)$, and let $\rho _2 \in \D (O_2)$.  Choose a
  double cone $O$ such that $(O_1\cup O_2)^{-}\subseteq O$.  Let $E$ be a projection
  in $\alg{A}(O_1)$.  By property B, there are isometries $V_1,V_2\in \alg{A}(O)$
  such that $V_1V_1^*+V_2V_2^*=I$.  Define $\rho :\alg{A}\to \alg{A}$ by
  $$ \rho (A)=V_1 \rho _1(A)V_1^*+V_2\rho _2(A)V_2^* ,\qquad \forall A\in \alg{A} .$$
  Since $V_iV_j=\d _{ij}I$, and $\sum _iV_iV_i^*=I$, it follows that $\rho$ is a
  morphism.  Since $\rho _1,\rho _2$ are localized in $O$, and $V_1,V_2\in
  \alg{A}(O)$, it follows that $\rho$ is localized in $O$.

  To see that $\rho$ is transportable, let $\widetilde{O}$ be another double cone.
  Since the $\rho _i$ are transportable, there are endomorphisms $\rho _i'$ localized
  in $\widetilde{O}$, and unitary operators $U_i\in \Hom (\rho _i,\rho _i')$.  As
  before, choose isometries $V_1',V_2'$ in $\alg{A}(\widetilde{O})$, and set $\rho
  '=V_1'\rho _1'{V_1'}^*+V_2'\rho _2'{V_2'}^*$.  Then $\rho '$ is localized in
  $\widetilde{O}$ and
  \begin{equation*} {V_1'}U_1V_1^* \in \Hom (\rho ,\rho ') ,\qquad {V_2'}U_2V_2^* \in
    \Hom (\rho ,\rho ') . \end{equation*} If we set
  $W={V_1'}U_1V_1^*+{V_2'}U_2V_2^*$, then $W\in \Hom (\rho ,\rho ')$ since it is a
  vector space.  Furthermore,
\begin{eqnarray*} W^*W &=&
  [{V_1'}U_1V_1^*+{V_2'}U_2V_2^*]^*[{V_1'}U_1V_1^*+{V_2'}U_2V_2^*]  \\
  &=&
  [V_1U_1^*{V_1'}^*+V_2U_2^*{V_2'}^*] [{V_1'}U_1V_1^*+{V_2'}U_2V_2^*] \\
  &=& V_1V_1^*+V_2V_2^* = I , \end{eqnarray*} and similarly for $WW^*$.  Therefore
$W$ is a unitary operator in $\Hom (\rho ,\rho ')$, showing that $\rho$ is
transportable.
\end{proof}

\begin{defn} If $\rho _1,\rho _2 \in \D$, we denote their direct sum by $\rho
  _1\oplus \rho _2$. \end{defn}

\begin{prop} The DHR category $\D$ has subobjects.  \end{prop}

\begin{proof} Let $\rho \in \D (O)$, and let $E$ be a
  projection in $\End (\rho )$; i.e.\ $E\rho (A)=\rho
  (A)E$, for all $A\in \alg{A}$.  Then for all $A\in
  \alg{A}(O')$,$$ EA =E\rho (A)=\rho (A)E=AE.  $$
  Therefore, by duality in the vacuum sector $E\in
  \alg{A}(O)$.  Choose $O_1$ such that $O^{-}\subseteq
  O_1$.  By property B, there is an isometry $V\in
  \alg{A}(O_1)$ such that $VV^*=E$.  Now define $\rho
  ':\alg{A}\to \alg{A}$ by
  \begin{equation*} \rho '(A)=V^*\rho (A)V ,\qquad \forall A\in \alg{A}
    .\end{equation*} The isometry $V$ embeds $\rho '$ into $\rho$.  Indeed,
  \begin{equation*} \rho '(A)V^*=V^*\rho (A)VV^*=V^*\rho (A)E =V^*\rho (A)
    .\end{equation*} and $V$ is an isometry in $\Hom (\rho ',\rho )$ such that
  $VV^*=E\in \End (\rho )$.

  To see that $\rho '$ is transportable, suppose that
  $O_2$ is an arbitrary double cone.  Choose a double
  cone $O_3$ such that $O_3^{-}\subseteq O_2$.  Since
  $\rho$ is transportable, there is a morphism $\sigma$
  localized in $O_3$ and a unitary $U\in \Hom (\rho
  ,\sigma)$.  It then follows that $U\End (\rho
  )U^*=\End (\sigma )$, thus $E'=UEU^*$ is a projection
  in $\End (\sigma )$.  Using property B, there is an
  isometry $V'\in \alg{A}(O_1)$ such that
  $V'{V'}^*=E'$.  Let $\sigma '={V'}^*\sigma V'$.
  Clearly $\sigma '$ is localized in $O_1$, and
  $W={V'}^*UV\in \Hom (\rho ',\sigma ')$.  Finally, $W$
  is unitary:
  \begin{eqnarray*} W^*W &=& V^*U^*V'{V'}^*UV \: = \: V^*U^*E'UV \\
    &= & V^*EV \:= \: V^*VV^*V \: =\: I ,\end{eqnarray*} and similarly for $WW^*$.
  Thus $\sigma '$ is equivalent to $\rho '$.  Since $O_2$ was an arbitrary double
  cone, $\rho ' \in \D$.
\end{proof}

\begin{defn} Suppose that $\2C$ is a $\7C$-linear category.  An object $X$ in $\2C$
  is said to be \emph{irreducible} if it is nonzero and $\End (X)=\7C \id _X$.
\end{defn}

\begin{note} Let $\iota$ be the identity endomorphism of $\alg{A}$.  Then $\iota \in
  \Obj (\D )$, and since the vacuum representation of $\alg{A}$ is irreducible,
  $\iota$ is an irreducible object.  \end{note}

\label{tensor}

We now define a bifunctor $\otimes =(\otimes ,\times )$
on the DHR category $\D$, and verify that $(\D ,\otimes
,\iota )$ is a tensor $*$-category.  But first we
recall the pertinent definitions.

\begin{defn} A \emph{bifunctor} on a category $\2C$
  consists of two mappings $F:\Obj (\2C)\times \Obj
  (\2C )\to \Obj(\2C )$ and $F:\Hom (\2C )\times \Hom
  (\2C )\to \Hom (\2C )$, such that for $s\in \Hom
  (X,Y)$ and $t\in \Hom (X',Y')$, $F(s,t)\in \Hom
  (F(X,X'),F(Y,Y'))$, and
  \begin{eqnarray*} 
    F(s_1\circ s_2,t) & = & F(s_1,t)\circ F(s_2,t),  \\
    F(s,t_1\circ t_2) & = & F(s,t_1)\circ F(s,t_2), \\
    F(\id _X ,\id_{X'}) & = & \id _{F(X,X')} .
\end{eqnarray*} 
If $\2C$ is a $*$-category, then a bifunctor $F$ is
also required to be bilinear and to commute with the
$*$-operation.  That is, for $s_i\in \Hom (X,X')$,
$t\in \Hom (Y,Y')$ and $c\in \7C$, we have
\begin{eqnarray*} 
  F(s_1+s_2,t) &=& F(s_1,t)+F(s_2,t) ,\\
  F(s,t_1+t_2) &=& F(s,t_1)+F(s,t_2) , \\
  F(cs,t) &=& cF(s,t) \quad =\quad F(s,ct) ,
\end{eqnarray*}
and
\begin{eqnarray*}
  F(s,t)^* &=& F(s^*,t^* ) .
\end{eqnarray*}
\end{defn}

\begin{defn}
  Let $\otimes =(\otimes ,\times )$ be a bifunctor on
  the category $\2C$, and let $\11 \in \Obj (\2C )$.
  Then $(\2C ,\otimes ,\11 )$ is said to be a
  \emph{tensor category} just in case $\otimes$ is
  associative up to a natural isomorphisms, and $\11$
  is a two sided identity up to natural isomorphisms.
  The object $\11$ is called the \emph{monoidal unit}.
  To be precise, to say that $\otimes$ is `associative
  up to a natural isomorphisms' means that for each
  $X,Y,Z\in \Obj (\2C)$, there is an isomorphism
  $\alpha _{X,Y,Z}:X\otimes (Y\otimes Z)\to (X\otimes
  Y)\otimes Z$ that is `natural' for all $X,Y,Z$; i.e.\
  if $s:X\to X'$ then
\begin{eqnarray}
  ((s\otimes \id _Y)\otimes \id _Z)\circ \alpha _{X,Y,Z} \:=\: \alpha _{X',Y,Z}\circ
  (s\otimes (\id _Y\otimes \id _Z)) ,
\end{eqnarray}
and similarly for $Y$ and $Z$.  Furthermore, $\alpha$ is required to make the
pentagonal diagram commute:

\begin{diagram}[grid=pentagon]
  &&&& (X\otimes Y)\otimes (Z\otimes Z') & \\
  &&& \relax\ruTo(4,2)^\alpha  && \relax\rdTo(4,2)^\alpha &\\
  X\otimes (Y\otimes (Z\otimes Z'))&&&&&&&&((X\otimes Y)\otimes Z)\otimes Z'&\\
  &\relax\rdTo_{\id _X \otimes \alpha }&&&&&&\relax\ruTo_{\alpha \otimes \id _{Z'}}&\\
  && X\otimes ((Y\otimes Z)\otimes Z') && \relax\rTo^\alpha && (X \otimes (Y\otimes
  Z))\otimes Z' &
\end{diagram} 

To say that $\11 \in \2C$ is a two sided identity up to natural isomorphisms means
that for each object $X\in \Obj (\2C )$, there are isomorphisms $\lambda _X\in \Hom
(\11 \otimes X,X)$ and $\rho _X\in \Hom (X\otimes \11,X)$ such that:
\begin{enumerate}
\item $\lambda _X$ and $\rho _X$ are natural in $X$; i.e.\ for any $s:X\to Y$,
  \begin{eqnarray}
    s\circ \lambda _X=\lambda _Y\circ (\id _{\11}\otimes s) ,\\
    s\circ \rho _X=\rho _Y\circ (s\otimes \id _{\11}) .\end{eqnarray} 
  In other words, the following two diagrams commute: \\

  \begin{tabular}{cc}
  \begin{diagram}
\11 \otimes X  & \rTo^{\lambda _X} & X \\ 
\dTo^{\id _{\11}\otimes s} & & \dTo_s \\
1\otimes Y & \rTo_{\lambda _Y} & Y  
\end{diagram}   &  
\begin{diagram}
  X\otimes \11  & \rTo^{\rho _X} & X \\
  \dTo^{s\otimes \id _{\11}} & & \dTo_s \\
  Y\otimes \11 & \rTo_{\rho _Y} & Y
\end{diagram}
\end{tabular}

\item $\lambda _X$ and $\rho _X$ make the triangular diagram commute:
\begin{diagram}
  X\otimes (\11 \otimes Y) & & \rTo^{\alpha } & & (X\otimes \11 )\otimes Y \\
  & \rdTo(4,2)_{\id _X\otimes \lambda _Y}  & & & \dTo_{\rho _X\otimes \id _Y} \\
  & & & & X\otimes Y
\end{diagram}
\end{enumerate}
If $\2C$ is also a $*$-category, there are two further requirements: (a.) the
bifunctor $\otimes$ must be compatible with the operations of $+$ and $*$ (as
required in the definition of bifunctor), and (b.) the monoidal unit $\11$ must be
irreducible, i.e.\ $\End (\11 )=\7C \id _{\11}$.  For a $C^*$-category $\2C$ we
require in addition that $\norm{s\times t}_{X\otimes Y,X'\otimes Y'}\leq
\norm{s}_{X,X'}\cdot \norm{t}_{Y,Y'}$.
\end{defn}

Mac Lane's coherence theorem shows that we can without danger ignore the natural
isomorphisms $\alpha ,\lambda$, and $\rho$.  That is, we can treat $X\otimes
(Y\otimes Z)$ and $(X\otimes Y)\otimes Z$ as the \emph{same} object, and we can treat
$X$, $\11 \otimes X$, and $X\otimes \11$ as the same object.  To be more precise, we
define:

\begin{defn} A tensor category $\2C$ is said to be
  \emph{strict} if $\a _{X,Y,Z} ,\lambda _X ,\rho _X$
  are identity morphisms for all $X,Y,Z\in \Obj (\2C
  )$.\end{defn}

For example, the tensor category
$(\mathrm{Vect},\otimes ,\7C)$ of vector spaces is
\emph{not} strict, since e.g.\ $V\otimes \7C$ is not
literally the same vector space as $V$.  On the other
hand, a commutative monoid $M$ can be thought of as a
strict tensor category with one object and with arrows
corresponding to elements of $M$.  The coherence
theorem can then be formulated as follows.

\begin{thm}[Coherence Theorem] Every tensor category is
  equivalent to a strict tensor
  category. \label{coherence}
\end{thm}

\begin{proof} See \cite{cwm}. \end{proof}

\begin{defn} If $\2C$ is a tensor category, then we let
  $\2C ^{\mathrm{st}}$ denote its
  strictification. \end{defn}

With these definitions in hand, we proceed now to define a bifunctor on $\D$, and to
verify that it satisfies all of the relevant properties.  Our product $\otimes$ of
objects in $\D$ will be just the composition of endomorphisms.

\begin{prop} If $\rho ,\sigma \in \Obj (\D )$ then $\rho \sigma \in \Obj (\D )$.
\end{prop}

\begin{proof} It is clear that if $\rho$ is localized in $O_1$ and $\sigma$ is
  localized in $O_2$, then $\rho \sigma$ is localized in any double cone that
  contains $O_1\cup O_2$.  

  To see that $\rho \sigma$ is transportable, let $O_3$
  be an arbitrary double done.  Since $\rho$ and
  $\sigma$ are transportable, there are $\rho ',\sigma
  '\in \D (O_3)$ and unitaries $U\in \Hom (\rho ,\rho
  ')$ and $V\in \Hom (\sigma ,\sigma ')$.  Then $\rho
  '\sigma '$ is localized in $O_3$ and $U\rho (V)$ is a
  unitary in $\Hom (\rho \sigma ,\rho '\sigma ')$.
  Therefore, $\rho \sigma$ is transportable.
\end{proof}

\begin{defn} Define $\otimes :\Obj (\D )\times \Obj (\D )\to \Obj (\D )$ by $\rho
  \otimes \sigma =\rho \sigma$.  \end{defn}  

The product $\times$ of arrows is slightly more complicated.

\begin{prop} If $S\in \Hom (\rho ,\rho ')$ and $T\in \Hom (\sigma ,\sigma ')$ then
  $S\rho (T)\in \Hom (\rho \otimes \sigma ,\rho '\otimes \sigma ')$. \end{prop}

\begin{proof} Since $S\rho (T)=\rho '(T)S$, it follows that for any $A\in \alg{A}$,
  \begin{eqnarray*} (S\rho (T))\rho \sigma (A) &=& S\rho (T\sigma (A)) \:= \: \rho
    '(T\sigma (A))S \:=\: \rho '(\sigma '(A)T)S \\
    &=& \rho '\sigma '(A)(\rho '(T)S) \: =\: \rho '\sigma '(A)(S\rho (T)).
  \end{eqnarray*} Therefore $S\rho (T)\in \Hom (\rho \sigma ,\rho' \sigma ')$.
\end{proof}

\begin{defn} Define $\times :\Hom (\D )\times \Hom (\D )\to \Hom (\D )$ by: for $S\in
  \Hom (\rho ,\rho ')$ and $T\in \Hom (\sigma ,\sigma ')$, we set $S\times T=S\rho
  (T)\in \Hom (\rho \otimes \sigma ,\rho '\otimes \sigma ')$. \end{defn}

In the remainder of this section, we verify that $(\D,
\otimes ,\iota )$ is a tensor $*$-category.

\bigskip \noindent \underline{$\otimes$ is a bifunctor on $\D$} 

\smallskip \begin{prop} For $S_1,S_2,T_1,T_2\in \Obj (\D )$, if the source of $T_i$ is the
  target of $S_i$ (so that $T_i\circ S_i$ is defined) then
$$ (T_1\times T_2)\circ (S_1\times S_2) = (T_1\circ S_1)\times (T_2\circ S_2) .$$
\end{prop}

\begin{proof} Straightforward calculation.  \end{proof}

We must now check that $\times$ is compatible with $*$

\begin{prop} For all $S,T\in \Hom (\D )$,
$$ (S\times T)^*=S^*\times T^*. $$ \end{prop}

\begin{proof} Straightforward calculation.
\end{proof}

\bigskip \noindent \underline{$\iota$ is a monoidal unit}

\smallskip For each $\rho \in \Obj (\D )$, $\iota \otimes \rho$ must be naturally isomorphic to
$\rho \otimes \iota$ and to $\rho$ (as expressed in the monoidal unit diagrams).  But
in the present case, $\iota \otimes \rho =\rho \otimes \iota =\rho$, so this natural
isomorphism holds trivially.

\bigskip \noindent \underline{Natural associativity of $\otimes$}

\smallskip Next, the product operation $\otimes =(\otimes ,\times )$ must be associative up to
natural isomorphisms, as expressed by the pentagonal diagram.  But this is trivial in
the present case, because associativity holds strictly; that is:

  \begin{prop} For all $\rho _1,\rho _2,\rho _3\in \Obj (\D )$, $$\rho _1\otimes
    (\rho _2\otimes \rho _3)=(\rho _1\otimes \rho _2)\otimes \rho _3 ,$$ and for all
    $T_1,T_2,T_3\in \Hom (\D)$, $$ T_1\times (T_2\times T_3)=(T_1\times T_2)\times
    T_3.  $$ \end{prop}

\begin{proof} The first claim follows trivially from the fact that composition of
  endomorphisms is associative.  The second claim can be verified by a
  straightforward calculation.  \end{proof}

\begin{lemma} $(\D ,\otimes ,\iota )$ is a $C^*$-tensor category with the norms
  inherited from $\alg{A}$. \end{lemma}

\begin{proof} We must verify that $\Hom (\rho ,\sigma )$ is closed in the norm on
  $\alg{A}$.  But this follows immediately from the fact that 
$$ \Hom (\rho ,\sigma )= \{ T\in \alg{A}:T\rho (A)=\sigma (A)T ,\, \forall A\in
\alg{A} \} .$$ 
It's clear that $\norm{s\circ t}\leq \norm{s}\norm{t}$.  Furthermore,
$$ \norm{S\times T}=\norm{S\rho (T)}\leq \norm{S}\cdot \norm{\rho (T)}\leq
\norm{S}\cdot \norm{T} .$$
\end{proof}

To this point we have shown that (i): $\D$ is a $C^*$-category, and
(ii): $\D$ is a tensor $*$-category.  The following five Subsections
are not linearly ordered.  Subection \ref{braided} shows how to define
the canonical braiding $\ve _{\rho _1,\rho _2}$ on $(\D ,\otimes
,\iota )$ such that it is a `braided' tensor $*$-category.  Then in
Subsection \ref{ss-dhr} we make good our claims about the motivation
for studying the category $\D$: we prove that there is a functoral
correspondence between $\D$ and the category of representations that
satisfy the DHR selection criterion.  We then pick up some technical
information about tensor $*$-categories that is essential for the
physical interpretation of the corresponding representations.  In
Subsection \ref{dimension} we see how to define a notion of the
`dimension' of an object in a tensor $*$-category, and we define the
notion of `conjugate' objects.  In Subsection \ref{ss-cov} we take a
detour to talk about the relation of spacetime symmetries to the DHR
representations.  Finally, in Subsection \ref{cat-stats} we give the
intrinsic statistical classification of objects of $\D _f$ that
corresponds to the intuitive distinction between Bosons and Fermions,
or Bose fields and Fermi fields.

\subsection{$\D$ is a braided tensor $*$-category}
\label{braided}

In this Subsection we define the canonical braiding on
$\D$; this gives us a grasp on what happens if we
change the order in products, say $\rho \otimes \sigma$
versus $\sigma \otimes \rho$.  We will also see that
there is a remarkable connection between spacetime
dimension and the properties of this braiding: if the
spacetime has three or more dimensions, the braiding is
a symmetry.  We first recall the pertinent definitions.

\begin{defn} If $(\2C ,\otimes ,\11 )$ is a tensor
  category then a \emph{braiding} on $\2C$ is a family
  of isomorphisms
$$ \bigl\{ c_{X,Y}\in \Hom (X\otimes Y,Y\otimes X): X,Y\in \Obj (\2C ) \bigr\} ,$$ 
satisfying the following two conditions:
\begin{enumerate}
\item $c_{X,Y}$ is natural in $X$ and $Y$; i.e.\ for any $f\in \Hom (X,X')$ and $g\in
  \Hom (Y,Y')$,
  \begin{equation} (g\times f)\circ c_{X,Y}= c_{X',Y'}\circ (f\times g) .
\end{equation} 
\item $c_{X,Y}$ makes the following two hexagonal diagrams commute:
\label{hexagon}

  \begin{diagram}
    (X\otimes Y)\otimes Z & \rTo^{c_{X\otimes Y,Z}} & Z\otimes (X\otimes Y) \\
    \dTo^{\alpha ^{-1}} & & \dTo_{\alpha} \\
    X\otimes (Y\otimes Z) & & (Z\otimes X)\otimes Y \\
    \dTo^{\id _{X}\otimes c_{Y,Z}} & & \dTo_{c_{Z,X}\otimes \id _{Y}} \\
    X\otimes (Z\otimes Y) & \rTo_{\alpha} & (X\otimes Z)\otimes Y \end{diagram}
 
\begin{diagram}
  X\otimes (Y\otimes Z) & \rTo^{c_{X,Y\otimes Z}} & (Y\otimes Z)\otimes X \\
  \dTo^{\alpha} & & \dTo_{\alpha ^{-1}} \\
  (X\otimes Y)\otimes Z & & Y\otimes (Z\otimes X) \\
  \dTo^{c_{X,Y}\otimes \id _Z} & & \dTo_{\id _{Y}\otimes c_{Z,X}} \\
  (Y\otimes X)\otimes Z & \rTo_{\alpha ^{-1}} & Y\otimes (X\otimes Z) \end{diagram}

That is, suppressing the associativity isomorphisms, $c_{X\otimes Y,Z}$ is expressed
in terms of $c_{X,Y}$ and $c_{X,Z}$ as:
$$ c_{X\otimes Y,Z} \: =\: (\id_Y\otimes c_{Z,X})^{-1}\circ (\id_X\otimes c_{Y,Z}) ,$$
and $c_{X,Y\otimes Z}$ is expressed in terms of $c_{X,Y}$ and $c_{Z,X}$ as:
$$ c_{X,Y\otimes Z} \: =\: (\id_Y\otimes c_{Z,X})^{-1}\circ (c_{X,Y}\otimes
\id_Z). $$
\end{enumerate}
\end{defn}

\begin{defn} A braiding $c_{X,Y}$ is called a \emph{symmetry} if
  $(c_{X,Y})^{-1}=c_{Y,X}$ for all $X,Y\in \Obj (\2C )$. \end{defn}

\begin{defn} A tensor category with a privileged
  braiding (symmetry) is called a \emph{braided
    (symmetric) tensor category}. \end{defn}

In order to find our braiding on $\D$, we will need the following technical lemma.

\begin{lemma} If $\rho \in \D (O_1)$ and $\sigma \in \D (O_2)$ where $O_1$ and $O_2$
  are spacelike separated, then $\rho \sigma =\sigma \rho$.  \label{commute}
\end{lemma}

\begin{proof} Since the union of $\{ \alg{A}(O):O_1\cup
  O_2\subseteq O \}$ is dense in $\alg{A}$, it suffices
  to show that $\rho \sigma (A)=\sigma \rho (A)$
  whenever $A\in \alg{A}(O)$ with $O_1\cup O_2\subseteq
  O$.  Choose $O_3,O_4$ that are spacelike to $O$ and
  such that $O_1\cup O_3$ is spacelike to $O_2\cup
  O_4$.  (This may always be done, even in two
  dimensional spacetime.)  Since $\rho ,\sigma$ are
  transportable, there are $\rho ',\sigma '$ localized
  in $O_3$ and $O_4$ respectively and unitary operators
  $U_1\in \Hom (\rho ,\rho ')$ and $U_2\in \Hom (\sigma
  ,\sigma ')$.  Then
$$ \sigma (A)=U_2\sigma '(A)U_2^*=U_2^*AU_2 .$$
Furthermore, $U_2\in \alg{A}(O_1')$ by duality in the vacuum sector.  Hence $\rho
(U_2)=U_2$, and
$$ \rho \sigma (A)=U_2U_1AU_1^*U_2^* .$$
Since $U_2U_1=U_1U_2$, it follows that $\rho \sigma (A)=\sigma \rho (A)$.
\end{proof}

We will not be able to define the braiding $\ve _{\rho _1,\rho _2}$ in one shot.
Rather, we first define arrows
$$\ve _{\rho _1,\rho _2}(U_1,U_2)\in \Hom (\rho _1\otimes \rho _2,\rho _2\otimes \rho
_1) ,$$ that depend on the choice of ``spectator morphisms'' $\wt \rho _1,\wt \rho
_2$, and unitary intertwiners $U_i\in \Hom (\rho _i,\wt \rho _i )$.  We will then
show that this definition is independent of the spectator morphisms and unitary
intertwiners.  (But, interestingly, when spacetime is two dimensional, the definition
depends on the choice of a spatial orientation.)

\begin{defn} Suppose that $\rho _1\in \D (O_1)$ and $\rho _2\in \D (O_2)$.  Let $\wt
  O_1$ and $\wt O_2$ be spacelike separated double cones.  Since $\rho _1$ and $\rho
  _2$ are transportable, there are $\wt \rho _i\in \D (\wt O_i)$ and unitary
  operators $U_i\in \Hom (\rho _i,\wt \rho _i)$.  Thus $U_1\times U_2 \in \Hom (\rho
  _1\otimes \rho _2 ,\wt\rho _1\otimes \wt\rho _2)$, and $U_2^*\times U_1^* \in \Hom
  (\wt\rho _2\otimes \wt\rho _1,\rho _2\otimes \rho _1)$.  Since $\wt O_1$ is
  spacelike to $\wt O_2$, Lemma \ref{commute} entails that $\wt\rho _1\otimes \wt\rho
  _2 =\wt\rho _2\otimes \wt\rho _1$.  Thus, we may define $\ve _{\rho _1,\rho
    _2}(U_1,U_2)\in \Hom (\rho _1\otimes \rho _2 ,\rho _2\otimes \rho _1 )$ by
  \begin{eqnarray}
    \ve _{\rho _1,\rho _2}(U_1,U_2) \: := \: (U_2\times U_1)^* \circ
    (U_1\times U_2) \: =\:  \rho _2 (U_1^*)U_2^*U_1\rho _1(U_2) .\label{def-braiding} \end{eqnarray}\end{defn}

\begin{note} Since endomorphisms preserve unitarity, $\ve _{\rho _1,\rho
    _2}(U_1,U_2)$ is unitary.
\end{note}

To show that $\ve _{\rho _1,\rho _2}(U_1,U_2)$ is independent of $U_1$ and $U_2$, we
need the following Lemma, which shows that $\ve _{\rho _1,\rho _2}(U_1,U_2)$ does not
change under certain ``perturbations'' of $U_1$ and $U_2$.

\begin{lemma} For $i=1,2$, let $\rho _i\in \D (O_i)$, let $\wt O_1$ and $\wt O_2$ be
  spacelike separated, let $\wt\rho _i\in \D (\wt O_i)$, and let $U_i\in \Hom (\rho
  _i,\wt\rho _i)$.  Then $\ve _{\rho _1,\rho _2}(U_1,U_2)$ depends only on
  neighborhoods of $U_1,U_2$ in the following sense: if $W_1,W_2$ are unitaries such
  that $W_1\in \alg{A}(\wt O_2')$, $W_2\in \alg{A}(\wt O_1')$, and $W_1W_2=W_2W_1$,
  then
$$ \ve _{\rho _1,\rho _2}(W_1U_1,W_2U_2)=\ve _{\rho _1,\rho _2}(U_1,U_2) .$$
\label{hunca-munca} \end{lemma}

\begin{proof} We must show that
  \begin{eqnarray} (W_2U_2\times W_1U_1)^*\circ (W_1U_1\times W_2U_2) = (U_2^*\times
    U_1^*)\circ (U_1\times U_2) .\label{long} \end{eqnarray} For any two unitary
  operators, $W_1,W_2\in \alg{A}$, we have
\begin{eqnarray*} W_iU_i\times W_jU_j &=& W_iU_i\rho _i(W_j)\rho _i(U_j) = W_i\rho
  _i'(W_j)(U_i\times U_j) .\end{eqnarray*} Since $W_1\in \alg{A}({{\wt O}_2}')$ and $\wt
\rho _2$ is localized in $\wt O_2$, $\wt \rho _2(W_1)=W_1$; and similarly, $\wt \rho
_1(W_2)=W_2$.  Hence, the left hand side of Eqn.\ \ref{long} becomes
\begin{eqnarray*}
  [(U_2\times U_1)^*\wt \rho _2(W_1^*)W_2^*]\, [W_1\wt \rho _1(W_2)(U_1\times U_2)]  &=& (U_2\times U_1)^*W_1^*W_2^*W_1W_2(U_1\times U_2) \\
  &=& (U_2\times U_1)^*(U_1\times U_2) ,\end{eqnarray*} where we used $W_1W_2=W_2W_1$
for the second equality. \end{proof}

\newcommand{\dt}[1]{\tilde{\tilde{#1}}}

\begin{lemma} Let $\rho _i\in \D (O_i)$, $i=1,2$, and let $T\in \Hom (\rho _1,\rho
  _2)$.  Then $T\in \alg{A}(O)$ for any double cone containing $O_1\cup O_2$.
  \label{hom-loc}
\end{lemma}

\begin{proof} Let $O$ be a double cone containing $O_1\cup O_2$, and let $A\in
  \alg{A}(O')$.  Then $\rho _1(A)=\rho _2(A)=A$, and so
  \begin{eqnarray*} TA=T\rho _1(A)=\rho _2(A)T =AT .\end{eqnarray*} Hence $T\in
  \alg{A}(O')'$, and by duality in the vacuum sector, $T\in \alg{A}(O)$. \end{proof}

Now we can show that $\ve _{\rho _1,\rho _2}(U_1,U_2)$ depends only on the
localization regions of the spectator morphisms.

\begin{prop} $\ve _{\rho _1,\rho _2}(U_1,U_2)$ is definable in terms of $\rho _1
  ,\rho _2$, and the regions $\wt O_1,\wt O_2$; and does not change if the latter are
  replaced by double cones $\dt{O}_1,\dt{O}_2$ such that $\wt O_1\subseteq \dt{O}_1$
  and $\wt O_2\subseteq \dt{O}_2$. \label{perturb}
\end{prop}

\begin{proof} (1.) We show first that for a given pair $(\wt O_1,\wt O_2)$ of
  spacelike separated double cones, the definition $\ve _{\rho _1,\rho _2}(U_1,U_2)$
  is independent of spectator morphisms $(\wt \rho _1,\wt \rho _2)$, and unitary
  intertwiners $(U_1,U_2)$.  So, suppose that $\dt{\rho}_i\in \D (\wt O_i)$, and
  $U'_i\in \Hom (\rho _i,\dt{\rho} _i )$.  Let $W_i=U_i'U_i^*\in \Hom (\wt \rho
  _i,\dt{\rho} _i)$, so that $U'_i=W_iU_i$.  Since $W_i$ has left and right support
  in $\wt O_i$, $W _i\in \alg{A}(\wt O_i)\subseteq \alg{A}(\wt O_j)'$.  Thus,
  $W_1W_2=W_2W_1$, and the hypotheses of Lemma \ref{hunca-munca} are satisfied.
  Therefore $\ve _{\rho _1,\rho _2}(U_1,U_2)=\ve _{\rho _1,\rho _2}(U_1',U_2')$.

  \bigskip (2.) Now let $\dt{O}_1$ and $\dt{O}_2$ be double cones such that
  $\dt{O}_1\perp \dt{O}_2$, and $\wt O_i\subseteq \dt{O}_i$, for $i=1,2$.  Choose
  $\dt{\rho}_i\in \D (\dt{O}_i)$, and unitaries $U_i'\in \Hom (\rho _i,\dt{\rho}_i)$.
  But we also have $\wt \rho _i\in \D (\wt O_i)\subseteq \D (\dt{O}_i)$.  And the
  first part of the proof shows that for fixed support regions $(\dt{O}_1,\dt{O}_2)$,
  the definition of $\ve _{\rho _1,\rho _2}$ is independent of the choice of
  spectator morphisms and unitary intertwiners.  Therefore $\ve _{\rho _1,\rho
    _2}(U_1,U_2)=\ve _{\rho _1,\rho _2}(U_1',U_2')$.
\end{proof}

\begin{note} We can always choose the spectator morphisms to be localized in strictly
  spacelike separated regions.  Indeed, given $\wt O_1$ and $\wt O_2$ that are
  spacelike separated, choose $\dt{O}_1$ such that $(\dt{O}_1)^{-}\subseteq \wt O_1$.
  But then the Lemma implies (by switching $\dt{O}_1$ with $\wt O_1$, and setting
  $\dt{O}_2=\wt O_2$) that we get the same definition from using $\dt{O}_1$ or $\wt
  O_1$.  More generally, since $\rho _1$ is transportable, the regions $\wt O_i$ can
  be chosen arbitrarily small.
\end{note}

\begin{note} The previous note shows that a definition of $\ve _{\rho _1,\rho
    _2}(U_1,U_2)$ is always equivalent to a definition using spectator morphisms
  localized in strictly spacelike separated regions.  That is, there is a
  neighborhood $N$ of zero such that $\wt O_1+x\subseteq \wt O_2'$ for all $x\in N$.
  Again, since $\wt O_1$ and $\wt O_1+x$ are contained in a double cone
  $\dt{O}_1\subseteq \wt O_2'$, the previous Lemma (applied twice) entails that the
  pairs $(\wt O_1,\wt O_2)$ and $(\wt O_1+x,\wt O_2)$ yield the same definition of
  $\ve _{\rho _1,\rho _2}$.

  By also shrinking $\wt O_2$ if necessary, and repeating the above construction, we
  see that for any vector $x$, the pairs $(\wt O_1+x,\wt O_2+x)$ and $(\wt O_1,\wt
  O_2)$ yield the same definition of $\ve _{\rho _1,\rho _2}$. \label{translate}
\end{note}

\begin{note} In what follows, by ``one dimensional
  spacetime'', we mean one dimensional space with zero
  dimensional time.  In this case, a double cone is
  just an open interval, and ``spacelike separated''
  means disjoint. \end{note}

\begin{prop} For spacetime of dimension two or less,
  $\ve _{\rho _1,\rho _2}(U_1,U_2)$ is definable in
  terms of $\rho _1,\rho _2$ and the spatial
  orientation of $\wt O_1$ with respect to $\wt O_2$.
  That is, $\ve _{\rho _1,\rho _2}(U_1,U_2)$ is
  independent of the choice of $(\wt O_1,\wt O_2)$,
  subject to the constraint of having the same spatial
  orientation.  \label{orient} \end{prop}

\begin{proof} Let $\wt O_i,\dt{O}_i$ be given such that $\wt O_1\perp \wt O_2$,
  $\dt{O}_1\perp \dt{O}_2$, and $\wt O_1$ is oriented with respect to $\wt O_2$ as
  $\dt{O}_1$ is with respect to $\dt{O}_2$.  Recall that translations of $(\wt
  O_1,\wt O_2)$ do not change $\ve _{\rho _1,\rho _2}(U_1,U_2)$; nor does replacement
  of $\wt O_i$ with a double cone either containing it or contained in it, and
  spacelike to $\wt O_j$.  But $(\wt O_1,\wt O_2)$ can be replaced by
  $(\dt{O}_1,\dt{O}_2)$ in a series of such moves.
\end{proof}

\begin{defn} For spacetime of two dimensions or less, fix a spatial orientation, and
  use $O_1<O_2$ to indicate that $O_1$ is to the left of $O_2$.  \end{defn} 

\begin{lemma} For spacetimes of dimension two or less, if the spatial orientation of
  $\wt O_1$ with respect to $\wt O_2$ is the opposite of the spatial orientation of
  $\dt{O}_1$ with respect to $\dt{O}_2$, then
$$ \ve _{\rho _1,\rho _2}(U_1,U_2) =[\ve _{\rho _2,\rho _1}(U_2',U_1')]^* .$$
\end{lemma}

\begin{proof} For defining $\ve _{\rho _1,\rho _2}(U_1,U_2)$, we can choose $\wt
  O_1=O_1$, $\wt O_2<O_1$, $\wt \rho _1=\rho _1$, and $U_1=I_{\rho _1}=I\in \Hom
  (\rho _1,\rho _1)$.  In this case, the definition simplifies to
$$ \ve _{\rho _1,\rho _2}(I,U_2) = U_2^*\rho _1(U_2),\qquad
\qquad (\wt O_2<\wt O_1) .$$ Using the same spectator morphisms, we have
$$ \ve _{\rho _2,\rho _1}(U_2,I) = \rho _1(U_2^*)U_2 ,\qquad \qquad (\wt O_2<\wt
O_1). $$ This latter expression uses the opposite spatial orientation.  By the
definability of $\ve _{\rho _1,\rho _2}(U_1,U_2)$ in terms of spatial orientation
(Prop.\ \ref{orient}), we see that $\ve _{\rho _1,\rho _2}(U_1,U_2)=[\ve _{\rho
  _2,\rho _1}(U_2',U_1')]^*$ when the opposite spatial orientations are used for the
two definitions.  \end{proof}

\begin{bxd}
  \begin{defn}[The Canonical Braiding on $\D$] For spacetimes of dimension two or
    less, we implement the convention that $\ve _{\rho _1,\rho _2}=\ve _{\rho _1,\rho
      _2}(U_1,U_2)$ with $\wt O_2<\wt O_1$.  The previous Lemma shows that if we
    define $\ol\ve _{\rho _1,\rho _2}$ with the opposite convention, then $\ol\ve
    _{\rho _1,\rho _2}=(\ve _{\rho _2,\rho _1})^*$.  For spacetimes of dimension
    three or more, we define $\ve _{\rho _1,\rho _2}=\ve _{\rho _1,\rho _2}(U_1,U_2)$
    with $\wt O_1$ and $\wt O_2$ spacelike separated.
\end{defn} \end{bxd}

We now verify that $\ve _{\rho _1,\rho _2}$ is a braiding on $(\D ,\otimes ,\iota )$.

\begin{prop} $\ve _{\rho ,\sigma}$ is a braiding on the DHR category $(\Delta
  ,\otimes ,\iota )$.
\end{prop}

\begin{proof} (1) We first show that $\ve _{\rho ,\sigma }$ is natural in $\rho$ and
  $\sigma$.  For this it suffices to show that if $T\in \Hom (\rho ,\rho ')$ then
\begin{eqnarray} 
  (I_{\sigma}\times T)\circ \ve _{\rho ,\sigma } &=& \ve _{\rho ',\sigma }\circ
  (T\times I_{\sigma}), \label{aaa} \\ 
  \ve _{\sigma ,\rho }\circ (I_{\sigma}\times T^*)  &=& (T^*\times I_{\sigma})\circ
  \ve _{\sigma ,\rho '} \label{nature}.
\end{eqnarray}
Let $O_1,O_2,O_3$ be double cones such that $\rho \in \D (O_1)$, $\sigma \in \D
(O_2)$, and $\rho '\in \D (O_3)$.  Choose a double cone $O_4$ the is spacelike to
$O_i$, $i=1,2,3$; and if the spacetime dimension is less than three, choose $O_4$ to
the left of all three.  Choose $\sigma '\in \D (O_4)$, and $U\in \Hom (\sigma ,\sigma
')$.  Then $\ve (\rho ,\sigma )=U^*\rho (U)$ and $\ve (\rho ',\sigma )=U^*\rho '(U)$.
Since $T\in \Hom (\rho ,\rho ')\subseteq \alg{A}(O_4')$, it follows that $\sigma
'(T)=T$.  Thus,
\begin{eqnarray*}
  \sigma (T)U^*\rho (U) = U^*\sigma '(T)\rho (U) =U^*T\rho (U)=U^*\rho '(U)T .
\end{eqnarray*}
This establishes Eqn.\ \ref{aaa}.  The second equation can be established by a
similar calculation.

\bigskip (2) Now we show that $\ve _{\rho ,\sigma}$ makes the hexagonal diagrams
commute.  Since $\D$ is strict monoidal, we can omit the associativity isomorphisms.
That is, it suffices to show that
\begin{eqnarray} 
  \ve _{\rho \otimes \sigma ,\tau} &=& \bigl( \ve _{\rho ,\tau }\times I_{\sigma}\bigr) \circ \bigl( I_{\rho }\times \ve
  _{\sigma ,\tau}\bigr) ,\label{bbb} \\
  \ve _{\rho ,\sigma \otimes \tau} &=& \bigl(I_{\sigma}\times \ve _{\rho ,\tau}
  \bigr) \circ \bigl( \ve _{\rho ,\sigma}\times I_{\tau} \bigr) \label{ccc} . 
\end{eqnarray}
Choose $\tau '\in \D$ such $\tau '$ is supported in a region that is spacelike to the
support regions of $\rho ,\sigma ,\tau$; for spacetimes of one or two dimensions,
choose the support region of $\tau '$ to the left.  Let $U\in \Hom (\tau ,\tau ')$.
Then $\ve _{\rho ,\tau}=U^*\rho (U)$, $\ve _{\sigma ,\tau}=U^*\sigma (U)$, and $\ve
_{\rho \otimes \sigma ,\tau}=U^*\rho \sigma (U)$.  Furthermore,
\begin{eqnarray*} U^*\rho (U)\rho [U^*\sigma (U)] = U^*\rho [UU^*\sigma (U)]=U^*\rho
  \sigma (U) ,\end{eqnarray*} establishing Eqn.\ \ref{bbb}.  The second equation is
proven analogously.
\end{proof}

\begin{prop} \label{uniqsymm}
For spacetimes of dimension two or less, $\ve _{\rho _1,\rho _2}$ is the
  unique braiding on $(\D ,\otimes ,\iota )$ such that $\ve _{\rho _1,\rho _2}=I$ when
  $\rho _i\in \D (O_i)$ with $O_2<O_1$.  For spacetimes of dimension three or more,
  $\ve _{\rho _1,\rho _2}$ is the unique braiding on $(\D ,\otimes ,\iota )$ such
  that $\ve _{\rho _1,\rho _2}=I$ when $\rho _i\in \D (O_i)$ with $O_1$ and $O_2$
  spacelike separated.  \end{prop}

\begin{proof} Choosing $\wt O_2$ to the left of $O_1$ we can set $\ve _{\rho _1,\rho
    _2}=\ve _{\rho _1,\rho _2}(I,U_2)=U_2^*\rho _1(U_2)$ where $U_2\in \Hom (\rho
  _2,\rho _2')$.  Now let $c _{\sigma _1,\sigma _2}$ be another braiding on $(\D
  ,\otimes ,\iota )$ such that $c _{\sigma _1,\sigma _2}=I$ whenever $\sigma _1$ is
  localized in a region to the right of the localization region of $\sigma _2$.  Then
  since $c_{\rho _1,\rho _2}$ is natural in $\rho _1$ and $\rho _2$ and $c _{\rho
    _1,\rho _2'}=I$,
$$ c_{\rho _1,\rho _2}=  (U_2^*\times I_{\rho _1})\circ c_{\rho _1,\rho _2'}\circ
(I_{\rho _1}\times U_2) = U_2^*\rho _1(U_2)=\ve _{\rho _1,\rho _2} .$$ The proof for
the higher dimensional case is structurally identical.
\end{proof}

\begin{prop} For spacetime of dimension three or more, $\ve _{\rho _1,\rho _2}=(\ve
  _{\rho _2,\rho _1})^{-1}$, hence $\ve _{\rho _1,\rho _2}$ is a symmetry on $(\D
  ,\otimes ,\iota )$.  \end{prop}

\begin{proof} We first show that $\ve _{\rho _1,\rho _2}(U_1,U_2)$ is independent of
  the choice $(\wt O_1,\wt O_2)$ of supports for the spectator morphisms.  (Compare
  the proof of Prop.\ \ref{orient}.)  Since the spacetime has at least three
  dimensions, there is a sequence of double cones $O_{i}$, $i=1,\dots ,n$, such that:
  $O_1=\wt O_2$, for each $i$, $O_{i}\cup O_{i+1}$ is contained in a double cone
  spacelike to $\wt O_1$, and $O_{n}$ has the opposite spatial orientation to $\wt
  O_1$ as did $\wt O_2$.  Applying Prop.\ \ref{perturb} repeatedly, we conclude that
  $\ve _{\rho _1,\rho _2}(U_1,U_2)=\ve _{\rho _1,\rho _2}(U_1,U_2')$, where
  $\dt{\rho}_2\in \D (O_n)$ and $U_2'\in \Hom (\rho _2,\dt{\rho}_2)$.  Thus, $\ve
  _{\rho _1,\rho _2}$ does not depend on the relative spatial orientation of $\wt
  O_1$ and $\wt O_2$.  Prop.\ \ref{orient} shows that $\ve _{\rho _1,\rho
    _2}(U_1,U_2)$ can depend on $(\wt O_1,\wt O_2)$ only through their relative
  spatial orientation.  Therefore, $\ve _{\rho _1,\rho _2}(U_1,U_2)$ is independent
  of $(\wt O_1,\wt O_2)$.

  We can choose $\wt O_1=O_1$, $\wt O_2\perp O_1$, $\wt \rho _1=\rho _1$, and
  $U_1=I_{\rho _1}=I\in \Hom (\rho _1,\rho _1)$ so that
$$ \ve _{\rho _1,\rho _2}=U^*\rho _1(U_2). $$
But given the independence of $\ve _{\rho _1,\rho _2}$ from the orientation of $(\wt
O_1,\wt O_2)$, we also have
$$ \ve _{\rho _2,\rho _1}=\rho _1(U_2)^*U_2 =(\ve _{\rho _1,\rho _2})^* .$$  
Since $\ve _{\rho _1,\rho _2}$ is unitary, $\ve _{\rho _1,\rho _2}=(\ve _{\rho
  _2,\rho _1})^{-1}$.  \end{proof}

\begin{note} The preceding Proposition is the first place where we invoked the
  dimension of the underlying spacetime.  We will be clear when subsequent results
  depend on assumptions about dimension.  \end{note}

\begin{defn} Let $\ve _\rho := \ve (\rho ,\rho )\in \mathrm{End}(\rho \otimes \rho)$.
\end{defn}

\subsection{Relation between localized endomorphisms
  and representations} \label{ss-dhr} While the
categories $\Delta$ and $\Delta_f$ defined in this
section have very remarkable properties, their physical
and philosophical relevance is certainly not
obvious. We therefore relate the category $\Delta$ to a
certain category of representations of the net
$\alg{\alg{A}}$:

\begin{defn} \label{def-dhr} Let $O\mapsto \alg{A}(O)$
  be a net of observables and $\pi_0:\alg{A}\rarr
  B(H_0)$ a vacuum representation.  Then a
  DHR-representation (w.r.t.\ the vacuum representation
  $\pi_0$) is a $*$-representation $\pi:\alg{A}\rarr
  B(H)$ such that
  $\pi|\alg{A}(O')\cong\pi_0|\alg{A}(O')$ for any
  double cone $O$.  I.e., upon restriction to
  $\alg{A}(O')$, the representations $\pi$ and $\pi_0$
  are unitarily equivalent. The category whose objects
  are DHR-representations of $\alg{A}$ with bounded
  intertwining operators is denoted by
  $DHR(\alg{A})$. It clearly is a $C^*$-category.
\end{defn}

\def\6#1{{\mathfrak #1}}

\begin{defn} \label{def-cov} Let $\6A$ be a net that is
  Poincar\'e covariant w.r.t.\ the positive energy
  representation $U_0:\2P\rarr\2U(H_0)$. A
  representation $(H,\pi)$ of $\6A$ is called covariant
  (with positive energy) if it is equipped with a
  strongly continuous unitary representation
  $U_\pi:\widehat{\2P}\rarr\2U(H)$ (with $\mathrm{spec}
  P^\mu\subset V_+$) of the universal covering of the
  Poincar\'e group such that
  $\mathrm{Ad}\,U_\pi(h)\circ\pi=\pi\circ\alpha_h$ for
  all $h\in\widehat{\2P}$, where, omitting the covering
  map $\widehat{P}\rarr\2P$ from the notation,
  $\alpha_h=\mathrm{Ad}\,U_0(h)$. \end{defn}

Note that the definition implies that the representation space $H$ of a DHR representation must have the same
dimension as the vacuum Hilbert space $H_0$.

\begin{prop} \label{prop-dhr} There is a functor $F:
  \D\rarr DHR(\alg{A})$ of $C^*$-categories such that
  $F(\rho)=\pi_0\circ\rho$ for the objects and
  $F(s)=\pi_0(s)$ for $s\in\Hom_\Delta(\rho,\sigma)$
  for morphisms. This functor is an equivalence.
\end{prop}

\begin{proof}
  We first note that these definitions make sense:
  $\rho\in\Obj(\D)$ maps $\al A$ into itself and can
  therefore be composed with the representation
  $\pi_0$, defining a new representation. Furthermore,
  if $S$ is an arrow in $\Delta$, then Lemma
  \ref{twine} gives $S\in\al A$, thus $F(S)=\pi_0(S)$
  makes sense. With $S\in\Hom_\D(\rho,\rho')$ we have
  \[
  F(S)F(\rho)(A)=\pi_0(S)\pi_0(\rho(A))=\pi_0(S\rho(A))=\pi_0(\rho'(A)S)=F(\rho')(A)F(S)\quad\forall
  A\in\alg{A},\] thus $F(s)\in\Hom(F(\rho),F(\rho'))$. Since
  $\id_\rho$ is the unit of $\alg{A}$ we have
  $F(\id_\rho)=I_{H_0}=\id_{F(\rho)}$. The property $F(s\circ
  t)=F(s)\circ F(t)$ is obvious. Since $\pi_0$ is faithful, $F$ is
  faithful. We must show that the representation
  $F(\rho)=\pi_0\circ\rho$ satisfies the DHR criterion. Since
  $\rho\in\D$ is transportable, for every double cone $O$ there exist
  $\rho_O\in\D$ localized in $O$ and a unitary $U_O:\rho\rarr\rho_O$.
  Since $\rho_O$ is localized in $O$, the representation
  $F(\rho_O)=\pi_0\circ\rho_O$ coincides with $\pi_0$ on
  $\alg{A}(O')$. Since $F(U_O):F(\rho)\rarr F(\rho_O)$ is unitary, we
  have
\[ F(\rho)| \alg{A}(O')\cong F(\rho')| \alg{A}(O')=\pi_0|\alg{A}(O'), \]
implying $F(\rho)=\pi_0\circ\rho\in DHR(\alg{A})$. Now let $\rho,\rho'\in\Obj(\Delta)$ and 
$\tilde{S}\in\Hom(F(\rho),F(\rho'))$. If $O$ is a double cone containing the localization regions of $\rho,\rho'$,
\[
\tilde{S}\pi_0(A)=\tilde{S}\pi_0(\rho(A))=\tilde{S}F(\rho)(A)=F(\rho')(A)S=\pi_0(\rho'(A))\tilde{S}
=\pi_0(A)\tilde{S} \] for all $A\in\alg{A}(O')$.  Therefore, by Haag
duality for $\pi_0$,
$\tilde{S}\in\pi_0(\alg{A}(O'))'=\pi_0(\alg{A}(O))$. Thus there exists
$s\in\Hom_\D(\rho,\rho')$ such that $\tilde{S}=F(S)$. This proves that
the functor $F$ is full. Finally, let $\pi\in DHR(\alg{A})$ be a DHR
representation on a Hilbert space $H$. Choose any double cone $O$.
Then the DHR criterion implies the existence of a unitary $U:H\rarr
H_0$ such that $U\pi(A)=\pi_0(A)U$ for all $A\in\alg{A}(O')$. Define a
new representation $\pi'$ of $\alg{A}$ on the vacuum Hilbert space
$H_0$ by $\pi'(\cdot)=U\pi(\cdot)U^*$. By the very definition, we have
$\pi'(A)=\pi_0(A)$ for all $A\in\alg{A}(O')$. If now $\widehat{O}$ is
any double cone containing $O$, and $A\in\alg{A}(\widehat{O}')$ and
$B\in\alg{A}(\widehat{O})$ then
\[ \pi'(B)\pi_0(A)=\pi'(BA)=\pi'(AB)=\pi'(A)\pi'(B)=\pi_0(A)\pi'(B), \]
implying $\pi'(\alg{A}(\widehat{O}))\subset\pi_0(\alg{A}(\widehat{O}'))'=\pi_0(\alg{A}(O))$ by Haag duality for $\pi_0$. 
Thus $\pi'$ maps the quasilocal algebra $\alg{A}$ into $\pi_0(\alg{A})$. Since $\pi_0$ is injective, we can define an 
endomorphism $\rho$ of $\alg{A}$ by $\rho=\pi_0^{-1}\circ\pi'$. By construction, $\rho$ is localized in $O$, and we
have $\pi'=\pi_0\circ\rho=F(\rho)$. Repeating the argument with a different double cone $\tilde{O}$, we see that $\rho$ is
transportable, thus $\rho\in\Delta$. Since $\pi\cong\pi'=F(\rho)$, we have proven that every DHR representation is unitarily
equivalent to one of the form $F(\rho)$ where $\rho\in\D$. Thus the functor $F$ is essentially
surjective, and therefore, cf.\ Appendix \ref{app-A} an equivalence of categories.
\end{proof}

\begin{note} \label{note-dhr} The significance of Proposition
  \ref{prop-dhr} is twofold. On the one hand, it provides an
  interpretation of the category $\D$ in terms of a class of
  representations of $\alg{A}$. If one accepts for a moment that the
  category $DHR(\alg{A})$ is worth studying, the above equivalence is
  a powerful tool. Namely, it permits us to pull the symmetric
  monoidal structure on $\D$ over to $DHR(\alg{A})$ -- which as
  defined initially is just a category -- such as to make the functor
  $F:\D\to DHR(\alg{A})$ an equivalence of symmetric tensor
  $C^*$-categories.  But once this is understood, it is much more
  convenient to work just with the category $\D$ rather than with
  $DHR(\alg{A})$, since the tensor structure on $DHR(\alg{A})$ will
  not be strict.

  As to the physical motivation of the DHR condition,
  we give three arguments:

\begin{enumerate}
\item By an increasing sequence of double cones we mean
  a sequence $O_1\subset O_2\subset \cdots $ of double
  cones such that $\cup_i O_i=\7R^d$ (typically
  $d=4$). In the appendix of \cite{dhr3}, the following
  result (the converse of which is trivial) is proven:

\begin{thm}
Let $\omega$ be a state on $\alg{A}$ such that 
\[ \lim_{n\to\infty}\|(\omega-\omega_0)|\alg{A}(O_n')\|=0 \]
and the GNS-representation $\pi_\omega$ associated with $\omega$ satisfies property B. Then there is
a double cone $O$ such that $\pi_\omega|\alg{A}(O')\cong\pi_0|\alg{A}(O')$.
\end{thm}

\item In Section \ref{sec-localfield} we will show that
  the DHR criterion is satisfied by superselection
  sectors that are connected to the vacuum sector by a
  field net satisfying Bose-Fermi commutation
  relations. (See Section \ref{sec-localfield} for
  precise definitions and statements.)  Conversely, in
  Section \ref{sec-DR} we will prove that every DHR
  representation with finite dimension arises in this
  way. Together these results imply that DHR
  superselection sectors are precisely those induced by
  (graded) local fields. We refer to Sections
  \ref{sec-localfield}-\ref{sec-DR} for further
  discussion.

\item Let $(H,\pi)$ be a Poincar\'e covariant representation (in the
sense of Definition \ref{def-cov}) of $\alg{A}$ such that $H$ is
separable and the spectrum $\mathrm{sp}\,P^\mu\subset\7R^d$ of the
momentum operator $P^\mu$ has an isolated mass shell $\{p \ | \
p^2=m^2\}$ at its bottom, where $m>0$. (Such a representation is
called a massive one-particle representation.) Then, as proven in
\cite{buch-fred}, for every `spacelike cone' $\2C$ one has a unitary
equivalence $\pi|\alg{A}(\2C')\cong\pi_0|\alg{A}(\2C')$. (For the
definition of spacelike cones cf.\ \cite{buch-fred}.) Despite the fact
that this localization property is weaker than the one imposed by the
DHR criterion, the category of representations that are localized in
spacelike cones still can be equipped with a braided monoidal
structure, cf.\ \cite{buch-fred}. (The purely representation theoretic
part of this theory was considerably simplified in \cite[Section
4]{dr2}.) In this theory, the dimension of spacetime must be $\ge 3+1$
in order for the braiding to be a symmetry! On the technical side the
mathematical treatment is more complicated for the following reason:
If $\pi$ is a representation such that
$\pi|\alg{A}(\2C')=\pi_0|\alg{A}(\2C')$, then Haag duality implies
$\pi(\alg{A}(\2C))\subset\pi(\alg{A}(\2C))''$, but due to the weak
closure the right hand side is not contained in the algebra $\alg{A}$.
The construction of a field net that we discuss in Section
\ref{sec-DR} can nevertheless be generalized to charges localized in
spacelike cones, cf.\ \cite[Section 5]{dr2}. On the grounds of the
cited results it seems evident that the cone-localized superselection
sectors are physically better motivated than the more restrictive DHR
sectors. The D(H)R theory expounded in Sections
\ref{sec-DHR}-\ref{sec-DR} remains useful as a technically easier
`mathematical laboratory'.
\end{enumerate}
\end{note}

\subsection{Dimension theory in tensor
  $*$-categories} \label{dimension}

For any tensor $*$-category, we can define a notion of
``conjugates.''  The following is a simplified version
of this definition for the case of a \emph{strict}
tensor $*$-category.  

\begin{defn} Let $\2C$ be a strict tensor $*$-category
  and $X\in\Obj (\2C )$. A \emph{solution of the
    conjugate equations} is a triple
  $(\overline{X},r,\overline{r})$, where
  $\overline{X}\in\Obj (\2C )$ and $r:\11\rarr
  \overline{X}\otimes X,\ \overline{r}:\11\rarr
  X\otimes\overline{X}$ satisfy
\begin{eqnarray*}
  (\overline{r}^*\otimes\id_X )\circ (\id_X\otimes r) &=&\id_X, \\
  (r^*\otimes\id_{\overline{X}})\circ (\id_{\overline{X}}\otimes\overline{r}) 
  &=&\id_{\overline{X}}.
\end{eqnarray*}
A strict tensor $*$-category $\2C$ \emph{has
  conjugates} if there is a solution of the conjugate
equations for every $X\in\2C$.
\end{defn}

\begin{example} The definition of conjugates is
  exemplified in the (strictification of the) category
  $\Rep _fG$ of finite dimensional representations of a
  compact group.  In particular, it is well known that
  for each representation $(H, \pi)$ of $G$, there is a
  conjugate representation $(\ol H ,\ol\pi )$ of $G$.
  (There are several different constructions of this
  conjugate representation; see e.g.\ \cite[p.\
  30]{simon}.)  In terms of universal properties, $(\ol
  H,\ol\pi )$ is the unique irreducible representation
  of $G$ such that $(H\otimes \ol H ,\pi \otimes \ol\pi
  )$ contains a copy of the trivial representation of
  $G$.
\end{example}

\begin{note} Suppose that $(\ol X,r,\ol r)$ is a
  conjugate for $X$, and that the tensor unit $\11$ is
  irreducible.  Then $r^*\circ r\in \End (\11 )=\7C \id
  _{\11}$.  Thus up to a scalar, $r$ is an isometry,
  and therefore $\11$ is a direct summand of $\ol
  X\otimes X$.  Furthermore, as can be shown using the
  conjugate equations, the map $\End (X)\to \Hom (\11
  ,\ol X\otimes X)$, defined by $s\mapsto (\id _{\ol
    X}\otimes s )\circ r$, is an isomorphism of vector
  spaces.  Therefore, if $X$ is irreducible, the direct
  summand $\11$ appears with multiplicity $1$ in $\ol
  X\otimes X$.
  \label{multiplicity}
\end{note}

\begin{defn} Let $\2C$ be a tensor $*$-category and
  $X\in \Obj (\2C )$.  A solution $(\ol X,r,\ol r)$ of
  the conjugate equations relative to $X$ is called
  \emph{normalized} if
$$ r^*\circ r = \ol r^*\circ \ol r ,$$
and \emph{standard} if
\begin{eqnarray*} 
  r^*\circ (\id _{\ol X}\otimes a)\circ r =\ol r^* \circ (a\otimes \id _{\ol X})\circ
  \ol r ,\end{eqnarray*}
for all $a\in \End (X)$.  
\end{defn}

\begin{note} If $X,Y$ have (standard) conjugates, then
  $X\otimes Y$ and $X\oplus Y$ also have (standard)
  conjugates.  If an object has a conjugate, then it
  has a standard conjugate.  For more details, see the
  appendix.
\end{note}

\begin{defn} If an object $X\in \Obj (\2C )$ has a standard conjugate $(\ol X,r,\ol r
  )$, we define its dimension $d(X)$ by
$$d(X)\id _{\11}=r^*\circ r . $$  
If an object $X$ does not have a conjugate, we formally say $d(X)=+\infty$.  \end{defn}

\begin{note} For all $X\in \Obj (\2C )$, $d(X)\geq 0$.
  Furthermore, if $X ,Y\in \Obj (\2C )$ have conjugates
  then
\begin{eqnarray*}
  d(\ol X)=d(X) ,\qquad d(X\otimes Y)=d(X)\cdot d(Y),\qquad d(X\oplus Y)=d(X)+d(Y)
  ,\end{eqnarray*}
and $d(\11 )=1$.  (See the appendix for the discussion of these facts.)  \end{note}

\begin{defn} Let $\D$ be the DHR category.  We define
  the full subcategory $\D _f$ of objects with finite
  dimension:
$$ \Obj (\D _f)=\{ \rho \in \Obj (\D ):d(\rho )<+\infty \} .$$ 
\end{defn}

\begin{note} 
By definition, $\D_f$ is a category with conjugates. It is closed under tensor products, direct sums
and subobjects. In any $C^*$-tensor category with conjugates, the dimension of any object takes
values in $[1,\infty)$, and in the interval $[1,2]$ only values of the form $2\cos(\pi/n), n\ge 3$
can appear, cf.\ \cite{LR}. In a symmetric $C^*$-tensor category, all dimensions are integers, as is
proven in the Appendix.
\end{note}

\begin{prop} \label{prop-fd} For each $X,Y\in \Obj (\D
  _f)$, $\Hom (X,Y)$ is a finite dimensional vector
  space.  Every object $X\in \Obj (\D _f)$ is a finite
  direct sum of irreducible objects; i.e.\ the category
  $\D _f$ is semisimple.  \label{semisimple}
\end{prop}

\begin{proof} See the appendix.  \end{proof}

\begin{note} \label{note-longo}
There is an important connection, discovered by Longo \cite{longo} and explored further in
\cite{LR},  between the dimension of a DHR sector $\rho\in\D$ and subfactor theory. Among many other
things, the latter associates  to any inclusion $N\subset M$ of factors an index
$[M:N]\in[1,\infty]$. In order to apply this theory to AQFT we need  to assume (or prove) that the
local von Neumann algebras $\alg{A}(O)$ are factors. (This is automatic, e.g., in conformally
covariant theories.) If $\rho\in\Delta$ is localized in $O$, it restricts to a normal
$*$-homomorphism  of $\alg{A}(O)$ into itself, giving rise to an inclusion
$\rho(\alg{A}(O))\subset\alg{A}(O)$. The index of this subfactor is related to the categorically
defined dimension $d(\rho)$ by  
\begin{equation}\label{eq-longo} [\alg{A}(O):\rho(\alg{A}(O))]=d(\rho)^2. \end{equation} 

Longo's result allows to give a very direct formula for the dimension of (the localized
endomorphisms associated to) a DHR representation. Namely, all endomorphisms $\rho\in\D$ for which
$\pi\cong\pi_0\circ\rho$ have the same categorical dimension, justifying to write $d(\pi)$, and for
any double cone $O$ we have 
\[ d(\pi)=[\pi(\alg{A}(O'))':\pi(\alg{A}(O))]^{1/2}. \]
This is seen as follows: $\pi$ is unitarily equivalent
to a representation $\pi'=\pi_0\circ\rho$, where
$\rho\in\D$ is localized in $O$. Then the inclusion
$\pi(\alg{A}(O))\subset\pi(\alg{A}(O'))'$ is unitarily
equivalent to
$\pi'(\alg{A}(O))\subset\pi(\alg{A}(O'))'$, which
equals
$\pi_0(\rho(\alg{A}(O)))\subset\pi_0(\alg{A}(O))$. Now
the claim follows by Eqn.\ (\ref{eq-longo}) and the
fact that the index is invariant under unitary
transformations: $[UMU^*:UNU^*]=[M:N]$.

Another comment seems in order: The categorical
definition of dimension of an object requires the
existence of a conjugate object. On the other hand,
assuming factoriality of the local algebras, the
expressions $[\alg{A}(O):\rho(\alg{A}(O))]$ (for an
endomorphism localized in $O$) and
$[\pi(\alg{A}(O'))':\pi(\alg{A}(O))]$ (whose
independence of $O$ follows from mild additional
axioms) do not presuppose the existence of a
conjugate. In fact, one can show that finiteness of
these subfactor indices implies the existence of a
conjugate DHR representation, cf.\ \cite{guido-longo}.
\end{note}

\subsection{Covariant representations} \label{ss-cov}
Since we decided to work with the category $\Delta$ of
localized transportable endomorphisms rather than
directly with DHR representations, we need the
following

\begin{defn} Let $\6A$ be a Poincar\'e covariant net
  with covariant vacuum representation
  $(H_0,\pi_0)$. An endomorphism $\rho\in\Delta(\6A)$
  is called covariant if there exists a strongly
  continuous positive energy representation $\pi_\rho:
  \widehat{\2P}\rarr U(H_0)$ such that
  \begin{equation} \label{eq-cov}
    \mathrm{Ad}\,U_\rho(h)\circ\pi_0\circ\rho=\pi_0\circ\rho\circ\beta_h\quad\quad\forall
    h\in\widehat{\2P}. \end{equation} The full
  subcategory of $\Delta(\6A)$ consisting of the
  covariant morphisms is denoted by $\Delta_c(\6A)$.
\end{defn}

\begin{note} \label{note-cov1} For $\rho\in\Delta,\
  h\in\2P$ we define
  $\rho_h=\beta_h\circ\rho\circ\beta_h^{-1}$. If $\rho$
  is localized in the double cone $O$ then $\rho_h$ is
  localized in $hO$. If $\rho\in\Delta_c$ then Eqn.\
  (\ref{eq-cov}) can be restated as
  \[ \mathrm{Ad}(U(h)U_\rho(h)^*)\circ\pi_0\circ\rho
  =\pi_0\circ\beta_h\circ\rho\circ\beta_h^{-1}
  =\pi_0\circ\rho_h \quad\quad\forall
  h\in\widehat{\2P}. \] Since $\rho$ and $\rho_h$ are
  both localized, it follows that $X_\rho(h)\equiv
  U(h)U_\rho(h)^*\in\Hom(\rho,\rho_h)$, thus
  $X_\rho(h)\in\6A$.  This $\6A$-valued cocycle is very
  convenient since expressions like $\rho(U(h))$ don't
  make sense, whereas $\rho(X_\sigma(h))$ does. It
  satisfies the following cocycle equation:
\begin{eqnarray*} X_\rho(gh) &=& U(gh)U_\rho(gh)^*=U(g)U(h)U_\rho(h)^*U_\rho(g)^* \\
   &=& \beta_g(U(h)U_\rho(h)^*)U(g)U_\rho(g)^* =  \beta_g(X_\rho(h))X_\rho(g). 
\end{eqnarray*}
The same computation implies that, if $\rho\in\Delta$
and $h\mapsto X_\rho(h)\in\6A$ satisfies
$X_\rho(gh)=\beta_g(X_\rho(h))X_\rho(g)$ for all
$g,h\in\2P$, then $U_\rho(h):=X_\rho(h)^*U(h)$ is a
representation of $\2P$ and Eqn.\ \ref{eq-cov} holds,
i.e.\ $\rho\in\Delta_c$.
\end{note}

\begin{prop} $\Delta_c$ is closed under tensor products, direct sums and subobjects.
\end{prop}

\begin{proof}
Let $\rho,\rho'\in\Delta_c$ with associated cocycles $X_\rho, X_{\rho'}$. Then
\begin{equation} \label{eq-Xtens}
  X_{\rho\rho'}(h)=X_{\rho}(h)\otimes
  X_{\rho'}(h)=X_{\rho}(h)\rho(X_{\rho'}(h))
  \in\Hom(\rho\otimes\rho',\rho_h\otimes\rho'_h)
\end{equation}
clearly satisfies the cocycle equation, thus
$\rho\rho'$ is covariant. The proof for direct sums and
subobjects is omitted, cf.\ \cite{rob-lec}.
\end{proof}

If $T\in\Hom(\rho,\rho')$ then 
\[
\beta_h(T)\rho_h(A)=\beta_h(T\rho\beta^{-1}_h(A))=\beta_h(\rho'\beta^{-1}_h(A)T)=\rho'_h(A)\beta_h(T), \]
thus $\beta_h(T)\in\Hom(\rho_h,\rho'_h)$.

Now we explore some consequences of finite dimensionality:

\begin{prop} \label{prop-intertw}
Let $\rho,\rho'\in\Delta_{fc}:=\Delta_f\cap\Delta_c$. Then 
\begin{itemize}
\item[(i)] If $T\in\Hom(\rho,\rho')$ then $TU_\rho(h)=U_{\rho'}(h)T$ for all $h\in\widehat{\2P}$. 
\item[(ii)] Every $\rho\in\Delta_{fc}$ is covariant w.r.t.\  a unique representation $U_\rho$.
\item[(iii)] If $\rho,\rho'\in\Delta_{fc}$ and $T\in\Hom(\rho,\rho')$ then 
\begin{equation}\label{eq-betaT} \beta_h(T)X_\rho(h)=X_{\rho'}(h)T\quad\quad\forall h\in\widehat{\2P}.
\end{equation}
\item[(iv)] $\Delta_{fc}$ is closed under conjugates.
\end{itemize}
\end{prop}

\begin{proof} (i) For $h\in\widehat{\2P}$ define $T_h=U_{\rho'}(h)TU_\rho(h)^*$. For any $A\in\6A$
we have 
\begin{eqnarray*} \lefteqn{  T_h\rho(A) = U_{\rho'}(h)TU_\rho(h)^*\rho(A)
   = U_{\rho'}(h)T\rho(\beta_h^{-1}(A))U_\rho(h)^* } \\
   &&=U_{\rho'}(h)\rho'(\beta_h^{-1}(A))TU_\rho(h)^*=\rho'(A)U_{\rho'}(h)TU_\rho(h)^*=T_h\rho'(A), 
\end{eqnarray*}
thus $T_h\in\Hom(\rho,\rho')$. By assumption, $\rho,\rho'$ have conjugates and thus
$\Hom(\rho,\rho')$ is finite dimensional by Proposition \ref{prop-fd}. Thus $(h,T)\mapsto T_h$ is a
finite dimensional representation of the Poincar\'e group $\2P$. The claim
$TU_\rho(h)=U_{\rho'}(h)T$ is equivalent to triviality of this representation. This triviality
follows from the non-existence of finite dimensional unitary representations of $\widehat{\2P}$ as
soon as one produces a positive definite $\widehat{\2P}$-invariant inner product on
$\Hom(\rho,\rho')$. For this last step we refer, e.g., to \cite{rob-lec}. 

(ii) Apply (i) to $\rho'=\rho,\ U_{\rho'}(h)=\widetilde{U}_\rho(h),\ T=\id_\rho=\11_{H_0}$ to conclude
$U_\rho=\widetilde{U}_\rho$. 

(iii) Using (i) we compute
\begin{eqnarray*} \beta_h(T)X_\rho(h) &=& (U(h)TU(h)^*)(U(h)U_\rho(h)^*)=U(h)TU_\rho(h^{-1}) \\
   &=& U(h)U_{\rho'}(h^{-1})T=X_{\rho'}(h)T, \end{eqnarray*}

(iv) See \cite{rob-lec}.
\end{proof}

\begin{note}  \label{note-cov2}
Under weak additional assumptions on the net $\6A$, it is shown in \cite[Theorem 5.2]{guido-longo}
that every localized endomorphism of finite dimension is automatically covariant with positive
energy! Equivalently, $\Delta_f\subset\Delta_c$, and therefore $\Delta_{fc}=\Delta_f$.
\end{note}

\subsection{Statistics in braided tensor $*$-categories} \label{cat-stats}

\begin{defn} Let $(\2C ,\otimes ,\11 )$ be a tensor
  $*$-category with unitary braiding $c_{X,Y}$, and
  suppose that each $X\in \Obj (\2C )$ has a conjugate.
  For each $X\in \Obj (\2C )$, we define the
  \emph{twist} of $X$, $\Theta _X\in \End (X)$, by
$$ \Theta _X = (\ol r^*\otimes \id _X)\circ (\id _{\ol X}\otimes c_{X,X})\circ
(\ol r\otimes \id _{X} ) ,$$ where $(\ol X,r,\ol r)$ is
a standard conjugate for $X$.
\end{defn}

\begin{note} For each $X\in \Obj (\2C )$, $\Theta _X$ is unitary.  When $X$ is
  irreducible, $\End (X)=\7C \id _X$, and so $\Theta _X =\om _X\id _X$, where $\om
  _X$ is a complex number of unit modulus (called the \emph{statistics phase}).  In
  the case that $c_{X,Y}$ is a symmetry, then $(c_{X,X})^*=c_{X,X}$, and so $(\Theta
  _X)^*=\Theta _X$.  Together with unitarity, this implies that $(\Theta _X)^2=\id
  _X$.
\end{note}

\begin{defn} Let $(\2C ,\otimes ,\11 )$ be a tensor $*$-category with unitary
  symmetry $c_{X,Y}$.  If $X\in \Obj (\2C )$ is irreducible, we say that $X$ is a
  \emph{Bosonic object} if $\om _X=1$, and we say that $X$ is a \emph{Fermionic
    object} if $\om _X=-1$.  \end{defn}

\begin{note} We give a number of justifications for our
  focus on the category $\D_f$ of DHR representations
  with finite dimension.

  (i): In the heuristic interpretation of the (unitary
  equivalence classes) of irreducible DHR
  representations as corresponding to the types of
  particles in a QFT, the conjugate $\ol{\rho}$ of a
  DHR representation $\rho$ corresponds to the
  antiparticle. It may happen that a particle is its
  own antiparticle, i.e.\ $\rho\cong\ol{\rho}$; but the
  existence of antiparticles seems to be an integral
  part of relativistic quantum field theories.

  (ii): The DHR sectors admitting a conjugate in the
  above sense are (rough) analogues in the operator
  algebraic approach to AQFT of Wightman fields with
  finitely many components. In the Wightman framework
  \cite{SW} it is well known that infinite components
  behave `pathologically' in that the PCT and
  spin-statistics theorems do not apply to them, and
  can in fact be violated. In algebraic QFT, these
  results are reflected in the fact that we cannot even
  define Bosonic and Fermionic objects that have
  dimension $\infty$, in the sense that they have no
  conjugates.

  (iii): In \cite{fred-anti} it was shown that every
  massive one-particle representation (cf.\ Note
  \ref{note-dhr}(iii)), which by the mentioned result
  of \cite{buch-fred} is localizable in space-like
  cones, has a conjugate in the $C^*$-tensor category
  of cone-localizable representations. It therefore
  seems natural to require existence of conjugates also
  in the more restrictive setting of double cone
  localizable representations.

  (iv): As pointed out in Note \ref{note-cov2}, DHR
  endomorphisms of finite dimension are automatically
  covariant, provided one accepts the additional
  conditions on the net $\6A$ needed for this result.
  Even if one doesn't wish to appeal to this result,
  finite dimensionality of the objects is needed (via
  finite dimensionality of the hom-sets) for the proof
  of Proposition \ref{prop-intertw}. The latter will be
  crucial for lifting the Poincar\'e action from $\6A$
  to the field theory $\6F$ in Section \ref{sec-DR}.
\end{note}

\section{From Fields to Representations}  \label{sec-localfield}

In the current section we take the `top down' approach to
superselection rules.  That is, we are given a field algebra $\al F$
and a gauge group $G$ acting concretely on a Hilbert space $\2H$.  We
then define the observables as the gauge invariant elements of $\al
F$.  The representation of $\al F$ on $\2H$ then gives us a preferred
set of representations of $\al A$; viz.\ those that can be `created
from the vacuum representation by the action of local fields.'  Our
main mathematical objective in the current section is to show that
these representations satisfy the DHR selection criterion.  Thus, all
superselection sectors that arise in the traditional way --- viz.\ by
acting on the vacuum with fields --- fall within the purview of DHR
superselection theory.  (But note: We are restricting attention to
\emph{local} fields.)

\begin{bxd} \begin{defn} \label{field-system} Let $\om _0$ be a state
    on $\al A$, and let $(\2H _0,\pi _0 )$ be the corresponding GNS
    representation.  A \emph{field system with gauge symmetry} for
    $(\al A,\om _0)$ is a quadruple $\fields$, where $(\2H ,\pi )$ is
    a representation of $\al A$, $O\mapsto \al F(O)$ is a net of von
    Neumann algebras acting irreducibly on $\2H$, $G$ is a strongly
    compact group of unitary operators on $\2H$, $k$ is a central
    element of $G$ such that $k^2=e$, and such that:
    \begin{list}%
 {\boldmath{\greek{bean})}}{\usecounter{bean}
 \setlength{\rightmargin}{\leftmargin}%
 \setlength{\itemindent}{-1em}  }
\item $(\2H _0,\pi _0)$ is a subrepresentation of $(\2H ,\pi )$, i.e.\
  there is an isometry $V:\2H _0\to \2H$ such that $V\pi _0=\pi V$; 
\item $V$ maps $\2H _0$ into the subspace of $G$-invariant vectors of
  $\2H$;
\item the $U\in G$ induce automorphisms that leave each
  $\al F(O)$ globally fixed, and $\pi (\al
  A(O))''\subseteq \al F(O)$ is the set of fixed points
  under the action of $G$ on $\al F(O)$;
\item for each $O\in \2K$, $V(\2H _0 )$ is cyclic for $\alg{F}(O)$;
\item the fields are local relative to the observables, i.e.\ $\alg{F}(O_1)$ and $\pi
  (\alg{A}(O_2))$ commute elementwise whenever $O_1$ and $O_2$ are spacelike
  separated. \end{list}

\end{defn}
\end{bxd}

A few remarks on the definition of a field system: the
fact that $\al F$ is generated by local algebras $\{
\al F (O):O\in \2K \}$ means that elements of $\al F$
represent \emph{local} fields --- i.e., fields whose
excitations can be localized within a bounded spacetime
region.  Furthermore:

\begin{itemize} 
\item[$\delta )$] is the Reeh-Schlieder Condition: it
  states that each local region $O$ carries a full set
  of fields in the sense that these local fields can
  reach each sector from the vacuum sector.  [But note
  that Condition $(\g )$ only guarantees that sectors
  in $\2H$ can be reached from the vacuum sector.  A
  stronger notion of completeness would rely on some
  intrinsic criterion for physical sectors of $\al A$,
  and would require that all these sectors be contained
  in $\2H$; see Definition \ref{complete}.]
\item[$\gamma )$] can be interpreted as saying that the
  group $G$ is an internal symmetry group of the field:
  it does not change the spacetime localization region
  of field operators.
\item[$\ve )$] is the Relative Locality Condition.  Since fields need
  not be observable, the field algebra is \emph{not} required to
  satisfy microcausality.  However, in the typical situation (i.e.\
  normal commutation relations), field operators localized in one
  spacetime region either commute or anticommute with field operators
  localized in a spacelike separated region.  Condition $(\ve )$ is a
  weakening of the requirement of normal (Bose/Fermi) commutation
  relations.
\end{itemize}

Since $G$ is a compact group of unitary operators
acting on $\2H$, we can apply all of the apparatus of
the theory of unitary representations of compact groups
(see e.g.\ \cite{folland}).  In particular, $\2H$
decomposes into a direct sum of orthogonal subspaces
$\2H _{\xi}$ which reduce the action of $G$.  Thus the
reduced unitary representation $U_{\xi}$ of $G$ on $\2H
_\xi$ is factorial, i.e.\ the von Neumann algebra
generated by the operators $\{ g|_{\2H _\xi}:g\in G\}$
is a factor.  The representation $U_{\xi}$ decomposes
into a direct sum of \emph{unitarily equivalent}
irreducible representations of $G$.  So, there is a
privileged direct sum decomposition of $\2H$:
$$ \2H = \bigoplus _{\xi} \2H _{\xi} ,$$
where the subspace $\2H _{\xi}$ is generated by the vectors in $\2H$
that transform according to the character $\xi$ (unitary equivalence
class of irreducible representations) of $G$.

In the present section our primary objectives are:
\begin{enumerate}
\item Show that the subspaces $\2H _{\xi}$ reduce the action of the
  observable algebra $\al A$.  So, the representation of $\al A$ on
  $\2H$ decomposes into a direct sum $\bigoplus _{\xi}\pi _{\xi}$ of
  representations on the subspaces $\2H _\xi$.
\item Show that each representation $(\2H _\xi ,\pi
  _{\xi})$ of $\al A$ is factorial, so that the
  irreducible subrepresentations of $(\2H _\xi ,\pi
  _\xi )$ are mutually equivalent.  (Hence each
  character $\xi$ of $G$ labels an equivalence class of
  irreducible representations of $\al A$.)
\item Show that $V(\2H _0)$ is precisely the subspace of $G$-invariant
  vectors in $\2H$.  (Hence the character $\11$ of $G$ labels the
  equivalence class of the vacuum representation of $\al A$.)
\item Show that each subrepresentation of $(\2H ,\pi )$
  is a DHR representation.  In slogan form, the sectors
  that can be reached from the vacuum by application of
  local fields correspond to DHR representations (i.e.\
  representations that are equivalent, modulo any local
  region, to the vacuum representation).
\end{enumerate}

Regarding objectives (1) and (2), it will suffice to show that $\pi
(\al A)''=G'$, because then the von Neumann algebras $\pi (\al A)''$
and $G''$ share the same central projections.

\begin{prop} If $\fields$ is a field system with gauge
  symmetry for $(\alg{A},\om _0)$ then $\pi (\al
  A)'=G''$.
\end{prop}

Our notation will henceforth be simplified if we use $g$ and $U(g)$
ambiguously to denote elements of the unitary group $G$ on $\2H$.
That is, we think of $g\to U(g)$ as the identity representation of $G$
on $\2H$.

\begin{proof} Define $M:\bh \to G'$ by
  \[ M(A)=\int _{G}U(g)AU(g)^{*}\, d\mu (g) ,\] where
  $\mu$ is the Haar measure on $G$.  Then $M$ is a
  faithful, normal projection of norm one from $\bh$
  onto $G'$.  Since $M$ is weakly continuous on the
  unit ball of $\bh$, we have
  \begin{eqnarray} G'=M(\bh )=M(\overline{\alg{F}})=\overline{M(\al
      F)} = \overline{\pi (\al A)}. \end{eqnarray} Thus, $G''=\pi
  (\alg{A})'$.
\end{proof}

It follows then that the factorial subrepresentations of the
representation $(\2H ,\pi )$ of $\al A$ are in one to one
correspondence with the factorial subrepresentations of the action of
$G$ on $\2H$.

\begin{note} Since $G$ is compact each irreducible
  representation of $G$ is finite dimensional.  Let
  $\hat{G}$ be the set of characters (equivalence class
  of irreducible representations) of $G$, and for $\xi
  \in \hat{G}$, let $d(\xi )$ be the dimension of the
  underlying Hilbert space.  Then the previous result
  gives a nice intuitive picture of the representation
  $(\2H ,\pi )$ of $\al A$.  For each $\xi \in
  \hat{G}$, select an irreducible subrepresentation
  $(\2H _\rho ,\pi _\rho )$ of the factorial
  representation $(\2H _\xi ,\pi _\xi )$.  Then we have
$$ \pi (A) = \bigoplus _{\xi \in \hat{G}} d(\xi )\pi _{\rho }(A) =
\bigoplus _{\xi \in \hat{G}}(\pi _{\rho}(A)\otimes
I_\rho ) ,$$ where $d(\xi )\pi _\rho (A)=\pi _\rho
(A)\oplus \cdots \oplus \pi _\rho (A)$, $d(\xi )$
times, and $I_\rho$ is the identity on an $d(\xi
)$-dimensional Hilbert space.
\end{note}

\begin{lemma} Let $\fields$ be a field system with gauge symmetry for
  $(\al A,\om _0 )$.  Then $\2H _0$ is separating for $\al
  F$.  \end{lemma}

\begin{proof} Let $F\in \al F$.  If $F\2H _0=\{ 0\}$
  then $\4E (F^*F)\2H _0=\{ 0\}$.  Since $\4E (F^*F)\in
  \pi (\al A)$ and $\pi _0$ is faithful, $\4E
  (F^*F)=0$.  Since $\4E$ is faithful, $F=0$.
  Therefore, $\2H _0$ is separating for $\al F$.
\end{proof}

To obtain further information about the field system $\fields$, we
identify ``tensors under the action of $G$'' in the field algebra $\al
F$.  To make sense of this idea, forget momentarily that $\al F$ has a
product operation, and consider it merely as a Banach space.  The map
$U\mapsto \mathrm{Ad}U$ is a (strongly) continuous representation of
the compact group $G$ in $\Aut \al F$, which is of course a subset of
the invertible linear operators on $\al F$.  As in the case of a
representation of a compact group on a Hilbert space $H$, the
representation of $G$ on $\al F$ decomposes into a direct sum of
disjoint representations.  An operator $F\in \al F$ is said to
\emph{transform according to a representation $\rho$ of $G$} just in
case it is contained in a linear subspace $H_\rho$ of $\al F$ carrying
the corresponding representation of $G$.  In fact, we will show that
the irreducible subspaces in $\al F$ have a special algebraic
property: they have support $I$.

\begin{lemma} Let $\fields$ be a field system with
  gauge symmetry for $(\al A,\om _0)$, and suppose that
  $\al A$ satisfies property B relative to $\om _0$.
  Then the net $O\mapsto \pi (\alg{A}(O))''$ satisfies
  property B.  \label{BB}
\end{lemma}

\begin{proof} We first establish that $\pi
  _0|_{\alg{A}(O')}$ is quasiequivalent to $\pi
  |_{\alg{A}(O')}$ for each double cone $O$.

  By the Relative Locality Condition $(\ve )$, $\al
  F(O)\subseteq \pi (\al A(O'))'$.  By the
  Reeh-Schlieder Condition $(\d )$, $\2H _0$ is a
  cyclic subspace for $\al F(O)$.  Thus,
$$ \2H =[\al F(O)\2H _0]\subseteq [\pi (\al A(O'))'\2H _0] .$$
Let $E_0$ be the orthogonal projection onto $\2H _0$.  The central
support of $E_0$ in $\pi (\al A(O'))'$ is the projection onto
$[\pi(\alg{A}(O'))'E_0(\2H )]$ \cite[Prop.\ 5.5.2]{kr}.  Thus $E_0$
has central support $I$ in $\pi (\al A(O'))'$, and therefore $(\pi
_0|_{\al A(O')},\2H _0)$ and $(\pi |_{\al A(O')},\2H )$ are
quasiequivalent \cite[Thm.\ 10.3.3]{kr}.

Let $O_1$ be a double cone whose closure is contained in $O$, and let
$E$ be a nonzero projection in $\pi (\alg{A}(O_1))''$.  Choose a
double cone $O_2$ that is spacelike separated from $O$.  The
preceding argument shows that there is a $*$-isomorphism $\f$ from
$\pi _0(\alg{A}(O_2'))''$ to $\pi (\alg{A}(O_2'))''$ such that $\f
(\pi _0(A))=\pi (A)$ for all $A\in \alg{A}$.  This isomorphism $\f$
preserves the net structure: $\f [\pi _0(\alg{A}(O_3))]=\pi
(\alg{A}(O_3))$ for any double cone $O_3$ contained in $O_2'$.
Further, since $\f$ is ultraweakly continuous \cite[Cor.\ 7.1.16]{kr},
$\f [\pi _0(\alg{A}(O_3))'']=\pi (\alg{A}(O_3))''$.  In particular,
$\f (E)$ is a projection in $\pi _0 (\alg{A}(O_1))''$.  By property B
for $\pi _0$, there is an isometry $V\in \pi _0(\alg{A}(O))''$ such
that $VV^*=\f (E)$.  Thus, $W:=\f ^{-1}(V)\in \pi (\alg{A}(O))''$ is
an isometry such that $WW^*=E$.  Therefore the net $O\mapsto \pi
(\alg{A}(O))''$ satisfies property B.
\end{proof}

\begin{defn} Consider the ordered $n$-tuple $(F_1,\dots ,F_n)$ of
  elements in $\alg{F}$.  We say that this $n$-tuple \emph{transforms
    under the action of $G$ according to character $\xi$} just in
  case:
  \begin{enumerate}
\item $F_{i}^*F_j=0$ if $i\neq j$; and 
\item $\a _g(F_i) =\sum _{j=1}^{n}u^{\xi}_{ij}(g)F_{j}$, where
  $u^{\xi}_{ij}$ is a set of matrix elements for $\xi$.  That
  is, for some representation $(H,\rho )$ of $G$ of class $\xi$,
  and orthonormal basis $\{ e_1,\dots ,e_n\}$ for $H$,
  $u_{ij}(g):=\langle e_{i},\rho (g)e_j\rangle _{H}$.
\end{enumerate}
\end{defn}

\begin{note} If $(F_1,\dots ,F_n)$ is a tensor in $\al F$ transforming
  according to $\xi$, then we can always replace the $F_i$'s with
  partial isometries $V_i$ with orthogonal ranges.  Indeed, let
  $V_i|F_i|$ be the polar decomposition of $F_i$.  When $i\neq j$,
  $F_i^*F_j=0$, and so $F^{*}_i$ and $F^*_j$ have orthogonal ranges.
  Recall that if $F=V|F|$, then $V$ annihilates the orthogonal
  complement of $\mathrm{r}(F^*)=\mathrm{r}(|F|)$ (see \cite[Thm.\
  6.1.2]{kr}).  Thus $V_j|F_i|=\d _{ij}F_j$, and $$ \Bigl( \sum
  _ju^{\xi}_{ij}(g)V_j \Bigr) |F_i| = \sum
  _{j}u^{\xi}_{ij}(g)V_{j}|F_j| = \sum _ju^{\xi}_{ij}(g)F_j =
  F_i .$$ By the uniqueness of the polar decomposition, $\sum
  _{j}u^{\xi}_{ij}(g)V_{j}=V_i$.  Hence $(V_1,\dots ,V_n)$ is a
  tensor transforming according to $\xi$.
\end{note}

\begin{defn} Given $\f ,\psi \in \2H _{\xi}$, define a map
  $M^{\xi}_{\f ,\psi}:\al F\to \al F$ by
$$ M^{\xi}_{\f,\psi }(F) =\int _G\langle \f ,U(g)\psi \rangle \a _g(F)
\,d\mu (g) ,$$ where $\mu$ is the Haar measure on $G$.
\end{defn}

\begin{fact} Due to the invariance of $\mu$ we have
  $\a _g \circ M^{\xi}_{\f ,\psi}(F)=M^{\xi}_{U(g)\f ,\psi
  }(F)$.  \end{fact}

\begin{lemma} Let $(F_1,\dots ,F_n)$ be a tensor in $\al F (O)$
  transforming as a unitary representation of class $\xi$.  Then
  $F_i(\2H _0)\subseteq \2H _{\xi}$, where $\2H _{\xi}$ is the
  subspace of vectors of $\2H$ that transform according to $\xi$.
  \label{crash}
\end{lemma}

\begin{proof}[Sketch of proof] Let $\f \in \2H _0$, and let $g\in G$.
  Then
  \begin{eqnarray*} U(g)[M^{\xi}_{\f ,\psi}(F)\f] = U(g)M^{\xi}_{\f
      ,\psi}(F)U(g)^*\f = M^{\xi}_{U(g)\f ,\psi}(F)\f
    . \end{eqnarray*} Then a straightforward calculation using matrix
  elements for $\xi$ establishes the result.
\end{proof}

\begin{lemma} Let $\xi$ be a character of $G$ that occurs
  nontrivially in the decomposition of the action of $G$ on $\2H$.
  Then for each double cone $O$, there is a tensor $(F_1,\dots ,F_n)$
  in $\al F(O)$ that transforms as a unitary representation of class
  $\xi$.
  \label{fullm}
\end{lemma}

\begin{proof}[Sketch of proof] Let $(\psi _1,\dots ,\psi _n)$ be an
  orthonormal basis from a $G$-irreducible subspace of $\2H _\xi$.
  Let $\f$ be a unit vector in this same subspace.  Since $\2H _0$ is
  cyclic for $\al F(O)$, and $\al F(O)$ is a von Neumann algebra,
  there is an $F\in \al F(O)$ and a vector $\f _0\in \2H _0$ such that
  $F\f _0=\f$.  Let $F_i=M^{\xi}_{\psi _i, \f } (F)$.  One verifies
  then that $(F_1,\dots ,F_n)$ is the required tensor.
\end{proof}

\begin{lemma} Let $F_1,\dots ,F_n\in \al F(O)$ such that $(F_1,\dots
  ,F_n)$ transforms according to the character $\xi$.  Then if $\ol
  O\subseteq O_1$, there are $X_1,\dots ,X_n\in \al A(O_1)$ such that
$(X_1,\dots ,X_n)$ transforms according to $\xi$ and
$$ \sum _{i=1}^{n}X_i^*X_i=I .$$  \label{support}
\end{lemma}

\begin{proof} First replace $F_1,\dots ,F_n$ with partial isometries
  $V_1,\dots ,V_n$, then let $V=\sum _{i=1}^{n}V_i$.  Since the $V_i$
  have orthogonal ranges, $V$ is a partial isometry, and $V^*V=\sum
  _{i=1}^{n}V_i^*V_i$.  A straightforward calculation shows that $\a
  _g(V^*V)=V^*V$ for all $g\in G$.  Thus, $E=V^*V$ is a projection in
  $\al F(O)\cap G'=\pi (\al A(O))''$.  By Lemma \ref{BB}, $O\mapsto
  \pi (\al A(O))''$ satisfies property B.  Thus, there is an isometry
  $W\in \pi (\al A (O_1))''$ with $WW^*=E$.  For $i=1,\dots ,n$, let
  $X_i=V_iW$.  Then the tensor $(X_1,\dots ,X_n)$ transforms according
  to $\xi$, and
$$\sum _{i=1}^{n}X_i^*X_i=W^*\Bigl( \sum _{i=1}^{n}V_i^*V_i \Bigr) W =I .$$
\end{proof}

\begin{lemma} Let $\2H _\xi\subseteq \2H$ be the range of a central
  projection in $\pi (\al A)''$.  Then for each double cone $O$, $\2H
  _\xi$ is cyclic for $\al F(O)$.  \label{sectors} \end{lemma}

\begin{proof} Let $O_1$ be a double cone such that $\ol O_1\subseteq
  O$.  By the Reeh-Schlieder Condition, $\2H _0$ is cyclic for $\al
  F(O_1)$.  By Lemma \ref{fullm}, there is a tensor $(F_1,\dots F_n)$
  in $\al F (O_1)$ that transforms according to the representation
  $(H,\rho )$ of $G$.  By Lemma \ref{support}, there is a tensor
  $(X_1,\dots ,X_n)$ in $\al F(O)$ that transforms that same way, and
  such that $\sum _{i=1}^{n}X_i^*X_i=I$.  Then
$$ \al F(O)\2H _0 \:=\: \al F(O)\sum _{i=1}^{n}X_i^*X_i\2H _0 \:\subseteq \:
\al F(O)\2H _{\xi} \, ,$$ where the final inclusion follows by Lemma
\ref{crash}.  Therefore $\2H _{\xi}$ is cyclic for $\al F(O)$.
\end{proof}

\begin{defn} Let $\Rep _{\al F}\al A$ be the category
  of subrepresentations of the representation $(\2H
  ,\pi )$ of $\al A$.  We mean to take $\Rep _{\al
    F}\al A$ as a full subcategory of the category of
  all representations of $\al A$, i.e.\ the hom-sets
  between representations in $\Rep _{\al F}\al A$ are
  the same as the hom-sets in the larger
  category.  \end{defn}

\begin{prop} Let $\fields$ be a field system with gauge
  symmetry for $(\al A,\om _0)$.  Then there is a
  faithful functor $F:\Rep _{\al F}\al A \to DHR(\al
  A)$.
  \label{uto}
\end{prop}

\begin{proof} Suppose that $(\2H ',\pi ')$ is an object of $\Rep _{\al
    F}\al A$.  That is, there is an isometry $V:\2H '\to \2H$ such
  that $V\pi '=\pi V$.  We subsequently identify $\2H '$ with its
  image in $\2H$, and treat $\pi '$ as mapping into $\bh$.  We must
  show that $(\2H ',\pi ')$ is in $DHR(\al A)$; that is, for any
  double cone $O$, $(\2H ',\pi '|_{\alg{A}(O')})$ is unitarily
  equivalent to $(\2H_0,\pi _0|_{\alg{A}(O')})$.  

  Let $\overline{\pi}=\pi |_{\alg{A}(O')}$.  Since
  $E_{\iota},E_{\xi}\in \pi (\alg{A})'\subseteq \pi
  (\alg{A}(O'))'$, $E_{\iota}$ and $E_{\xi}$ reduce
  $\overline{\pi}$.  We first establish that
  $E_{\iota}$ and $E_{\xi}$ have the same central
  support in $\pi (\alg{A}(O'))^{-}$, from which it
  follows that $E_{\iota}\ol\pi$ and $E_{\xi}\ol\pi$
  are quasiequivalent \cite[Thm.\ 10.3.3]{kr}.

  By the Relative Locality Condition $(\boldmath{\delta})$,
  $\alg{F}(O)\subseteq \pi (\alg{A}(O'))'$.  By the Reeh-Schlieder
  Condition $(\boldmath{\gamma})$, $E_{\iota}\2H$ is a cyclic subspace
  for $\alg{F}(O)$.  Thus,
$$ \2H =[\alg{F}(O)E_{\iota}(\2H )]\subseteq [\pi (\alg{A}(O'))'E_{\iota}(\2H )] .$$  
Similarly, Lemma \ref{sectors} entails that
$E_{\xi}\2H$ is a cyclic subspace for $\alg{F}(O)$, and
so $[\pi (\alg{A}(O'))'E_{\xi}(\2H )]=\2H$.  However,
the central support of $E_\iota$ in $\pi
(\alg{A}(O'))'$ is the projection onto
$[\pi(\alg{A}(O'))'E_0(\2H )]$, and similarly for
$E_{\xi}$ \cite[Prop.\ 5.5.2]{kr}.  Thus, $E_{\iota}$
and $E_{\xi}$ have central support $I$ in $\pi
(\alg{A}(O'))'$.  Therefore, $(\pi
_{0}|_{\alg{A}(O')},\2H _0)$ and $(\pi
_{\xi}|_{\alg{A}(O')},\2H _{\xi})$ are quasiequivalent,
i.e.\ there is a $*$-isomorphism $\f :\pi
_0(\alg{A}(O'))\to \pi _{\xi}(\alg{A}(O'))$ such that
$\f (\pi _0(A))=\pi _{\xi}(A)$ for all $A\in
\alg{A}(O')$.

The previous reasoning also shows (by replacing $O$ with a spacelike
separated double cone) that for each double cone $O$, $(\pi
_{0}|_{\alg{A}(O)},\2H_0)$ is quasiequivalent to $(\pi
_{\xi}|_{\alg{A}(O)},\2H _{\xi})$.  Thus, in particular, since the net
$O\to \pi _0(\alg{A}(O))''$ of von Neumann algebras satisfies property
B (by assumption), so does the net $O\to \pi _{\xi}(\alg{A}(O))''$.

To establish that $(\pi _{0}|_{\alg{A}(O')},\2H_0)$ and $(\pi
_{\xi}|_{\alg{A}(O')},\2H_{\xi})$ are unitarily equivalent, we will
use the following result (\cite[Theorem 7.2.9]{kr}):
\begin{quote}
  \textit{Let $\alg{R}_j$, $j=1,2$, be von Neumann
    algebras acting on Hilbert spaces $\2H _j$
    respectively.  Suppose that for $j=1,2$, there is a
    vector $x_j\in \2H _j$ that is cyclic and
    separating for $\alg{R}_j$.  If $\a :\alg{R}_1\to
    \alg{R}_2$ is a $*$ isomorphism then there is a
    unitary operator $U$ from $\2H _1$ to $\2H _2$ that
    implements $\a$.}
\end{quote}
in conjunction with the fact (\cite[Exercise 9.6.32]{kr}):
\begin{quote}
  \textit{If $\alg{R}$ is a von Neumann algebra acting on a separable Hilbert space
    $\2H$, and if $\alg{R}'$ is properly infinite, then there is vector $x\in \2H$
    that is cyclic and separating for $\alg{R}$.}
\end{quote}
By Proposition \ref{infinite}, $\pi _0 (\alg{A}(O'))'$ and $\pi
_{\xi}(\alg{A}(O'))'$ are properly infinite.  By assumption, $\2H _0$
is separable.  Thus, it will suffice to show that $\2H _{\xi}$ is
separable.  Since $\pi _{\xi}$ is irreducible, each non-zero vector
$x\in \2H _{\xi}$ is cyclic for $\pi _{\xi}(\alg{A})$.  Thus, $\2H
_{\xi}$ is the closure of the union of $\pi _{\xi}(\alg{A}(O_n))x$ for
an increasing sequence $O_n$ of double cones.  Hence it suffices to
show that $\pi _{\xi}(\alg{A}(O))x$ is separable for each $O\in \2K$.
Since $\2H _0$ is separable, the unit ball of $\alg{B}(\2H _0)$ is
compact metrizable \cite[Thm.\ 5.1.3; Exercise 5.7.7]{kr}.  Since the
unit ball of $\pi _0(\alg{A}(O))''$ is a closed subset of the unit
ball of $\alg{B}(\2H _0)$, it is also compact metrizable.  But $\pi
_{\xi}(\alg{A}(O))''$ is $*$ isomorphic, hence ultraweakly
homeomorphic, to $\pi _{0}(\alg{A}(O))''$.  Therefore, the unit ball
of $\pi _{\xi}(\alg{A}(O))''$ is compact metrizable, hence separable,
in the weak operator topology.  It follows that $\pi
_{\xi}(\alg{A}(O))''x$ is separable.
\end{proof}

In Proposition \ref{prop-dhr} it was shown that there is a faithful,
essentially surjective functor $F'$ from the category $DHR(\al A)$ of
DHR representations to the category $\D$ of localized transportable
morphisms of $\al A$.  So, the previous Proposition entails that
$F'\circ F$ is a faithful functor from $\Rep _{\al F}\al A$ into $\D$.
We subsequently replace $F'\circ F$ with just $F$.

Recall that $\D _f$ is the full subcategory of $\D$ of objects with
conjugates.  The final thing we need to show in this Section is that
the image of each object in $\Rep _{\al F}\al A$ under $F$ is
isomorphic to an object in $\D _f$.  That is, we need to show that the
image object has a conjugate.

\begin{proof}[Sketch of proof] One shows that the subrepresentations
  of $G$ on $\2H$ are closed under taking conjugates.  This can be
  proven by noting a correspondence between the action of $G$ on $\2H$
  and the action of $G$ on $\al F$.  Then use the fact that $\al F$ is
  a $*$-algebra.  Thus, for each irreducible subrepresentation $\pi
  _\rho $ of $\pi$, there is an irreducible subrepresentation $\pi
  _{\ol\rho}$ of $\pi$.  Verify that $(F'\circ F)(\pi _{\ol\rho})$ is
  a conjugate for $(F'\circ F)(\pi _\rho )$. \end{proof}

\noindent Therefore, $F'\circ F$ is a faithful functor
from $\Rep _{\al F}\al A$ into $\D _f$.  So we have
shown:
\begin{quote} \textit{Each representation of $\al A$ that arises from
    its being taken as the gauge invariant part of a field algebra is
    a representation of the form $\pi _0\circ \rho$ with $\rho \in
    \Obj (\D _f)$.}  \end{quote} Thus, the study of $\D _f$
encompasses the study of representations that arise from the approach
that begins with a field algebra.

We said above that in the ``normal'' situation, field operators in
$\al F(O_1)$ with either commute or anticommute with field operators
in $\al F(O_2)$ when $O_1$ and $O_2$ are spacelike separated.  To be
more precise, we would expect for a Bose field operator to commute
both with other Bose field operators, as well as with Fermi field
operators; and we would expect for a pair of Fermi field operators to
anticommute.  But what are Bose and Fermi field operators?  The
distinction between the two is defined in terms of the privileged
element $k$ of the gauge group $G$.  

\begin{defn} If $\a _k(F)=F$ then $F$ is said to be a \emph{Bose field
    operator}; and if $\a _k(F)=-F$ then $F$ is said to be a
  \emph{Fermi field operator}. \end{defn}

We define a Bosonic sector in $\2H$ to be a subspace $\2H _{\xi}$ such
that $U(k)|_{\2H _{\xi}}=I$, and a Fermionic sector in $\2H$ to be a
subspace $\2H _{\xi}$ such that $U(k)|_{\2H_{\xi}}=-I$.  It then
follows that Bosonic field operators create Bosonic sectors from the
vacuum, and Fermionic field operators create Fermionic sectors from
the vacuum.  

We can now make sense of the notion of normal commutation relations:
Bose field operators should commute with each other and commute with
Fermionic field operators.  Fermionic field operators should
anticommute with each other.

\begin{bxd}
  \begin{defn} \label{normal-commutation} A local operator algebra
    system of fields $(\pi ,(G,k),\alg{F})$ is said to \emph{satisfy
      normal commutation relations} just in case the local field
    algebras satisfy \emph{graded local commutativity:} If $O_1$ and
    $O_2$ are spacelike, and $F_\sigma\in \alg{F}(O_1)$,
    $F_{\sigma}'\in \alg{F}(O_2)$ are such that $\alpha
    _k(F_{\sigma})=\sigma F_{\sigma}$ and $\alpha
    _k(F'_{\sigma})=\sigma F_{\sigma}$, ($\sigma =\pm$), then
    \[ F_+F_+'=F_+'F_+,\qquad F_+F_-'=F_-'F_+,\qquad F_-F_-'=-F_-'F_- .\]
\end{defn}
\end{bxd}


\section{From Representations to Fields} \label{sec-DR}

The preceding section derives properties of
representations of $\al A$, given that these
representations are created from the vacuum
representation by the action of local fields on the
vacuum.  But such an approach will seem at best
heuristic to the Algebraic Imperialist.  From the
Imperialist's point of view, the entire content of the
theory is contained in the abstract net $\al A$ of
observable algebras.  

On the one hand, the Imperialist might be an eliminativist about
fields and gauge group.  On the other hand, the Imperialist might
claim that the fields and gauge group are physically significant, but
only because they can be `reconstructed' from the net of observable
algebras.  In order to justify this latter stance, the Imperialist
would need to accomplish the following:
\begin{quote} \textbf{Task:} Try to reconstruct, in a
  mathematically rigorous fashion, the entire apparatus
  of QFT --- fields, gauge groups, etc. --- from the
  net of observable algebras.
\end{quote}
A quixotic task indeed! For one, philosophers seemed to
have settled that theory is always underdetermined by
data; and so we should not expect to be able to find
the full theoretical apparatus hidden within the net of
observable algebras.  But there is a surprise in store:
the Task was undertaken, and was achieved.  The DR
Reconstruction Theorem shows in a fully rigorous and
precise way that the DHR category encodes all the
information needed to reconstruct, uniquely, the fields
and the gauge group.  This section provides the details
of the reconstruction.

\begin{bxd}
  \begin{defn} A field system with gauge symmetry
    $\fields$ for $(\al A,\om _0)$ is said to be
    \emph{complete} if the representation $\pi$ of $\al
    A$ contains copies of all representations in the
    DHR category $DHR(\al A)$ of $\al
    A$.  \label{complete} \end{defn}
\end{bxd}

\begin{bxd}
  \label{equiv-fields}
  \begin{defn} Two field systems with gauge symmetry
    $(\pi _1, \2H _1 ,\al F_1, G _1)$ and $(\pi _2,\2H
    _2 ,\al F_2 ,G_2)$ for $(\alg{A},\om _0)$ are said
    to be \emph{equivalent} if there exists a unitary
    operator $W:\2H_1\to \2H_2$ such that:
\begin{eqnarray*}
  W\pi _1(A)&=&\pi _2(A)W, \quad \forall A\in \alg{A} ,\\
  WU(G _1)&=&U(G _2)W ,\\
  W\alg{F}_1(O)&=&\alg{F}_2(O)W , \ \mbox{for each double cone} \ O .\end{eqnarray*}
\end{defn}
\end{bxd}

\noindent \bigskip \doublebox{\begin{minipage}{0.9\textwidth}
    \begin{DR} Let $(\al A,\om _0)$ be a net of
      observable algebras satisfying duality and
      property B relative to a privileged `vacuum'
      state $\om _0$.  Then there exists a field system
      with gauge symmetry $\fields$ for $(\al A,\om
      _0)$ that is complete, and that has normal
      commutation relations.  Any complete, normal
      field system for $(\alg{A},\om _0)$ is equivalent
      to $\fields$.
    \end{DR}
  \end{minipage} } \bigskip

The proof of the reconstruction theorem is contained in \cite{DR} and
\cite{dr2}.  In this article, we give an alternative proof, based on
Deligne's embedding theorem \cite{del}, and results obtained by
Roberts \cite{problem} prior to obtaining the full proof of the
reconstruction theorem.

In outline, the theorem shows first --- as was
essentially established in \cite{dhr3} --- that the DHR
superselection sectors naturally have the structure of
a braided tensor $*$-category with conjugates --- and
when the spacetime dimension is three or greater, we
can replace ``braided'' with ``symmetric.''  Now, until
the late 1980's, this first result was merely
suggestive: it is known that the category $\Rep _f G$
of representations of a compact group $G$ on finite
dimensional (super) Hilbert spaces is a symmetric
tensor $*$-category with conjugates.  Hence, the
category of DHR superselection sectors seems to have
all the structure of $\Rep _fG$ for some compact $G$.
By the classical Tannaka-Krein duality theorem, it is
possible to reconstruct $G$ from $\Rep _fG$.
Furthermore, Roberts \cite{problem} proved the
conditional claim that \emph{if} the category of
superselection sectors was equivalent to the category
$\Rep _fG$ for some compact $G$, then the field algebra
$\alg{F}$ could be reconstructed.

But there is a crucial difference between the category of
superselection sectors and the category $\Rep _fG$.  The category
$\Rep _fG$ is ``concrete'' --- it comes with an embedding into the
category of Hilbert spaces, namely the forgetful functor, and hence
its objects can be regarded as structured sets.  It is also precisely
the existence of such an embedding that is needed to construct a field
algebra, because one needs the objects in the category to have
``internal structure,'' as, for example, an object in the category
$\2H$ of Hilbert spaces is a structured set.  Before we state the
embedding theorem, whose proof is given in the Appendix, we need some
preparatory definitions concerning `supermathematics'.

\subsection{Supermathematics and the embedding theorem}

\begin{defn} A \emph{super vector space}, alternatively
  a $\7Z_2$-\emph{graded vector space}, is a vector
  space $V$ with a distinguished decomposition
  $V=V_+\oplus V_-$.  The subspace $V_+$ is called the
  \emph{even} subspace, and $V_-$ is called the
  \emph{odd} subspace.  Elements of $V_+\cup V_-=:V_h$
  are called \emph{homogeneous}.  Define the parity
  function $\om$ on the homogeneous elements by setting
  $\om (v)=\pm 1$ if $v\in V_{\pm}$.  A morphism
  between two super vector spaces is a linear mapping
  $T:V\to W$ such that $T(V_\pm )\subseteq W_\pm $.  We
  let $\SVect$ denote the category of super vector
  spaces.  A \emph{super Hilbert space} is a super
  vector space $V$ with a positive definite inner
  product such that $V_-\perp V_+$.  We use $\2S\2H$ to
  denote the category of super Hilbert spaces.
\end{defn}

We now define operations that make $\SVect$ into a
symmetric tensor category.  It is straightforward to
verify that the set $\Mor (V,W)$ of morphisms between
two super vector spaces is a linear subspace of
$B(V,W)$.  Thus, $\SVect$ is a linear category.

If $V$ and $W$ are super vector spaces, then their
direct sum is the vector space $V\oplus W$ with even
subspace $V_+\oplus W_+$ and odd subspace $V_-\oplus
W_-$.  We define the monoidal product in $\SVect$ as
the vector space $V\otimes W$ whose even and odd
subspaces are defined by
\[ (V\otimes W)_{\sigma }=\bigoplus _{\sigma '\sigma
  ''=\sigma}V_{\sigma '}\otimes W_{\sigma ''} ,\] where
$\sigma =\pm$.  Thus,
\begin{eqnarray*} (V\otimes W)_+ &=& (V_+\otimes W_+)\oplus (V_-\otimes W_-)  ,\\
  (V\otimes W)_- &=& (V_+\otimes W_-)\oplus (V_-\otimes
  W_+) .\end{eqnarray*} The monoidal unit is $\7C$,
with even subspace $\7C$.

\begin{defn} For two super vector spaces $V,W$, we
  define the symmetry isomorphism
  \[ c_{V,W}:V\otimes W\to W\otimes V ,\] by setting
  \[ c_{V,W}(v\otimes w)=(-1)^{(1-\om (v))(1-\om
    (w))/4} w\otimes v ,\qquad \forall v\in V_h,\forall
  w\in W_h. \] on homogeneous simple tensors, and then
  by extending linearly.
\end{defn}

\begin{prop} Both $(\SVect ,\otimes ,\7C,c_{V,W})$ and $(\2S\2H ,\otimes ,\7C
  ,c_{V,W})$ are symmetric tensor $*$-categories.
\end{prop}

\begin{note} By the coherence theorem $\2S\2H$ is equivalent to a strict symmetric
  tensor $*$-category, which we will also denote by $\2S\2H$.  
\end{note}

\begin{defn} A \emph{supergroup} is a pair $(G ,k )$ where $G$ is a group and $k$ is
  a central element in $G$ such that $k\gp k=e$.  A morphism between two supergroups
  $(G _1,k _1)$ and $(G _2,k_2)$ is a group homomorphism $\f:G _1\to G_2$ such that
  $\f(k_1)=k_2$.
\end{defn}

\begin{defn} A \emph{(unitary) representation} $\pi$ of
  a supergroup $(G,k)$ is a super Hilbert space
  $V=V_+\oplus V_-$ together with a (unitary)
  representation $\pi$ of $G$ on $V$ such that
  $\pi(k)|_{V_\pm}=\pm \id _{V_{\pm}}$.  The
  representations $\Rep (G,k)$ of $(G,k)$ form a
  symmetric tensor $*$-category with tensor product and
  symmetry inherited from $\2S\2H$, and monoidal unit
  the trivial representation of $(G,k)$ on
  $\7C$.  \end{defn}

\begin{note} Let $\2S\2H_f$ be the full subcategory of
  finite dimensional super Hilbert spaces.  For a
  supergroup $(G,k)$ we denote by $\Rep _f(G,k)$ the
  full subcategory of finite dimensional
  representations of $(G,k)$.  The categories $\2S\2H
  _f$ and $\Rep _f(G,k)$ are semisimple and have
  conjugates (see the Appendix for more on this
  terminology).  Also, there is a canonical forgetful
  functor $K:\Rep _f(G,k)\to \2S\2H_f$.  \end{note}

We now move on to the statement of the Embedding
Theorem, which will be required for the reconstruction
of the field algebra and gauge group.  For more on
supermathematics, we refer the reader to
\cite{vara,super}.  (But note that DHR superselection
theory is not concerned with supersymmetry in the sense
of a symmetry transforming Bosonic and Fermionic fields
into each other.  Also, our definition of a supergroup
is idiosyncratic.)

\begin{bxd} \begin{emb} Let $\2S\2H_f$ be the category of
    finite-dimensional super Hilbert spaces over $\7C$.  Let $(\2C
    ,\otimes ,\11 ,c_{X,Y})$ be a tensor $C^*$-category with unitary
    symmetry $c_{X,Y}$, conjugates, direct sums, subobjects, and
    irreducible monoidal unit $\11$. (Such a category is called an
    $STC^*$ in the Appendix.)  Then
\begin{enumerate}
\item There is a faithful symmetric tensor $*$-functor $E:\2C\to \2S\2H_f$.
\item There is a compact supergroup $(G,k)$, where $G$ is the group of unitary
  natural monoidal transformations of $E$, and an equivalence $F:\2C \to \Rep
  _f(G,k)$ of symmetric tensor $*$-categories such that $E=F\circ K$, where $K:\Rep
  _f(G,k)\to \2S\2H_f$ is the forgetful functor.
\end{enumerate}
\end{emb}
\end{bxd}

\begin{note}
  The embedding theorem is proven in Appendix B.  In its proof we
  assume the tensor category $\2C$ to be strict and we will work with
  the strictification $\2S\2H$ of the category of super Hilbert
  spaces. In view of the coherence theorem for symmetric tensor
  categories the strictness assumptions do not limit the generality of
  the result. The tensor functor $F:\2C\rarr\2S\2H_f$ that we
  construct will, however, not be a strict tensor functor. In the
  construction of the field net below we do pretend for notational
  simplicity that $F$ is strict. We will comment on this issue again
  at the end of this section.
\end{note}

\subsection{Construction of the field net, algebraic} \label{ss-Falg}

We now apply the Embedding Theorem to the case of the DHR category $\D
_f$ of localized transportable morphisms with finite dimension.  In
particular, we show that given an embedding $E:\D _f\to \2S\2H_f$, it
is possible to construct a local system of field algebras $\fields$.
This strategy of reconstruction is based on the unpublished manuscript
\cite{problem}, which assumes the existence of an embedding (or fiber)
functor.  The actual existence theorem for the embedding functor ---
which is based on the work of Tannaka and Deligne, but incorporates
more recent simplifications --- can be found in the Appendix.

\begin{defn} As a set, the field algebra $\alg{F}_0 $
  consists of equivalence classes of triples $(A,\rho
  ,\psi)$, with $A\in \alg{A}$, $\rho \in \Obj (\D
  _f)$, and $\psi \in E(\rho )$, modulo the equivalence
  relation
$$ (AT,\rho ,\psi )=(A,\rho ',E(T)\psi ) ,$$
for $T\in \Hom (\rho ,\rho ')$.  Since $E(\lambda \id _{\rho})=\lambda \id _{E(\rho
  )}$ we have $(\lambda A,\rho ,\psi )=(A,\rho ,\lambda \psi )$.  Subsequently, we do
not distinguish notationally between a triple $(A,\rho ,\psi )$ and its equivalence
class.
\end{defn}

\begin{prop} $\alg{F}_0 $ is a complex vector space under the operations:
  \begin{equation} \lambda (A,\rho ,\psi )\: := \:(\lambda A,\rho ,\psi ) ,\qquad
    \lambda \in \7C ,\end{equation} and
  \begin{equation} (A_1,\rho _1,\psi_1)+(A_2,\rho _2,\psi _2)\: :=
    \:(A_1W_1^*+A_2W_2^*,\rho ,E(W_1)\psi _1+E(W_2)\psi _2), \label{addition}
  \end{equation} where $\psi _i\in E(\rho _i)$ and $W_i\in \Hom (\rho _i ,\rho )$ are
  isometries with
  \begin{equation} W_1W_1^*+W_2W_2^*=\id _\rho .\end{equation} In addition,  
  \begin{eqnarray*} (A_1,\rho ,\psi )+(A_2,\rho ,\psi )&=&(A_1+A_2,\rho ,\psi ) \\
    (A,\rho ,\psi _1)+(A,\rho ,\psi _2) &=&(A,\rho ,\psi _1+\psi _2) .\end{eqnarray*}
  Therefore, identifying $\alg{A}$ with $\{ (A,\iota ,1):A\in \alg{A}, \, 1\in
  \7C\equiv E(\iota )\}$, $\alg{A}$ becomes a linear subspace of $\alg{F}_0 $; and
  identifying $E(\rho)$ with $\{ (I,\rho ,\psi ):\psi \in E(\rho )\}$, $E(\rho )$
  becomes a linear subspace of $\alg{F}_0$.
\label{linear-space}
\end{prop}

\begin{proof} We first verify that the operations are well defined.
  Scalar multiplication is well defined since for any $T\in \Hom (\rho
  ,\rho')$, $(\lambda A,\rho ',E(T)\psi )=((\lambda A)T,\rho ,\psi
  )=(\lambda (AT),\rho ,\psi )$.  To show that addition is well
  defined, we first establish that Eqn.\ (\ref{addition}) is
  independent of the choice of $W_1$ and $W_2$.  If $W_i'\in \Hom
  (\rho _i,\rho ')$ is another such choice then
  \begin{eqnarray*}
    \lefteqn{ \Bigl( A_1W_1'^*+A_2W_2'^*,\rho ',E(W_1')\psi _1+E(W_2')\psi _2 \Bigr) } \\
    &=& \Bigl( A_1W_1^*+A_2W_2^*(W_1W_1'^*+W_2W_2'^*),\rho ',E(W_1')\psi _1+E(W_2')\psi _2 \Bigr)  \\
    &=& \Bigl(A_1W_1^*+A_2W_2^*,\rho ,E(W_1W_1'^*+W_2W_2'^*)E(W_1')\psi _1+E(W_2')\psi _2 \Bigr) \\
    &=& \Bigl(A_1W_1^*+A_2W_2^*,\rho ,E(W_1)\psi _1+E(W_2)\psi _2 \Bigr) . \end{eqnarray*}  
  To see that addition is independent of equivalence classes, let $T_i\in \Hom (\rho
  _i,\rho _i')$, $W_i$ isometries in $\Hom (\rho _i,\rho )$, and $W_i'$ isometries in
  $\Hom (\rho _i',\rho )$.  Then, 
\begin{eqnarray*}
  \lefteqn{   (A_1T_1,\rho _1,\psi _1)+(A_2T_2,\rho _2,\psi _2) } \\
  &=& \Bigl( A_1T_1W_1^*+A_2T_2W_2^*,\rho ,E(W_1)\psi
  _1+E(W_2)\psi _2\Bigr) \\
  &=& \Bigl( (A_1{W'_1}^*+A_2{W'_2}^*)(W'_1T_1W_1^*+W'_2T_2W_2^*),\rho ,E(W_1)\psi _1+E(W_2)\psi _2\Bigr)
  \\
  &=& \Bigl( A_1{W'_1}^*+A_2{W'_2}^*,\rho ,E(W_1')E(T_1)\psi _1+E(W_2')E(T_2)\psi _2 \Bigr) \\
  &=& ( A_1,\rho _1', E(T_1)\psi _1)+(A_2,\rho _2',E(T_2)\psi _2 ) .\end{eqnarray*}
To prove additivity in the first argument, choose $\sigma =\rho \oplus \rho$, and
$W_i\in \Hom (\rho ,\sigma )$ the corresponding isometries.  Then  
\begin{eqnarray*}
  \lefteqn{ (A_1,\rho ,\psi )+(A_2,\rho ,\psi ) } \\
  & =& (A_1W_1^*+A_2W_2^*,\sigma ,(E(W_1)+E(W_2))\psi )  \\ 
  &=& (A_1W_1^*+A_2W_2^*,\sigma ,E(W_1+W_2)\psi ) \\
  &=& ((A_1W_1^*+A_2W_2^*)(W_1+W_2),\rho ,\psi ) \\
  &=& (A_1+A_2,\rho ,\psi ).\end{eqnarray*}  
Finally, to prove additivity in the second argument, choose $\sigma =\rho \oplus
\rho$, and $W_i\in \Hom (\rho ,\sigma )$ the corresponding isometries.  Then
\begin{eqnarray*}
  \lefteqn{ (A,\rho ,\psi _1)+(A,\rho ,\psi _2) } \\
  & =& (AW_1^*+AW_2^*,\sigma ,E(W_1)\psi _1+E(W_2)\psi _2 )  \\ 
  &=& (A(W_1^*+W_2^*),\sigma ,E(W_1)\psi _1+E(W_2)\psi _2 ) \\
  &=& (A,\rho ,E(W_1^*+W_2^*)(E(W_1)\psi _1+E(W_2)\psi _2 ) ) \\
  &=& (A,\rho ,\psi _1+\psi _2).\end{eqnarray*}  
\end{proof}

\begin{prop} The complex linear space $\alg{F}_0 $ becomes an algebra if we define
  \begin{equation} (A_1,\rho _1,\psi _1)(A_2,\rho _2,\psi _2)\: := \: \Bigl( A_1\rho
    _1(A_2),\rho _1\otimes \rho _2,\psi _1\otimes \psi _2 \Bigr) ,\label{multiply}
  \end{equation} where $\psi _i\in E(\rho _i)$, $i=1,2$.  Furthermore, $\alg{A}$ is a
  subalgebra of $\alg{F}_0 $, and the equivalence class of $(I,\iota ,1)$ is a
  multiplicative identity, where $I$ is the multiplicative identity of $\alg{A}$,
  and $1\in E(\iota )=\7C$.  \end{prop}

\begin{proof} We first verify that Eqn.\ (\ref{multiply}) is well-defined on $\alg{F}_0
  $.  Let $T_i\in \Hom (\rho _i,\rho _i')$.  Recalling that $T_1\times T_2=\rho
  _1'(T_2)T_1$, we have
\begin{eqnarray*}
  (A_1T_1,\rho _1,\psi _1)(A_2T_2,\rho _2,\psi _2) &=& (A_1T_1\rho _1(A_2T_2),\rho
  _1\otimes \rho _2,\psi _1\otimes \psi
  _2) \\
  &=& (A_1\rho _1'(A_2T_2)T_1,\rho _1\otimes \rho _2,\psi _1\otimes \psi _2) \\
  &=& (A_1\rho _1'(A_2)\rho _1'(T_2)T_1,\rho _1\otimes \rho _2,\psi _1\otimes \psi
  _2) \\
  &=& (A_1\rho _1'(A_2)(T_1\times T_2),\rho _1\otimes \rho _2,\psi _1\otimes \psi _2) \\
  &=& (A_1\rho _1'(A_2),\rho _1'\otimes \rho _2',E(T_1\times T_2)(\psi _1\otimes \psi _2)) \\
  &=& (A_1\rho _1'(A_2),\rho _1'\otimes \rho _2',E(T_1)\psi _1\otimes E(T_2)\psi _2) \\
  &=& (A_1,\rho _1',E(T_1)\psi _1)(A_2,\rho _2',E(T_2)\psi _2) .\end{eqnarray*} 
A straightforward calculation shows that multiplication is associative.  For
distributivity, let $W_i\in \Hom (\rho _i,\rho )$.  Then,
\begin{eqnarray*}
  \lefteqn{ \Bigl[ (A_1,\rho _1,\psi _1)+(A_2,\rho _2,\psi _2)\Bigr] (A_3,\rho _3,\psi _3) } \\
  &=& \bigl( (A_1W_1^*+A_2W_2^*)\rho
  (A_3),\rho \otimes \rho _3,(E(W_1)\psi _1+E(W_2)\psi _2)\otimes \psi _3 \bigr) \\
  &=& \bigl( A_1\rho _1(A_3)W_1^{*}+A_2\rho _2(A_3)W_2^*,\rho \otimes \rho _3,(E(W_1)\psi _1)\otimes \psi
  _3+(E(W_2)\psi _2)\otimes \psi _3 \bigr) \\
  &=& \bigl( A_1\rho _1(A_3)(W_1^{*}\times 1_{\rho _3})+A_2\rho _2(A_3)(W_2^*\times 1_{\rho
    _3}), \rho \otimes \rho _3,(E(W_1)\psi _1)\otimes \psi _3+(E(W_2)\psi _2)\otimes \psi _3 ) \bigr) \\
  &=& (A_1,\rho _1,\psi _1)(A_3,\rho _3,\psi _3)+(A_2,\rho _2,\psi _2)(A_3,\rho _3,\psi _3) .\end{eqnarray*}
\end{proof}

We will need the following basic lemma from linear algebra.

\begin{defn} If $H,H'$ are Hilbert spaces and $S\in \Hom (H\otimes H',\7C )$, then we
  define an antilinear mapping $\2JS:H\to H'$ by setting
  \[ ((\2JS)x,x')=S(x\otimes x') ,\qquad \qquad \forall x\in H,\forall
    x'\in H'.\]
\end{defn}

\begin{lemma} \mbox{}
  \begin{enumerate}
  \item $\2J$ is antilinear: $\2J(\lambda S)=\overline{\lambda}(\2JS)$, and
    $\2J(S_1+S_2)=\2JS_1+\2JS_2$.
  \item \label{magic} If $T\in \Hom (H',H)$ then \begin{eqnarray*}
      T\circ (\2JS) &=& \2J(S\circ (I_{H}\otimes T^{*})) ,\\
      (\2JS)\circ T &=& \2J(S\circ (T\otimes I_{H'})) . \end{eqnarray*}
  \item If $S'\in \Hom (H'\otimes H'', \7C )$ then $(\2JS')\circ (\2JS)=(S\otimes
    1_{H''})\circ (1_{H}\otimes S'^*)$.
   \item Let $S_1\in \End (H_1\otimes H_1',\7C )$ and $S_2\in \End (H_2\otimes
     H_2',\7C )$.  Then
   \begin{eqnarray*} (\2JS_2 \otimes \2JS_1)\circ \Sigma _{H_1,H_2} &=&\2J[S_1\circ
     (1_{H_1}\otimes S_2\otimes 1_{H_1'})] .
   \end{eqnarray*}
\end{enumerate} \label{trivial}
\end{lemma}

\begin{proof} Straightforward.  A nice exercise in basic linear
  algebra. \end{proof}

\begin{note} We will apply the previous Lemma to super Hilbert spaces.
  But we will take $\Sigma _{H,H'}$ to be the ordinary symmetry on the
  category $\2H _f$ of finite dimensional Hilbert spaces. \end{note}

\begin{lemma} Let $T\in \Hom (\rho ,\rho ')$ and pick solutions
  $(\overline{\rho},R,\overline{R})$ and ($\overline{\rho}',R',\overline{R}')$ of the
  conjugate equations with respect to $\rho$ and $\rho'$; that is, $R\in \Hom (\iota
  , \overline{\rho}\otimes \rho )$, $\overline{R}\in \Hom (\iota ,\rho \otimes
  \overline{\rho})$ such that $(\overline{R}^*\times I_\rho)\circ (I_\rho \times
  R)=I_{\rho}$, $(R^*\times I_{\overline{\rho}})\circ (I_{\overline{\rho}}\times
  \overline{R})=I_{\overline{\rho}}$, and analogously for $R'$ and $\overline{R}'$.
  Set
\begin{eqnarray*} 
  \overline{T} &:= & (I_{\overline{\rho}}\times \overline{R}'^*)\circ
  (I_{\overline{\rho}}\times T\times 1_{\overline{\rho}'})\circ (R\times
  1_{\overline{\rho}'}) \: =\: \ol\rho (\ol R'^*T)R .\end{eqnarray*}
Then $\overline{T}\in \Hom (\overline{\rho}',\overline{\rho})$ and 
\begin{eqnarray}
  (I_{\overline{\rho}}\times T)\circ R &=& (\overline{T}\times I_{\rho '})\circ R'
  , \label{bill} \\
  (I_\rho \times \overline{T}^*)\circ \overline{R} &=& (T^*\times
  I_{\overline{\rho}'})\circ \overline{R}' \label{bob} .\end{eqnarray}
\end{lemma}

\begin{proof} 
  For Eqn.\ (\ref{bill}), we have
\begin{eqnarray*}
  (\overline{T}\times I_{\rho'})\circ R' &=& \overline{T}R'
  \: =\: \overline{\rho}(\overline{R}'^*T)RR' \: =\:
  \overline{\rho}(\overline{R}'^*T\rho (R'))R \\
  &=& \overline{\rho}(\overline{R}'^*\rho '(R')T)R ,\end{eqnarray*}
where we used the definition of $\times$ for the first equality, the definition of
$\overline{T}$ for the second equality, $R\in \Hom (\iota ,\overline{\rho}\otimes \rho )$ for
the third equality, and $T\in \Hom (\rho ,\rho ')$ for the fourth equality. 
But by the conjugate equations, $\overline{R}'^*\rho '(R')=(\overline{R}'^*\times I_{\rho'})\circ (I_{\rho '}\times
R')=I_{\rho '}=1$, and hence $(\overline{T}\times I_{\rho'})\circ
R'=\overline{\rho}(T)R=(I_{\overline{\rho}}\times T)\circ R$.  For Eqn.\ (\ref{bob}),
we have 
\begin{equation}
  (I_{\rho}\times \overline{T}^*)\circ \overline{R} \: =\: \rho
  (\overline{T}^*)\overline{R} \: =\: \rho (R^*)\rho
  \overline{\rho}(T^*\overline{R}')\overline{R} \: =\: \rho
  (R^*)\overline{R}T^*\overline{R}' ,\end{equation}  
where we used the definition of $\overline{T}$ for the second equality, and
$\overline{R}\in \Hom (\iota ,\rho \otimes \overline{\rho})$ for the third equality.  But by the
conjugate equations $\rho (R^*)\overline{R}=(I_{\rho}\times R^*)\circ
(\overline{R}\times I_{\rho})=I_{\rho}$, and hence $(I_{\rho}\times
\overline{T}^*)\circ \overline{R}=T^*\overline{R}'=(T^*\times
I_{\overline{\rho}'})\circ \overline{R}'$.  
\end{proof}

\begin{prop} \label{prop-starF}
The algebra $\alg{F}_0 $ becomes a $*$-algebra if we define
  \begin{equation} (A,\rho ,\psi )^* \: := \: (R^*\overline{\rho}(A)^*,\ol\rho
    ,\2JE(\overline{R}^*)\psi ) ,\end{equation} where $\psi \in E(\rho )$, and
  $(\overline{\rho},R,\overline{R})$ is a conjugate to $\rho$.
\end{prop}

\begin{proof} We first show that the definition of $*$ is independent of the choice
  of conjugate to $\rho$.  For this, let $(\overline{\rho}_1,R_1,\overline{R}_1)$ be
  any other choice.  Define $W\in \Hom (\ol\rho ,\ol\rho _1)$ by
  \begin{eqnarray} W &:=& (R^*\times I_{\overline{\rho} _1})\circ
    (I_{\overline{\rho}}\times \ol R_1) \: =\: R^*\overline{\rho}(\ol R_1)
  \end{eqnarray} we have by the conjugate equations
\begin{eqnarray*}
  W^{-1} & := & (R_1^*\times I_{\ol\rho}) \circ (I_{\ol\rho_1}\times \ol R) \:
  =\: R_1^*\ol\rho_1 (\ol R) .\end{eqnarray*} Moreover,
\begin{eqnarray*}
  (R_1^*\ol\rho _1(A)^*,\ol\rho _1,\2JE(\ol R_1^*)\psi ) &=& (R^*W^{-1}\ol\rho
  _1(A)^*,\ol\rho _1,\2JE(\ol
  R^*(I_{\rho}\times W^*))\psi )   \\ 
  &=&   (R^*\ol\rho (A)^*,\ol\rho ,E(W^{-1})\2JE(\ol R^*(1_{\rho}\times W^*))\psi ) \\
  &=& (R^*\ol\rho (A)^*,\ol\rho ,\2JE(\ol R^*)\psi ),\end{eqnarray*}
where we used Lemma \ref{trivial}.3 for the final equality.  

To see that the definition of $*$ is independent of equivalence classes, suppose that
$T\in \Hom (\rho ,\rho ')$ and $\psi \in E(\rho )$.  Then
\begin{eqnarray*} (AT,\rho ,\psi )^* &=& (R^*\ol\rho (T^*A^*),\ol\rho ,\2JE(\ol R^*)\psi ) \\
  &=& (R'^*\ol T^*\ol\rho (A^*),\ol\rho ,\2JE(\ol R^*)\psi ) \\
  &=& (R'^*\ol\rho '(A^*)\ol T^*,\ol\rho ,\2JE(\ol R^*)\psi ) \\
  &=& (R'^*\ol\rho '(A^*),\ol\rho ',E(\ol T^*)\2JE(\ol R^*)\psi ) \\
  &=& (R'^*\ol\rho '(A^*),\ol\rho ',\2JE(\ol R^*\circ (I_{\rho }\times \ol T))\psi )\\
  &=& (R'^*\ol\rho '(A^*),\ol\rho ',\2JE(\ol R'^*\circ (T\times I_{\ol\rho '}))\psi ) \\
  &=& (R'^*\ol\rho '(A^*),\ol\rho ',\2JE(\ol R'^*)E(T)\psi ) \\
  &=& (A,\rho ',E(T)\psi )^* , \end{eqnarray*} where we used Eqn.\ (\ref{bill}) for the
second equality, the fact that $\ol{T}^*\in \Hom (\ol\rho ,\ol\rho ')$ for the third
equality, Lemma (\ref{trivial}.\ref{magic}) for the fifth equality, and Eqn.\ (\ref{bob})
for the sixth equality.

We verify that $*$ is involutive:
\begin{eqnarray*}
  (A,\rho ,\psi )^{**} &=& (R^*\ol\rho (A)^*,\ol\rho ,\2JE(\ol R^*)\psi )^* \\
  &=& (\ol R^*\rho (\ol\rho (A)R),\rho,\2JE(R^*)\2JE(\ol R^*)\psi ) \\
  &=& (A\ol R^*\rho (R),\rho ,\2JE(R^*)\2JE(\ol R^*)\psi ) \\
  &=& (A,\rho ,E((R^*\times I_{\ol\rho})(1_{\ol\rho}\times \ol R))\psi ) \\
  &=& (A,\rho ,\psi ) ,
\end{eqnarray*}
where we used Lemma \ref{trivial}.3 for the penultimate equality, and the conjugate
equations for the final equality.

To verify that $*$ is antilinear, let $W_i\in \Hom (\rho _i,\rho )$.  Then,
\begin{eqnarray} \lefteqn{ \bigl[ (A_1,\rho _1,\psi _1)+(A_2,\rho _2,\psi _2)\bigr]
    ^* \:=\: \bigl(
    A_1W_1^*+A_2W_2^*,\rho ,E(W_1)\psi _1+E(W_2)\psi _2 \bigr) ^* } \nonumber \\
  &=& (R^*\ol\rho (W_1A_1^*+W_2A_2^*),\ol\rho ,\2JE(\ol R^*)(E(W_1)\psi _1+E(W_2)\psi
  _2)) .\label{inv} \end{eqnarray} But we may take $R=(\ol W_1\times W_1)\circ
R_1+(\ol W_2\times W_2)\circ R_2$, $\ol R=(W_1\times \ol W_1)\circ \ol R_1+(W_2\times
\ol W_2)\circ \ol R_2$, where $\ol W_i\in \Hom (\ol\rho _i,\ol\rho )$ are isometries,
$\ol W_1\ol W_1^*+\ol W_2\ol W_2^*=I_{\ol\rho}$.  Then Eqn.\ (\ref{inv}) becomes
\begin{eqnarray*} 
  \lefteqn{  \left[ (A_1,\rho _1,\psi _1)+(A_2,\rho _2,\psi _2)\right] ^* } \\
  &=& (R_1^*\ol\rho
  _1(A_1^*)\ol W_1^*+R_2^*\ol\rho _2(A_2^*)\ol W_2^*,\ol\rho ,\2JE(\ol R_1^*(I_{\rho _1}\times
  \ol W_1^*))\psi _1+\2JE(\ol R_2^*(I_{\rho _2}\times \ol W_2^*))\psi _2) \\
  &=& (R_1^*\ol \rho _1(A_1^*)\ol W_1^*+R_2^*\ol \rho _2(A_2^*)\ol W_2^*,\ol\rho ,E(\ol
  W_1)\2JE(\ol R_1^*)\psi _1+E(\ol W_2)\2JE(\ol R_2^*)\psi _2) \\
  &=& (A_1,\rho _1,\psi _1)^*+(A_2,\rho _2,\psi _2)^* ,\end{eqnarray*} using
Lemma \ref{trivial} for the second equality.

Finally, we show that $[(A_1,\rho _1,\psi _1)(A_2,\rho _2,\psi _2)]^*=(A_2,\rho
_2,\psi _2)^*(A_1,\rho _2,\psi _1)^*$.  If $\rho =\rho _1\otimes \rho _2$ and $\rho
'=\rho _1'\otimes \rho _2'$ then we may take $R=(I_{\ol\rho _2}\times R_1\times
I_{\rho _2})\circ R_2$ and $\ol R=(I_{\rho _1}\times \ol R_2\times I_{\ol\rho
  _1})\circ \ol R_1$.  Thus,
\begin{eqnarray*}
  \lefteqn{ \left[ (A_1,\rho _1,\psi _1)(A_2,\rho _2,\psi _2)\right] ^*  \: =\: (A_1\rho
    _1(A_2),\rho _1\otimes \rho _2,\psi _1\otimes \psi  _2)^* } \\
  &=& (R_2^*\ol\rho _2(R_1^*)\ol\rho _2\ol\rho _1(\rho
  _1(A_2^*)A_1^*),\ol\rho _2\otimes \ol\rho _1,\2JE(\ol
  R_1^*\circ (I_{\rho _1}\times \ol R_2^*\times I_{\ol\rho _1}))\psi _1\otimes \psi
  _2 ) \\
  &=& (R_2^*\ol\rho _2(A_2^*)\ol\rho _2(R_1^*\ol\rho _1(A_1^*)),\ol\rho _2\otimes
  \ol\rho _1,\2JE(\ol R_1^*\circ
  (I_{\rho _1}\times \ol R_2^*\times I_{\ol\rho _1}))\psi _1\otimes \psi _2) \\
  &=& (R_2^*\ol\rho _2(A_2^*)\ol\rho _2(R_1^*\ol\rho _1(A_1^*)),\ol\rho _2\otimes \ol\rho_1,\2JE(\ol R_2^*)\psi
  _2\otimes \2JE(\ol R_1^*)\psi _1) \\
  &=& (R_2^*\ol\rho _2(A_2^*),\ol\rho _2 ,\2JE(\ol R_2^*)\psi _2)(R_1^*\ol \rho
  _1(A_1^*), \ol\rho _1,\2JE(\ol
  R_1^*)\psi _1) \\
  &=& (A_2,\rho _2,\psi _2)^*(A_1,\rho _1,\psi _1)^* ,\end{eqnarray*}
where the third equality follows from the fact that $R_1^*\in \Hom (\ol\rho _1\otimes
\rho _1,\iota )$, and the fourth equality follows by Lemma \ref{trivial}.4. 
\end{proof}

\begin{prop} Let $E:\D _f\to \2S\2H_f$ be the embedding functor from the DHR
  category $\D _f$ into the strictified category $\2S\2H_f$ of finite dimensional
  super Hilbert spaces.  Then the formula 
  \begin{equation} \a _g(A,\rho ,\psi )\: =\: (A,\rho
    ,g_\rho \psi ),\qquad \qquad A\in \alg{A},\psi \in
    E(\rho ). \label{automorphism} \end{equation}
  defines a group isomorphism $g\mapsto \a _g$ from the
  intrinsic group $G$ of $E$ into $\Aut
  _{\alg{A}}\alg{F}_0$, the group of $*$-automorphisms
  of $\alg{F}_0$ leaving $\alg{A}$ pointwise
  fixed.  \label{group-isomorphism} \end{prop}

\begin{proof} Since $g$ is a natural monoidal transformation, $g _\iota =\id
  _{E(\iota )}=\id _{\7C }$.  For any $g\in G$, $\alpha _g$ is well defined on
  $\alg{F}_0$ since for $S\in \Hom (\rho ,\rho ')$,
  \begin{eqnarray*} \alpha _g(AS,\rho ,\psi )&=& (AS,\rho ,g_\rho \psi ) \:=\:
    (A,\rho ', E(S)g_\rho \psi ) \\
    &=& (A,\rho ', g(\rho ')E(S)\psi ) \: =\: \alpha _g(A,\rho ',E(S)\psi ).
  \end{eqnarray*} Since $g _\iota =\id _{\7C}$, $\alpha _g$ leaves $\alg{A}\subset
  \alg{F}_0$ pointwise fixed.  Each $g_\rho $ is linear so $\alpha _g$ is linear.
  \begin{eqnarray*} (A_1,\rho _1,g _{\rho _1}\psi _1)(A_2,\rho _2,g_{\rho _2}\psi _2)
    &=& \Bigl( A_1\rho _1(A_2),\rho _1\otimes \rho _2,(g _{\rho _1}\otimes g_{\rho
      _2})(\psi _1\otimes \psi _2) \Bigr) ,\end{eqnarray*} but $g_{\rho _1\otimes
    \rho _2}=g_{\rho _1}\otimes g_{\rho _2}$ so
  \begin{eqnarray*} (A_1,\rho _1,g_{\rho _1}\psi _1)(A_2,\rho _2,g_{\rho _2}\psi _2)
    &=& \Bigl( A_1\rho _1(A_2),\rho _1\otimes \rho _2,g_{\rho _1\otimes \rho _2}(\psi
    _1\otimes \psi _2) \Bigr) .\end{eqnarray*} Thus,
  \begin{equation} \label{homo} \alpha _g(F_1)\alpha _g(F_2)=\alpha _g(F_1F_2)
    .\end{equation}

To show that $\alpha _g$ is a $*$-homomorphism, recall that
\begin{equation} (A,\rho ,g_\rho \psi )^*=(R^*\ol\rho (A)^*,\ol\rho ,\2JE(\ol
  R^*)g_\rho \psi ).\label{get} \end{equation} If $\ol\psi \in E(\ol\rho )$ then
$E(\ol R^*)(g_\rho \psi \otimes g_{\ol\rho }\ol\psi )=E(\ol R^*)((g_\rho \otimes
g_{\ol\rho})(\psi \otimes \ol\psi ))$.  Furthermore,
$$ E(\ol R^*)(g_\rho \psi \otimes g _{\ol\rho }\ol\psi )=g _\iota E(\ol R^*)(\psi
\otimes \ol\psi )=E(\ol R^*)(\psi \otimes \ol\psi ).$$ Hence $g _{\ol \rho
}^*\2JE(\ol R^*)g_\rho =\2JE(\ol R^*)$ and since $g _{\ol\rho }$ is unitary we get
from (\ref{get}),
$$ (A,\rho ,g_\rho \psi )^*=(R^*\ol\rho (A)^*,\ol\rho ,\2JE(\ol R^*)g_\rho \psi )=(R^*\ol\rho
(A)^*,\ol\rho ,g _{\ol\rho }\2JE(\ol R^*)\psi ) ,$$ so \begin{equation} \alpha
  _g(F^*)=\alpha _g(F)^* ,\qquad F\in \alg{F}. \label{star} \end{equation}

Equations (\ref{homo}), (\ref{star}) show that $\alpha _g$ is a $*$-homomorphism, its
inverse is clearly $\alpha _{g^{-1}}$ so $\alpha _g$ defined by Eqn.\
(\ref{automorphism}) is an element of $\Aut _{\alg{A}}\alg{F}$.  The mapping $g\mapsto
\alpha _g$ is clearly a homomorphism.  

Since $G$ is a compact group, for every $g\neq e$, there exists a $(H, \pi ) \in \Rep
_fG$ such that $\pi (g)\neq \id _H$.  Since the functor $E$ is an equivalence, in
particular essentially surjective, there exists a $\rho \in \Obj (\D _f)$ such that
$E(\rho )$ is isomorphic to $(H,\pi )$.  Thus there exists $\psi \in E(\rho )$ such
that $$ \pi (g)\psi =g_{\rho}\psi \neq \psi .$$ Defining $F=(I,\rho ,\psi )$, we have
$\a _g(F)\neq F$.  This proves injectivity of $g\mapsto \a _g$. 

It remains to show that $G\mapsto \Aut _{\alg{A}}\alg{F}_0$ is onto.  Let $\alpha \in
\Aut _{\alg{A}}\alg{F}_0$, $A\in \alg{A}$ and $\psi \in E(\rho )\subset \alg{F}_0$.
Let $\Psi =(I,\rho ,\psi )$.  Then
$$ (\alpha (\Psi ))A = \alpha (\Psi A) = \alpha (\rho (A)\Psi )=\rho (A)\alpha (\Psi
).$$ It is easily checked that this implies that $\alpha (\Psi )$ is of the form
$(I,\rho ,\psi ')$ with $\psi '\in E(\rho )$.  Thus $\psi \mapsto \psi '$ is a linear
map of $E(\rho )$ into $E(\rho )$ which we denote by $g_\rho $, and it remains to
show that $g=(g_\rho )_{\rho \in \D _f}$ is monoidal natural transformation of $E$.
For $S\in \Hom (\rho ,\rho ')$, we have
$$ (S,\rho ,g_\rho \psi )=\alpha (S,\rho ,\psi )=\alpha (I,\rho ',E(S)\psi )=(I,\rho' ,g_{\rho '}E(S)\psi
) .$$ Hence
$$ E(S)g_\rho \psi =g_{\rho '}E(S)\psi ,\qquad \psi \in E(\rho ).$$
That is,
$$ E(S)g_\rho =g_{\rho '}E(S),$$
and $g\in \Nat E$.  To check monoidality, choose arbitrary $\psi _i\in E(\rho _i)$
and let $\Psi _i=(I,\rho _i,\psi _i)$.  Then,
\begin{eqnarray*} g_{\rho _1\otimes \rho _2}(\psi _1\otimes \psi _2) &=& \alpha (\Psi
  _1\Psi _2)\:=\:\alpha (\Psi _1)\alpha (\Psi _2)\:=\:(g_{\rho _1}\otimes g_{\rho
    _2})(\psi _1\psi _2 ) .
\end{eqnarray*} Thus,
$g\in \Nat _{\otimes}E$.  It remains to show that $g$ is unitary.  For $\psi ,\psi '\in
E(\rho )$ and $\Psi =(I,\rho ,\psi ),\Psi '=(I,\rho ,\psi ')$ we have
$$ \big\langle g_\rho \psi ,g_\rho \psi '\big\rangle _{E(\rho )}I =\alpha (\Psi )^*\alpha (\Psi ') =\alpha (\Psi
^*\Psi ') =\bigl\langle \psi ,\psi '\bigr\rangle _{E(\rho )}I ,$$ where the first and
last equalities follow from Prop.\ \ref{ip}.  Hence $g_\rho $ is unitary for each
$\rho \in \Obj (\Delta _f)$.  Therefore every $\a \in \Aut _{\alg{A}}\alg{F}_0$ is of
the form $\a _g$ with $g\in G=\Nat _{\otimes }E$.
\end{proof}

\begin{defn} Given a double cone $O$, we define $\alg{F}_0 (O)$ to consist of those
  elements $F$ in $\alg{F}_0 $ such that there exists $A\in \alg{A}(O)$, $\rho \in
  \Obj (\D _f )$ localized in $O$, and $\psi \in E(\rho )$ with $F=(A,\rho ,\psi )$.
\end{defn}

\begin{prop} $\alg{F}_0 (O)$ is a $*$-subalgebra of $\alg{F}_0 $.  \end{prop}

\begin{proof} Let $F_1=(A_1,\rho _1,\psi _1)$ and $F_2=(A_2,\rho _2,\psi _2)$ be in
  $\alg{F}_0 (O)$.  Thus, the $A_i$ can be chosen from $\alg{A}(O)$ and the $\rho _i$
  can be chosen localized in $O$.  Since $\rho _1(\alg{A}(O))\subseteq \alg{A}(O)$,
  it follows that 
$$ F_1F_2 = (A_1\rho _1(A_2),\rho _1\otimes \rho _2,\psi _1\otimes \psi _2) ,$$
is also in $\alg{F}_0 (O)$.  By transportability, $\ol\rho$ can be chosen localized in
$O$, and in this case $\ol\rho \otimes \rho$ is localized in $O$.  By Lemma
\ref{twine}, $R\in \alg{A}(O)$.  Hence,
$$ F^*=(R^*\ol\rho (A)^*,\ol\rho ,\mathcal{J}E(\ol R^*)\psi ) ,$$
is in $\alg{F}_0 (O)$.  Similarly, $\alg{F}_0 (O)$ is closed under the addition defined in
Prop.\ \ref{linear-space} since $\rho$ can also be chosen localized in $O$, and then
the isometries $W_1,W_2$ are in $\alg{A}(O)$ (by Lemma \ref{twine}).
\end{proof}

\begin{prop} The action of $G$ on $\alg{F}_0$ leaves $\alg{F}_0(O)$ globally fixed.
\end{prop}

\begin{proof} If $F\in \alg{F}_0(O)$ then $F=(A,\rho ,\psi )$ for some $A\in
  \alg{A}(O)$ and $\rho$ localized in $O$.  Then clearly $\a _g(F)=(A,\rho ,g_\rho
  \psi )$ is in $\alg{F} _0(O)$.  \end{proof}

\begin{note} Having defined an action of the supergroup $(G,k)$, the element $k\in G$
  induces a $\7Z _2$ grading on $\alg{F}_0$ and on the local algebras
  $\alg{F}_0(O)$.  

\end{note}

\begin{prop} The field net $\alg{F}_0$ satisfies normal commutation relations.  That
  is, if $O_1$ and $O_2$ are spacelike, and $F_i\in \alg{F}(O_i)$ are such that
  $$ \alpha _k(F_{i})=\sigma _iF_i ,$$ then
  $$ F_1F_2= (-1)^{(1-\sigma _1)(1-\sigma _2)/4}F_2F_1 .$$ \label{normality}
\end{prop}

\begin{proof} Choose $F_i=(A_i,\rho _i,\psi _i)$ with $A_i\in \alg{A}(O_i)$ and $\rho
  _i$ localized in $O_i$.  Then $A_1A_2=A_2A_1$, $\rho _1(A_2)=A_2$, $\rho
  _2(A_1)=A_1$, and $\ve _{\rho _1,\rho _2}=\id _{\rho _1\otimes \rho _2}$.  In view
  of the way $G$ acts on $\alg{F}_0$ we have 
  \begin{eqnarray*} \sigma _i(A_i,\rho _i,\psi _i) =\a _k(A_i,\rho _i,\psi _i) =
    (A_i,\rho _i,k_{\rho _i}\psi _i ) ,\end{eqnarray*} and hence $k_{\rho _i}\psi
  _i=\sigma _i\psi _i$.  That is, $\psi _i$ is homogeneous and $\om (\psi _i)=\sigma
  _i$.  Furthermore, since $E$ is a symmetric functor $E(\ve _{\rho _1,\rho
    _2})=\Sigma _{E(\rho _1),E(\rho _2)}$, where $\Sigma _{H,H'}$ is the symmetry on
  $\2S\2H_f$ and therefore
$$  \Sigma _{H,H'}(\psi _1\otimes \psi _2)=(-1)^{(1-\sigma _1)(1-\sigma _2)/4}(\psi
_2\otimes \psi _1) .$$ Hence
\begin{eqnarray*} F_1F_2 &=& (A_1\rho _1(A_2),\rho _1\otimes \rho _2,\psi _1\otimes  \psi _2)  \\ 
  &=& (A_1A_2 \ve _{\rho _2,\rho _1}, \rho _1\otimes \rho _2 ,\psi _1\otimes \psi_2) \\
  &=& (A_1A_2,\rho _2\otimes \rho _1,E(\ve _{\rho _2,\rho _1})(\psi _1\otimes \psi_2)) \\
  &=& (A_2\rho _2(A_1),\rho _2\otimes \rho _1, E(\ve _{\rho _2,\rho _1})(\psi_1\otimes \psi _2)) \\
  &=& (A_2\rho _2(A_1) ,\rho _2\otimes \rho _1,\Sigma _{E(\rho _2),E(\rho _1)}(\psi _1\otimes \psi _2)) \\
  &=& (-1)^{(1-\sigma _1)(1-\sigma _2)/4}(A_2\rho _2(A_1),\rho _2\otimes \rho _1,\psi_2\otimes \psi _1 ) \\
  &=& (-1)^{(1-\sigma _1)(1-\sigma _2)/4} F_2F_1 . \end{eqnarray*}
\end{proof}

\begin{prop} For all $\Psi =(I,\rho ,\psi),\Psi '=(I,\rho ,\psi ')$ with $\psi ,\psi
  '\in E(\rho)$ we have
  \begin{eqnarray}
    \Psi A = \rho (A)\Psi ,   \label{induce} \\
    \Psi ^*\Psi ' = (\psi ,\psi ')I .  
    \label{cuntz}  \end{eqnarray}
  For any orthonormal basis $\{ \psi _i:i=1,\dots, n\}$ of $E(\rho )$, we have 
  \begin{equation} \sum _{i=1}^{n}\Psi _i\Psi _i^* = I .
  \label{basisz} \end{equation}
  \label{ip} \end{prop}

\begin{proof} 
$$   (I,\rho ,\psi )(A,\iota ,1)=(\rho (A),\rho ,\psi )=(\rho (A),\iota ,1)(I,\rho ,\psi ) ,$$ whence
(\ref{induce}).  For (\ref{cuntz}), we check:
\begin{eqnarray*} (I,\rho ,\psi )^*(I,\rho ,\psi ') &=& (R^*,\ol\rho \otimes \rho ,(\2JE(\ol R^*)\psi )\otimes \psi ') \\
  &=& (I,\iota, E(R^*)((\2JE(\ol R^*)\psi )\otimes \psi ')) .\end{eqnarray*} Since
$\2JE(\ol R^*):E(\rho )\to E(\ol \rho)$ and $E(R^*):E(\ol\rho )\otimes E(\rho )\to
\7C$, it follows that $E(R^*)((\2JE(\ol R^*)\psi )\otimes \psi ')$ is a complex
number.  In fact, by the definition of $\2J$ and Lemma \ref{trivial}.3,
\begin{eqnarray*} E(R^*)((\2JE(\ol R^*)\psi )\otimes \psi ') &=& \bigl\langle
  \2JE(R^*)\circ \2JE(\ol R^*)\psi ,\psi
  '\bigr\rangle _{E(\rho )} \\
  &=& \bigl\langle \2JE((R^*\times I_\rho )\circ (I_\rho \times \ol R^*))\psi ,\psi
  '\bigr\rangle _{E(\rho )} \\
  & =& \bigl\langle \psi ,\psi '\bigr\rangle _{E(\rho )} ,\end{eqnarray*} where the
final equality follows by the conjugate equations.  So, combining the previous two
equations we have
$$ (I,\rho ,\psi )^*(I,\rho ,\psi ') \: = \: \bigl( I,\iota ,\langle \psi ,\psi
'\rangle _{E(\rho )}) \: = \: \langle \psi ,\psi '\rangle _{E(\rho )} \,\bigl(
I,\iota ,1\bigr) .$$ For Eqn.\ (\ref{basisz}), we have
\begin{eqnarray*} \lefteqn{ \sum _i(I,\rho ,\psi _i )(I,\rho ,\psi _i)^* = \left(
      \rho (R)^*,\rho \otimes \ol\rho ,\sum
      \psi _i\otimes \2JE(\ol R^*)\psi _i \right) } \\
  &&= (\rho (R)^*,\rho\otimes \ol\rho ,E(\ol R)1 ) = (\rho (R)^*\ol R,\iota, 1) \:
  =\: (I,\iota ,1) ,\end{eqnarray*} where the second equality follows from the
definition of $\2J$ and the final equality follows by the conjugate equations.
\end{proof}

\subsection{Completion of the field net}

We now construct a representation $(\2H ,\pi )$ of the
$*$-algebra $\alg{F}_0 $, and show that $\pi
|_{\alg{A}}$ has a nontrivial subrepresentation
equivalent to the GNS representation induced by the
vacuum state $\om _0$.  We do so by extending the state
$\om _0$ from $\alg{A}$ to $\alg{F}_0 $, and then by
taking the GNS representation.  In order to extend the
state $\om _0$ from $\alg{A}$ to $\alg{F}_0 $, it
suffices to show that there is a positive linear map
$m:\alg{F}_0 \to \alg{A}$.

\begin{note} Let $\rho \in \Obj (\D _f)$.  Since $\D _f$ is semisimple (see Prop.\
  \ref{semisimple}), $\rho$ is a finite direct sum $\rho =\rho _1\oplus \cdots \oplus
  \rho _n$ of irreducible objects in $\Obj (\D _f)$.  Therefore, there is a
  projection $P_{\iota}^{\rho}\in \End (\rho )$ onto the direct sum of those
  irreducibles in this decomposition that are isomorphic to $\iota$.  \end{note}

\begin{prop} Given $(A,\rho ,\psi )\in \alg{F}_0$, define
  \begin{equation} m(A,\rho ,\psi ) \: := \: (AP_{\iota}^{\rho},\rho ,\psi ) .
  \end{equation} Then $m:\alg{F}_0\to \alg{A}$ is a faithful positive linear
  projection from $\alg{F}_0$ onto $\alg{A}$.  Further,
  \begin{eqn} m(AF)=Am(F) ,\qquad A\in \alg{A},F\in \alg{F}_0  . \label{module}
  \end{eqn} \label{conditional} \end{prop}

\begin{proof} We first show that $m$ is well defined.  If $T\in \Hom (\rho ,\rho ')$
  then $TP_{\iota}^{\rho}=P_{\iota}^{\rho '}TP_{\iota}^{\rho}=P_{\iota}^{\rho'}T$,
  hence
$$ m(AT,\rho ,\psi )=(ATP_{\iota}^{\rho},\rho ,\psi )=(AP_{\iota}^{\rho '}T,\rho ,\psi )=(AP_{\iota}^{\rho'},\rho ',E(T)\psi
)=m(A,\rho ',E(T)\psi ) ,$$ as required.  $m$ is clearly linear and satisfies Eqn.\
(\ref{module}).  We now show that $m$ is positive.  First, since $\rho$ has finite
dimension, $\rho$ contains at most finitely many copies of the vacuum representation.
Thus, $P_{\iota}^{\rho}=\sum _iS_iS_i^*$ where $S_i\in \Hom (\iota ,\rho )$ and
$S_i^*S_j=\delta _{ij}\id _\iota $.  Thus,
$$ m(A,\rho ,\psi )=(AP_{\iota}^{\rho},\rho ,\psi )=\sum _i(AS_i,\iota ,E(S_i^*)\psi) .$$
However, $E(S_i^*)\psi =\lambda _i1$ so that
$$ m(A,\rho ,\psi )=\sum _i\lambda _i(AS_i,\iota ,1) \in \alg{A} .$$

Since each $\rho \in \Obj (\D _f )$ is a finite direct sum of irreducible objects
(Prop.\ \ref{semisimple}), any $F\in \alg{F}_0 $ may be written as a finite sum
$F=\sum _iF_i$, $F_i=(A_i,\rho _i,\psi _i)$, where $\psi _i\in E(\rho _i)$ with $\rho
_i$ irreducible and pairwise inequivalent.  Thus,
$$ m(F^*F)= \sum _{i,j}m(F_i^*F_j)=\sum _im(F_i^*F_i) .$$
Hence, to show that $m$ is positive and faithful, it suffices to consider $m(F^*F)$
with $F=(A,\rho ,\psi )$, $\psi \in E(\rho)$ and $\rho$ irreducible.  In this case,
$$ (A,\rho ,\psi )^*(A,\rho ,\psi )=\bigl( R^*\ol\rho (A^*A),\ol\rho \otimes \rho ,\2JE(\ol R^*)(\psi \otimes \psi )\bigr).$$
Using $P^{\ol\rho \otimes \rho }_{\iota}=\norm{RR^*}^{-1}RR^*=d(\rho )^{-1}RR^*$, we
have
\begin{eqnarray*} d(\rho )m(F^*F) &=& (R^*\ol\rho (A^*A)RR^*,\ol\rho \otimes \rho
  ,\2JE(\ol R^*)\psi \otimes \psi ) \\
  &=& (R^*\ol\rho (A^*A)R,\iota ,E(R^*)\2JE(\ol R^*)\psi \otimes \psi )
  .\end{eqnarray*} Now,
$$ E(R^*)\2JE(\ol R^*)(\psi \otimes \psi)=\bigl\langle \2JE(R^*)\2JE(\ol R^*)\psi
,\psi \bigr\rangle _{E(\rho )} ,$$ hence by Lemma \ref{trivial},
\begin{eqnarray*} d(\rho )m(F^*F) &=& R^*\ol \rho (A^*A)R \, \bigl\langle E((\ol
  R^*\times I _{\rho})\circ
  (I_{\rho}\times R))\psi ,\psi \bigr\rangle _{E(\rho )} \\
  &=& R^*\ol \rho (A^*A)R \,\bigl\langle \psi ,\psi \bigr\rangle _{E(\rho )}
  .\end{eqnarray*} Thus, $m(F^*F)\geq 0$ and $m(F^*F)=0$ implies $\psi =0$ or $\ol
\rho (A)R=0$.  But $\ol \rho (A)R=0$ only if
$$ 0=\ol R^*\rho \ol \rho (A)\rho (R) = A\ol R^*\rho (R) = A.$$
Thus $m(F^*F)=0$ implies $F=0$, and $m$ is a faithful positive linear projection from
$\alg{F}_0 $ onto $\alg{A}$.  \end{proof}

\begin{lemma} Let $P_0^{\rho}$ be the projection in $\End (E(\rho ))$ onto the
  subspace of $G$ invariant vectors with respect to the action $\pi _\rho
  (g)=g_\rho$.  Then $E(P^{\rho }_{\iota})=P_0^{\rho}$.  Furthermore, the conditional
  expectation $m$ is $G$-invariant, i.e.\ $m(\a _g(F))=m(F)$ for all $g\in G$ and
  $F\in \alg{F}_0$.  \label{invariant} \end{lemma}

\begin{proof} Recall that if $(H, \pi )$ is an irreducible representation of a
  compact group $G$ and $\pi$ is not the trivial representation, then $H$ contains no
  $G$-invariant vectors.  If $\rho =\bigoplus \rho _i$ with $\rho _i$ irreducible,
  then the previous observation implies that the $G$-invariant vectors in $E(\rho )$
  are precisely those in the image of $E(P_{\iota}^{\rho})$.  Thus
  $E(P_{\iota}^{\rho})=P_0^{\rho}$, implying $m(F)=\a _g(m(F))$.  Furthermore,
  \begin{eqnarray*} m\alpha _g(A,\rho ,\psi ) &=& m(A,\rho ,g_\rho \psi )\:=\:
    (AP_{\iota}^{\rho},\rho ,g_\rho \psi ) =(A,\rho ,P_{0}^{\rho}g_\rho \psi  ) \\
    &=& (A,\rho ,g_\rho P_{0}^{\rho}\psi ) = (A,\rho ,P_0^{\rho}\psi ) \\ &=&
    (AP_{\iota}^{\rho},\rho ,\psi ) = m(A,\rho ,\psi ).\end{eqnarray*}
\end{proof}

In view of Prop.\ \ref{conditional}, $\om _0\circ m$ is
a faithful state on the $*$-algebra $\al F_0$.  Let
$(\2H ,\pi )$ be the GNS representation of $\al F$
induced by $\om _0\circ m$, let $\al F$ be the norm
closure of $\pi (\al F_0)$, and let $\al F(O)$ be the
\emph{weak} closure of $\pi (\al F_0(O))$.  It is clear
that $\al F$ is the $C^*$-inductive limit of the net
$O\mapsto \al F(O)$.  Since $\om _0\circ m$ is
$G$-invariant by Lemma \ref{invariant}, there is a
unitary representation $U$ of $G$ on $\2H$ implementing
the automorphisms $\a _g$ of $\alg F_0$:
\[ \pi (\a _g(F))= U(g)\pi (F)U(g)^* ,\qquad g\in
G,F\in \alg{F}_0 ,\] and therefore it extends to $\al
F$.  Since $g\mapsto \a _g$ is injective, $U$ is
injective.

\begin{defn} \label{cond-exp}
Let $\sigma \in \hat{G}$ be an irreducible character of $G$.  Define a
  map $\4E _{\sigma}$ on $\bh$ by
\[ \4E _{\sigma}(A)=\int _{G}\overline{\sigma (g)}U(g)AU(g)^{*}\, d\mu (g) ,\] where
$\mu$ is the Haar measure on $U(G)$.  \end{defn}

\begin{note} Let $F=(A,\rho ,\psi )\in \alg{F}_0$.  Since the $U(g)$ implements $\a
  _g$ we have
  \begin{eqnarray} \4E _{\sigma}(\pi (F)) &=& \int _{G} \overline{\sigma (g)}\Bigl(
    \pi (\a _g(A, \rho ,\psi ))\Bigr)d\mu (g) = \int _G \overline{\sigma (g)}\pi
    (A,\rho ,g_\rho \psi )d\mu (g) \nonumber \\ &=& \pi (A,\rho
    ,P^{\rho}_{\sigma}\psi ) ,\label{eqn-projection} \end{eqnarray} where
  $P_{\sigma}^{\rho}\in \End (E(\rho ))$ is the orthogonal projection onto the
  subspace transforming according to the irreducible representation $\sigma$.  Since
  $G$ is compact, $\4E _\sigma$ is strongly continuous. Note that $\4E _{0}(\pi
  (F))=\pi [m(A,\rho ,\psi )]$. \end{note}

\begin{lemma} $\alg{F}_0(O)\alg{A}=\alg{F}_0$. \label{cyclic} \end{lemma}

  \begin{proof} Let $(A,\rho ,\psi )\in \alg{F}_0$.  Since $\rho$ is transportable,
    there is a unitary $T\in \Hom (\rho ,\rho ')$ with $\rho '$ localized in $O$.
    Then
$$ (A,\rho ,\psi )=(AT^*,\rho ',E(T)\psi )=(AT^*,\iota ,1)(I,\rho ',E(T)\psi )=BF ,$$
where $B\in \alg{A}$ and $F\in \alg{F}_0(O)$. Hence $\alg{A}\alg{F}_0(O)=\alg{F}_0$.
Since $\alg{A},\alg{F}_0(O)$ and $\alg{F}_0$ are $*$-algebras,
$\alg{F}_0(O)\alg{A}=\alg{F}_0$.  \end{proof}

\begin{thm} $\fields$ is a field system with gauge
  symmetry for $(\alg{A},\om _0)$ with normal
  commutation relations (in the sense of Definitions
  \ref{field-system} and
  \ref{normal-commutation}). \end{thm}

\begin{proof} It is obvious that $\al F(O)$ is a
  $G$-stable von Neumann subalgebra of $\al F$.  Also
  the net $O\mapsto \al F(O)$ satisfies normal
  commutation relations.  We now run through the
  individual conditions in Definition
  \ref{field-system}.

  \begin{enumerate}

  \item[($\g$)] We need to show that the fixed
    point algebra of $\al F(O)$ under the $G$ action is
    $\pi (\al A(O))$.  First note that $\4E (\pi (\al
    F_0(O)))=\pi (m(\al F_0(O)))$.  Thus,
\[ \al F(O)^{G}=\4E (\al F(O))=\4E \Bigl(\overline{\pi
  (\al F_0(O))}\Bigr)=\overline{\4E (\pi (\al
  F_0(O)))}= \overline{\pi (m(\al F_0(O)))
}=\overline{\pi (\al A(O))} .
\] The third equality follows by the normality of
$\4E$, and the last equality is due to the fact that
$m$ is a conditional expectation from $\al F_0$ to $\al
A$.


\item[($\d$)] Let $j:\al F_0\to \2H$ be the
  inclusion mapping derived from the GNS representation
  of $\om _0\circ m$.  Since $\overline{j(\alg{A})}=\2H
  _0$ we have
$$ \overline{\alg{F}(O)\2H _0}=\overline{\pi (\alg{F}_0(O))\2H
  _0}=\overline{\pi
  (\alg{F}_0(O))j(\alg{A})}=\overline{j(\alg{F}_0(O)\alg{A})}=\overline{j(\alg{F}_0)}=\2H
.$$

\item[($\ve$)] Let $O_1$ and $O_2$ be spacelike separated.  The
  subalgebra $\alg{A}(O _1)$ of $\alg{F}_0$ is pointwise invariant
  under the gauge transformations. In particular, $\a _k(A)=A$ for all
  $A\in \alg{A}(O)$, i.e.\ elements of $\alg{A}(O_1)$ are purely
  Bosonic.  Therefore relative locality follows by normality of the
  commutation relations (Prop.\ \ref{normality}).

\end{enumerate}

Now we claim that $\Aut _{\alg{A}}\alg{F}=G$.  By Eqn.\
(\ref{eqn-projection}), $\4E _\sigma (\pi (\alg{F}_0))$
is isomorphic as a Banach space to $\alg{A}\otimes
P_{\sigma}^{\rho}E(\rho )$, and so is a closed subspace
of $\alg{F}$, and so
$$ \4E _\sigma (\alg{F})=\4E _\sigma (\overline{\pi (\alg{F}_0)})=\overline{\4E
  _{\sigma}(\pi (\alg{F}_0))}= \4E _\sigma (\pi (\alg{F}_0)) .$$ Since for any $F\in
\alg{F}$ we have $F=\sum _{\sigma \in \hat{G}}\4E _\sigma (F)$, and $\4E _\sigma
(F)\in \pi (\alg{F}_0)$, it follows that an element $F\in \alg{F}$ is in $\pi
(\alg{F}_0)$ if and only if $\4E _{\sigma}(F)\neq 0$ for only finitely many $\sigma
\in \hat{G}$.  Together with linearity of $\a$, this implies that $\a (\pi
(\alg{F}_0))\subseteq \pi (\alg{F}_0)$.  Thus there exists a $g\in G$ such that $\a
|_{\pi (\alg{F}_0)}=\a _g$ (by Prop.\ \ref{group-isomorphism}).  Since $\a _g$ is
continuous and $\pi (\alg{F}_0)$ is dense in $\alg{F}$, $\a$ is the unique extension
of $\a _g$ to $\alg{F}$.
\end{proof}

\subsection{Poincar\'e covariance of the field net}
Covariance considerations have played no prominent role
in the DHR theory of Section \ref{sec-DHR} or in the
above reconstruction of a field net $\6F$. We now show
that the latter is Poincar\'e covariant if the
underlying DHR sectors are. (Recall from Remark
\ref{note-cov2} that under favorable circumstances we
have $\Delta_{fc}=\Delta_f$.)

\begin{thm} If in the construction of the field net $\6F$ we start from the category $\Delta_{fc}$
instead of $\Delta_f$, the field net constructed above is covariant under an automorphic action of
$\widehat{P}$. This action is implemented by a positive energy representation on the GNS
representation space of $\6F$ corresponding to the state $\omega_0\circ m$.  
\end{thm}

\begin{proof} Let $\beta_h=\mathrm{Ad}\,U(h)$ be the action of $\2P$ on $\6A$.
Recall from Note \ref{note-cov1} that $\rho_h=\beta_h\circ\rho\circ\beta_h^{-1}$ and 
$X_\rho(h)\equiv U(h)U_\rho(h)^*\in\Hom(\rho,\rho_h)$ for all $h\in\widehat{\2P}$.
We define an action $\widehat{\beta}$ of $\widehat{\2P}$ on $\6F_0$ by  
\begin{eqnarray} \widehat{\beta}_h((A,\rho,\psi)) &\equiv& (\beta_h(A),\rho_h,E(X_\rho(h))\psi) \nonumber\\
   &=& (\beta_h(A)X_\rho(h),\rho,\psi) =(U(h)AU_\rho(h)^*,\rho,\psi). \label{eq-beta}
\end{eqnarray}
Let $\rho,\rho'\in\Delta_{fc}$ and $T\in\Hom(\rho,\rho')$. Then $\beta_h(T)\in\Hom(\rho_h,\rho'_h)$,
and $TU_\rho(h)=U_{\rho'}(h)T$, cf.\ Section \ref{ss-cov}. Thus,
\begin{eqnarray*} \beta_h(T)X_\rho(h) &=& (U(h)TU(h)^*)(U(h)U_\rho(h)^*)=U(h)TU_\rho(h)^* \\
   &=& U(h)U_{\rho'}(h)T=X_{\rho'}(h)T, \end{eqnarray*}
Using this equation, we compute
\begin{eqnarray*} \widehat{\beta}_h((AT,\rho,\psi)) &=& (\beta_h(AT),\rho_h,E(X_\rho(h))\psi) 
   =  (\beta_h(A),\rho'_h,E(\beta_h(T)X_\rho(h))\psi)  \\
   &=&  (\beta_h(A),\rho'_h,E(X_{\rho'}(h)T)\psi) =  \widehat{\beta}_h((A,\rho',E(T)\psi)),
\end{eqnarray*}
thus $\widehat{\beta}_g$ is well defined. Let $i:A\mapsto(A,\iota,\11)$ be the inclusion of $\6A$ in
$\6F$. Then $\widehat{\beta}_h\circ i=i\circ\beta_g$, thus $\widehat{\beta}_g$ extends $\beta_g$.
If $F\in\6F(O)$ then there exists a representation
$F=(A,\rho,\psi)$ with $A\in\6A(O)$ and $\rho\in\Delta(O)$. Now it is evident from the definition
that $\widehat{\beta}_h(F)\in\6F(hO)$. That $g\mapsto\widehat{\beta}_g$ is a group homomorphism
is obvious from the r.h.s.\ of Eqn. (\ref{eq-beta}). Now,
\begin{eqnarray*} \lefteqn{ \widehat{\beta}_g((A_1,\rho_1,\psi_1)(A_2,\rho_2,\psi_2)) =
   \widehat{\beta}_g((A_1\rho_1(A_2),\rho_1\rho_2,\psi_1\otimes\psi_2)) } \\
  &&= (U(h)A_1\rho_1(A_2)U_{\rho_1\rho_2}(h)^*,\rho_1\rho_2,\psi_1\otimes\psi_2)  \\
  &&= (\beta_h(A_1)\rho_{1,h}(\beta_h(A_2))U(h)U_{\rho_1\rho_2}(h)^*,\rho_1\rho_2,\psi_1\otimes\psi_2) \\
  &&= (\beta_h(A_1)\rho_{1,h}(\beta_h(A_2))X_{\rho_1\rho_2}(h),\rho_1\rho_2,\psi_1\otimes\psi_2) \\
  &&= (\beta_h(A_1)\rho_{1,h}(\beta_h(A_2))X_{\rho_1}(h)\rho_1(X_{\rho_2}(h)),\rho_1\rho_2,\psi_1\otimes\psi_2) \\
  &&= (\beta_h(A_1)X_{\rho_1}(h)\rho_1(\beta_h(A_2)X_{\rho_2}(h)),\rho_1\rho_2,\psi_1\otimes\psi_2) \\
  &&= (U(h)A_1U_{\rho_1}(h)^*\rho_1(U(h)A_2U_{\rho_2}(h)^*),\rho_1\rho_2,\psi_1\otimes\psi_2) \\
  &&= (U(h)A_1U_{\rho_1}(h)^*,\rho_1,\psi_1)(U(h)A_2U_{\rho_2}(h)^*,\rho_2,\psi_2) \\
  &&= \widehat{\beta}_g((A_1,\rho_1,\psi_1))\widehat{\beta}_g((A_2,\rho_2,\psi_2)),
\end{eqnarray*}
where the fifth equality is due to Eqn.\ (\ref{eq-Xtens}). Thus $\widehat{\beta}_g$ is an algebra
homomorphism.   

Let $\rho\in\Delta_{fc}$ and choose a conjugate $(\ol{\rho},R,\ol{R})$. Since the trivial morphism
$\iota$ is covariant with $X_\iota=\id_\iota$, applying Eqn.\ (\ref{eq-beta}) with
$T=R^*\in\Hom(\ol{\rho}\rho,\iota)$ we get
$R^*=\beta_h(R^*)X_{\ol{\rho}\rho}(h)=\beta_h(R^*)X_{\ol{\rho}}(h)\ol{\rho}(X_\rho(h))$, 
where we used Eqn.\ (\ref{eq-Xtens}) again. This is equivalent to
\begin{equation}\label{eq-abc} R^*\overline{\rho}(X_\rho(h)^*)=\beta_h(R^*)X_{\overline{\rho}}(h),
\end{equation}
which will be used below. Now
we compute
\begin{eqnarray*}
 (\widehat{\beta}_h(A,\rho,\psi))^* &=&  (U(h)AU_\rho(h)^*,\rho,\psi)^* \\
   &=& (R^*\overline{\rho}(U(h)AU_\rho(h)^*)^*,\ol\rho,\2JE(\overline{R}^*)\psi) \\
   &=& (R^*\overline{\rho}(U_\rho(h)A^*U(h)^*),\ol\rho,\2JE(\overline{R}^*)\psi) \\
   &=& (R^*\overline{\rho}(U_\rho(h)U(h)^*\beta_h(A^*)),\ol\rho,\2JE(\overline{R}^*)\psi) \\
   &=& (R^*\overline{\rho}(X_\rho(h)^*\beta_h(A^*)),\ol\rho,\2JE(\overline{R}^*)\psi) \\
   &=& (\beta_h(R^*)X_{\overline{\rho}}(h)\overline{\rho}(\beta_h(A))^*,\ol\rho,\2JE(\overline{R}^*)\psi) \\
   &=& (U(h)R^*U_{\overline{\rho}}(h)^*\overline{\rho}(\beta_h(A))^*,\ol\rho,\2JE(\overline{R}^*)\psi) \\
   &=& (U(h)R^*\overline{\rho}(A)^*U_{\overline{\rho}}(h)^*,\ol\rho,\2JE(\overline{R}^*)\psi) \\
   &=& \widehat{\beta}_h((R^*\overline{\rho}(A)^*,\ol\rho,\2JE(\overline{R}^*)\psi)) \\
   &=& \widehat{\beta}_h((A,\rho,\psi)^*),
\end{eqnarray*}
thus $\widehat{\beta}_h$ is a $*$-homomorphism. (In the sixth equality we used Eqn.\ (\ref{eq-abc}).)

In view of 
\begin{eqnarray*} \widehat{\beta}_h((A,\rho,\psi))  &=&  (U(h)AU_\rho(h)^*,\rho,\psi) \\
    \alpha_g((A,\rho,\psi)) &=& (A,\rho,\pi_{E(\rho)}(g)\psi) \end{eqnarray*}
it is clear that $\widehat{\beta}_h\circ\alpha_g=\alpha_g\circ\widehat{\beta}_h$ for all 
$g\in G, h\in\widehat{\2P}$. In view of $\pi\circ m=\2E_0\circ\pi$, we have
$\omega_0\circ\beta_h\circ m=\omega_0\circ m$. 
Thus the vacuum state of $\6F$ is $\widehat{\2P}$-invariant, and $\widehat{\2P}$ is unitarily
implemented in the GNS representation. 
\end{proof}

\subsection{Uniqueness of the field net}
In the present section we have shown that, given a fiber functor $E:\Delta_f(\6A)\rarr\2S\2H$, there
exists a field net with normal commutation relations that is complete, i.e.\ creates all
representations in $\D_f(\6A)$ from the vacuum. We call this the Roberts field net and denote it by
$\6F^R_E$. We first consider the dependence of this construction on the functor $E$.

\begin{prop}
Let $E_1,E_2:\Delta_f\rarr\2H$ be two fiber functors. Then the Roberts field nets
$\6F^R_{E_1},\6F^R_{E_2}$ constructed from them are unitarily equivalent.
\end{prop}

\begin{proof} By Theorem \ref{theor-uniq1} from the
  appendix, there exists a unitary monoidal natural
  isomorphism $\alpha: E_1\rarr E_2$. Based on this we
  define a map $\gamma:\6F^R_{0,1}\rarr\6F^R_{0,2}$ by
  $\gamma:(A,\rho,\psi)\mapsto(A,\rho,\alpha_\rho\psi)$. This
  makes sense since $\psi\in E_1(\rho)$ and
  $\alpha\in\Hom(E_1(\rho),E_2(\rho))$. $\gamma$ is
  well defined since, for $T\in\Hom(\rho,\rho')$, we
  have
\begin{eqnarray*} \gamma(AT,\rho,\psi)&=&(AT,\rho,\alpha_\rho\psi)=(A,\rho',E_2(T)\circ\alpha_\rho\psi)\\
  &=& (A,\rho',\alpha_\rho\circ E_1(T)\psi))=\gamma(A,\rho',E_1(T)\psi). \end{eqnarray*}
That $\gamma$ is an algebra homomorphism follows from
\begin{eqnarray*} \lefteqn{ \gamma((A_1,\rho_1,\psi_1))\gamma((A_2,\rho_2,\psi_2)) =
   (A_1,\rho_1,\alpha_{\rho_1}\psi_1)(A_2,\rho_2,\alpha_{\rho_2}\psi_2) } \\
   &&= (A_1\rho_1(A_2),\rho_1\otimes\rho_2,\alpha_{\rho_1}\psi_1\otimes\alpha_{\rho_2}\psi_2) 
   = (A_1\rho_1(A_2),\rho_1\otimes\rho_2,\alpha_{\rho_1\otimes\rho_2}(\psi_1\otimes\psi_2)) \\
   &&= \gamma((A_1\rho_1(A_2),\rho_1\otimes\rho_2,\psi_1\otimes\psi_2)) 
   = \gamma((A_1,\rho_1,\psi_1)(A_2,\rho_2,\psi_2)),
\end{eqnarray*}
where we have used monoidality $\alpha_{\rho_1\otimes\rho_2}=\alpha_{\rho_1}\otimes\alpha_{\rho_2}$
of $\alpha$. Since an inverse can be obtained using the natural isomorphism $\alpha^*$, $\gamma$ is
an isomorphism between the field algebras $\6F^R_{0,1}$ and $\6F^R_{0,2}$. It clearly respects the
local structure, i.e.\ maps $\6F^R_{0,1}(O)$ to $\6F^R_{0,2}(O)$.

Next we claim that $m_2\circ\gamma=\gamma\circ m_1$, where $m_1,m_2$ are the projections defined
earlier. Namely, 
\[ m_2\circ\gamma((A,\rho,\psi)) = m_2((A,\rho,\alpha_\rho\psi))
   =(AP^\rho_\iota,\rho,\alpha_\rho\psi)  =\gamma((AP^\rho_\iota,\rho,\psi))
  =\gamma\circ m_1((A,\rho,\psi)).
\]
This implies that the states $\omega_0\circ m_1$ and $\omega_0\circ m_2\circ\gamma$ on $\6F^R_{0,1}$
coincide, and therefore the isomorphism $\gamma:\6F^R_{0,1}\rarr\6F^R_{0,2}$ extends to a unitary
equivalence of the norm completions in the GNS representations.
\end{proof}

In order to study an arbitrary complete normal field net $\6F$, not a priori of the form
$\overline{\6F^R}^{\|\cdot\|}$, we use the following

\begin{prop}  \label{prop-FtoE}
Let $\6F$ be a complete normal field net for the observable net $\6A$. Then there exists a strict
tensor functor $E_\6F:\D_f(\6A)\rarr\2S\2H_f$ to the category of finite dimensional super Hilbert
spaces. On the objects, $E_\6F$ is given by the vector space 
\[ E_\6F(\rho):=\{ F\in\6F\ | \ F\pi_0(A)=\pi_0(\rho(A))F \ \ \forall A\in\6A\}. \]
The inner product is given by $\langle F,F'\rangle\11=F^*F'$ and the $\7Z_2$-grading by the action
of $k\in G$. For irreducible $\rho,\rho'\in\D_f$,
we have $E(\ve(\rho,\rho'))=\pm \sum_{i,j}\psi'_i\psi_j{\psi'_i}^*\psi_j^*$,
where  $\{\psi_i, i=1,\ldots,d(\rho)\}$ and $\{\psi'_i, i=1,\ldots,d(\rho')\}$ are orthonormal
bases of $E(\rho)$ and $E(\rho')$, respectively, and the minus sign appears iff $\rho$ and $\rho$
are both fermionic. 
\end{prop}

\begin{proof} (In this proof we write $E$ instead of $E_\6F$.) For $s\in\Hom(\rho,\rho')$ we define
$E(s)=\pi_0(s)\in\6F$. For $F\in E(\rho)$ we have  
$\pi_0(s)F\pi_0(A)=\pi_0(s)\pi_0(\rho(A))F=\pi_0(s\rho(A))F=\pi_0(\rho'(A)s)F=\pi_0(\rho'(A))\pi_0(s)F$
for all $A\in\6A$, thus $\pi_0(s)F\in E(\rho')$ and $E$ is a functor. If $F,F'\in E(\rho)$ then
$F^*F'\in\6F\cap\6A'=\7C\11$, allowing us to define $\langle F,F'\rangle\11=F^*F'$.
Let $s\in\Hom(\rho,\rho')$ and $F\in E(\rho),F'\in E(\rho')$. Then 
\[ \langle F',E(s)F\rangle=\langle F',\pi_0(s)F\rangle={F'}^*\pi_0(s)F
  =(\pi_0(s)^*F')^*F =\langle\pi_0(s^*)F',F\rangle=\langle E(s^*)F',F\rangle, \]
where we have used that $\pi_0$ is $*$-preserving, shows that $E$ is $*$-preserving.
By Section \ref{sec-localfield} we have $E(\rho)E(\rho')=E(\rho\otimes\rho')$. If
$S_i\in\Hom(\rho_i,\rho_i'), \ F_i\in E(\rho_i)$ then 
\[ E(S_1\times
S_2)F_1F_2=\pi_0(S_1\rho_1(S_2))F_1F_2=\pi_0(S_1)F_1\pi_0(S_2)F_2
\in E(\rho_1'\otimes\rho_2'), \] thus $E(S_1\times
S_2)=E(S_1)\otimes E(S_2)$, thus $E$ is a strict tensor
functor.  Completeness of the field net together with
the discussion in Section \ref{sec-localfield} implies
that $E$ is faithful and satisfies $\dim
E(\rho)=d(\rho)$. (The latter follows also by
Proposition \ref{prop-stand2} of the appendix.)
Finally, let $F\in E(\rho), F'\in E(\rho')$ be of norm
one. Now let $\rho,\rho'\in\D_f$, and let $\psi_i,
i=1,\ldots,d(\rho)$ and $\psi'_i, i=1,\ldots,d(\rho')$
be orthonormal bases of $E(\rho)$ and $E(\rho')$,
respectively. Then
\[ \tilde{c}(\rho,\rho')=\sum_{i,j}\psi'_i\psi_j{\psi'_i}^*\psi_j^* \]
is in $\6F^G$ and independent of the chosen bases. Furthermore,
$\tilde{c}(\rho,\rho')\in\Hom(\rho\otimes\rho',\rho'\otimes\rho)$. The functoriality of $E$ that was
proven above implies that $\tilde{c}(\rho,\rho')$ is natural in both arguments. If now
$\{\rho''\in\D_f\}$ and $\psi_k'', k=1,\ldots,d(\rho'')$ is an orthonormal basis in $E(\rho'')$,
then $\{\psi'_j\psi''_k\}$ is an orthonormal  basis in $E(\rho'\otimes\rho'')$, thus 
\begin{eqnarray*}
  \tilde{c}(\rho,\rho'\otimes\rho'')&=&
\sum_{i,j,k}\psi'_j\psi_k''\psi_j{\psi''_k}^*{\psi'_j}^*\psi_i^* \\
  &=&\sum_{i,j,m,k,l} \psi'_m (\psi''_i\psi_j{\psi''_i}^*\psi_j^*) {\psi'_m}^*
    (\psi'_k\psi_l{\psi'_k}^*\psi_l^*) \\
   &=& \id_{\rho'}\otimes\tilde{c}(\rho,\rho'')\circ\tilde{c}(\rho,\rho')\otimes\id_{\rho''},
\end{eqnarray*}
which is one of the braid relations. One easily sees that
$\tilde{c}(\rho,\rho')\tilde{c}(\rho',\rho)=\11$, thus $\tilde{c}(\cdot,\cdot)$ is a symmetry for
the tensor category $\Delta_f$. If $\rho$ and $\rho'$ are irreducible and localized spacelike to
each other, the normal commutation relations of the corresponding fields imply that
$\tilde{c}(\rho,\rho)=\pm 1$, where the minus sign occurs iff $\rho$ and $\rho''$ are
fermionic. Now, for irreducible $\rho,\rho'$ define $c(\rho,\rho')=\pm c(\rho,\rho')$, where we take
the minus sign iff $\rho$ and $\rho'$ are fermionic, and extend $c$ to reducible objects by
naturality. Then $c(\rho,\rho')=1$ whenever $\rho,\rho'$ are localized spacelike. Now it follows
from the uniqueness result Proposition \ref{uniqsymm} that $E(\ve(\rho,\rho'))=c(\rho,\rho')$. 
Thus $E_\6F$ is a symmetric tensor functor in the sense that it maps the symmetry $\ve$ of $\D_f$ to
the symmetry $c$ of the category $\2H$ of Hilbert spaces. Equivalently, $E$ is a symmetric tensor
functor into the category of super Hilbert spaces equipped with the symmetry $\widetilde{c}$.
\end{proof}

Thus every complete normal field net $\6F$ gives rise to a strict symmetric $*$-preserving fiber
functor $E_\6F$. Denoting by $\6F^R_{E_{\6F}}$ the Roberts field net associated to the latter, our
aim is to construct an isomorphism $\6F\cong\6F^R_{E_{\6F}}$. 

\begin{thm} Let $\6F$ be a complete normal field net for $\6A$ and $E_\6F:\Delta_f\rarr\2S\2H$ the fiber
functor from Proposition \ref{prop-FtoE}. Then there is a unitary equivalence
$\6F_{E_{\6F}}\rarr\6F$ of field nets.
\end{thm}

\begin{proof} By Proposition \ref{prop-FtoE}, there is
  a symmetric $*$-preserving fiber functor
  $E_\6F:\Delta_f\rarr\2S\2H$. By the concrete Tannaka
  theorem (Theorem \ref{theor-T2} of the appendix), the
  compact group $G_{E_\6F}$ of unitary monoidal natural
  transformations of $E$ is unitarily represented on
  the spaces $E_\6F(\rho)$. On the other hand, the
  compact group $G$ coming with our field net $\6F$
  also acts on these spaces, providing a homomorphism
  $G\rarr G_{E_\6F}$. This homomorphism is injective
  since $G$ is concretely given as a group of unitaries
  on the Hilbert space $H$ where $\6F$ lives. It is
  also surjective, since otherwise $\pi|\6A$ would
  contain representations that are not in $\Delta_f$,
  contradicting the assumption that $\6F$ is a complete
  field net. Thus the given group $G$ can be identified
  with the one reconstructed from the fiber functor
  $E_\6F$. For every $\sigma\in\widehat{F}$ we define a
  projection $\2E_\sigma$ on $\6F$ as in Definition
  \ref{cond-exp}. We denote by $\6F_0$ the algebraic
  direct sum
  $\oplus_{\sigma\in\widehat{G}}\2E_\sigma(\6F)$, which
  is the same as $\{ F\in\6F\ | \ \2E_\sigma(F)=0\
  \mbox{for almost all}\ \sigma\in\widehat{G}\}.$

We now define a map $\gamma:\6F^R_{E_{\6F},0}\rarr\6F$ by 
$\gamma:(A,\rho,\psi)\mapsto\pi_0(A)\psi$. At first sight, this formula looks strange, but it makes
perfect sense since $\psi\in E_\6F(\rho)$, where $E_\6F(\rho)$ by definition is a subspace of $\6F$.
As usual, $\gamma$ is well defined since, for $T\in\Hom(\rho,\rho')$,
\[ \gamma((AT,\rho,\psi))=\pi_0(AT)\psi=\pi_0(A)E_\6F(T)\psi   =\gamma((A,\rho',E_\6F(T)\psi)). \]
Furthermore,
\begin{multline*}  \gamma((A_1,\rho_1,\psi_1)(A_2,\rho_2,\psi_2)) =
  \gamma((A_1\rho_1(A_2),\rho_1\rho_2,\psi_1\otimes\psi_2)) 
   =\pi_0(A_1\rho_1(A_2))\psi_1\psi_2  \\ 
    = \pi_0(A_1)\psi_1\pi_0(A_2)\psi_2 
   = \gamma((A_1,\rho_1,\psi_1))\gamma((A_2,\rho_2,\psi_2)), 
\end{multline*}
where we have used $\psi_1\in E_\6F(\rho_1)=\{F\in\6F\ | \ F\pi_0(A)=\pi_0(\rho(A))F\}$. Thus
$\gamma$ is an algebra homomorphism. This, together with $(A,\rho,\psi)=(A,\iota,1)(\11,\rho,\psi)$
implies that $\gamma$ is a $*$-homomorphism provided $\gamma(F^*)=\gamma(F)^*$ for
$F=(\11,\rho,\psi)$. Now, using the $*$-operation on $\6F^R$ defined in Proposition
\ref{prop-starF}, we have 
\[ \gamma((\11,\rho,\psi)^*)=\gamma((R^*,\ol\rho,(\2JE(\overline{R}^*))\psi))
  = \pi_0(R^*)(\2JE(\overline{R}^*)\psi). \]
On the other hand, $\gamma((1,\rho,\psi))^*=\psi^*$, thus $\gamma$ is a $*$-homomorphism provided 
$\psi^*=R^*(\2JE(\overline{R}^*))\psi$ holds for all $\psi\in E(\rho)$.

Now, for any $\ol{\psi}\in E(\ol{\rho})$, we have
$R^*\ol{\psi}\rho(A)=R^*\ol{\rho}\rho(A)\ol{\psi}=AR^*\ol{\psi}$, 
thus $(R^*\ol{\psi})^*\in E(\rho)$. Applying this to
$\ol{\psi}=\2JE(\overline{R}^*)\psi\in E(\ol{\rho})$, we see that 
$\psi^*=R^*(\2JE(\overline{R}^*))\psi$ holds iff
$\psi^*\psi'=R^*(\2JE(\overline{R}^*)\psi)\psi'$ for all $\psi'\in E(\rho)$.

By Proposition \ref{prop-stand2} of the Appendix, $(E(\rho),E(R),E(\ol{R}))$ is a conjugate of
$E(\rho)$ in the category of Hilbert spaces. (Or super Hilbert spaces. This doesn't matter since we
don't use the symmetry.) Thus there are bases $\{e_i\},\{f_i\}$ of $E(\rho)$ and $E(\ol{\rho})$,
respectively, with dual bases $\{\widehat{e_i}\},\{\widehat{f_i}\}$ in
$\widehat{E(\rho)},\widehat{E(\ol{\rho})}$ such that  
\[ E(R) = \sum_i f_i\otimes e_i, \quad E(\ol{R})=\sum_i e_i\otimes f_i, \quad 
  E(R)^* = \sum_i \widehat{f_i}\otimes\widehat{e_i}, \quad E(\ol{R})^*=\sum_i\widehat{e_i}\otimes\widehat{f_i}.
\]
Thus, for $\psi\in E(\rho),\ol{\psi}\in E(\ol{\rho})$, we have
\[ \langle\2JE(\ol{R}^*)\psi,\ol{\psi}\rangle
   =\left(\sum_i\widehat{e_i}\otimes\widehat{f_i}\right)(\psi\otimes\ol{\psi})
   =\sum_i \widehat{e_i}(\psi)\widehat{f_i}(\ol{\psi})
\]
and therefore $\2JE(\ol{R}^*)\psi=\sum_i\overline{ {\widehat{e_i}(\psi)}} f_i$. Thus
\[ E(R)^*((\2JE(\ol{R}^*)\psi)\otimes\psi')=(\sum_i\widehat{f_i}\otimes\widehat{e_i})(
   \sum_j\ol{  {\widehat{e_j}(\psi)}}f_j\otimes\psi')=\sum_i\ol{ {\widehat{e_i}(\psi)}}\widehat{e_i}(\psi')
  =\langle\psi,\psi'\rangle.
\]
Now, in $\6F$, the left hand side equals $R^*(\2JE(\ol{R}^*)\psi)\psi'$ and the right hand side
equals $\psi^*\psi'$, proving the desired identity $\psi^*=R^*(\2JE(\overline{R}^*))\psi$.

Now, for $(A,\rho,\psi)\in \2F^R_{E_{\6F},0}$ is is clear that $\gamma((A,\rho,\psi))$ is contained
in a finite dimensional $G$-stable subspace of $\6F$ and thus in $\6F_0$. Every $F\in\6F_0$ is a sum
of finitely many terms of the form $\2E_\sigma(F)$ with $\sigma\in\widehat{G}$. Picking an
irreducible subspace $H_\sigma$ of isometries in $\6F_0$ transforming according to the class
$\sigma$, there is an endomorphism $\rho\in\Delta_f$ induced by the subspace $H_\sigma$. 
Since every $F\in \2E_\sigma(\6F)$ is a linear combination $\sum_i A_i\psi_i$ with 
$A_i\in\6A, \psi_i\in H_\sigma$, we have $F=\gamma(\sum_i(A,\rho,\psi_i)$, proving
$\gamma(\6F^R_{E_\6F,0})=\6F_0$. 

Let $(A,\rho,\psi)\in\6F^R_0$. By construction of $\6F_0$, we have a finite sum representation
$(A,\rho,\psi)=\sum_i(A_i,\rho_i,\psi_i)$, where the $\rho_i$ are irreducible and mutually
non-isomorphic. Now $\gamma((A,\rho,\psi))=\sum_i A_i\psi_i$, where the spaces
$E(\rho_i)\subset\6F$ transform under mutually inequivalent irreducible representations of $G$.
Thus $\gamma((A,\rho,\psi))=0$ iff $A_i\psi_i=0$ for all $i$. 
isometries transforming according to a representation in the class $\sigma$.
Since by harmonic analysis, every $F\in\6F$ has a unique representation of the form 
$F=\sum_\sigma A_{\sigma,i}\psi^\sigma_i$, this implies that for each $i$ we have
$(A_i,\rho_i,\psi_i)=0$. Thus $\gamma$ is injective.

We have thus proven that $\gamma:\6F^R_{E_\6F,0}\rarr\6F_0$ is an isomorphism. Since the vacuum state
$\omega_0^\6F=(\Omega_0,\cdot\Omega)$ of $\6F$ is by assumption gauge invariant, the states
$\omega_0^\6F\circ \gamma$ and $\omega_0^\6A\circ m$ on $\6F^R_{E_\6F,0}$ coincide, implying that
the completed nets are unitarily equivalent in their GNS representations.
\end{proof}

\begin{cor}
Every complete normal field net $\6F$ is unitarily equivalent to a Roberts field net $\6F^R_E$,
where it doesn't matter which fiber functor $E$ we use. 
\end{cor}

\begin{note}
As promised, we return to the issue of strictness of the functor $F:\2C\rarr\2S\2H_f$ that was
assumed in the construction of the field net, but not proven in the appendix. In the latter, we
constructed a non-strict fiber functor, i.e.\ a functor $E:\2C\rarr\2S\2H_f$ together with natural
isomorphisms $d^E_{\rho,\rho'}: E(\rho)\otimes E(\rho')\rarr E(\rho\otimes\rho')$ and
$e^E:\11_{\2S\2H}\rarr E(\iota_\Delta)$ satisfying Eqns.\ (\ref{eq-A1}), (\ref{eq-A2}).
The construction of the (algebraic) field algebra $\6F_0$ in Subsection \ref{ss-Falg} can easily be
generalized to this situation: The product of fields is defined by
\[ (A_1,\rho _1,\psi _1)(A_2,\rho _2,\psi _2)\: := \: \Bigl(
A_1\rho_1(A_2),\rho _1\otimes \rho _2, d^E_{\rho_1,\rho_2}(\psi
_1\otimes \psi _2) \Bigr) \] and the unit is
$(\11,\iota,e^E1_\7C)$. Now associativity and the unit property are
obvious consequences of Eqns.\ (\ref{eq-A1}), (\ref{eq-A2}). The rest
of the constructions and proofs goes through as before, just carrying
the unitaries $d^E, e^E$ along. An interesting consequence of this and
of Proposition \ref{prop-FtoE} is that we can prove the existence of a
strict fiber functor $E':\Delta_f\rarr\2S\2H'$, where $\2S\2H'_f$ is a
strictification of the category of finite dimensional super Hilbert
spaces. This is consistent with strictification results in category
theory. (Strictification of tensor categories is nicely treated in
\cite[Chap.\ XI]{kas}, but for strictification of tensor functors the
best reference remains \cite[Sect.\ 1]{stjo}.)
\end{note}

\subsection{Further relations between $\alg{A}$ and $\alg{F}$, and a Galois interpretation}
In Section \ref{sec-localfield} we have discussed at
length the structure of the superselection sectors of a
net $\alg{A}$ of observables in relation to the
harmonic analysis of the action of a (global) gauge
group on a field net $\alg{F}$. Note that we did not
claim that all DHR representations of the fixed point
net $\alg{A}=\alg{F}^G$ are connected to the vacuum
representation by the fields in $\alg{F}$. In order to
see that this is in general false, consider a theory
$\alg{A}$ with non-trivial DHR-category and take
$\alg{F}:=\alg{A}$ as `field net', acted upon by the
trivial group $G=\{e\}$. Obviously, all DHR
representations of $\alg{A}$ are not created by the
action of $\alg{F}$ on $H_0$. In the special case where
$\alg{F}$ is Bosonic and itself satisfies all the
requirements on an observable net, it may have
non-trivial DHR sectors. Restricting a DHR
representation $\pi$ of $\alg{F}$ with $d(\pi)<\infty$
to $\alg{A}$, one obtains a DHR representation of
$\alg{A}$ of the same dimension, which therefore
decomposes into a finite direct sum of irreducibles. If
$\pi$ is irreducible and inequivalent to the vacuum
representation $\pi_0$ of $\alg{F}$, then all the
irreducible representations of $\alg{A}$ obtained in
this way are disjoint from those contained in
$\pi_0|\alg{A}$. We refrain from a further analysis of
this issue.  We do, however, wish to point out that one
can specify conditions on a net $\alg{F}$ implying that
all DHR representations of $\alg{A}$ are contained in
$\pi_0|\alg{A}$. This involves the net-cohomology or
local 1-cohomology developed by J.E. Roberts and
reviewed, e.g., in \cite[\S 3.4]{rob-lec}. We refrain
from even attempting to give precise statements and
only say the following: If $\alg{F}$ has `quasi-trivial
1-cohomology' and is acted upon by a compact group $G$
of global gauge symmetries, then the equivalent (by
Proposition \ref{prop-dhr}) categories
$DHR_f(\alg{A})\simeq\D_f(\alg{A})$ are equivalent, as
symmetric tensor categories to $\Rep_fG$. In
\cite{goldstone} it is shown, e.g., that the theory of
a free massive scalar field has quasi-trivial
1-cohomology. Thus, if one takes $\alg{F}$ to be the
direct product of $N$ copies of such fields (of the
same mass) then $SO(N)$ acts on $\alg{F}$. Therefore,
$\D_f(\alg{A})\simeq\Rep_fG$ whenever $G\subset SO(N)$
is a closed subgroup and $\alg{A}=\alg{F}^G$. In
\cite{dop-groups} this observation is combined with a
limit construction to prove that every (second
countable) compact group arises as a DHR gauge
group. In a similar fashion, one shows that if
$\alg{F}$ is the theory of a massive Fermion with its
canonical $\7Z/2$-symmetry, then
$\D_f(\alg{F}^{\7Z/2})\simeq\Rep_f\7Z/2$.

There are results in the opposite direction, i.e.\ from
the superselection structure of $\alg{A}$ to that of
$\alg{F}$. By \cite[Theorem 3.6]{dr2}, which we have
not covered entirely in Section \ref{sec-localfield},
the field net reconstructed in \cite{dr2} and in
Section \ref{sec-DR} above satisfies `twisted Haag
duality'. In particular, if $\alg{A}$ has no Fermionic
representations then $\alg{F}$ satisfies Haag
duality. In this case, one can study the categories
$DHR(\alg{F})$ or $\D(\alg{F})$. In \cite{CDR}, the
following has been proven:

\begin{thm} \label{thm-cdr} Let $\alg{A}$ be a net of
  observables such that there are at most countably
  many DHR representations of finite dimension, all of
  which are Bosonic. Then the complete field net
  $\alg{F}$ has no non-trivial DHR representations of
  finite dimension.
\end{thm}

Rather than trying to comment on the many other known
results related to those treated in the preceding
sections, we close this section by commenting on a very
satisfactory {\it mathematical} interpretation of
DHR/DR theory. We are referring to the evident analogy
between this theory and the Galois theory of algebraic
field extensions. (It should be clear that in the
latter context, by `field' we mean the algebraic
structure of which $\7Q,\7R,\7C$ are examples, not the
theory of classical or quantum fields.) A field $\7F$
in the latter sense is called algebraically closed if
every polynomial $P(x)$ with coefficients in $\7F$ has
a zero in $\7F$. ($P$ then is a product of linear
factors $x-a$.) Every field $\7F$ is a subfield of an
essentially unique algebraically closed field
$\overline{\7F}$ that is an algebraic extension of
$\7F$. The latter means that $\overline{\7F }$ is
obtained by adjoining, typically transfinitely,
solutions of polynomial equations to $\7F$. The group
$G_\7F=\Aut_\7F(\overline{\7F})$ is compact, and one
has a bijective correspondence between intermediate
fields $\7F'\subset\overline{\7F},\7F'\supset\7F$ and
closed subgroups $H\subset G$. (The correspondence is
given by $H\mapsto \overline{\7F}^H,\
\7F'\mapsto\Aut_{\7F'}(\overline{\7F})$.) A similar
Galois correspondence holds in AQFT, cf.\ e.g.\
\cite{CDR,carpi-conti}. In view of Theorem
\ref{thm-cdr}, the construction of the complete DR
field net is entirely analogous to that of the
algebraic closure and can be considered as the passage
to a simpler or better behaved theory. Conversely, just
as taking the fixed field of an algebraically closed
field $\7F$ under the action of a closed subgroup
$G\subset\Aut\7F$ will result in an algebraically
non-closed field $\7F^G$, taking the $G$-fixed subnet
of a net $\alg{F}$ with trivial category
$\D_f(\alg{F})$ (more precisely, quasi-trivial
1-cohomology) will result in a net with non-trivial
category $\D_f(\alg{A})$.  Thus the `complication'
manifested by a non-trivial DHR-category
$\D_f(\alg{A})$ indicates that the theory $\alg{A}$
`really' is just a subtheory of a simpler one.

Physically, however, it is not at all clear whether the
`observable' net $\alg{A}=\alg{F}^G$ with its
non-trivial representation category $\D_f(\alg{A})$ or
the `field net' $\alg{F}$ with trivial $\D_f(\alg{F})$
but non-trivial global symmetry group $G$ is more
fundamental -- at least when $\alg{F}$ is Bosonic. In
\cite{haag} it is argued that the `right' description
of the physical situation should be in terms of a net
without any global symmetries. (On the other hand, in
\cite[Section III.4.2]{haag} one finds a tentative
postulate on a `good' net $\alg{A}$ of observables that
implies triviality of $\D_f(\alg{A})$. As the above
discussion shows, it will be very hard to find a theory
that has both a trivial DHR category $\D_f$ and trivial
global symmetry group! The theory of a single free
massive Bose field is one of the rare examples.)
Whether or not one subscribes to these views, from a
mathematical point of view, both nets $\6A$ and $\6F$
contain the same amount of information. This
equivalence is in fact a useful tool, since it permits
to view many problems from different angles. For
example, while a spin statistics theorem can be proven
in a `field' framework, its physical interpretation may
be clearer in the `observable' setting.

\subsection{Spontaneous symmetry breaking} \label{ssb}

So far, our entire analysis has presupposed the axiom of Haag duality for the theory
$\alg{A}$. Haag duality played an important r\^ole in our analysis of the category
$\D(\alg{A})$, but is needed also to establish the equivalence between the
$\D(\alg{A})$ and the representations satisfying the a priori physically motivated
DHR criterion (Definition \ref{def-dhr}). Thus, while it seems that the study of DHR
representations is physically motivated also for non-Haag dual nets, our mathematical
analysis soon gets stuck. We will therefore briefly comment on an approach to resolve
this issue, which turns out to have a profound physical interpretation.

\begin{defn} Let $O\mapsto\alg{R}(O)$ be a net of von Neumann algebras on a vacuum Hilbert space $H_0$.
The \emph{dual net} $\alg{R}^d$ of $\alg{R}$ is the assignment $O\mapsto \alg{R}(O')'$.
\end{defn}

If we have $O_1\subset O_2$ then $O_2'\subset O_1'$, thus
$\alg{R}(O_2')\subset\alg{R}(O_1')$, and therefore
$\alg{R}^d(O_1)\subset\alg{R}^d(O_2)$. Thus the dual net really satisfies isotony.
Microcausality of $\alg{R}$ is equivalent to
$\alg{R}(O)\subset\alg{R}(O')'=\alg{R}^d(O)$, or briefly $\alg{R}\subset\alg{R}^d$,
and Haag duality of $\alg{R}$ is equivalent to $\alg{R}=\alg{R}^d$. If
$\alg{A}_1\subset\alg{A}_2$ (in the sense of an inclusion for every $O$) then
$\alg{A}^d_2\subset\alg{A}^d_1$, thus $\alg{A}\subset\alg{A}^{dd}$, and a standard
argument shows $\alg{A}^{d}=\alg{A}^{ddd}$. Note, however, that microcausality of
$\alg{R}$ does not imply microcausality of $\alg{R}^d$! This motivates the following

\begin{defn} A net $O\mapsto\alg{R}(O)\subset B(H_0)$ satisfies \emph{essential
    duality} if both $\alg{R}$ and the dual net $\alg{R}^d$ (both indexed by double
  cones) satisfy microcausality.
\end{defn}

\begin{lemma} If $\alg{R}$ satisfies essential duality then $\alg{R}^d=\alg{R}^{dd}$, i.e.\
$\alg{R}^d$ satisfies Haag duality. 
\end{lemma} 

\begin{note} Essential duality can be difficult to verify. Fortunately, essential
  duality follows from \emph{wedge duality}, to wit $\alg{R}(W')'=\alg{R}(W)$ for all
  wedge regions (the regions obtained from the standard wedge $W_0=\{ x\in\7R^{1+s}\
  | \ x_0\ge|x_1| \}$ by Poincar\'e transformations). Besides being much easier to
  verify than essential duality, wedge duality is a very fundamental property that
  might well be required of any `reasonable' QFT.
\end{note}

Assuming that $\alg{R}$ satisfies essential duality,
$\alg{R}^d$ satisfies Haag duality and the D(H)R
analysis applies to it. Thus we obtain a symmetric
tensor $*$-category with conjugates
$\D_f^d(\alg{R}):=\D_f(\alg{R}^d)\simeq
DHR_f(\alg{R}^d)$, and we can construct the complete DR
field net $\al F$ associated with
$(\alg{R}^d,\D_f(\alg{R}^d))$. One thus has an
inclusion $\alg{R}\subset\alg{R}^d\subset\alg{F}$ of
nets. The DR gauge group acts on $\alg{F}$ and we have
$\alg{F}^G=\alg{R}^d$ and also
$G=\Aut_{\alg{R}^d}(\alg{F})$. Since the group $G$ is
implemented by unitaries that leave the vacuum vector
fixed, $G$ consists of `unbroken symmetries'. One can
now define a larger group
\[ \widehat{G}=\Aut_{\alg{R}}(\alg{F}) \] and topologize it suitably.  Now
$G\subset\widehat{G}$ consists precisely of the elements of $\widehat{G}$ that are
unitarily implemented. The point is that the net $\alg{R}$ acts irreducibly on $H_0$,
thus a unitary whose adjoint action leaves all algebras $\alg{R}(O)$ pointwise fixed
must be a multiple of the identity also on $\alg{R}^d$.

Concerning the categories associated with $\alg{R}$, little can be said about the
category $\D(\alg{R})$, but Roberts proved the existence of an extension functor
$K:DHR(\alg{R})\rarr DHR(\alg{R}^d)$ such that $K(\pi)|\alg{R}=\pi$ for every $\pi\in
DHR(\alg{R})$, cf.\ \cite[\S 3.4]{rob-lec}. (Again, a crucial r\^ole is played by the
theory of local 1-cohomology. Furthermore, this result breaks down in less than three
spacetime dimensions due to the phenomenon of solitonic representations.) This
functor actually is an equivalence, thus spontaneous symmetry breakdown doesn't
manifest itself in the superselection structure.

For the detailed analysis we refer to \cite{roberts-broken,rob-lec}
and the remarkable paper \cite{goldstone}, in which the Goldstone
phenomenon is analyzed in the context of algebraic QFT.

\bigskip \noindent {\small \textit{Notes:} DHR superselection theory
  originates in a four-paper series: \cite{dhr1} starts with a field
  algebra and gauge group and then derives properties of the
  superselection sectors.  \cite{dhr2} reconstructs the field algebra
  and gauge group from the category of representations in the special
  case where the objects are all one dimensional (i.e.\ the
  equivalence classes of objects of $\D _f$ form an abelian group with
  the monoidal product and conjugates as inverses).  \cite{dhr3}
  defines the symmetry $\ve _{\rho _1,\rho _2}$, and uses it to give
  the statistical classification of objects of $\D$.}

{\small For surveys of DHR theory in general, see
  \cite{rob-struc,rob-lec,rob-more,fred-review,fred}, \cite[Ch.\
  6]{araki}, and \cite[Ch.\ IV.2]{haag}.}

{\small The full proof of the DR reconstruction theorem
  is distributed over \cite{dhr4,dr0,DR} and
  \cite{dr2}.  The alternative approach to the
  reconstruction theorem that we use in this paper is
  based on \cite{problem} and \cite{del}, incorporating
  simplifications of the latter due to \cite{bichon}
  and ourselves.}

{\small For informal expositions of the DR econstruction theorem, see
  \cite{dop87,dop-abstract,dop-progress,dop-duals,dop-cats}.  For an
  interesting description of the goal of reconstructing fields and a
  gauge group, written before a solution was obtained, see
  \cite{rob7}.}

\section{Foundational Implications of the Reconstruction Theorem}
\label{last-DHR}

We now return to the foundational questions (Section \ref{inter}) that
motivated our investigation.  We also point out a few other cases
where discussions in the philosophical literature (e.g.\ about
permutation symmetry and the identity of particles) might benefit from
the study of superselection theory.

\subsection{Algebraic Imperialism and Hilbert Space Conservatism}

DHR superselection theory sheds light on some questions that
philosophers have asked about the role of inequivalent representations
of the algebra of observables.  But it will not answer all of our
questions.  We first bracket off those questions for which DHR theory
provides little help, and then we proceed to develop a case for the
relevance of DHR theory to foundational questions.

The DHR analysis requires that we fix a vacuum state $\om _0$, and
hence a base representation $(\2H_0 ,\pi _0)$.  Inequivalent DHR
representations do not correspond to different vacuua; rather, they
correspond to different local excitations of one and the same vacuum
state.  So, DHR theory effectively ignores the question of how to
choose a vacuum representation.  (But note that the power of the DHR
analysis strongly suggests --- against the Algebraic Imperialist ---
that representations \emph{are} essential components of the physical
content of the theory.)

Second, in some of the most familiar cases --- e.g., the free Boson
field --- the DHR category is trivial.  That is, $DHR(\al A)=\{ \pi _0
\}$, and so $\al F=\al A$.  In this case, the vacuum representation is
the \emph{only} DHR representation (relative to itself).  Thus, in
such cases, the elaborate apparatus of DHR superselection theory seems
to provide little insight into the physical importance of inequivalent
representations.  (However, if we are able to find a physical reason
for choosing a preferred vacuum representation, then the DHR analysis
suggests that no other representations are relevant for explaining the
phenomena.)

Finally, even in cases where $DHR(\al A)$ is nontrivial, the field
algebra itself has inequivalent representations.  (After all, it's
just another large $C^*$-algebra.)  And one might worry that the same
Conservative versus Imperialist debate will arise with respect to the
field algebra.  

But DHR theory has something to say about inequivalent DHR
representations, and about representations of the field algebra.
First, the field algebra $\al F$ is constructed concretely as
operators on a Hilbert space $\2H$; i.e.\ $\al F$ comes with a
preferred representation.  (Recall that the preferred representation
of $\al F$ is on a Hilbert space $\2H$ that decomposes into a direct
sum of superselection sectors for $\al A$.)  Of course, we could
consider other representations of $\al F$.  But in other
representations of $\al F$, we no longer have the intuitive
interpretation of elements of $\al F$ as intertwiners between the DHR
sectors of $\al A$.  If the physically meaningful quantities have to
have some connection to observable phenomena, then the interpretation
of elements of $\al F$ in terms of $\al A$ might be thought to be a
necessary condition for interpretation; and so the given
representation of $\al F$ might be preferred for these reasons.

So, DHR theory suggests that the issue of inequivalent representations
does not come up for the field algebra.  Regarding the issue of
inequivalent representation of the observable algebra, we can divide
the problem into several questions: 

\begin{enumerate}
\item Is there a physically privileged vacuum state/representation?
  What features pick it out?
\item Are all physical representations in the DHR category of some
  vacuum state?  (We are ignoring for the time being theories with
  long range forces (see \cite{buch-fred}).  In more general settings,
  we expect the \emph{form} of the question to remain the same: do we
  require physical states to be reachable from a fixed vacuum state by
  the action of an appropriate set of fields?)
\item If the answer to the second question is No, then how should we
  compare representations that are not connected to a vacuum
  representation by fields to representations that are?
\end{enumerate}

Let's suppose that the first question has been answered in the
affirmative, and that the vacuum representation $(\2H _0,\pi _0 )$ is
fixed.  Suppose also that $DHR(\al A)$ is nontrivial.  Then how should
we think about the inequivalent representations in $DHR(\al A)$?  A
naive transcription of Hilbert Space Conservatism to the current
context would tell us the following: the representations in $DHR(\al
A)$ are analogous to competing theories: one is correct, and the
others are surplus structure.  The naive transcription of Algebraic
Imperialism to the current context would say: the representations in
$DHR(\al A)$ are surplus structure; the physical content of the theory
is in $\al A$, the abstract algebra of observables.

Both Conservatism and Imperialism are based on an
oversimplified view of the formalism: it is supposed
that the elements of reality correspond to operators in
the abstract algebra or in some Hilbert space
representation thereof, and that the possible states
are either all states on the abstract algebra or some
particular folium of states.  But the fundamental
insight of DHR theory is that the set of
representations itself has structure, and it is
\emph{this} structure that explains phenomena.  So, a
more adequate position would take all the
representations more seriously.  Hence, we propose that
according to \emph{Representation Realism}, the content
of the theory is given by: (i) the net $O\mapsto \al
A(O)$, (ii) the dynamics on the quasilocal algebra
(i.e.\ the representation of the translation group in
$\Aut \al A$), and (iii) the symmetric tensor
$*$-category $DHR(\al A)$ of DHR representations.

Recall that the Conservative claims to have the
advantage in terms of explanatory power: more structure
(provided by choosing a representation) provides more
elements of reality, and so more satisfying
explanations.  But DHR superselection theory shows that
this claimed advantage is misleading: the focus on one
representation ignores the most interesting structure,
namely the relations between representations.  Indeed,
if we committed ourselves to one representation, and
ignored others, we would have no field operators, no
gauge group, no definition of Bose and Fermi fields, no
definition of antiparticles, etc..

And yet there is a strong \emph{prima facie} objection to
Representation Realism: since the Hamiltonian is always an observable,
no possible dynamical evolution can take us from a state in one
representation to a state in an inequivalent representation.  So,
inequivalent representations are dynamically isolated from each other,
and abstract relations between them cannot explain the features of
states in the representation that best describes our universe.

The fact that the Hamiltonian is an observable --- hence cannot map
states from one sector to states in another --- raises a further
problem for our interpretation of field operators.  Recall that we
speak of ``creating'' states in a sector $\2H _\rho$ by acting on the
vacuum with elements from the field algebra.  That is, we can choose
$F\in H_\rho \subseteq \al F$ such that $F\Om \in \2H _\rho$, where
$(\2H _\rho ,\pi _\rho )$ is disjoint from the vacuum representation
$(\2H _0,\pi _0)$.  The talk of ``creation'' here suggests that we are
talking about some sort of dynamical process.  On the one hand, $F\in
\al F$ can be chosen unitary, so structurally the map $\Om \mapsto
F\Om$ looks like dynamics.  But since the Hamiltonian is an
observable, the transition $\Om \mapsto F\Om$ is not dynamically
allowable.  So, in what sense are states in $\2H _\rho$ accessible
from the vacuum?  Is the key insight behind superselection rules that
there are two notions of dynamic accessibility?  If so, then how are
we to understand the differences between these two notions?

\subsection{Explanatory relations between representations}

If we consider a $C^*$-algebra $\alg{A}$ with no
further structure, then the mathematically definable
relations between representations (and hence, between
states in the folia of these representations) are
exhausted by the following table: 
\begin{table}[h] \begin{center}
\begin{tabular}{||l||}
  \hline \hline $\pi _1$ and $\pi _2$ are equivalent   \\ 
  \hline $\pi _1$ and $\pi _2$ are quasiequivalent  \\
  \hline $\pi _1$ and $\pi _2$ are disjoint   \\
  \hline $\pi _1$ and $\pi _2$ are weakly equivalent  \\
  \hline \hline \end{tabular} \caption{Relations Between Representations
  of $\alg{A}$} \label{coarse} \end{center} \end{table}

\noindent Outside of the fourth relation (which makes
special reference to the topology of the state space),
these relations are precisely those that can be defined
in an arbitrary $*$-category $\2C$ with subobjects.
Two objects $X,Y$ in $\2C$ are equivalent if there is a
unitary $u\in \Hom (X,Y)$; are quasiequivalent if there
is an isometry $v\in \Hom (X,Y)$; and are disjoint just
in case they are not quasiequivalent.

Consider now the normal state space $K$ of a $C^*$-algebra $\al A$.
The GNS theorem provides a map $\om \mapsto (\2H _\om ,\pi _\om )$
from $K$ into the representation category of $\al A$.  We then use
this map to induce relations corresponding to those in Table
\ref{coarse} on $K$: we talk about equivalent, quasiequivalent, and
disjoint states.  Furthermore, the individual folia (sets of states
whose GNS representations are quasiequivalent) have a rich geometrical
structure which corresponds exactly to the normal state space of $\bh$
for some Hilbert space $\2H$.  Thus, within a folium we have a notion
of ``transition probability'' between pure states (corresponding to
rays in $\2H$), and a three place relation ``$\om$ is a superposition
of $\om _1$ and $\om _2$.''  However, if two states lie in disjoint
folia, then these relations trivialize.  The transition probability
between disjoint states is zero, and no state can be a superposition
of states from a different folia.  It seems that the only physically
suggestive thing we can say about states from different folia is that
they are ``orthogonal.''

It is precisely the preceding considerations that have lead
philosophers to worry about inequivalent representations.  The worry
is based on the fact that disjoint representations seem to be
\emph{competitors}, and nothing more.  In order to alleviate worries
about the competition between representations, some philosophers
\cite{rindler,hans} go so far as to claim that these representations
are ``complementary'' descriptions of the phenomena (in the sense of
Bohr).  The word ``complementarity'' is of course meant to suggest
that the representations are not \emph{merely} competitors, and the
choice of one does not need to be seen as completely ruling out the
relevance of another.

We wish to replace suggestive --- and possibly misleading ---
terminology with some real facts about the relationships between
inequivalent representations.  To illustrate what we mean by this,
consider the case of \emph{group} representations: let $\Rep _fG$ be
the category of unitary representations of a compact group $G$.  $\Rep
_fG$ is not only a $*$-category, but it has a monoidal product and
conjugates.  That is, for objects $X,Y$ in $\Rep _fG$, there is a
product object $X\otimes Y$, and a conjugate object $\ol X$.  For our
purposes, this is the crucial difference between group representations
and the representations of an arbitrary $C^*$-algebra $\al A$.  For an
arbitrary $C^*$-algebra $\al A$, there is no product of
representations, or conjugate of a representation.  

In the case of compact group representations, typically $X\in \Rep
_fG$ will be disjoint from both $X\otimes Y$ and $\ol X$.  But in this
case, we are not tempted to see $X$ is \emph{merely} as a competitor
of $X\otimes Y$, or of $\ol X$; there are some interesting relations
between these representations.  Roughly speaking, information about
$X$ gives us information about $X\otimes Y$ and $\ol X$.  Thus,
although these representations are technically ``disjoint,'' they are
not completely unrelated to each other.\footnote{But note also:
  Philosophers of physics have so far not worried about inequivalent
  \emph{group} representations as competing descriptions of reality.
  And for good reason, because group elements are not observables, and
  groups do not have states.  Another insight of DHR theory is to show
  that physicist's intuitions about group representations are not
  totally baseless, because in fact the interesting (DHR)
  representations of the observable algebra correspond to
  representations of a compact group.}

One of the main accomplishments of the DHR analysis and DR
reconstruction theorem is to show that the category $\D _f$ of
physical representations is a tensor $*$-category with conjugates;
indeed the Embedding Theorem (see the Appendix) shows that $\D _f$ is
equivalent to the category $\Rep _fG$ for some compact group $G$.  The
obvious question then is whether these additional relations on the
category of representations can help us get past the idea that
disjoint representations are merely competing descriptions.

An analogy to states might be helpful.  Consider a pair $H_1,H_2$ of
Hilbert spaces, and let $\psi _i\in H_i$ be unit vectors.  Now
consider the following two ``descriptions of reality'': 
\begin{enumerate}
\item The state is $\psi _1$.
\item The state is $\psi _1\otimes \psi _2$.  
\end{enumerate} 
What do we say here: are these competing descriptions?
In one sense, (1) and (2) are competitors, because they
cannot both be fully adequate descriptions of reality
at the same time.  However, (1) and (2) are not
competitors in the same sense that, say, two orthogonal
vectors in a single Hilbert space are competitors.  The
two state descriptions are not merely competitors,
because there is an interesting sense in which $\psi
_1$ is a ``part'' of $\psi _1\otimes \psi _2$.  Indeed,
information about $\psi _1$ (e.g.\ expectation values
it assigns to observables) does in fact give us
information about $\psi _1\otimes \psi _2$ because of
the canonical mappings between $H_1$ and $H_1\otimes
H_2$.

Now let $\pi _1,\pi _2$ be objects in the DHR category
$\D _f$, and suppose (as will often be the case) that
the representations $\pi _1$ and $\pi _1\otimes \pi _2$
are disjoint.  Are these competing descriptions?
Again, $\pi _1$ and $\pi _1\otimes \pi _2$ are
competitors in the sense that if the state of an object
(or of the universe?) is in $\2H _{(\pi _1\otimes \pi
  _2)}$ then it is not in $\2H _{\pi _1}$.
Nonetheless, $\pi _1$ and $\pi _1\otimes \pi _2$ are
not merely competitors, because in one sense $\pi _1$
is ``part'' of $\pi _1\otimes \pi _2$.

But two words of caution should be issued here.  First,
we must be cautious with the use of the ``part''
metaphor.  For example, $\D _f$ can have a nontrivial
representation $\pi$ such that $\pi \otimes \pi$ is
equivalent to the vacuum representation.  Then it is
not so clear that we should say that ``$\pi$ is part of
$\pi \otimes \pi$.''  Second, there is one significant
disanalogy between the case of states $\psi _1$ and
$\psi _1\otimes \psi _2$ and the case of
representations $\pi _1$ and $\pi _1\otimes \pi _2$:
the two representations are GNS representations of
states on a \emph{single} $C^*$-algebra $\al A$.  Hence
we can directly compare the expectation values that
these states assign to observables in $\al A$, and they
will disagree significantly (indeed, for any $\e >0$
there is an observable $A\in \al A$ such that
$\norm{A}\leq 1$ and $\norm{\om _1(A)-\om
  _2(A)}>2-\e$).  Thus, there is a clear, empirically
verifiable sense in which states in $\pi _1$ are
competitors with states in $\pi _1\otimes \pi _2$.

Finally, there is an interesting physical relation between a DHR
representation $\pi$ and its conjugate $\ol \pi$, even though $\pi$
and $\ol \pi$ are typically disjoint.  In short, $\ol \pi$ is like an
inverse of $\pi$: if $\pi$ is irreducible, then $\ol\pi$ is the unique
irreducible representation such that $\pi \otimes \ol\pi$ contains a
copy of the vacuum representation.  In fact, when $\pi =\pi _0\circ
\rho$ where $\rho$ is a dimension $1$ element of $\D _f$, $d(\rho
)=1$, then this is the exact relation: $\rho$ is an automorphism and
$\ol \rho =\rho ^{-1}$.  In terms of field operators, if $F$ creates
the charge $\xi$, then $\ol F$ annihilates the charge $\xi$.
Furthermore, when $\pi$ admits a particle interpretation, then the
states in the folium of $\ol \pi$ are the antiparticle states of the
states in the folium of $\pi$ \cite{dhr2}.

\subsection{Fields as theoretical entities, as surplus structure}

From the standpoint of superselection theory, there is a sharp
distinction between observable and unobservable fields, namely, a
field operator is an observable iff it is invariant under all gauge
transformations.  To what extent does this distinction between fields
and observables match up with the philosopher of science's distinction
between theoretical and observational components of a theory?  Even if
the two notions are not exactly the same, the connection is
suggestive.  In particular, it seems interesting to ask about the
extent to which the field plus gauge part of QFT is fixed by the
observable algebra.

First, the notion of equivalent systems of field operators seems a
fairly close analogue of the philosopher's notion of ``theoretical
equivalence.''
\begin{defn} Let $\alg{F}_1=(\alg{F}_1,\2H_1,G _1)$ and
  $\alg{F}_2=(\alg{F}_2,\2H_2,G_2)$ be local field systems with gauge
  symmetry for $(\al A,\om )$.  (See Defn.\ \ref{field-system} on p.\
  \pageref{field-system}.) Then $\alg{F}_1$ and $\alg{F}_2$ are
  \emph{theoretically equivalent} just in case they are unitarily
  equivalent as systems of local field operators.  (See Defn.\
  \ref{equiv-fields} on p.\ \pageref{equiv-fields}.)  \end{defn}

\begin{rema} (i) This definition is not fully adequate, because it
  does not make reference to dynamics.  For example, this definition
  entails that the free Bose field nets for different positive masses
  are theoretically equivalent.  For a fully adequate definition, it
  would probably be necessary to require that the unitary mapping
  $W:\2H _1\to \2H _2$ also intertwines the dynamical groups on the
  two Hilbert spaces. (ii) If $\al F_1$ and $\al F_2$ are
  theoretically equivalent, then they are equivalent in \emph{all}
  physically relevant senses (modulo dynamics): they have the same
  type of commutation relations (either both have normal or abnormal
  commutation relations), they have isomorphic gauge groups, etc..
\end{rema}

The working analogy also suggests that we define ``observational
equivalence'' between two theories in terms of some equivalence
between their nets of observable algebras.  There are a myriad number
of ways we could explicate the notion of observational equivalence in
this setting; philosophers have their work cut out for them.  The
following two definitions give extremely weak notions of observational
equivalence that do not take into account a representation of the
algebra of observables.

\begin{defn}
  Let $\al F_1$ and $\al F_2$ be two local field systems with gauge
  symmetry, and let $\al A_1$ and $\al A_2$ be the fixed point
  algebras; i.e., 
$$ \al A_i =\{ A\in \al F_i :\a _g(A)=A,\; \mbox{for all}\; g\in
G_i \} .$$ Then we say that $\al F_1$ and $\al F_2$ are \emph{weakly
  observationally equivalent} just in case there is a $*$-isomorphism
$\a$ from the algebra $\al A_1$ onto the algebra $\al A_2$.
\end{defn}

\begin{defn}
  Let $\al F_1$ and $\al F_2$ be two local field systems with gauge
  symmetry, and let $\al A_1$ and $\al A_2$ be their fixed point nets;
  i.e.\ for each double cone $O$,
$$ \al A_i(O) =\{ A\in \al F_i(O):\a _g(A)=A,\; \mbox{for all}\; g\in
G_i \} .$$ Then we say that $\alg{F}_1$ and $\alg{F}_2$ are
\emph{observationally equivalent} just in case there is a net
isomorphism $\a :\al A_1\to \al A_2$ (see Defn.\ \ref{net-morphism} on
p.\ \pageref{net-morphism}).
\end{defn}

\begin{rema} The first definition is weaker because it does not
  require that the net structure be preserved by the $*$-isomorphism
  $\a$.  \end{rema}

Again, the definitions omit reference to dynamics, which would be an
important component of a fully adequate treatment of observational
equivalence.  Nonetheless, even with these definitions, we can make
some sense of remarks about underdetermination of fields by
observables, or about the physical equivalence of different field
theories.

\begin{enumerate}
\item (Construction of observationally equivalent theories) The DR
  reconstruction theorem provides a general, nontrivial recipe for
  constructing non-equivalent theories that are observationally
  equivalent: If $(\alg{A},\om _0)$ has nontrivial DHR superselection
  sectors, then it can be embedded into two nonequivalent field
  algebras $\alg{F}_1$ and $\alg{F}_2$.  Indeed, $\alg{A}$ is always a
  field algebra over itself (but incomplete), but the field algebra
  $\alg{F}$ from the DR reconstruction theorem is complete.
  \item (Elimination of parafields) It has long been thought that
  parafields are a theoretical artifact.  More precisely, it has been
  claimed that every parafield theory is ``physically equivalent'' to
  a theory with normal commutation relations (see \cite{araki-para}).
  The DR reconstruction theorem partially validates this claim by
  showing that every parafield theory is \emph{observationally
    equivalent} to a theory with normal commutation relations.
  Indeed, suppose that $\alg{F}_1$ is a parafield theory.  Then we can
  extract the observable algebra $\alg{A}$ contained in $\alg{F}_1$,
  and apply the DR reconstruction theorem to construct a field algebra
  $\alg{F}_2$ with normal commutation relations.  Since $\alg{F}_1$
  and $\alg{F}_2$ have the same net of local observable algebras, they
  are observationally equivalent.
\item \noindent Some have claimed that the relation between quantum
  fields (the field algebra $\al F$) and observables (the observable
  algebra $\al A$) is analogous to relation between coordinates and a
  manifold in differential geometry.  However, the DR reconstruction
  theorem shows that (subject to normal commutation relations), there
  is a \emph{unique} field net $\alg{F}$ and gauge group $G$
  compatible with the observable algebra $(\alg{A},\om _0)$.  Thus,
  there is a strong disanalogy between the two cases, since there
  seems to be no sense in which one coordinate system of a manifold is
  a better representation of reality than another coordinate system.
\end{enumerate}

Finally, we are in a position to get a clear picture of the
interrelations between questions about inequivalent representations
and questions about gauge invariance.

According to a common interpretive stance, if two states of a physical
system can be identified under a gauge symmetry, then those two states
are different descriptions of the same underlying reality.  So, for
the purposes of counting states, we should look to the quotient of the
state space under gauge orbits.  Similarly, it is thought that the
``real'' quantities of a theory are the gauge invariant quantities
(see \cite{ear}).

In the setting of DHR superselection theory, the algebra of
  observables $\al A$ is precisely the gauge invariant part of the
  field algebra $\al F$, that is,
$$ \al A = \{ A\in \al F:\a _g(A)=A, \,\mbox{for all}\; g\in G \} ,$$ where $G$ is the
gauge group.  This of course means that for any observable $A\in \al
A$, there is no difference between a state $\psi$ and the gauge
transformed state $U(g)\psi$.  (Of course, if $\psi$ is a state vector
in the vacuum representation, then $U(g)\psi =\psi$, since the
representation of the gauge group there is trivial.)  So, if the
common interpretive stance is correct, the physical content of the
theory is in the observable algebra $\al A$; the fields are
``descriptive fluff.''

So suppose that we ignore the field algebra $\al F$, and just look to
the observable algebra $\al A$ to provide the physical content of the
theory.  But what should we say about the representations of $\al A$?
Are they just descriptive fluff?  If not, then is there one correct
representation of $\al A$, or do we somehow need inequivalent
representations in order to account for all of the physical facts? 

The DR reconstruction theorem shows that the preceding two sets of
questions --- regarding the status of gauge variant quantities on the
one hand, and representations on the other hand --- are tightly
intertwined.  The full structure of the theory, field algebra $\al F$
and gauge group $G$, is uniquely recoverable (modulo completeness, and
normal commutation relations) from the structure of the category
$DHR(\al A)$ of representations.  The ontological significance of the
gauge variant fields is closely aligned with the ontological
significance of inequivalent representations.  (We will revisit this
question in the next section when we discuss permutation symmetry.)

Of course, there is a crucial disanalogy between the global gauge
symmetries in DHR superselection theory, and the local gauge
symmetries of electromagnetism or general relativity.  But it is not
clear that this disanalogy makes the DR reconstruction theorem any
less interesting for understanding the relation between gauge symmetry
and superselection rules.

\subsection{Statistics, permutation symmetry, and identical particles}

Philosophers have taken an active interest in the differences between
the Maxwell-Boltzmann statistics of classical physics, and the
Bose-Fermi statistics of quantum physics.  Indeed, it has been
provocatively claimed that Bose-Fermi statistics is explained by
permutation invariance --- i.e.\ there are no physical differences
between states with permuted particle labels --- and that this entails
that quantum particles are not ``individuals'' in the same sense that
classical particles are.  (See \cite{french2,french} for an account of
the argument.)

But such discussions can be hampered by an overly simplistic
formalism.  In particular, states are usually identified with unit
vectors (or at best with rays) in a single Hilbert space, and no
account is given of the status of non-permutation invariant operators.
It might be helpful then to look at the issue from the perspective of
a more flexible formalism that allows a distinction between fields and
observables, and in which states can be represented by vectors in
inequivalent representations of the algebra of observables.

There is another reason why the issue of permutation invariance should
be revisited within the context of QFT.  Some of the literature
suggests that the metaphysical problem about the individuality of
particles is \emph{aufgehoben} in the transition to QFT, because: (i)
QFT is about fields and not particles, and (ii) the Fock space
formalism of QFT already identifies permuted states, and so rules out
a notion of particles as individuals.  We have already noted that it
is not so clear how to make sense of the idea that QFT is about fields
as opposed to particles.  Furthermore, the DR reconstruction theorem
shows precisely how to make sense of non-permutation invariant
quantities and states in a theory that is manifestly permutation
invariant.

It is not surprising that DHR theory is relevant for the issue of
permutation invariance and statistics: one of the original goals of
DHR was to clarify the role of statistics in QFT.  Riding the crest of
a wave of mathematical success, Roberts made the following bold claim
about the DHR analysis of statistics:
\begin{quote}
  One of the insights provided by the study of
  superselection sectors concerns the origin of what is
  termed the `statistics' of a particle. \dots Now just
  as a theory should determine its particle states so
  should it determine the statistics of these
  particles.  Ordinary quantum mechanics ignores this
  challenge saying in effect that the statistics of
  particles is one of the parameters determining the
  theory, the one telling you what symmetry the
  $n$-particle states have.  QFT does a little better:
  it says that the statistics of the particles in the
  theory is determined by the commutation relations of
  fields at spacelike separations.  \ldots In adopting
  the philosophy that the local observables determine
  the theory, we are forced to meet this challenge in
  full.  \cite[p.\ 203]{rob-stat}
\end{quote} In the remainder of the paper from which the quote is
taken, Roberts shows how Bose-Fermi particle statistics emerges
naturally from the DHR analysis of physical representations of the
algebra of observables.

Roberts' claim is of crucial relevance to the philosophical debate
about statistics and identical particles.  The philosophers have
asked, ``what explains Bose-Fermi statistics?''  Roberts' answer is
that the explanation comes from the structure of the category of
representations of the observable algebra.

Let us recall then how the Bose-Fermi distinction is supposed to
emerge from the DHR analysis.  In Section \ref{dimension}, an object
$\rho$ of the category $\D$ is shown to have an intrinsic dimension
$d(\rho )$.  The dimension is finite iff $\rho$ has a conjugate; in
this case we define a unitary operator $\Theta _\rho \in
\End (\rho )$ called the \emph{twist} of $\rho$.  If $\rho$ is
irreducible, then $\Theta _{\rho}=\om _\rho \id _{\rho}$ where $\om
_\rho =\pm 1$.  We then stipulate that a ``Bosonic'' object is one
with $\om _\rho =1$ and a ``Fermionic'' object is one with $\om _\rho
=-1$.  

Of course, $\rho$ is not the sort of thing that the philosophers have
been calling Bosonic or Fermionic --- it is not a wavefunction.  To
connect the two pieces of formalism, recall that an object of $\D _f$
(endomorphisms of the algebra of observables) corresponds to a
representation $\pi _0\circ \rho$ of the algebra of observables.  So,
we call the representation $\pi _0\circ \rho$ Bosonic when $\rho$ is
Bosonic, and Fermionic when $\rho$ is Fermionic.  Finally, we then
call a state (``wavefunction'') Bosonic if it is in the folium of a
Bosonic representation, and Fermionic if it is in the folium of a
Fermionic representation.  The claims to be assessed are: (i) does
this stipulative definition adequately reproduce the distinction
between Bosonic and Fermionic wavefunctions made in elementary
nonrelativistic QM; and if so, (ii) what does this tell us about
permutation invariance?

\subsubsection{The Bose-Fermi distinction in nonrelativistic QM}

In nonrelativistic QM, the state space of $n$ identical particles is
the tensor product $H\otimes \cdots \otimes H$ of $n$ copies of a
Hilbert space $H$.  The Hilbert space $H\otimes \cdots \otimes H$ is
spanned by \emph{product states}, i.e.\ states of the form
$$ \psi _1\otimes \cdots \otimes \psi _n ,$$
with $\psi _1,\dots ,\psi _n\in H$.  

\begin{defn} We define the natural action of the permutation group
  $S_n$ on $H\otimes \cdots \otimes H$ as follows.  Let $\{ \psi
  _1,\dots ,\psi _m\}$ be an orthonormal basis for $H$, and define for
  each permutation $\sigma$,
$$ U(\sigma )(\psi _{i_1}\otimes \cdots \otimes \psi _{i_m}) = \psi
_{\sigma (i_1)}\otimes \cdots \otimes \psi _{\sigma (i_m)} ,$$
and extend $U(\sigma )$ by linearity to all of $H$.  
\end{defn}
If $\dim H>1$, then the representation $U$ of $S_n$ is reducible.  It
contains copies of the two one-dimensional representations of $S_n$,
namely the identity representation $S_n\to \11$, and the alternating
representation.  The subspace of vectors of $H\otimes \cdots \otimes
H$ transforming according to the identity representation is called the
\emph{symmetric} subspace, and the subspace of vectors transforming
according to the alternating representation is called the
\emph{antisymmetric} subspace.  Vectors in the symmetric subspace are
called Bosonic, and vectors in the antisymmetric subspace are called
Fermionic.  These traditional definitions have served as the basis of
discussions of permutation invariance in the philosophical literature.

In QM, states of $n$ particles that differ only by permuting labels
--- for example, $\psi _1\otimes \psi _2$ versus $\psi _2\otimes \psi
_1$ --- should not be counted separately.  For the purposes of
statistical weighting, these two symbols represent one state.  This
has been stated as the principle of Permutation Invariance.
\begin{quote}
  \textbf{Permutation Invariance (PI):} Let $\al A$ be the observables
  for the $n$ particle system.  Then for each state $\psi$, and for
  each permutation $\sigma \in S_n$, we have
$$ \langle U(\sigma)\psi ,AU(\sigma )\psi \rangle = \langle \psi
,A\psi \rangle .$$ 
\end{quote}
Permutation Invariance is sometimes also called the
\emph{Indistinguishability Postulate}: two states that differ by a
permutation are indistinguishable (i.e.\ there is no measurement that
can distinguish between the two).  It has been claimed that PI entails
that state has Bose or Fermi statistics, no states with
``parastatistics'' are allowed.
\begin{quote}
  \textbf{Dichotomy:} For each state vector $\psi$ and permutation
  $\sigma$, we have $U(\sigma )\psi =\pm \psi$.
\end{quote}
(See \cite[pp.\ 389ff]{bas} for an account of attempts to prove
Dichotomy from PI.  See \cite{jb} for further details.)  In other
words, the states that are not in the symmetric or antisymmetric
subspaces are surplus structure.

This leaves us with a number of puzzles.  First, what do we say about
the vectors in $H\otimes \cdots \otimes H$ that are not in the
symmetric or antisymmetric subspaces?  Are they surplus structure?
Are they possibilities that are contingently not realized in nature?
More generally, not all vectors in $H\otimes \cdots \otimes H$ are of
a definite symmetry type; and even among those that are of a definite
symmetry type, not all are totally symmetric or totally antisymmetric.
For any irreducible representation $\xi$ of $S_n$ we say that a
wavefunction $\psi$ in $H\otimes \cdots \otimes H$ is of symmetry type
$\xi$ just in case $\psi$ is contained in the subspace corresponding
to the representation $\xi$.  Then $H\otimes \cdots \otimes H$ is the
direct sum of subspaces of definite symmetry type vectors. But now the
principle of plenitude suggests that there should be particles of
every symmetry type.  Why do we not see them?

\subsubsection{An intrinsic characterization of symmetric and
  antisymmetric subspaces}

We began with the full $n$-particle Hilbert space $H\otimes \cdots
\otimes H$, and then we reduced to the symmetric and antisymmetric
subspaces.  We were then left wondering what to do with the remaining
elements of $H\otimes \cdots \otimes H$.  

The intrinsic description of the symmetric and antisymmetric subspaces
is that they are representations of the group $S_n$.  (In fact, they
are quasiequivalent to the one dimensional irreducible representations
of $S_n$.)  So we can also work backwards.  That is, if we are given a
representation $(H,\pi )$ of $S_n$, we can ask after the intrinsic
characterization of this representation.  Recall that the irreducible
representations of $S_n$ are in one-to-one correspondence with Young
tableaux with $n$ boxes (see \cite{simon}).  There is a natural
grouping of representations of $S_n$ into para-Bose and para-Fermi
categories: we specify the representation $(H,\pi )$ by a pair of
numbers $(d(\pi ),\om _\pi )$, with $d(\pi )\in \{ 1,2,\dots ,n\}$ and
$\om _\pi =\pm 1$.
\begin{enumerate}
\item For $(d,+1)$, all Young tableaux whose columns have length less
  than or equal $d$. (In this case, we say that $\pi$ has
  \emph{Para-Bose statistics} of order $d$.)
\item For $(d,-1)$, all Young tableaux whose rows have length less
  than or equal $d$.  (In this case, we say that $\pi$ has
  \emph{para-Fermi statistics} of order $d$.)
\end{enumerate}
Clearly the division of representations into para-Bose and para-Fermi
is mutually exclusive, but not exhaustive.  (e.g., there are
representations of $S_n$ that contain copies of both the $\11$
representation and the alternating representation.)

Now suppose that we are in the following position
(described vividly by Roberts in the opening quote): we
are given a pure state $\om$ of the algebra of
observables $\al A$ and we asked whether its
``intrinsic'' statistics of its states is Bosonic or
Fermionic.  What can we do?  First we construct the GNS
representation $(\2H ,\pi )$ induced by $\om$.  At
least this makes things more concrete.  But the Hilbert
space $\2H$ is not itself a tensor product, and so
there is no natural representation of $S_n$ on $\2H$.
Nor would it help to construct tensor products of
elements of $\2H$ in the normal way, because for $\psi
\otimes \cdots \otimes \psi$ is trivially Bosonic.  So,
the obvious approach does not seem to tell us anything
about the \emph{intrinsic} symmetry type of elements of
$\2H$.

The key insight again comes from the DHR analysis: the
representation $(\2H ,\pi )$ is naturally isomorphic to
an object $\rho$ of a symmetric tensor $*$-category,
namely the category $\D _f$ of localized transportable
endomorphisms.  Since $\D _f$ has products we can
construct $\rho \otimes \rho$, and the symmetry $\ve
_{\rho ,\rho }$ gives us notion of permuting $\rho
\otimes \rho$.  [Recall that $\ve _{\rho ,\rho }\in
\Hom (\rho \otimes \rho )$.]  As we will see in the
following section, this gives us a natural
representation $u$ of $S_n$ in $\End (\rho \otimes \rho
)$.  Furthermore, the pair $(d(\rho ), \om _\rho )$,
where $d(\rho )$ is the dimension of $\rho$, and $\om
_\rho$ is the statistical phase of $\rho$, coincide
with the classification of $u$ as a para-Bose or
para-Fermi representation of $S_n$.  We will also see
that this natural representation $u$ of $S_n$
corresponds to a permutation of wavefunctions in the
``larger'' Hilbert space of the field algebra $\al F$.

\subsubsection{Representations of $S_n$ in a symmetric tensor
  $*$-category}

Unitary representations of the permutation group $S_n$ arise naturally
in tensor $*$-categories with a unitary symmetry.  Let $(\2C ,\otimes,
\11)$ be a tensor $*$-category with unitary symmetry $c_{X,Y}$.  Fix
an object $X\in \Obj (\2C )$, and define a map $u:S_2\to \End
(X\otimes X)$ by setting
\begin{eqnarray*}
  u((1)) = \id _{X\otimes X} ,\qquad \qquad u((1,2))= c_{X,X}
  .\end{eqnarray*}
Since $(c_{X,X})^{2}=\id _{X\otimes X}$, $u$ is a unitary
representation of $S_2$ in $\End (X\otimes X)$.  This construction can be iterated: define a map $u:S_n\to \End
(X\otimes \cdots \otimes X)$ by setting
\begin{eqnarray*} u((i,i+1))= \id _X\otimes \cdots \otimes
  c_{X,X}\otimes \cdots \otimes \id _X .\end{eqnarray*} It is easy to
verify that $u$ extends uniquely to a unitary representation of $S_n$
in $\End (X\otimes \cdots \otimes X)$.

\begin{fact} Let $\2C$ be a tensor $*$-category with
  unitary symmetry and conjugates.  Then for each
  irreducible object $X\in \Obj (\2C )$ the induced
  unitary representation $u$ of $S_n$ in $\End
  (X\otimes \cdots \otimes X)$ is para-Bose of order
  $d(X)$ if $\om _X =+1$, and is para-Fermi of order
  $d(X)$ if $\om _X=-1$.  Furthermore, the statistical
  phase $\om _X$ is the trace of $u((1,2))=c_{X,X}$.
  (See Appendix \ref{App-B} for more details.)
  \label{risque}
\end{fact}

The physical interpretation becomes more clear in the presence of
field operators.  (Of course, the point of the Reconstruction Theorem
is to show that such field operators are always available.)  Let $(\2H
,\al F,(G,k))$ be a field system with gauge symmetry for the
observable algebra $\al A$ and vacuum state $\om$.  Let $O_1,\dots
,O_n$ be mutually spacelike separated regions.  Let $\rho$ be an
irreducible object in $\D _f$.  Then using the transportability of
$\rho$ we can choose $F_i\in \al F(O_i)$ such that $F_i\Om$ is in the
sector $\2H _{\hat{\rho}}$.  (Recall that sectors are labeled by
unitary equivalent classes $\hat{\rho}$ of objects in $\D _f$.)  In
other words, $F_i$ creates the charge $\hat{\rho}$ in the region
$O_i$.  Let $\sigma$ be a permutation of $\{ 1,\dots ,n \}$ and
consider the following two state vectors in $\2H$:
\begin{eqnarray}
  \psi _1\times \psi _2\times \cdots \times \psi _n &\equiv & F_1F_2\cdots
  F_n\Om  ,\label{ident} \\
  \psi _{\sigma (1)}\times \psi _{\sigma (2)}\times \cdots \times \psi _{\sigma
    (n)} & \equiv & F_{\sigma (1)}F_{\sigma (2)}\cdots F_{\sigma
    (n)}\Om . \label{permu}
\end{eqnarray}
These two vectors are typically distinct.  In fact, if the field net
has normal commutation relations then we can compute the difference
between the two vectors.  Supposing that $\sigma$ only permutes two
numbers, the two vectors will be the same if $\rho$ is Bosonic, and
the two vectors will differ by a sign if $\rho$ is Fermionic.
However, the two vectors always induce the same state on the algebra
of observables $\pi (\al A)$.  Indeed, if $\rho _i\in \D _f(O_i)$ are
the corresponding morphisms, then the states induced by the two
vectors, respectively, are
\begin{eqnarray}
  \om \circ (\rho  _1\otimes \rho _2\otimes \cdots \otimes \rho _n) &=& \om \circ (\rho _1\rho
  _2\cdots \rho _n) ,\\
  \om \circ \bigl( \rho _{\sigma (1)}\otimes \rho _{\sigma (2)}\otimes
  \cdots \otimes \rho _{\sigma (n)}\bigr) & = & \om \circ (\rho _{\sigma (1)}\rho
  _{\sigma (2)}\cdots \rho _{\sigma (n)}) .
\end{eqnarray}
Since endomorphisms that are localized in spacelike separated regions
commute, these two states are equal.  Thus, permutation invariance
holds for the observables, but not for the fields.

The interpretive upshot of the the DHR treatment of statistics is as
follows: permutation invariance is a gauge symmetry in the sense that
it leaves all observables' values invariant, but changes the values
assigned to field operators.  Are two states related by a permutation
the same or different?  Of course, the answer to the mathematical
question is clear: the states of the observable algebra are the same,
the states of the field algebra are different.  So, whether or not we
take permutations to correspond to ``real'' changes depends on what we
think about the status of the field operators.  So the issue of
permutation invariance is just a special version of the issue of gauge
invariance, and accordingly is tightly connection to the question of
the status of inequivalent representations.

\subsubsection{Parastatistics and nonabelian gauge groups}

The abstract Tannaka Theorem (Appendix B) shows that
each symmetric tensor $*$-category (STC$^*$) $\2C$ is
equivalent to the representation category $\Rep
_f(G,k)$ of a compact supergroup $(G,k)$.  Applied to
our current topic, the theorem shows that the category
$\D _f$ of localized transportable morphisms is
equivalent to the representation category of the gauge
group.  Furthermore, Section \ref{saggy} shows that
each object $X$ of an STC$^*$ gives rise naturally to a
unitary representation of the symmetric group $S_n$ in
$\End (X\otimes \cdots \otimes X)$, and this
representation corresponds to the intrinsic statistical
characterization of $X$.  Now, we know that the
categorical dimension of a representation $(H,\pi )$ of
$(G,k)$ corresponds to the dimension of the underlying
Hilbert space $H$.  Hence:

\begin{lemma} The category $\Rep _f(G,k)$ has irreducible objects of
  dimension greater than $1$ iff $G$ is nonabelian.  \end{lemma}

\begin{proof}[Sketch of proof] The set of irreducible representations
  of $G$ separates the elements of $G$.  Hence for $gh\neq hg$, there
  is an irreducible representation $(H,\pi )$ such that $\pi (g)\pi
  (h)\neq \pi (h)\pi (g)$.  Therefore $\dim H\geq 2$.  \end{proof}
\noindent It immediately follows from Fact \ref{risque}, in
conjunction with the fact that the embedding functor preserves
dimension, that:

\begin{prop} There is an irreducible object $X$ of $\2C \simeq \Rep
  _f(G,k)$ with parastatistics iff the corresponding group $G$ is nonabelian.
\end{prop}

Applied to our current case, this means that there are representations
and states with parastatistics iff the gauge group $G$ is
nonabelian.\footnote{But there is an ambiguity in ``parastatistics.''
  We mean para-Bose or para-Fermi statistics, not mixtures of Bose and
  Fermi statistics.}  But we have good reasons to think that the case
of nonabelian gauge groups is physically relevant.  So, the DHR
approach ignores worries about the supposed nonexistence of
paraparticle states, and undermines claims that there is a proof of
Dichotomy.

\subsubsection{Braid group statistics}

Recall from Section \ref{braided} that when spacetime
is dimension $2$, then $\ve _{\rho _1,\rho _2}$ is not
necessarily a symmetry on $\D _f$, but only a braiding.
In this case, objects in $\D _f$ are not classified
according to representations of the symmetric group
$S_n$; rather, objects in $\D _f$ are classified in
terms of representations of the braid group $B_n$.  In
physical terms, states might not be permutation
invariant, but satisfy the more general braid group
statistics.

\begin{defn} The \emph{braid group} $B_n$ on $n$
  strands is the group generated by the set $\{ \sigma
  _1,\dots ,\sigma _{n-1} \}$ satisfying the equations
  \[ \begin{array}{llll} (1) & \sigma _i\sigma _j =\sigma _j\sigma _i
    , & &
    \abs{i-j}\geq 2 ,\vspace{4pt} \\
    (2) & \sigma _{i+1}\sigma _i\sigma _{i+1} =\sigma _i\sigma
    _{i+1}\sigma _i .\end{array} \]
\end{defn}
The braid group on $n$ strands can be given the
following heuristic, geometric description: a braid
consists of two sets $\{ a_1,\dots ,a_n\}$ and $\{
b_1,\dots ,b_n\}$ of base points, and $n$ strands of
yarn, where each yarn strand has one end attached to
$a_i$, and the other end attached to $b_j$, and each
base point touches only one strand of yarn.  We
identify two configurations if one can be transformed
into the other without disconnecting the strands from
the base points.  In this picture, the identity braid
has $a_i$ connected to $b_i$, and no twists.  The
generating element $\sigma _i$ can be thought of as the
simple braid where: $a_i$ is connected to $b_{i+1}$,
$a_{i+1}$ is connected to $b_i$, and these two strands
are twisted only once.  (Otherwise, the $\sigma _i$
braid looks just like the identity braid.)  Under this
interpretation, the product $gh$ of two braids is the
braid formed by attaching the ending points of $g$ to
the beginning points of $h$.

\begin{prop} For each $n\in \7N$, the mapping
  \begin{eqnarray*} \ve ^{(n)}_{\rho}(\sigma _i) & = &
    \rho ^{i-1}(\ve _\rho) \quad =\quad I_\rho \times
    \cdots \times I_\rho \times \ve _{\rho} \times
    I_\rho \times \cdots \times
    I_{\rho},\end{eqnarray*} defines a unitary
  representation of the braid group $B_n$ in
  $\mathrm{End}\left( \rho \otimes \cdots \otimes \rho
  \right)$.  For each $i,j\in \7N$ with $i\leq j$ we
  have
$$ \ve ^{(i)}_{\rho}(g) =\ve ^{(j)}_{\rho}(\f _{ij}(g)) ,\qquad \qquad \forall g\in
S_i .$$ \label{thm-braid}
\end{prop}

(A proof of this Proposition can be found in \cite{rehr}.
Alternatively, it is obvious given the considerations in Appendix
\ref{App-B}.)  In other words, the product object $\rho \otimes \cdots
\otimes \rho$ carries a unitary representation of the braid group,
which is induced by the unitary operators of the form:
$$I_\rho \times \cdots \times \ve _{\rho ,\rho} \times \cdots
\times I_\rho .$$ This represents an elementary permutation of the
$i$-th and $(i+1)$-st copy of $\rho$.

There is a natural homomorphism of the braid group onto the symmetric
group.  This is obvious when we recall that the definition of the
symmetric group is exactly the same as the definition of the braid
group with the additional condition that each generator is its own
inverse.  Hence, van Dyck's theorem \cite[p.\ 78]{hung} entails that
the obvious map $f$ on generators extends uniquely to a group
homomorphism $\ol f:B_n\to S_n$.  So, each representation $\pi$ of
$S_n$ yields a representation $\pi \circ \ol f$ of $B_n$. In slogan
form: a system that obeys permutation statistics also obeys braid
statistics.

Recall now the worrisome argument for the existence of paraparticles: 
There should be particles corresponding to all irreducible
representations of $S_n$.  For $n\geq 3$, there are non Bose or Fermi
representations of $S_n$, so there should be paraparticles.  

Now we can see that either something is wrong with this
argument, or the problem is much more severe than we
thought.  Since any system that has $S_n$ as a symmetry
group also has $B_n$ as a symmetry group, the argument
commits us to predicting the existence of particles
corresponding to all irreducible representations of
$B_n$.  But $B_n$ has infinitely many irreducible
representations.  (Indeed, its representations have so
far resisted classification.)  Furthermore, we could
now repeat the argument for \emph{any} group $K$ that
can be mapped homomorphically onto $B_n$, and there is
an infinite chain of larger groups with this property.
Therefore, the principle of plenitude applied to group
representations predicts more particles than we could
ever possibly describe.

\bigskip \noindent {\small \textit{Notes:} For discussions of
  statistics of DHR representations, see \cite{rob-stat,dop-stat}.}

\addcontentsline{toc}{section}{Bibliography}

\bibliographystyle{named} {\small \bibliography{dr} }

\newpage

\appendix

\noindent {\Large \textbf{Appendix (by Michael M{\"u}ger})}
\addcontentsline{toc}{section}{Appendix (by Michael M{\"u}ger)}

\def\endexem{\hfill{ }\medskip}

\theoremstyle{change}

\newcommand{\be}{\begin{equation}}
\newcommand{\ee}{\end{equation}}
\newcommand{\bea}{\begin{eqnarray}}
\newcommand{\eea}{\end{eqnarray}}
\newcommand{\bean}{\begin{eqnarray*}}
\newcommand{\eean}{\end{eqnarray*}}

\newcommand{\bdefin}{\begin{defn}}
\newcommand{\blemma}{\begin{lemma}}
\newcommand{\bprop}{\begin{prop}}
\newcommand{\btheor}{\begin{thm}}
\newcommand{\bcoro}{\begin{cor}}
\newcommand{\bconj}{\begin{conj}}
\newcommand{\bdefprop}{\begin{defprop}}
\newcommand{\bexam}{\begin{example}}
\newcommand{\edefin}{\end{defn}}
\newcommand{\elemma}{\end{lemma}}
\newcommand{\eprop}{\end{prop}}
\newcommand{\etheor}{\end{thm}}
\newcommand{\ecoro}{\end{cor}}
\newcommand{\econj}{\end{conj}}
\newcommand{\brem}{\begin{rema}}
\newcommand{\erem}{\endexem\end{rema}}
\newcommand{\edefprop}{\end{defprop}}
\newcommand{\eexam}{\endexem\end{example}}

\def\1#1{{\bf #1}}
\def\2#1{{\cal #1}}
\def\3#1{{\sl #1}}
\def\4#1{{\tt #1}}
\def\5#1{{\sf #1}}
\def\6#1{{\mathfrak #1}}
\def\7#1{{\mathbb #1}}

\def\qed{\hfill{$\Box$}\medskip}
\def\altem{\slshape}

\def\id{\mathrm{id}}
\newcommand{\Mod}{\mathrm{Mod}}
\renewcommand{\ker}{\mathrm{ker}}
\newcommand{\coker}{\mathrm{coker}}
\newcommand{\im}{\mathrm{im}}

\def\implies{\Rightarrow}
\renewcommand{\restr}{\upharpoonright}
\renewcommand{\rarr}{\rightarrow}
\newcommand{\lrarr}{\longrightarrow}
\renewcommand{\impl}{\Rightarrow}
\renewcommand{\oli}{{\overline{\imath}}}
\renewcommand{\ve}{\varepsilon}
\renewcommand{\op}{{\mbox{\scriptsize op}}}

\renewcommand{\DS}{\displaystyle}

\def\ip#1#2{{(\,#1\, , \, #2\,)}}
\def\m#1{{\vert #1 \vert}} \def\n#1{\Vert #1 \Vert}
\def\prf{\noindent \emph{Proof.\ }}

\newarrow{Congruent} 33333

\vspace{2em} \noindent Not much in these two appendices
is new. (Theorem \ref{theor-F} probably is, and see
Remark \ref{rem-new}.) However, this seems to be the
first exposition of the reconstruction theorem for
symmetric tensor categories that gives complete and
streamlined proofs, including a short and transparent
proof of Tannaka's classical theorem. In the first
section we provide the necessary concepts and results
of category theory to the extent that they don't
involve the notion of fiber functor, whereas the second
section is concerned with the Tannaka theory
proper. Our main reference for category theory is
\cite{cwm}, preferably the second edition. The reader
having some experience with categories is advised to
skip directly to Section \ref{App-B}, using the present
section as a reference when needed.

\section{Categorical Preliminaries} \label{cats} \label{app-A}

\subsection{Basics}
\bdefin \label{def-cat} \index{category}
A \emph{category} $\2C$ consists of:
\begin{itemize}
\item A class $\Obj (\2C )$ of \emph{Objects}.  We
  denote the objects by capital letters $X,Y,\ldots$.
\item For any two objects $X,Y$ a set $\Hom_\2C(X,Y)$
  of \emph{arrows} (or morphisms); we write $f:X\to Y$
  to indicate that $f\in \Hom_\2C(X,Y)$, and we omit
  the subscript $\2C$ whenever there is no risk of
  confusion.
\item For any object $X$ a distinguished arrow $\id _X\in \End (X)=\Hom(X,X)$.
\item For each $X,Y,Z\in \Obj (\2C )$, a function
  $\circ :\Hom(Y,Z)\times \Hom(X,Y)\to \Hom (X,Z)$ such
  that:
  \[ h\circ (g\circ f)=(h\circ g)\circ f ,\] and
  \[ \id _Y\circ f=f ,\qquad g\circ \id _Y=g ,\] whenever $f\in \Hom
  (X,Y)$, $g\in \Hom (Y,Z)$, and $h\in \Hom (Z,W)$.
\end{itemize} \edefin

\bdefin \index{isomorphism} A morphism $f\in \Hom (X,Y)$ is an
\emph{isomorphism} iff it is invertible, i.e.\ there is a $g\in \Hom
(Y,X)$ such that $g\circ f=\id_{X}$ and $f\circ g=\id_Y$. If an
isomorphism $X\rarr Y$ exists, we write $X\cong Y$.  \edefin

\bdefin \index{subcategory} If $\2C$ is a category, then a
\emph{subcategory} $\2D\subset\2C$ is defined by a subclass
$\Obj\,\2D\subset\Obj\,\2C$ and, for every $X,Y\in\Obj\,\2D$, a subset
$\Hom_\2D(X,Y)\subset\Hom_\2C(X,Y)$ such that $\id_X\in\Hom_\2D(X,X)$
for all $X\in\Obj\,\2D$ and the morphisms in $\2D$ is closed under the
composition $\circ$ of $\2C$. A subcategory $\2D\subset\2C$ is
\emph{full} if $\Hom_\2D(X,Y)=\Hom_\2C(X,Y)$ for all
$X,Y\in\Obj\,\2D$.  \edefin

\bdefin \index{functor}
A (covariant) \emph{functor} $F$ from category $\2C$ to category $\2D$
  maps objects of $\2C$ to objects of $\2D$ and arrows of $\2C$ to arrows of $\2D$
  such that $F(g\circ f)=F(g)\circ F(f)$, and $F(\id _X)=\id _{F(X)}$.  A
  \emph{contravariant functor} is just like a covariant functor except that it
  reverses the order of arrows. \edefin

  \bdefin \index{functor!faithful} \index{functor!full} A functor
  $F:\2C\to \2D$ is \emph{faithful}, respectively \emph{full}, if the
  map
\[ F_{X,Y}:\Hom _{\2C}(X,Y)\to \Hom _{\2D}(F(X),F(Y)), \]
is injective, respectively surjective, for all $X,Y\in \Obj (\2C)$.  
\edefin

\bdefin \index{functor!essentially surjective}
A functor $F:\2C\to \2D$ is \emph{essentially surjective} if for every $Y\in\Obj\,\2D$ there
is an $X\in\Obj\,\2C$ such that $F(X)\cong Y$.
\edefin

\bdefin \index{natural transformation} \label{nat-trans}
If $F:\2C\to \2D$ and $G:\2C\to \2D$ are functors, then a \emph{natural
    transformation} $\eta$ from $F$ to $G$ associates to every $X\in\Obj (\2C )$ a
  morphism $\eta _X \in \Hom _{\2D }(F(X),G(X))$ such that 
\[ \begin{diagram} F(X) & \rTo^{F(s)} & F(Y) \\
   \dTo^{\eta_X} && \dTo_{\eta_Y} \\
   G(X) & \rTo_{G(s)} & G(Y) \end{diagram} \]
commutes for every arrow $f\in \Hom_{\2C }(X,Y)$.  If $\eta _X$ is an isomorphism
for every $X\in \Obj (\2C )$, then $\eta$ is said
  to be a \emph{natural isomorphism}. 
\edefin

\bdefin \index{category!equivalence} \index{equivalence!of categories}
A functor $F:\2C\to \2D$ is an \emph{equivalence of categories} if there exist a functor
$G:\2D\to \2C$ and natural isomorphisms $\eta:FG\to \id _{\2D}$ and $\ve :\id _{\2C}\to GF$. Two
categories are \emph{equivalent}, denoted $F\simeq G$, if there exists an equivalence
$F:\2C\rarr\2D$. 
\edefin

\bdefin A category is \emph{small} if $\Obj\,\2C$ is a set (rather than just a class). A category is
\emph{essentially small} if it is equivalent to a small one, i.e.\ $\Obj\,\2C/\cong$ is a set. 
\edefin

\brem Without wanting to go into foundational technicalities we point out that the category of a
`all representations' of a group is a huge object. However, considered modulo equivalence the
representations are of reasonable cardinality, i.e.\ are a set.
\erem

\subsection{Tensor categories and braidings}
\bdefin \index{category!product}
Given two categories $\2C,\2D$, the product category $\2C\times\2D$ is defined by
\bean \Obj(\2C\times\2D) &=& \Obj\,\2C\times\Obj\,\2D, \\
   \Hom_{\2C\times\2D}(X\times Y,Z\times W) &=& \Hom_\2C(X,Z)\times\Hom_\2D(Y,W), \\
  \id_{X\times Y} &=& \id_X\times\id_Y \eean
with the obvious composition $(a\times b)\circ(c\times d):=(a\circ c)\times(b\circ d)$.
\edefin

\bdefin \index{category!monoidal} \index{tensor!category}
A strict tensor category (or strict monoidal category) is a category $\2C$ equipped with a
distinguished object $\11$, the tensor unit, and a functor $\otimes:\2C\times\2C\rarr\2C$ such that:
\begin{enumerate}
\item $\otimes$ is associative on objects and morphisms, i.e.\ $(X\otimes Y)\otimes
Z=X\otimes(Y\otimes Z)$ and $(s\otimes t)\otimes u=s\otimes(t\otimes u)$ for all
$X,Y,Z,X',Y',Z'\in\Obj\,\2C$ and all $s:X\rarr X', t:Y\rarr Y', u:Z\rarr Z'$.
\item The unit object behaves as it should: $X\otimes\11=X=\11\otimes X$ and
$s\otimes\id_\11=s=\id_\11\otimes s$ for all $s:X\rarr Y$.
\item The interchange law 
\[ (a\otimes b)\circ(c\otimes d)=(a\circ c)\otimes(b\circ d) \]
holds whenever $a\circ c$ and $b\circ d$ are defined.
\end{enumerate}
\edefin

\brem
Many categories with tensor product are not strict in the above sense. A tensor category is a
category equipped with a functor $\otimes:\2C\times\2C\rarr\2C$, a unit $\11$ and natural
isomorphisms $\alpha_{X,Y,Z}:(X\otimes Y)\otimes Z\rarr X\otimes(Y\otimes Z)$, 
$\lambda_X:\11\otimes X\rarr X$, $\rho_X:X\otimes\11\rarr X$ satisfying certain identities.
The notions of braiding, monoidal functor and monoidal natural transformation generalize to such
categories. The generality achieved by considering non-strict categories is only apparent: By the
coherence theorems, every (braided/symmetric) tensor category is monoidally naturally equivalent to
a strict one. See \cite{cwm,JS2} for all this. 

Strictly speaking (pun intended) the categories of vector spaces and Hilbert spaces are not strict. 
However, the coherence theorems allow us to pretend that they are, simplifying the formulae
considerably. The reader who feels uncomfortable with this is invited to insert the isomorphisms
$\alpha,\lambda,\rho$ wherever they should appear.
\erem

\bdefin \index{subcategory!monoidal} \index{tensor!subcategory}
A (full) tensor subcategory of a tensor category $\2C$ is a (full) subcategory
$\2D\subset\2C$ such that $\Obj\,\2D$ contains the unit object $\11$ and is closed under the tensor
product $\otimes$.
\edefin

\bdefin \label{def-mon-func} \index{tensor!functor} \index{functor!monoidal}
Let $\2C, \2D$ be strict tensor categories. A tensor functor (or a monoidal functor) is a functor
$F: \2C\rarr\2D$ together with isomorphisms $d^F_{X,Y}: F(X)\otimes F(Y)\rarr F(X\otimes Y)$ for all
$X,Y\in\2C$ and a morphism $e^F: \11_\2D\rarr F(\11_\2C)$ such that
\begin{enumerate}
\item The morphisms $d^F_{X,Y}$ are natural w.r.t.\ both arguments.
\item For all $X,Y,Z\in\2C$ the following diagram commutes:
\be \label{eq-A1}
\begin{diagram}
F(X)\otimes F(Y)\otimes F(Z) & \rTo^{d^F_{X,Y}\otimes\id_{F(Z)}} & F(X\otimes Y)\otimes F(Z) \\
\dTo^{\id_{F(X)}\otimes d^F_{Y,Z}} & & \dTo_{d^F_{X\otimes Y,Z}} \\
F(X)\otimes F(Y\otimes Z) & \rTo_{d^F_{X, Y\otimes Z}} & F(X\otimes Y\otimes Z)
\end{diagram}
\end{equation}
\item The following compositions are the identity morphisms of $F(X)$
\be \label{eq-A2} 
\begin{array} {c}
\begin{diagram}
F(X) \equiv F(X)\otimes \11_\2D & \rTo^{\id_{F(X)}\otimes e^F} & F(X)\otimes F(\11_\2C) & \rTo^{d_{X,\11}} &
   F(X\otimes\11_\2C) \equiv F(X)
\end{diagram} \\
\begin{diagram}
F(X) \equiv \11_\2D\otimes F(X) & \rTo^{e^F\otimes\id_{F(X)}} & F(\11_\2C)\otimes F(X) & \rTo^{d_{\11,X}} &
   F(\11_\2C\otimes X) \equiv F(X)
\end{diagram}
\end{array} \ee
for all $X\in\2C$.
\end{enumerate}
If $\2C,\2D$ are tensor $*$-categories and $F$ is $*$-preserving, the isomorphisms $e, d_{X,Y}$ are 
required to be unitary. 
\edefin

\bdefin \label{def-monnat}\index{natural transformation!monoidal}
Let $\2C,\2D$ be strict tensor categories and $F,G:\2C\rarr\2D$ tensor functors. A natural
transformation $\alpha:\2C\rarr\2D$ is monoidal if 
\[ \begin{diagram} F(X)\otimes F(Y) & \rTo^{d^F_{X,Y}} &  F(X\otimes Y) \\
  \dTo^{\alpha_X\otimes\alpha_Y} && \dTo_{\alpha_{X\otimes Y}} \\
   G(X)\otimes G(Y) & \rTo_{d^G_{X,Y}} &  G(X\otimes Y) 
\end{diagram} 
\]
commutes for all $X,Y\in\2C$ and the composite
$\11_\2D\stackrel{e^F}{\longrightarrow}F(\11)\stackrel{\alpha_\11}{\longrightarrow}G(\11)$
coincides with $e^G$.
\edefin

\brem A tensor functor between strict tensor categories is called strict if all the isomorphisms
$d_{X,Y}$ and $e$ are identities. However, it is not true that every tensor functor is equivalent to
a strict one! 
\erem

\bdefin \index{equivalence!of tensor categories}
A tensor functor $F:\2C\rarr\2D$ is an \emph{equivalence} (of tensor categories) if there
exist a tensor functor $G:\2D\rarr\2C$ and monoidal natural isomorphisms $GF\rarr\id_\2C$ and
$FG\rarr\id_\2C$. 
\edefin

\begin{prop} \label{prop-equiv}
A functor $F:\2C\to \2D$ is an equivalence iff $F$ is faithful, full and essentially surjective. A
tensor functor $F:\2C\to \2D$ of (strict) tensor categories is an equivalence of tensor categories
iff $F$ is faithful, full and essentially surjective. 
\end{prop}

\prf For the first statement see \cite[Theorem 1, p.\ 91]{cwm} and for the second \cite{SR}.
\qed


\bdefin \index{braiding} \index{category!braided monoidal} \index{tensor!category, braided}
\index{symmetry}\index{category!symmetric monoidal} \index{tensor!category, symmetric}
A braiding for a strict tensor category $\2C$ is a family of isomorphisms 
$c_{X,Y}:X\otimes Y\rarr Y\otimes X$ for all $X,Y\in\Obj\,\2C$ satisfying
\begin{enumerate}
\item Naturality: For every $s:X\rarr X',\ t:Y\rarr Y'$, the diagram
\[ \begin{diagram} X\otimes Y & \rTo^{c_{X,Y}} & Y\otimes X \\
  \dTo^{s\otimes t} && \dTo_{t\otimes s} \\
  X'\otimes Y' & \rTo^{c_{X',Y'}} & Y'\otimes X' 
\end{diagram}\]
commutes.
\item The `braid equations' hold, i.e.\ the diagrams
\[ \begin{diagram} X\otimes Y\otimes Z & \rTo^{c_{X,Y}\otimes\id_Z} & Y\otimes X\otimes Z \\
   &\rdTo_{c_{X,Y\otimes Z}} & \dTo_{\id_Y\otimes c_{X,Z}} \\ && Y\otimes Z\otimes X
\end{diagram}
\quad\quad\quad\quad
\begin{diagram} X\otimes Y\otimes Z & \rTo^{\id_X\otimes c_{Y,Z}} & X\otimes Z\otimes Y \\
   &\rdTo_{c_{X\otimes Y, Z}} & \dTo_{c_{X,Z}\otimes\id_Y} \\ && Z\otimes X\otimes Y
\end{diagram}
\]
commute for all $X,Y,Z\in\Obj\,\2C$.
\end{enumerate}
If, in addition, $c_{Y,X}\circ c_{X,Y}=\id_{X\otimes Y}$ holds for all $X,Y$, the braiding is called
a symmetry. 

A strict braided (symmetric) tensor category is a strict tensor category equipped with a braiding
(symmetry).
\edefin

\bdefin \index{functor!braided monoidal}\index{functor!symmetric monoidal}
If $\2C,\2D$ are strict braided (symmetric) tensor categories, a tensor functor
$F:\2C\rarr\2D$ is braided (symmetric) if 
\[ F(c_{X,Y})=c_{F(X),F(Y)} \quad \forall X,Y\in\Obj\,\2C. \]
(Note that on the l.h.s., respectively r.h.s, $c$ is the braiding of $\2C$, respectively $\2D$.
\edefin

There is no additional condition for a monoidal natural transformation to  be braided/symmetric.


\subsection{Graphical notation for tensor categories}
We will on some occasions use the so-called `tangle
diagrams' for computations in strict (braided) tensor
categories, hoping that the reader has seen them
before. By way of explanation (for much more detail see
e.g.\ \cite{kas}) we just say that identity morphisms
(equivalently, objects) are denoted by vertical lines,
a morphism $s:X\rarr Y$ by a box with lines
corresponding to $X$ and $Y$ entering from below and
above, respectively.  Compositions and tensor products
of morphisms are denoted by vertical and horizontal
juxtaposition, respectively. Braiding morphisms are
represented by a crossing and the duality morphisms
$r,\ol{r}$ by arcs:
\[ \Hom(X,Y)\ni s \equiv\quad \begin{tangle}\hstep\object{Y}\\ \hh\hstep\id\\ \frabox{s}\\
    \hh\hstep\id\\  \hstep\object{X}\end{tangle}\quad\quad
   c_{X,Y}\equiv\quad\begin{tangle}\object{Y}\Step\object{X}\\ \xx \\
    \object{X}\Step\object{Y}\end{tangle}  \quad\quad
   c^{-1}_{Y,X}\equiv\quad\begin{tangle}\object{Y}\Step\object{X}\\ \x \\
    \object{X}\Step\object{Y}\end{tangle}  \] 

\[ r_X\equiv\quad \begin{tangle}\object{\ol{X}}\Step\object{X}\\ \ev \end{tangle} \quad\quad
    \ol{r}_X\equiv\quad \begin{tangle}\object{X}\Step\object{\ol{X}}\\ \ev \end{tangle} 
\] 
(If $c$ is a symmetry, both lines are drawn unbroken.)
The reason for using this diagrammatic representation is that even relatively simple formulae in
tensor categories become utterly unintelligible as soon as morphisms with `different numbers of in-
and outputs' are involved, like $s: A\rarr B\otimes C\otimes D$. This gets even worse when braidings
and duality morphisms are involved. Just one example of a complete formula: The interchange law
$s\otimes\id_W\circ\id_X\otimes t=\id_Y\otimes t\circ s\otimes\id_Z$ for $s:X\rarr Y,\ t:Z\rarr W$
is drawn as
\[ \begin{tangle} \hstep\object{Y}\Step\object{W} \\ \hh \hstep\id\Step\id \\ 
   \hh\frabox{s}\step[1.5]\id \\ \hh \hstep\id\Step\id \\ 
   \hh\hstep\id\step[1.5]\frabox{t}\\ \hh\hstep\id\Step\id\\ \hstep\object{X}\Step\object{Z}  \end{tangle}
\quad\quad=\quad\quad
\begin{tangle} \hstep\object{Y}\Step\object{W} \\ \hh \hstep\id\Step\id \\ 
   \hh\hstep\id\step[1.5]\frabox{t} \\ \hh \hstep\id\Step\id \\ 
   \hh\frabox{s}\step[1.5]\id\\ \hh\hstep\id\Step\id\\ \hstep\object{X}\Step\object{Z}  \end{tangle}
\]
The diagram (correctly!) suggests that we have may pull morphisms alongside each other.


\subsection{Additive, $\7C$-linear and $*$-categories}
\bdefin \label{def-abcat} \index{category!Ab-}
A category is an Ab-category if all hom-sets are abelian groups and the composition $\circ$ is
bi-additive. 
\edefin

\bdefin \label{def-dirsum}\index{direct sum}
Let $X,Y,Z$ be objects in a Ab-category. Then $Z$ is called a \emph{direct sum} of $X$ and $Y$,
denoted $Z\cong X\oplus Y$, if there are morphisms $u:X\rarr Z, u':Z\rarr X,v:Y\rarr Z, v':Z\rarr Y$
such that $u'\circ u=\id_X,\ v'\circ v=\id_Y$ and $u\circ u'+v\circ v'=\id_Z$. (Note that every
$Z'\cong Z$ also is a direct sum of $X$ and $Y$. Thus direct sums are defined only up to
isomorphism, which is why we don't write $Z=X\oplus Y$.) We say  that $\2C$ \emph{has direct sums}
if there exists a direct sum $Z\cong X\oplus Y$ for any two object $X,Y$.
\edefin

\bdefin
An object $\10$ in a category $\2C$ is called a zero object if, for every $X\in\2C$, the sets
$\Hom(X,\10)$ and $\Hom(\10,X)$ both contain precisely one element. A morphism to or from a zero
object is called a zero morphism. 
\edefin

It is immediate from the definition that any two zero objects are isomorphic. If a category doesn't
have a zero object it is straightforward to add one. If $z$ is a zero morphism and $f$ is any
morphism, then $z\circ f, f\circ z, z\otimes f, f\otimes z$ are zero morphisms (provided they make sense).

\bdefin \index{category!additive}  
An \emph{additive} category is an Ab-category that has a zero object and direct sums.
\edefin

\bexam The category of abelian groups (with the trivial group as zero).
\eexam

\bdefin \index{category!$\7C$-linear}
A category $\2C$ is called $\7C$-linear if $\Hom(X,Y)$ is a $\7C$-vector space for all
$X,Y\in\Obj\,\2C$ and the composition map $\circ: (f,g)\mapsto g\circ f$ is bilinear. If $\2C$ is a
tensor category we require that also $\otimes: (f,g)\mapsto g\otimes f$ is bilinear.
Functors between $\7C$-linear category are always assumed to be $\7C$-linear, i.e.\
$\Hom_\2C(X,Y)\rarr\Hom_\2D(F(X),F(Y))$ must be $\7C$-linear.
\edefin

\bdefin \index{$*$-operation}\index{$*$-category}
A \emph{positive $*$-operation} on a $\7C$-linear category is a family of maps that to every
morphism $s\in\Hom(X,Y)$ associates a morphism $s^*\in\Hom(Y,X)$. This map must be antilinear,
involutive ($(s^*)^*=s$) and positive in the sense that $s^*\circ s=0$ implies $s=0$. A $*$-category
is a $\7C$-linear category equipped with a positive $*$-operation. A \emph{tensor $*$-category} is a tensor
category with a positive $*$-operation such that $(s\otimes t)^*=s^*\otimes t^*$ for all $s,t$.
We consider only unitary braidings (symmetries) of tensor $*$-categories.
\edefin

\bdefin \index{isometry}\index{projector}
A morphism $v:X\rarr Y$ in a $*$-category is called an \emph{isometry} if $v^*\circ v=\id_X$. 
It is called a \emph{unitary} if it satisfies $v^*\circ v=v\circ v^*=\id_Y$. A morphism
$p\in\End\,X$ is called a projector if $p=p\circ p=p^*$. We say that $\2C$ \emph{has subobjects} if
for every projector $p\in\End\,X$ there exists an isometry $v:Y\rarr X$ such that $v\circ v^*=p$.
In a $*$-category we strengthen Definition \ref{def-dirsum} by requiring that $u'=u^*, v'=v^*$,
i.e.\ $u,v$ must be isometries.
\edefin

\bdefin \index{functor!$*$-preserving}
A functor $F$ between $*$-categories is \emph{$*$-preserving} if $F(s^*)=F(s)^*$ for every morphism
$s$. The isomorphisms $d_{X,Y}, e$ coming with a functor between tensor $*$-categories  coming with
a functor of tensor $*$-categories are required to be unitaries. 
\edefin

\bdefin \label{def-conj} \index{conjugate}
Let $\2C$ be a tensor $*$-category and $X\in\Obj\,\2C$. An object $\ol{X}\in\Obj\,\2C$ is called a
\emph{conjugate} object of $X$ if there exist morphisms $r:\11\rarr\ol{X}\otimes X$ and 
$r:\11\rarr X\otimes\ol{X}$ satisfying the `conjugate equations'
\bean \id_X\otimes r^*\mcirc \ol{r}\otimes\id_X &=& \id_X\, \\
  \id_{\ol{X}}\otimes\ol{r}^*\mcirc r\otimes\id_{\ol{X}} &=& \id_{\ol{X}}.
\eean
We say that $(\ol{X},r,\ol{r})$ is a \emph{conjugate} of $X$.
If every non-zero object of $\2C$ has a conjugate then we say that $\2C$ has conjugates.
\edefin

Note also that a zero object cannot have a conjugate.
If $(\ol{X},r,\ol{r}),(\ol{X}',r',\ol{r}')$ both are conjugates of $X$ then one easily verifies
that $\id_{\ol{X}'}\otimes\ol{r}^*\mcirc r'\otimes\id_{\ol{X}}: \ol{X}\rarr\ol{X}'$ is unitary. Thus
the conjugate is unique up to unitary equivalence. 

\bdefin \index{object!irreducible}
An object in a $\7C$-linear category is \emph{irreducible} if $\End\,X=\7C\id_X$.
\edefin

\bdefin \index{TC$^*$} \index{STC$^*$} \index{BTC$^*$}
A $TC^*$ is an tensor $*$-category with finite dimensional hom-sets, with conjugates, direct sums, 
subobjects and irreducible unit $\11$. A $BTC^*$ is a TC$^*$ with a unitary braiding. An $STC^*$ is
a TC$^*$ with a unitary symmetry.
\edefin

\bexam \label{exam-H}
The tensor $*$-category $\2H$ of finite dimensional Hilbert spaces is a $STC^*$.
The symmetry $c_{H,H'}: H\otimes H'\rarr H'\otimes H$ is given by the flip isomorphism 
$\Sigma: x\otimes y\mapsto y\otimes x$. The conjugate of an object $H$ is the Hilbert space dual
$\ol{H}$. Picking a basis $\{e_i\}$ of $H$ with dual basis $\{f_i\}$, the conjugation morphisms are
given by 
\[ r=\sum_i f_i\otimes e_i, \quad \ol{r}=\sum_i e_i\otimes f_i. \]
In the same way one sees that the category $\Rep_fG$ of finite dimensional unitary representations
of a compact group $G$ is an $STC^*$.
\eexam

\blemma \label{lem-semisim} 
A TC$^*$ is semisimple, i.e.\ every object is a finite direct sum of irreducible objects.
\elemma

\prf For every $X\in\2C$, $\End\,X$ is a finite dimensional $\7C$-algebra with a positive
involution. Such an algebra is semisimple, to wit a multi matrix algebra. Thus $\id_X$ is a sum of
projections $p_i$ that are minimal in the sense that $p_i\End\,Xp_i\cong\7C$. Since $\2C$ has
subobjects, there are objects $X_i$ corresponding to the $p_i$, which are irreducible by minimality
of the $p_i$. Clearly, $X\cong\oplus_iX_i$.
\qed

\bdefin \index{conjugate!standard}
A solution $(X,r,\ol{r})$ of the conjugate equations is called standard if
\[ r^* \mcirc \id_{\ol{X}}\otimes s\mcirc r= \ol{r}^*\circ s\otimes\id_{\ol{X}}\circ\ol{r} \]
for all $s\in\End\,X$. In this case, $(\ol{X},r,\ol{r})$ is called a standard conjugate.
\edefin

\blemma \label{lem-stand1}
Let $\2C$ be a $TC^*$ and $(\ol{X},r,\ol{r})$ a conjugate for $X\in\2C$. Let $v_i:X_i\rarr X$,
$w_i:\ol{X_i}\rarr\ol{X}$ be isometries effecting the direct sum decomposition of $X,\ol{X}$ into
irreducibles. Then $(\ol{X},r,\ol{r})$ is a standard conjugate iff
$(\ol{X_i},w_i^*\otimes v_i^*\circ r,v_i^*\otimes w_i^*\circ\ol{r})$ 
is a standard conjugate for $X_i$ for all $i$. Every object admits a standard conjugate.
\elemma

\prf For the equivalence claim, see \cite{LR}, in particular Lemma 3.9. (Note that in \cite{LR},
standardness is defined by the property in the statement above.) We use this to prove that every
objects admits a standard conjugate. If $X$ is irreducible, we have $\End\,X=\7C\id_X$. Therefore
the standardness condition reduces to $r^*\circ r=\ol{r}^*\circ\ol{r}$, thus a conjugate
$(\ol{X},r,\ol{r})$ can be made standard by rescaling $r,\ol{r}$. In the general case, we use
semisimplicity to find a direct sum decomposition of $X$ into irreducibles $X_i$.  
Let $(\ol{X_i},r_i,\ol{r}_i)$ be standard conjugates of the $X_i$ and put $\ol{X}=\oplus\ol{X}_i$.
Let $v_i:X_i\rarr X,\ w_i:\ol{X}_i\rarr\ol{X}$ be the isometries effecting the direct sums. Defining
$r=\sum_i w_i\otimes v_i\circ r_i$ and $\ol{r}=\sum_i v_i\otimes w_i\circ\ol{r}_i$, the criterion in
the first part of the lemma applies and gives standardness of $(\ol{X},r,\ol{r})$.
\qed

\blemma \label{lem-conj}
Let $(\ol{X},r,\ol{r})$ be a (standard) conjugate of $X$, let $p\in\End\,X$ a projection and define
$\ol{p}=r^*\otimes\id_{\ol{X}}\mcirc\id_{\ol{X}}\otimes p\otimes\id_{\ol{X}}\mcirc\id_{\ol{X}}\otimes\ol{r}\in\End\,\ol{X}$.
If $v: Y\rarr X$, $w:\ol{Y}\rarr\ol{X}$ are isometries such that $v\circ v^*=p, w\circ w^*=\ol{p}$ then
$(\ol{Y},w^*\otimes v^*\mcirc r,v^*\otimes w^*\mcirc\ol{r})$ is a (standard) conjugate for $Y$.
\elemma

\prf Omitted. For the easy proof see \cite{LR} or \cite{mue06}.
\qed

\blemma \label{lem-mult}
If $(\ol{X},r,\ol{r}),(\ol{Y},r',\ol{r}')$ are (standard) conjugates of $X,Y$, respectively, then
$(\ol{Y}\otimes\ol{X}, r'',\ol{r}'')$, where $r''=\id_{\ol{Y}}\otimes r\otimes\id_Y\circ r'$,
$\ol{r}''=\id_X\otimes\ol{r'}\otimes\id_{\ol{X}}\circ\ol{r})$ 
is a (standard) conjugate for $X\otimes Y$.
\elemma

\prf That $(\ol{Y}\otimes\ol{X}, r'',\ol{r}'')$ is a conjugate is an easy computation. Standardness
is less obvious since the map $\End\,X\otimes\End\,Y\rarr\End\,X\otimes Y$ need not be
surjective. However, it follows using the alternative characterization of standardness given in
Lemma \ref{lem-stand1}.
\qed

\bprop \label{prop-trace} \index{trace}
Let $\2C$ be a TC$^*$. Let
$X\in\2C$ and let $(\ol{X}, r,\ol{r})$ be a standard conjugate. Then the map
\[ Tr_X: \ \End\,X\rarr\7C,\quad s\mapsto r^*\mcirc \id_{\ol{X}}\otimes s\mcirc r \]
is well defined, i.e.\ independent of the choice of $(\ol{X}, r,\ol{r})$. It is called the trace. It
satisfies
\bean Tr_X(s\mcirc  t) &=& Tr_Y(t\mcirc  s) \quad\forall s: Y\rarr X, \ t: X\rarr Y,  \\
  Tr_{X\otimes Y}(s\otimes t) &=& Tr_X(s)\, Tr_Y(t) \quad\forall s\in\End\,X,\ t\in\End\,Y. 
\eean
\eprop

\prf Easy exercise. \qed

\bdefin \index{dimension}
Let $\2C$ be a TC$^*$ and $X\in\2C$. The \emph{dimension} of $X$ is defined by
$d(X)=Tr_X(\id_X)$, i.e.\ $d(X)=r^*\circ r$ for any standard conjugate $(\ol{X},r,\ol{r})$. 
\edefin

\blemma \label{lem-dim}
The dimension is additive ($d(X\oplus Y)=d(X)+d(Y)$) and multiplicative ($d(X\otimes Y)=d(X)d(Y)$).
Furthermore, $d(\ol{X})=d(X)\ge 1$ for every object, and $d(X)=1$ implies that $X\otimes\ol{X}\cong\11$,
i.e.\ $X$ is invertible.
\elemma

\prf Additivity is immediate by the discussion of standard conjugates. Multiplicativity of the
dimension follows from Lemma \ref{lem-mult}.

If $(\ol{X},r,\ol{r})$ is a standard conjugate for $X$, then $(X,\ol{r},r)$ is a standard conjugate
for $\ol{X}$, implying $d(\ol{X})=d(X)$. The positivity of the $*$-operation implies that
$d(X)=r^*\circ r>0$. Since $X\otimes\ol{X}$ contains $\11$ as a direct summand, we have
$d(X)^2\ge 1$, thus $d(X)\ge 1$. Finally, if $d(X)=1$, $\11$ is the only direct summand of
$X\otimes\ol{X}$, to wit $X\otimes\ol{X}\cong\11$. Similarly, $\ol{X}\otimes X\cong\11$.
\qed

\bdefin \label{def-twist} \index{twist}
Let $\2C$ be a $BTC^*$. For every $X\in\2C$ pick a conjugate $\ol{X}$ and a standard solution
$r, \ol{r}$ of the conjugate equations. Define the twist $\Theta(X)\in\End\,X$ by
\[ \Theta(X)=r^*\otimes\id_X\circ\id_{\ol{X}}\otimes c_{X,X}\circ r\otimes\id_X. \]
\edefin

\blemma \label{lem-twist}
Let $\2C$ be a $BTC^*$. Then
\begin{enumerate}
\item[(i)] $\Theta(X)$ is well defined, i.e.\ does not depend on the choice of $(\ol{X},r,\ol{r})$. 
\item[(ii)] For every morphism $s: X\rarr Y$ we have $\Theta(Y)\circ s=s\circ\Theta(X)$. (I.e.,
$\Theta$ is a natural transformation of the identity functor of $\2C$.) 
\item[(iii)] $\Theta(X)$ is unitary.
\item[(iv)] $\Theta(X\otimes Y)=\Theta(X)\otimes\Theta(Y)\mcirc c_{Y,X}\mcirc c_{X,Y}$ for all
$X,Y$. 
\item[(v)] If $\2C$ is an $STC^*$, this simplifies to $\Theta(X)^2=\id_X$ and
$\Theta(X\otimes Y)=\Theta(X)\otimes\Theta(Y)$ for all $X,Y\in\2C$ (i.e., $\Theta$ is a
\underline{monoidal} natural transformation of the identity functor of $\2C$). If $X,Y$ are
irreducible, we have $\omega(X)=\pm 1$ and $\omega_Z=\omega_X\omega_Y$ for all irreducible direct
summands $Z\prec X\otimes Y$. 
\end{enumerate}
\elemma

\prf (i) is proven as Proposition \ref{prop-trace}. The other verifications are not-too-difficult
computations, for which we refer to \cite{LR} or \cite{mue06}. We just comment on (v): In an $STC^*$
we have $c_{X,X}^*=c_{X,X}^{-1}=c_{X,X}$, implying $\Theta(X)^*=\Theta(X)$. Together with unitarity
this gives $\Theta(X)^2=\id_X$. Multiplicativity of $\Theta$ in an $STC^*$ follows from 
$c_{Y,X}\mcirc c_{X,Y}=\id$. If $X,Y$ are irreducible, we have
$\Theta(X)=\omega_X\id_X, \Theta(Y)=\omega_Y\id_Y$ and thus 
$\Theta(X\otimes Y)=\omega_X\omega_Y\id_{X\otimes Y}$. Now $\omega(Z)=\omega_X\omega_Y$ for
irreducible $Z\prec X\otimes Y$ follows by naturality of $\Theta$.
\qed

The following is a reworking of Propositions 4.4 and 4.5 in \cite{LR}.

\bprop \label{prop-stand2}
Let $\2C,\2D$ be  $BTC^*$s and $E:\2C\rarr\2D$ a $*$-preserving braided tensor functor. If
$(\ol{X},r,\ol{r})$ is a standard conjugate of $X\in\2C$, then 
$(E(\ol{X}),(d^E_{\ol{X},X})^{-1}\circ E(r)\circ e^E,(d^E_{X,\ol{X}})^{-1}\circ E(\ol{r})\circ e^E)$ is a
standard conjugate for $E(X)$. In particular, 
\[ d(E(X))=d(X), \quad\quad\quad \Theta(E(X))=E(\Theta(X)) \quad\quad\forall X\in\2C. \]
\eprop

\prf We assume for a while that the functor $E$ is strict and that $X$ is irreducible. Let
$(\ol{X},r,\ol{r})$ be a standard conjugate. Since $E$ preserves the conjugate equations,
$(E(\ol{X}),E(r),E(\ol{r}))$ is a conjugate for $E(X)$, but if $E$ is not full, standardness
requires proof. We begin with
\[ \begin{tangle} \object{\ol{X}}\step\object{X}\\ 
  \hh\id\step\id\\ \hx\\ \hh\hev\obj{\ol{r}} \end{tangle}
  \quad=\quad
\begin{tangle} \object{\ol{X}}\step[3]\object{X}\\ 
  \id \step\hcoev\obj{\ol{r}^*}\step\id\\
  \hx\step\id\step\id\\
  \hh\hev\obj{\ol{r}}\step\id\step\id\\
  \hh\Step\hev\obj{r}
\end{tangle}
\quad=\quad
\begin{tangle} \object{\ol{X}}\step[3]\object{X}\\ 
  \hh\id \step\hcoev\obj{\ol{r}^*}\step\id\\
  \hx\step\id\step\id\\
  \id\step\hx\step\id\\
  \id\step\hxx\step\id\\
  \hh\hev\obj{\ol{r}}\step\id\step\id\\
  \hh\Step\hev\obj{r}
\end{tangle}
\quad=\quad
\begin{tangle} \Step\object{\ol{X}}\step\object{X}\\ 
  \hcoev\obj{\ol{r}^*}\step\id\step\id\\
  \id\step\hxx\step\id\\
  \hh\hev\obj{\ol{r}}\step\id\step\id\\
  \hh\Step\hev\obj{r}
\end{tangle}
\quad=\quad \omega_{\ol{X}}\ 
\begin{tangle} \object{\ol{X}}\step\object{X}\\ \hh\id\step\id\step\\ \hh \hev\obj{r} \end{tangle}
\]
Thus $c_{\ol{X},X}^*\circ\ol{r}=\omega_{\ol{X}}\cdot r$, which is equivalent to 
$c_{\ol{X},X}\circ r=\ol{\omega}_{\ol{X}}\ol{r}$. Now we let $s\in\End\,E(X)$ and compute
\bean E(r^*)\mcirc\id_{E(\ol{X})}\otimes s\mcirc E(r) &=&
 E(r^*)\mcirc c_{E(\ol{X}),E(X)}^*\mcirc c_{E(\ol{X}),E(X)}\mcirc \id_{E(\ol{X})}\otimes s\mcirc E(r) \\
  &=&(c_{E(\ol{X}),E(X)}\mcirc E(r))^*\mcirc c_{E(\ol{X}),E(X)}\mcirc \id_{E(\ol{X})}\otimes s\mcirc E(r) \\
  &=&(c_{E(\ol{X}),E(X)}\mcirc E(r))^* \mcirc s\otimes\id_{E(\ol{X})}\mcirc c_{E(\ol{X}),E(X)}\mcirc E(r)\\
  &=& E(c_{\ol{X},X}\mcirc r)^* \mcirc s\otimes\id_{E(\ol{X})}\mcirc E(c_{\ol{X},X}\mcirc r) \\
  &=& E(\ol{\omega}_{\ol{X}}\ol{r})^* \mcirc s\otimes\id_{E(\ol{X})}\mcirc E(\ol{\omega}_{\ol{X}}\ol{r}) \\
  &=& E(\ol{r})^* \mcirc s\otimes\id_{E(\ol{X})}\mcirc E(\ol{r}),
\eean
which means that $(E(\ol{X}),E(r),E(\ol{r}))$ is a standard conjugate for $E(X)$. (We have used
unitarity of the braiding, the fact that $E$ is $*$-preserving and braided, 
$c_{\ol{X},X}\mcirc r=\ol{\omega}_{\ol{X}}\ol{r}$ and $|\omega_{\ol{X}}|=1$.) 

Now let $X$ be reducible, $(\ol{X},r,\ol{r})$ a standard conjugate and let $v_i:X_i\rarr X$,
$w_i:\ol{X}_i\rarr\ol{X}$ be isometries effecting the decompositions into irreducibles. Defining 
$r_i=w_i^*\otimes v_i^*\circ r$, $\ol{r}_i=v_i^*\otimes w_i^*\circ\ol{r})$,
$(\ol{X_i},r_i,\ol{r}_i)$ is standard by Lemma \ref{lem-stand1}. Thus 
$(E(\ol{X_i}),E(r_i),E(\ol{r}_i))$ is standard by the first half of this proof. In view of
$E(r)=E(\sum_i w_i\otimes v_i\mcirc r_i)=\sum_i E(w_i)\otimes E(v_i)\mcirc E(r_i)$ and similarly for
$E(\ol{r})$, it follows that $(E(\ol{X}),E(r),E(\ol{r}))$ is standard (since it is a direct sum of
standard conjugates). 

If $E$ is not strict, we have to insert the unitaries $d^E_{X,Y}: E(X)\otimes E(Y)\rarr E(X\otimes Y)$,
$e^E:\11\rarr E(\11)$ at the obvious places in the above computations, but nothing else
changes. That $E$ preserves dimensions follows since the dimension is defined in terms of a standard
conjugate. Finally, standardness of $(E(\ol{X}),E(r),E(\ol{r}))$ together with
$E(c_{X,Y})=c_{E(X),E(Y)}$ imply $\Theta(E(X))=E(\Theta(X))$.
\qed

We close this subsection by commenting on the relation of $*$-categories with the more general notion
of $C^*$-tensor categories of \cite{DR,LR}.  

\bdefin \label{def-cstar} \index{category!$C^*$-}
A $C^*$-category is a $\7C$-linear category with a positive $*$-operation, where $\Hom(X,Y)$ is a
Banach space for all $X,Y$ and
$\|s\circ t\|_{\Hom(X,Z)}\le\|s\|_{\Hom(X,Y)}\cdot\|t\|_{\Hom(Y,Z)}$ for all $s:X\rarr Y,\ t:Y\rarr Z$
and $\|s^*\circ s\|_{\End\,X}=\|s\|^2_{\Hom(X,Y)}$ for all $s:X\rarr Y$. (Thus each $\End\,X$ is a
$C^*$-algebra.) A $C^*$-tensor category is a $C^*$-category and a tensor category such that 
$\|s\otimes t\|\le\|s\|\cdot\|t\|$ for all $s,t$.
\edefin

\bprop \cite{LR} Let $\2C$ be a $C^*$-tensor category with direct sums and irreducible unit. If $X,Y\in\2C$
admit conjugates then $\dim\Hom(X,Y)<\infty$. Thus a $C^*$-tensor category with direct sums,
subobjects, conjugates and irreducible unit is a $TC^*$. Conversely, given a TC$^*$, there are
unique norms on the spaces $\Hom(X,Y)$ rendering $\2C$ a $C^*$-tensor category. 
\eprop

\prf Assume that $X\in\2C$ has a conjugate $(X,r,\ol{r})$. Then the map
$\End\,X\rarr\Hom(\11,\ol{X}\otimes X),\ s\mapsto\id_{\ol{X}}\otimes s\mcirc r$ is an isomorphism of
vector spaces since $t\mapsto\ol{r}^*\otimes\id_X\mcirc\id_X\otimes t$ is its inverse, as is
verified using the conjugate equations. Now, $\Hom(\11,\ol{X}\otimes X)$ is a pre-Hilbert space w.r.t.\
the inner product $\langle a,b\rangle\id_\11=a^*\circ b$, and it is complete because $\2C$ is a
$C^*$-tensor category. Choose an orthogonal basis $(e_i)_{i\in I}$ in $\Hom(\11,\ol{X}\otimes X)$.
Then each $e_i:\11\rarr\ol{X}\otimes X$ is an isometry and $e_i^*\circ e_j=0$ for $i\ne j$, implying
that $\ol{X}\otimes X$ contains $\#I$ copies of $\11$ as direct summands. Since $X$ has a conjugate,
so does $\ol{X}\otimes X$, but this is impossible if $\#I$ is infinite. Thus 
$\Hom(\11,\ol{X}\otimes X)$ and therefore $\End\,X$ is finite dimensional.

Given arbitrary $X,Y$ having conjugates, pick a direct sum $Z\cong X\oplus Y$ with isometries
$u:X\rarr Z, v:Y\rarr Z$. Then also $Z$ has a conjugate, cf.\ Lemma \ref{lem-stand1}, and therefore
$\dim\End\,Z<\infty$. Now, the map $\Hom(X,Y)\rarr\End\,Z$ given by $s\mapsto v\circ s\circ u^*$ is
injective since it has $t\mapsto v^*\circ t\circ u$ as inverse. This implies $\dim\Hom(X,Y)<\infty$.

We omit the proof of the implication $TC^*\impl C^*$-tensor category, since it will not be used in
the sequel. It can be found in \cite{mue06}.
\qed

This result shows that the assumptions made in Appendix B are equivalent to those of \cite{DR},
formulated  in terms of $C^*$-tensor categories.


\subsection{Abelian categories} \label{ss-abelian}
In the second half of Appendix B, which is of a purely algebraic nature, we will need some basic
facts from the theory of abelian categories. Good references are, e.g., \cite{gab} and \cite[Chapter
VIII]{cwm}.

\bdefin \index{monic}\index{epic}
A morphism $s:X\rarr Y$ is called monic if $s\circ t_1=s\circ t_2$ implies $t_1=t_2$, whenever
$t_1,t_2$ are morphisms with target $X$ and the same source.
A morphism $s:X\rarr Y$ is called epi if $t_1\circ s=t_2\circ s$ implies $t_1=t_2$, whenever
$t_1,t_2$ are morphisms with source $Y$ and the same target.
\edefin

\bdefin \index{kernel}\index{cokernel}
Let $\2C$ be an additive category. Given a morphism $f:X\rarr Y$, a morphism $k: Z\rarr X$ is a
kernel of $f$ if $f\circ k=0$ and given any morphism $k': Z'\rarr X$ such that $f\circ k'=0$, there
is a unique morphism $l:Z'\rarr Z$ such that $k'=k\circ l$.

A cokernel  of $f:X\rarr Y$ is a morphism $c: Y\rarr Z$ if $c\circ f=0$ and given any morphism 
$c': Y\rarr Z'$ such that $c'\circ f=0$, there is a unique $d: Z\rarr Z'$ such that $c'=d\circ c$.
\edefin

It is an easy consequence of the definition that every kernel is monic and every cokernel is epic.

\bdefin \label{def-abelian} \index{category!abelian}
An additive category $\2C$ is \emph{abelian} if
\begin{enumerate}
\item Every morphism has a kernel and a cokernel.
\item Every monic morphism is the kernel of some morphism.
\item Every epic morphism is the cokernel of some morphism.
\end{enumerate}
\edefin

\bprop \label{prop-me} Let $\2C$ be an abelian category. Then
\begin{itemize}
\item[(i)] Every monic is the kernel of its cokernel and every epi is the cokernel of its kernel.
\item[(ii)] A morphism is an isomorphism iff it is monic and epic. (`Only if' is trivial.)
\item[(iii)] Every morphism $f:X\rarr Y$ in an abelian category admits a factorization $f=m\circ e$,
where  $e:X\rarr Z$ is epi and  $m:Z\rarr Y$ is monic. Given another epi $e':X\rarr Z'$ and monic
$m':Z'\rarr Y$ such that $f=m'\circ e'$, there exists an isomorphism $u:Z\rarr Z'$ such that
$e'=u\circ e$ and $m=m'\circ u$.  
\end{itemize}
\eprop

\prf See \cite[Chapter VIII]{cwm}. Concerning (iii): Defining $m=\mathrm{ker}(\mathrm{coker}(f))$,
$m$ is monic. In view of $(\mathrm{coker}\,f)\circ f=0$, $f$ factors as $f=m\circ e$ for a unique
$e$. Next one proves that $e$ is epi and $e=\mathrm{coker}(\mathrm{ker}(f))$. For the details cf.\
e.g.\ \cite{cwm}. \qed 

\bdefin \index{image} 
The \emph{image} of a morphism $f:X\rarr Y$ in an abelian category is the monic $m:Z\rarr Y$ (unique
up to isomorphism) in the monic-epic factorization $X\stackrel{e}{\rarr}Z\stackrel{m}{\rarr}Y$ of
$f$.
\edefin

In a concrete abelian category, the object $Z$ is isomorphic to the usual image of $f$, which is a
subset of $Y$, whence the terminology.

\bdefin \label{def-proj} \index{object!projective}
An object $P$ in an abelian category is \emph{projective} if, given any epimorphism
$p:A\rarr B$ and any morphism $b:P\rarr B$ there is a morphism $a:P\rarr A$ such that $b=p\circ a$. 
\edefin

\blemma \label{lem-abel}
Any $TC^*\ \2C$ that has a zero object is abelian.
\elemma

\prf It is clear that $\2C$ is additive. The other requirements of Definition \ref{def-abelian}
follow with a little work from semisimplicity, cf.\ Lemma \ref{lem-semisim}.
\qed

\subsection{Commutative algebra in abelian symmetric tensor categories} \label{ss-comm-alg}
A considerable part of the well known algebra of commutative rings, their ideals and modules (living
in the category $\mathrm{Ab}$ of abelian groups) can be generalized to other abelian symmetric or
even braided tensor categories. We state just those facts that will be needed, some of which seem
to be new.

\bdefin \label{def-monoid} \index{monoid}
Let $\2D$ be a strict tensor category. Then a monoid in $\2D$ is a triple $(Q,m,\eta)$,
where $Q\in\2D$ and $m: Q\otimes Q\rarr Q$ and $\eta: \11\rarr Q$ are morphisms satisfying
\[ m\circ(m\otimes\id_Q)=m\circ(\id_Q\otimes m),\quad \quad
   m\circ\eta\otimes\id_Q=\id_Q=m\circ\id_Q\otimes\eta. \]
If $\2D$ is braided then the monoid is called commutative if $m\circ c_{Q,Q}=m$.
\edefin

\bdefin \label{def-module}\index{module} \index{monoid!module}
Let $(Q,m,\eta)$ be a monoid in the strict tensor category $\2D$. Then a $Q$-module (in
$\2D$) is a pair $(M,\mu)$, where $M\in\2D$ and $\mu:Q\otimes M\rarr M$ satisfy
\[ \mu\circ\id_Q\otimes \mu = \mu\circ m\otimes\id_M, \quad\quad \mu\circ\eta\otimes\id_M=\id_M. \]
A morphism $s: (M,\mu)\rarr(R,\rho)$ of $Q$-modules is a morphism $s\in\Hom_\2D(M,R)$ satisfying
$s\circ\mu=\rho\circ\id_Q\otimes s$. The $Q$-modules in $\2D$ and their morphisms form a category
$Q-\Mod_\2D$. If $\2D$ is $k$-linear then $Q-\Mod_\2D$ is $k$-linear. The hom-sets in the category
$Q-\Mod$ are denoted by $\Hom_Q(\cdot,\cdot)$.
\edefin

\brem 1. The preceding definitions, which are obvious generalizations of the corresponding notions
in $\Vect$, generalize in a straightforward way to non-strict tensor categories. 

2. If $(M,\mu)$ is a $Q$-module and $X\in\2D$ then $(Q\otimes X,\mu\otimes\id_X)$ is a $Q$-module.

3. If $\2D$ has direct sums, we can define the direct sum $(R,\rho)$ of two $Q$-modules 
$(M_1,\mu_1),(M_2,\mu_2)$. Concretely, if $v_i: M_i\rarr R, i=1,2$ are the isometries corresponding
to $R\cong M_1\oplus M_2$ then 
$\rho=v_1\circ\mu_1\circ\id_Q\otimes v_1^*+v_2\circ\mu_2\circ\id_Q\otimes v_2^*$ provides a
$Q$-module structure. 

4. Given a monoid $(Q,m,\eta)$ in $\2D$, we have an obvious $Q$-module $(Q,m)$, and for any
$n\in\7N$ we can consider $n\cdot(Q,m)$, the direct sum of $n$ copies of the $Q$-module $(Q,m)$.
\erem

\bdefin Let $\2D$ be a strict tensor category with unit $\11$ and let $(Q,m,\eta)$ be a monoid in
$\2D$. We define a monoid $\Gamma_Q$ in the category of sets by $\Gamma_Q=\Hom(\11,Q)$, the
multiplication being given by $s\bullet t=m\circ t\otimes s$ and the unit 
by $\eta$. If $\2D$ is braided and $(Q,m,\eta)$ commutative then $\Gamma_Q$ is commutative.
\edefin

\blemma \label{lem-end}
Let $\2D$ be a strict tensor category and $(Q,m,\eta)$ a monoid in $\2D$. Then there is an
isomorphism of monoids $\gamma: \End_Q((Q,m))\rarr (\Gamma_Q,\bullet,\eta)$ given by
\[ \begin{array}{rrcll}
 \gamma: & \End_Q((Q,m)) & \rarr & \Hom(\11,Q),&  \quad u\mapsto u\circ\eta, \\
  \gamma^{-1}: & \Hom(\11,Q) & \rarr & \End_Q((Q,m)), &\quad s\mapsto m\circ\id_Q\otimes s. 
\end{array}\]
If $\2D$ (and thus $Q-\Mod_\2D$) is $k$-linear then $\gamma$ is an isomorphism of $k$-algebras.
If $\2D$ is braided and the monoid $(Q,m,\eta)$ is commutative then the monoid ($k$-algebra)
$(\Gamma_Q,\bullet,\eta)$, and therefore also $\End_Q((Q,m))$, is commutative.
\elemma

\prf That $(\Gamma_Q,\bullet,\eta)$ is a monoid (associative $k$-algebra) is immediate since
$(Q,m,\eta)$ is a monoid. For $s\in\Hom(\11,Q)$ we have 
$\gamma(\gamma^{-1}(s))=m\circ\id_Q\otimes s\circ\eta=s$ by the monoid axioms. On the other hand,
for $u\in\End_Q((Q,m))$ we have 
\[ \gamma^{-1}(\gamma(u))=m\circ\id_Q\otimes (u\circ\eta)=m\circ\id_Q\otimes u\circ\id_Q\otimes\eta
   =u\circ m\circ\id_Q\otimes\eta=u, \]
where the third equality is due to the fact that $s$ is a $Q$-module map (cf.\ Definition
\ref{def-module}). Clearly $\gamma(\id_Q)=\eta$. Furthermore,
\bean  \gamma^{-1}(s)\circ\gamma^{-1}(t) &=& (m\circ\id_Q\otimes s)\circ(m\circ\id_Q\otimes t)
   =m\circ m\otimes\id_Q\circ\id_Q\otimes t\otimes s  \\
  &=& m\circ \id_Q\otimes m\circ\id_Q\otimes t\otimes s = \gamma^{-1}(s\bullet t).
\eean
If $\2D$ is braided and the monoid $(Q,m,\eta)$ is commutative then 
\[ s\bullet t=m\circ t\otimes s=m\circ c_{Q,Q}\circ s\otimes t=m\circ s\otimes t=t\bullet s, \]
where we used naturality of the braiding and commutativity of the monoid.
\qed

\brem
1. We have seen that a monoid $(Q,m,\eta)$ in any abstract tensor category gives rise to a
monoid $(\Gamma_Q,\bullet,\eta)$ that is concrete, i.e.\ lives in the category Sets. The latter has
the cartesian product as a tensor product and any one-element set is a tensor unit $\11$. Thus for
any $X\in\mathrm{Sets}$, $\Hom(\11,X)$ is in bijective correspondence to the elements of
$X$. Therefore, if $\2D=Sets$ then the monoids $(Q,m,\eta)$ and $(\Gamma_Q,\bullet,\eta)$ are
isomorphic. For this reason, we call $\Gamma_Q$ the monoid of \emph{elements of $Q$} even when $\2D$
is an abstract category. 

2. The commutativity of $\End_Q((Q,m))$ in the case of a commutative monoid $(Q,m,\eta)$ in a
braided tensor category $\2D$ has a very natural interpretation: If $\2D$ has coequalizers, which
holds in any abelian category, then the category $Q-\Mod_\2D$ is again a tensor category and the
Q-module $(Q,m)$ is its unit object. In any tensor category with unit $\11$, $\End\,\11$ is a
commutative monoid  (commutative $k$-algebra if $\2D$ is $k$-linear). This is the real reason why
$\End_Q((Q,m))$ is commutative. 
More is known: If $\2D$ is symmetric and $Q$ abelian, then the tensor category $Q-\Mod_\2D$ is again
symmetric. (In the braided case this need not be true, but $Q-\Mod_\2D$ always has a distinguished
full subcategory that is braided.)
\erem

We now specialize to abelian categories.

\bprop
Let $(Q,m,\eta)$ be a monoid in an abelian strict tensor category $\2D$. Then the category
$Q-\Mod_\2D$ is abelian.
\eprop

\prf Omitted. (This is a nice exercise on abelian categories.)
\qed

\bdefin \index{ideal}
Let $\2D$ be an abelian strict symmetric tensor category.
An \emph{ideal} in a commutative monoid $(Q,m,\eta)$ is a monic $j:(J,\mu_J)\rarr(Q,m)$ in the
category $Q-\Mod$. An ideal $j:(J,\mu_J)\rarr(Q,m)$ is called \emph{proper} if $j$ is not an
isomorphism (i.e.\ not epi). If $j:(J,\mu_J)\rarr(Q,m))$ and $j':(J',\mu_{J'})\rarr(Q,m)$ are ideals
then $j:(J,\mu_J)\rarr(Q,m)$ is \emph{contained} in $j':(J',\mu_{J'})\rarr(Q,m)$, denoted
$j\prec j'$, if there exists a monic $i\in\Hom_Q((J,\mu_J),(J',\mu_{J'})$ such that $j'\mcirc i=j$. 
A proper ideal $j:(J,\mu_J)\rarr(Q,m)$ in $(Q,m,\eta)$ is called \emph{maximal} if every proper
ideal $j':(J',\mu_{J'})\rarr(Q,m)$ containing $j:(J,\mu_J)\rarr(Q,m)$ is isomorphic to
$j:(J,\mu_J)\rarr(Q,m)$.  
\edefin

\blemma \label{l-maxid}
Let $\2D$ be an essentially small abelian strict symmetric tensor category, $(Q,m,\eta)$ a
commutative monoid in $\2D$. Then every proper ideal $j:(J,\mu_J)\rarr(Q,m)$ in $(Q,m,\eta)$ is
contained in a maximal ideal $\widetilde{j}:(\widetilde{J},\widetilde{\mu})\rarr(Q,m)$.
\elemma

\prf The ideals in $(Q,m,\eta)$ do not necessarily form a set, but the isomorphism classes do, since
$\2D$ is assumed essentially small. The relation $\prec$ on the ideals in $(Q,m,\eta)$ gives rise to
a partial ordering of the set of isomorphism classes of ideals. The maximal elements w.r.t.\ this
partial order are precisely the isomorphism classes of maximal ideals. Now we can apply Zorn's Lemma
to complete the proof as in commutative algebra.
\qed

As in the category $R$-mod, we can quotient a commutative monoid by an ideal:

\blemma \label{l-quot} \index{monoid!quotient}
Let $\2D$ be an abelian strict symmetric tensor category, $(Q,m,\eta)$ a commutative monoid and
$j:(J,\mu_J)\rarr(Q,m)$ an ideal. Let $p=\coker\,j:(Q,m)\rarr(B,\mu_B)$. Then there exist unique
morphisms $m_B: B\otimes B\rarr B$ and $\eta_B: \11\rarr B$ such that  
\begin{enumerate}
\item $(B,m_B,\eta_B)$ is a commutative monoid,
\item $p\mcirc m=m_B\mcirc p\otimes p$,
\item $p\circ\eta=\eta_B$. 
\end{enumerate}
The monoid $(B,m_B,\eta_B)$ is called the quotient of $(Q,m,\eta)$ by the ideal
$j:(J,\mu_J)\rarr(Q,m)$. It is nontrivial ($B$ is not a zero object) iff the ideal is proper.

Furthermore, the map $p_\Gamma:\Gamma:\Gamma_Q\rarr\Gamma_B$ given by $s\mapsto p\circ s$ is a
homomorphism of commutative algebras, which is surjective if the unit $\11\in\2D$ is a projective
object.
\elemma

\prf The construction of $m_B,\eta_B$ proceeds essentially as in commutative algebra, despite the
fact that the absence of elements makes it somewhat more abstract. Since $p:(Q,m)\rarr(B,\mu_B)$ is
the cokernel of $j$, $B$ is non-zero iff $j$ is not epi, to wit if the ideal is proper. The
equations $p\mcirc m=m_B\mcirc p\otimes p$ and $p\circ\eta=\eta_B$ imply that $p_\Gamma$ is a unital
homomorphism. 
If $\11$ is projective then the very Definition \ref{def-proj} implies that for every $s:\11\rarr B$
there is $t:\11\rarr Q$ such that $s=p\circ t$, thus $p_\Gamma$ is surjective.
\qed

\blemma \label{lem-corr}
Let $\2D$ be an essentially small abelian strict symmetric tensor category.
Let $(Q,m,\eta)$ be a commutative monoid in $\2D$ and $j:(J,\mu)\rarr(Q,m)$ an ideal. Let
$(B,m_B,\eta_B)$ be the quotient monoid. Then there is a bijective correspondence between
equivalence classes of ideals in $(B,m_B,\eta_B)$ and equivalence classes of ideals
$j':(J',\mu')\rarr(Q,m)$ in $(Q,\mu,\eta)$ that contain $j:(J,\mu)\rarr(Q,m)$. 

In particular, if $j$ is a maximal ideal then all ideals in $(B,m_B,\eta_B)$ are either zero or
isomorphic to $(B,m_B)$.
\elemma

\prf As in ordinary commutative algebra.
\qed

\blemma \label{l-field}
Let $k$ be a field and $(Q,m,\eta)$ a commutative monoid in the strict symmetric abelian $k$-linear
category $\2D$. If every non-zero ideal in $(Q,m,\eta)$ is isomorphic to $(Q,m)$ then the
commutative unital $k$-algebra $\End_Q((Q,m))$ is a field.
\elemma

\prf 
Let $s\in\End_Q((Q,m))$ be non-zero. Then $\mathrm{im}\,s\ne 0$ is a non-zero ideal in $(Q,m)$, thus 
must be isomorphic to $(Q,m)$. Therefore $\mathrm{im}\,s$ and in turn $s$ are epi. Since $s\ne 0$,
the kernel $\ker\,s$ is not isomorphic to $(Q,m)$ and therefore it must be zero, thus $s$ is monic. 
By Proposition \ref{prop-me}, $s$ is an isomorphism. Thus the commutative $k$-algebra $\End_Q((Q,m))$
is a field extending $k$.
\qed

The following lemma is borrowed from \cite{bichon}:

\blemma \label{l-epiiso}
Let $\2D$ be an abelian strict symmetric tensor category and $(Q,m,\eta)$ a commutative monoid in
it. Then every epimorphism in $\End_Q((Q,m))$ is an isomorphism. 
\elemma

\prf Let $g\in\End_Q((Q,m))$ be an epimorphism and let $j:(J,\mu_J)\rarr(Q,m)$ be an ideal in
$(Q,m,\eta)$. Now, $Q-\Mod$ is a tensor category whose unit is $(Q,m)$, thus there is an isomorphism
$s\in\Hom_Q((J,\mu_J),(Q\otimes_Q J, \mu_{Q\otimes_Q J}))$. Let $h\in\End_Q((J,\mu_J))$ be the
composition 
\begin{diagram}
(J,\mu_J) & \rTo^{s} & (Q\otimes_Q J, \mu_{Q\otimes_Q J}) & \rTo^{g\otimes\id_J} & (Q\otimes_Q J,
  \mu_{Q\otimes_Q J})   & \rTo{s^{-1}} & (J,\mu_J).
\end{diagram}
Since the tensor product $\otimes_Q$ of $Q-\Mod$ is right-exact, $g\otimes\id_J$ is epi. Now,
$j\mcirc h=g\mcirc j$, and if we put $(j:(J,\mu_J)\rarr(Q,m))=\ker\,g$ we have $j\mcirc h=0$ and
thus $j=0$ since $h$ is epi. Thus $g$ is monic and therefore an isomorphism.
\qed


\subsection{Inductive limits and the Ind-category}  \label{ss-ind}
We need the categorical version of the concept of an inductive  limit. For our purposes, inductive
limits over $\7N$ will do, but in order to appeal to existing theories we need some definitions.

\bdefin If $\2I,\2C$ are categories and $F:\2I\rarr\2C$ a functor, then a \emph{colimit} (or 
\emph{inductive limit}) of $F$ consists of an object $Z\in\2C$ and, for every $X\in\2I$, of a
morphism $i_X:F(X)\rarr Z$ in  $\2C$ such that 
\begin{enumerate}
\item $i_Y\circ F(s)=i_X$ for every morphism $s:X\rarr Y$ in $\2I$.
\item Given $Z'\in\2C$ and a family of morphisms $j_X:F(X)\rarr Z'$ in $\2C$ such that 
$j_Y\circ F(s)=j_X$ for every morphism $s:X\rarr Y$ in $\2I$, there is a unique morphism 
$\iota:Z\rarr Z'$ such that $j_X=\iota\circ i_X$ for all $X\in\2I$.
\end{enumerate}
\edefin

The second property required above is the universal property. It implies that any two colimits of
$F$ are isomorphic. Thus the colimit is essentially unique, provided it exists.

\bdefin \label{def-filtered}
A category $\2I$ is \emph{filtered} if it is non-empty and 
\begin{enumerate}
\item For any two objects $X,Y\in\2I$ there is an $Z\in\2Z$ and morphisms $i:X\rarr Z, j:Y\rarr Z$.
\item For any two morphisms $u,v:X\rarr Y$ in $\2I$ there is a morphism $w:Y\rarr Z$ such that
$w\circ u=w\circ v$.
\end{enumerate}
\edefin

Note that any directed partially ordered set $(I,\leq )$ is a filtered category if we take the
objects to be the elements of $I$, and the arrows are ordered pairs $\{ (i,j):i\leq j\}$.

\bdefin Let $\2C$ be a category. Then the category
$\mathrm{Ind}\,\2C$ is defined as the functor category
whose objects are all functors $F:\2I\rarr\2C$, where
$\2I$ is a small filtered category.  For
$F:\2I\rarr\2C, F':\2I'\rarr\2C$, the hom-set is
defined by
\[ \Hom_{\mathrm{Ind}\,\2C}(F,F')=\lim_{\longleftarrow\atop X}\lim_{\longrightarrow\atop Y}
   \Hom_\2C(F(X),F'(Y)). \]
(An element of the r.h.s.\ consists of a family 
$(f_{X,Y}: F(X)\rarr F'(Y))_{X\in\2I, Y\in\2I'}$ satisfying $F'(s)\circ f_{X,Y}=f_{X,Y'}$ for every
$s:Y\rarr Y'$ in $\2I'$ and $f_{X',Y}\circ F(t)=f_{X,Y}$ for every $t:X\rarr X'$ in $\2I$.) We leave
it as an exercise to work out the composition of morphisms.
\edefin

Some properties of $\mathrm{Ind}\,\2C$ are almost obvious. It contains $\2C$ as a subcategory: To
every $X\in\2C$ we assign the functor $F:\2I\rarr\2C$, where $\2I$ has only one object $*$ and
$F(*)=X$. This embedding clearly is full and faithful. If $\2C$ is an 
Ab-category / additive / $\7C$-linear then so is $\mathrm{Ind}\,\2C$. If $\2C$ is a strict
(symmetric) tensor category then so is $\mathrm{Ind}\,\2C$: The tensor product of $F:\2I\rarr\2C$
and $F:\2I'\rarr\2C$ is defined by $\2I''=\2I\times\2I'$ (which is a filtered category) and 
$F\otimes F': \2I''\ni X\times Y\mapsto F(X)\otimes F'(Y)$. For the remaining results that we need,
we just cite \cite{SGA4}, to which we also refer for the proof:

\btheor 
$\mathrm{Ind}\,\2C$ has colimits for all small filtered index categories $\2I$.
If $\2C$ is an abelian category $\2C$ then $\mathrm{Ind}\,\2C$ is abelian.
\etheor

Thus every abelian (symmetric monoidal) category is a full subcategory of an abelian (symmetric monoidal)
category that is complete under filtered colimits. For us this means that in $\mathrm{Ind}\,\2C$ we
can make sense of infinite direct sums indexed by $\7N$, defining $\bigoplus_{i\in\7N} X_i$ as the
colimit of the functor $F:\2I\rarr\2C$, where $\2I$ is the poset $\7N$ interpreted as a filtered
category, and $F(n)=\bigoplus_{i=1}^n X_i$ together with the obvious morphisms $F(n)\rarr F(m)$ when
$n\le m$. 

\blemma \label{lem-proj}
If $\2C$ is a $TC^*$ then every object $X\in\2C$ is projective as an object of $\mathrm{Ind}\,\2C$.
\elemma

\prf First assume that $X$ is irreducible and consider $s:X\rarr B$. Given an epi $p:A\rarr B$ in
$\mathrm{Ind}\,\2C$, we have $A=\lim_{\longrightarrow}A_i$ with $A_i\in\2C$ and similarly for $B$.
Furthermore, $\Hom(A,B)=\lim_{\longleftarrow}\lim_{\longrightarrow}\Hom_\2C(A_i,B_j)$ and
$\Hom(X,B)=\lim_{\longrightarrow}\Hom_\2C(X,B_j)$. Since $X$ is irreducible and $\2C$ is semisimple,
$X$ is a direct summand of $B_j$ whenever $s_j:X\rarr B_j$ is non-zero. Since $p:A\rarr B$ is epi, the
component $A_i\rarr B_j$ is epi for $i$ sufficiently big. By semisimplicity of $\2C$, $s_j$ then
lifts to a morphism $X\rarr A_i$. Putting everything together this gives a morphism
$\widehat{s}:X\rarr A$ such that $p\circ\widehat{s}=s$.

Now let $X$ be a finite direct sum of irreducible $X_i$ with isometries $v_i:X_i\rarr X$ and $s:X\rarr B$.
Defining $s_i=s\circ v_i:X_i\rarr B$, the first half of the proof provides $\widehat{s_i}:X_i\rarr A$
such that $p\circ\widehat{s_i}=s_i$. Now define $\widehat{s}=\sum_i \widehat{s_i}\circ v_i^*:X\rarr A$.
We have 
\[ p\circ\widehat{s}=\sum_i p\circ\widehat{s_i}\circ v_i^*=\sum_i s_i\circ v_i^* 
   =\sum_i s\circ v_i\circ v_i^* = s, \]
proving projectivity of $X$.
\qed



\section{Abstract Duality Theory for Symmetric Tensor $*$-Categories} \label{App-B}
In the first two subsections we give self-contained statements of the results needed for the AQFT
constructions. Some of the proofs are deferred to the rest of this appendix, which hurried (or less
ambitious) or readers may safely skip. 

\subsection{Fiber functors and the concrete Tannaka theorem. Part I} \label{ss-concrete1}
Let $\Vect_\7C$ denote the $\7C$-linear symmetric tensor category of finite dimensional $\7C$-vector
spaces and $\2H$ denote the $STC^*$ of finite dimensional Hilbert spaces. We pretend that both
tensor categories are strict, which amounts to suppressing the associativity and unit isomorphisms
$\alpha,\lambda,\rho$ from the notation. Both categories have a canonical symmetry $\Sigma$, the
flip isomorphism $\Sigma_{V,V'}: V\otimes V'\rarr V'\otimes V$.  

\bdefin \index{functor!fiber}\index{fiber functor}
Let $\2C$ be an $STC^*$. A fiber functor for $\2C$ is a faithful $\7C$-linear tensor functor
$E:\2C\rarr\Vect_\7C$. A $*$-preserving fiber functor for $\2C$ is a faithful functor $E: \2C\rarr\2H$
of tensor $*$-categories. $E$ is symmetric if $E(c_{X,Y})=\Sigma_{E(X),E(Y)}$, i.e.\ the symmetry of  
$\2C$ is mapped to the canonical symmetry of $\Vect_\7C$ or $\2H$, respectively.
\edefin

A symmetric tensor category equipped with a symmetric $*$-preserving fiber functor is called
concrete, since it is equivalent to a (non-full!) tensor subcategory of the category $\2H$ of
Hilbert spaces. Our main concern in this appendix  are (1) Consequences of the existence of a fiber
functor, (2) Uniqueness of fiber functors, and (3) Existence of fiber functors. As to (2) we will prove:

\btheor \label{theor-uniq1} \index{fiber functor!uniqueness}
Let $\2C$ be an $STC^*$ and let $E_1,E_2:\2C\rarr\2H$ be $*$-preserving symmetric fiber
functors. Then $E_1\cong E_2$, i.e.\ there exists a unitary monoidal natural isomorphism  
$\alpha: E_1\rarr E_2$.
\etheor

We now {\it assume} a symmetric $*$-preserving fiber functor for the $STC^*\ \2C$ to be given. Let
$G_E\subset\Nat_\otimes E$ denote the set of unitary monoidal natural transformations of $E$ (to 
itself). This clearly is a group with the identical natural transformation as unit. $G_E$ can be
identified with a subset of $\prod_{X\in\2C}\2U(E(X))$, where $\2U(E(X))$ is the compact group of
unitaries on the finite dimensional Hilbert space $E(X)$. The product of these groups is compact by
Tychonov's theorem, cf.\ e.g.\ \cite[Theorem 1.6.10]{ped}, and since $G_E$ is a closed subset, it is
itself compact. The product and inverse maps are continuous, thus $G_E$ is a compact topological
group. By its very definition, the group $G_E$ acts on the Hilbert spaces $E(X), X\in\2C$ by unitary
representations $\pi_X$, namely $\pi_X(g)=g_X$ where $g_X$ is the component at $X$ of the natural
transformation $g\in G_E$.  

\bprop
There is a faithful symmetric tensor $*$-functor $F: \2C\rarr\Rep_fG_E$ such that $K\circ F=E$,
where $K:\Rep_fG_E \rarr\2H$ is the forgetful functor $(H,\pi)\mapsto H$.
\eprop 

\prf We define $F(X)=(E(X),\pi_X)\in\Rep_fG_E$ for all $X\in\2C$ and $F(s)=E(s)$ for all
$s\in\Hom(X,Y)$. For $s:X\rarr Y$ we have
\[ F(s) \pi_X(g)=F(s)g_X=g_YF(s)=\pi_Y(g)F(s) \]
since $g:E\rarr E$ is a natural transformation. Thus $F$ is a functor, which is obviously
$*$-preserving and faithful. In view of $g_\11=\id_{E(\11)}$ for every $g\in G_E$, we have
$F(\11_\2C)=(\7C,\pi_0)=\11_{\Rep_fG_E}$, where $\pi_0$ is the trivial representation. In order to
see that $F$ is a functor of tensor $*$-categories we must produce unitaries 
$d^F_{X,Y}: F(X)\otimes F(Y)\rarr F(X\otimes Y),\ X,Y\in\2C$ and 
$e:\11_{\Rep_fG_E}\rarr F(\11_\2C)$ satisfying (\ref{eq-A1}) and (\ref{eq-A2}), respectively.
We claim that the choice $e^F=e^E, \ d^F_{X,Y}=d^E_{X,Y}$ does the job, where the $e^E$ and
$d^E_{X,Y}$ are the unitaries coming with the tensor functor $E:\2C\rarr\2H$. It is obvious that  
$e^E$ and $d^E_{X,Y}$ satisfy (\ref{eq-A1}) and (\ref{eq-A2}), but we must show that they are
morphisms in $\Rep_fG_E$. For $d^E_{X,Y}$ this follows from the computation
\[ d^F_{X,Y}\mcirc (\pi_X(g)\otimes\pi_Y(g))=d^E_{X,Y}\mcirc g_X\otimes g_Y
   =g_{X\otimes Y}\mcirc d^E_{X,Y}=\pi_{X\otimes Y}(g)\mcirc d^F_{X,Y}, \]
where we have used that $g$ is a monoidal natural transformation. Now, by the definition of a
natural monoidal transformation we have $g_\11=\id_{E(\11)}$ for all $g\in G_E$, i.e.\
$F(\11)=(E(\11),\pi_\11)$ is the trivial representation. If the strict unit $\11_\2H=\7C$ is in the
image of $E$ then, by naturality, it also carries the trivial representation, thus $e^F$ in fact is
a morphism of representations. (In case $\11_\2H\not\in E(\2C)$, we equip $\11_\2H$ with the trivial
representation by hand.) Since the symmetry of $\Rep_fG_E$ is by definition given by
$c((H,\pi),(H',\pi'))=c(H,H')$, where the right hand side refers to the category $\2H$, and since
$E$ respects the symmetries, so does $F$. $K\circ F=E$ is obvious.
\qed

The proof of the following proposition is postponed, since it requires further preparations.

\bprop \label{prop-dense}
Let $\2C$ be an $STC^*$ and $E:\2C\rarr\2H$ a symmetric $*$-preserving fiber functor. Let $G_E$ and 
$F:\2C\rarr\Rep_fG_E$ as defined above. Then the following hold:
\begin{itemize}
\item[(i)] If $X\in\2C$ is irreducible then $\mathrm{span}_\7C\{ \pi_X(g),\ g\in G_E\}$ is dense in
  $\End\,E(X)$. 
\item[(ii)] If $X,Y\in\2C$ are irreducible and $X\not\cong Y$ then 
$\mathrm{span}_\7C\{ \pi_X(g)\oplus\pi_Y(g),\ g\in G_E\}$ is dense in $\End\,E(X)\oplus\End\,E(Y)$.
\end{itemize}
\eprop

\btheor \label{theor-equiv}
Let $\2C$ be an $STC^*$ and $E:\2C\rarr\2H$ a symmetric $*$-preserving fiber functor. Let $G_E$ and 
$F:\2C\rarr\Rep_fG_E$ as defined above. Then $F$ is an equivalence of symmetric tensor
$*$-categories. 
\etheor

\prf We already know that $F$ is a faithful symmetric tensor functor. In view of Proposition
\ref{prop-equiv} it remains to show that $F$ is full and essentially surjective. 

Since the categories $\2C$ and $\Rep_fG_E$ are semisimple, in order to prove that $F$ is full it is
sufficient to show that (a) $F(X)\in\Rep_fG_E$ is irreducible if $X\in\2C$ is irreducible and (b)
if $X,Y\in\2C$ are irreducible and inequivalent then $\Hom(F(X),F(Y))=\{0\}$. Now, (i) of
Proposition \ref{prop-dense} clearly implies that $\End(F(X))=\7C\,\id$, which is the desired
irreducibility of $F(X)$. Assume now that $X,Y\in\2C$ are irreducible and non-isomorphic and let 
$s\in\Hom(F(X),F(Y))$, to wit $s\in\Hom(E(X),E(Y))$ and $s\pi_X(g)=\pi_Y(g)s$ for all $g\in G_E$. 
Then (ii) of Proposition \ref{prop-dense} implies $su=vs$ for any $u\in\End\,E(X)$ and
$v\in\End\,E(Y)$. With $u=0$ and $v=1$ this implies $s=0$, thus the irreps $F(X)=(E(X),\pi_X)$ and
$F(Y)=(E(X),\pi_Y)$ are non-isomorphic. This proves that $F$ is full.

Therefore, $F$ is an equivalence of $\2C$ with a full tensor subcategory of $\Rep_fG_E$. If 
$g\in G_E$ is nontrivial, it is immediate by the definition of $G_E$ that there is an $X\in\2C$ such
that $g_X\ne\id_{E(X)}$ -- but this means $\pi_X(g)\ne\11$. In other words, the representations 
$\{F(X), X\in\2C\}$ separate the points of $G_E$. But it is a well known consequence of the
Peter-Weyl theorem that a full monoidal subcategory of $\Rep_fG_E$ separates the points of $G_E$ iff
it is in fact equivalent to $\Rep_fG_E$. Thus the functor $F$ is essentially surjective, and we are
done. 
\qed

Since they so important, we restate Theorems \ref{theor-uniq1} and \ref{theor-equiv} in a self
contained way:

\btheor \label{theor-T2} \index{Tannaka}
Let $\2C$ be an $STC^*$ and $E: \2C\rarr\2H$ a $*$-preserving symmetric fiber functor. Let $G_E$ be 
the group of unitary monoidal natural transformations of $E$ with the topology inherited from 
$\prod_{X\in\2C}\2U(E(X))$. Then $G_E$ is compact and the functor 
$F: \2C\rarr\Rep_fG_E,\ X\mapsto(E(X),\pi_X)$, where $\pi_X(g)=g_X$, is an equivalence of $STC^*$s. 
If $E_1,E_2: \2C\rarr\2H$ are $*$-preserving symmetric fiber functors then $E_1\cong E_2$ and
therefore $G_{E_1}\cong G_{E_2}$.
\etheor

\brem The preceding theorem is essentially a reformulation in modern language of the classical
result of Tannaka \cite{tannaka}. It can be generalized, albeit without the uniqueness part, to a
setting where $\2C$ is only braided or even has no braiding. This leads to a (concrete) Tannaka
theory for quantum groups, for which the interested reader is referred to the reviews \cite{JS1} and
\cite{MRT}. 
\erem

Before we turn to proving Theorem \ref{theor-uniq1} (Subsection \ref{ss-unique}) and Proposition
\ref{prop-dense} (Subsection \ref{ss-concrete2}) we identify a necessary condition for the existence
of fiber functors, which will lead us to a generalization of Theorem \ref{theor-T2}.


\subsection{Compact supergroups and the abstract Tannaka theorem} 
According to Theorem \ref{theor-T2}, an $STC^*$ admitting a symmetric $*$-preserving fiber functor
is equivalent, as a symmetric tensor $*$-category, to the category of finite dimensional unitary
representations of a compact group $G$ that is uniquely determined up to isomorphism. Concerning the
existence of fiber functors it will turn out that the twist $\Theta$ (Definition \ref{def-twist})
provides an obstruction, fortunately the only one.

\bdefin \index{$STC^*$!even}
An $STC^*$ is called even if $\Theta(X)=\id_X$ for all $X\in\2C$.
\edefin

\bexam A simple computation using the explicit formulae for $r,\ol{r},c_{X,Y}$ given in Example
\ref{exam-H} shows that the $STC^*\ \2H$ of finite dimensional Hilbert spaces is even. The same
holds for the category $\Rep_fG$ of finite dimensional unitary representations of a compact group
$G$.
\eexam

This suggests that an $STC^*$ must be even in order to admit a fiber functor. In fact:

\bprop If an $STC^*$ $\2C$ admits a $*$-preserving symmetric fiber functor $E$ then it is even.
\eprop

\prf By Proposition \ref{prop-stand2}, we have $E(\Theta(X))=\Theta(E(X))$. Since $\2H$ is even,
this equals $\id_{E(X)}=E(\id_X)$. Since $E$ is faithful, this implies $\Theta(X)=\id_X$.  
\qed

Fortunately, this is the only obstruction since, beginning in the next subsection, we will prove:

\btheor \label{theor-C} \index{fiber functor!existence}
Every even $STC^*$ admits a $*$-preserving symmetric fiber functor $E: \2C\rarr\2H$. 
\etheor

Combining this with Theorem \ref{theor-T2} we obtain:

\btheor \label{theor-B}
Let $\2C$ be an even $STC^*$. Then there is a compact group $G$, unique up to isomorphism, such that
there exists an equivalence $F: \2C\rarr\Rep_fG$ of $STC^*$s. 
\etheor

Theorem \ref{theor-B} is not yet sufficiently general for the application to quantum field theory,
which is the subject of this paper. Making the connection with DHR theory, we see that the twist of
an irreducible DHR sector is $\pm 1$, depending on whether the sector is bosonic or fermionic. Since
in general we cannot a priori rule out fermionic sectors, we cannot restrict ourselves to even
$STC^*$s. What we therefore really need is a characterization of all $STC^*$s. This requires a
generalization of the notion of compact groups: 

\bdefin \index{super!-group}
A (compact) supergroup is a pair $(G,k)$ where $G$ is a (compact Hausdorff) group and $k$ is an
element of order two in the center of $G$. An isomorphism
$\alpha:(G,k)\stackrel{\cong}{\rarr}(G',k')$ of (compact) supergroups is an isomorphism $\alpha:G\rarr G'$
of (topological) groups such that $\alpha(k)=k'$.
\edefin

\bdefin \index{super!-Hilbert space}
A (finite dimensional, unitary, continuous) representation of a compact supergroup $(G,k)$
is just a (finite dimensional, unitary, continuous) representation $(H,\pi)$ of $G$. 
Intertwiners and the tensor product of representations are defined as for groups, thus
$\Rep_{(f)}(G,k)\cong\Rep_{(f)}G$ as $C^*$-tensor tensor categories. (Since $k$ is in the center of
$G$, morphisms in $\Rep_{(f)}(G,k)$ automatically preserve the $\7Z_2$-grading induced by $\pi(k)$.
$\Rep_{(f)}(G,k)$ is equipped with a symmetry $\Sigma_k$ as follows: For every $(H,\pi)\in\Rep(G,k)$
let $P_\pm^\pi=(\id+\pi(k))/2$ be the projector on the even and odd subspaces of a representation
space $H$, respectively. Then 
\[ \Sigma_k((H,\pi),(H',\pi'))= \Sigma(H,H')(\11 - 2 P_-^\pi\otimes P_-^{\pi'}), \]
where $\Sigma(H,H'): H\otimes H'\rarr H'\otimes H$ is the usual flip isomorphism 
$x\otimes y\mapsto y\otimes x$. Thus for homogeneous $x\in H, y\in H'$ we have
$\Sigma_k((H,\pi),(H',\pi')): x\otimes y\mapsto\pm y\otimes x$, where the minus sign occurs iff
$x\in H_-$ and $y\in H'_-$. In the case $(G,k)=(\{e,k\},k)$, we call $\Rep_f(G,k)$ the category
$\2S\2H$ of super Hilbert spaces.
\edefin

\brem Note that the action of $k$ induces a $\7Z_2$-grading on $H$ that is stable under the
$G$-action. Since the symmetry $\Sigma_k$ defined above is precisely the one on the category
$\2S\2H$ of finite dimensional super Hilbert spaces, we see that there is a forgetful symmetric
tensor functor $\Rep_f(G,k)\rarr\2S\2H$.
\erem

\blemma
$\Sigma_k$ as defined above is a symmetry on the category $\Rep(G,k)$. Thus $\Rep_f(G,k)$ is a
$STC^*$. For every object $X=(H,\pi)\in\Rep_f(G,k)$, the twist $\Theta(X)$ is given by $\pi(k)$.
\elemma

\prf Most of the claimed properties follow immediately from those of $\Rep_fG$. 
It is clear that $\Sigma_k((H,\pi),(H',\pi'))\circ\Sigma_k((H',\pi'),(H,\pi))$ is the identity
of $H'\otimes H$. We only need to prove naturality and compatibility with the tensor product. This
is an easy exercise. The same holds for the identity $\Theta((H,\pi))=\pi(k)$.
\qed

We need a corollary of (the proof of) Theorem \ref{theor-B}:

\bcoro \label{coro-twist}
For any compact group $G$, the unitary monoidal natural transformations of the identity functor on
$\Rep_fG$ form an abelian group that is isomorphic to the center $Z(G)$.
\ecoro

\prf If $k\in Z(G)$ and $(H,\pi)\in\Rep_fG$ is irreducible then $\pi(k)=\omega_{(H,\pi)}\id_H$, where  
$\omega_{(H,\pi)}$ is a scalar. Defining $\Theta((H,\pi))=\omega_{(H,\pi)}\id_{(H,\pi)}$ and
extending to reducible objects defines a unitary monoidal natural isomorphism of $\Rep_fG$.
Conversely, let $\{\Theta((H,\pi))\}$ be a unitary monoidal isomorphism of the identity functor of
$\Rep_fG$ and $K:\Rep_fG\rarr\2H$ the forgetful functor. Then the family 
$(\alpha_{(H,\pi)}=K(\Theta((H,\pi))))$ is a unitary monoidal natural isomorphism of $K$. By Theorem
\ref{theor-T2}, there is a $g\in G$ such that $\alpha_{(H,\pi)}=\pi(g)$ for all
$(H,\pi)\in\Rep_fG$. Since $\pi(g)$ is a multiple of the identity for every irreducible $(H,\pi)$,
$g$ is in $Z(G)$ by Schur's lemma. Clearly the above correspondence is an isomorphism of abelian
groups. 
\qed

Modulo Theorem \ref{theor-C} we can now can prove the Main Result of this appendix:

\begin{bxd}
\btheor \label{theor-A} \index{Doplicher} \index{Roberts} \index{super!-fiber functor}
Let $\2C$ be an $STC^*$. Then there exist a compact supergroup $(G,k)$, unique up to isomorphism,
and an equivalence $F:\2C\rarr\Rep_f(G,k)$ of symmetric tensor $*$-categories. In particular, if
$K:\Rep_f(G,k)\rarr\2S\2H$ is the forgetful functor, the composite $E=K\circ F:\2C\rarr\2S\2H$ is a
`super fiber functor', i.e.\ a faithful symmetric $*$-preserving tensor functor into the $STC^*$ of
super Hilbert spaces.
\etheor
\end{bxd}

\prf We define a new $STC^*\ \widetilde{\2C}$ (the `bosonization' of $\2C$) as follows. As a 
tensor $*$-category, $\widetilde{\2C}$ coincides with $\2C$. The symmetry $\tilde{c}$ is defined by
\[ \tilde{c}_{X,Y}=(-1)^{(1-\Theta(X))(1-\Theta(Y))/4} c_{X,Y} \]
for irreducible $X,Y\in\obj\,\2C=\obj\,\tilde{\2C}$, and extended to all objects by naturality. It is
easy to verify that $(\tilde{\2C},\tilde{c})$ is again a symmetric tensor category, in fact an even 
one. Thus by Theorem \ref{theor-B} there is a compact group $G$ such that $\tilde{\2C}\simeq\Rep_fG$
as $STC^*$s. Applying Corollary \ref{coro-twist} to the category $\tilde{\2C}\simeq\Rep_fG$ and the
family $(\Theta(X))_{X\in\2C}$, as defined in the original category $\2C$ proves the existence of an
element $k\in Z(G), k^2=e$, such that $\Theta((H,\pi))=\pi(k)$ for all
$(H,\pi)\in\tilde{\2C}\simeq\Rep_fG$. Clearly $(G,k)$ is a supergroup. We claim that
$\2C\simeq\Rep_f(G,k)$ as $STC^*$s. Ignoring the symmetries this is clearly true since
$\Rep_f(G,k)\simeq\Rep_fG$ as tensor $*$-categories. That $\2C$ and $\Rep_f(G,k)$ are equivalent
as $STC^*$s, i.e.\ taking the symmetries into account, follows from the fact that $\2C$ is related
to $\tilde{\2C}$ precisely as $\Rep_f(G,k)$ is to $\Rep_fG$, namely by a twist of the symmetry
effected by the family  $(\Theta((H,\pi))=\pi(k))$. To conclude, we observe that the uniqueness result
for $(G,k)$ follows from the uniqueness of $G$ in Theorem \ref{theor-B} and that of $k$ in Corollary
\ref{coro-twist}. 
\qed

\brem Theorem \ref{theor-A} was proven by Doplicher and Roberts in \cite[Section 7]{DR} exactly as
stated above, the only superficial difference being that the terminology of supergroups wasn't
used. (Note that our supergroups are not what is usually designated by this name.) As above, the
proof was by reduction to even categories and compact groups. Independently and 
essentially at the same time, a result analogous to Theorem \ref{theor-C} for (pro)algebraic groups
was proven by Deligne in \cite{del}, implying an algebraic analogue of Theorem \ref{theor-B} by
\cite{SR,DM}. Recently, Deligne also discussed the super case, cf.\ \cite{del3}.
\erem

This concludes the discussion of the main results of this appendix. We now turn to proving
Theorem \ref{theor-uniq1}, Proposition \ref{prop-dense} and Theorem \ref{theor-C}.


\subsection{Certain algebras arising from fiber functors} \label{ss-algebras}
Let $\2C$ be a $TC^*$ and $E_1,E_2: \2C\rarr\Vect_\7C$ fiber functors. Recall that they come with
natural isomorphisms $d^i_{X,Y}: E_i(X)\otimes E_i(Y)\rarr E_i(X\otimes Y)$ and 
$e^i: \11_{\Vect}=\7C\rarr E_i(\11_\2C)$. Consider the $\7C$-vector space
\[ A_0(E_1,E_2)= \bigoplus_{X\in\2C}\Hom(E_2(X),E_1(X)). \]
For $X\in\2C$ and $s\in\Hom(E_2(X),E_1(X))$ we write $[X,s]$ for the element of $A_0(E_1,E_2)$ which
takes the value $s$ at $X$ and is zero elsewhere. Clearly, $A_0$ consists precisely of the finite
linear combinations of such elements. We turn $A_0(E_1,E_2)$ into a $\7C$-algebra by defining
$[X,s]\cdot[Y,t]=[X\otimes Y, u]$, where $u$ is the composite
\[ \begin{diagram}          
 E_2(X\otimes Y) & \rTo^{(d^2_{X,Y})^{-1}} & E_2(X)\otimes E_2(Y) & \rTo^{s\otimes t} &
   E_1(X)\otimes E_1(Y) & \rTo^{d^1_{X,Y}} & E_1(X\otimes Y)
\end{diagram} \]
Since $\2C$ is strict, we have $(X\otimes Y)\otimes Z=X\otimes (Y\otimes Z)$ and 
$\11\otimes X =X=X\otimes\11$. Together with the 2-cocycle type equation (\ref{eq-A1}) satisfied by
the isomorphisms $d^i_{X,Y}$ this implies that $A_0(E_1,E_2)$ is associative. 
The compatibility (\ref{eq-A2}) of $d^i_{X,Y}$ with $e^i$ for $i=1,2$ implies that
$[\11,e^1\circ(e^2)^{-1}]$ is a unit of the algebra $A_0(E_1,E_2)$.

\blemma \label{lem-ideal}
Let $\2C$ be a $TC^*$ and $E_1,E_2: \2C\rarr\Vect_\7C$ fiber functors. 
The subspace  
\[ I(E_1,E_2)=\mathrm{span}_\7C\{ [X, a\circ E_2(s)]-[Y,E_1(s)\circ a ] \ | \ 
    s:X\rarr Y,\ a\in\Hom(E_2(Y),E_1(X)) \} \]
is a two-sided ideal. 
\elemma

\prf  To show that $I(E_1,E_2)\subset A_0(E_1,E_2)$ is an ideal, let 
$s:X\rarr Y,\ a\in\Hom(E_2(Y),E_1(X))$, thus $[X,a\circ E_2(s)]-[Y,E_1(s)\circ a]\in I(E_1,E_2)$,
and let $[Z,t]\in A_0(E_1,E_2)$. Then 
\bean \lefteqn{  ([X, a\circ E_2(s)]-[Y,E_1(s)\circ a ]) \cdot [Z,t] } \\
  &&=  [X\otimes Z, d^1_{X,Z}\circ (a\circ E_2(s))\otimes t\circ (d^2_{X,Z})^{-1}] -
   [Y\otimes Z, d^1_{Y,Z}\circ (E_1(s)\circ a)\otimes t\circ (d^2_{Y,Z})^{-1}] \\
  &&=  [X\otimes Z, d^1_{X,Z}\circ a \otimes t\circ (d^2_{Y,Z})^{-1}\circ E_2(s\otimes\id_Z)] -
   [Y\otimes Z, E_1(s\otimes\id_Z)\circ d^1_{X,Z}\circ a\otimes t\circ (d^2_{Y,Z})^{-1}] \\
  &&=  [X', a'\circ E_2(s')] - [Y', E_1(s')\circ a'] \ \in I(E_1,E_2), 
\eean
where in the second equality we used naturality of $d^i$, and in the last line we wrote 
$X'=X\otimes Z, Y'=Y\otimes Z, s'=s\otimes\id_Z: X'\rarr Y'$ and
$a'=d^1_{X,Z}\circ a \otimes t\circ (d^2_{Y,Z})^{-1}\in\Hom(E_2(Y'),E_1(X')$ in order 
to make clear that the result is in $I(E_1,E_2)$. This proves that the latter is a  left ideal in
$A_0(E_1,E_2)$. Similarly, one shows that it is a right ideal. 
\qed

We denote by $A(E_1,E_2)$ the quotient algebra $A_0(E_1,E_2)/I(E_1,E_2)$. It can also be understood
as the algebra generated by symbols $[X,s]$, where $X\in\2C, s\in\Hom(E_2(X),E_1(X))$, subject to
the relations $[X,s]+[X,t]=[X,s+t]$ and $[X, a\circ E_2(s)]=[Y,E_1(s)\circ a ]$ whenever 
$s:X\rarr Y,\ a\in\Hom(E_2(Y),E_1(X))$. Therefore it should not cause confusion that we denote the image 
of $[X,s]\in A_0(E_1,E_2)$ in $A(E_1,E_2)$ again by $[X,s]$.

\bprop \label{prop-comm}
Let $\2C$ be an $STC^*$ and $E_1,E_2: \2C\rarr\Vect_\7C$ fiber functors. If $E_1,E_2$ are symmetric
then $A(E_1,E_2)$ is commutative.   
\eprop

\prf Assume $\2C$ is symmetric and the fiber functors satisfy $E_i(c_{X,Y})=\Sigma_{E_i(X),E_i(Y)}$. 
Let $[A,u],[B,v]\in A_0(E_1,E_2)$, thus $A,B\in\2C$ and $u: E_2(A)\rarr E_1(A), v: E_2(B)\rarr E_1(B)$.
Then 
\[ [A,u]\cdot [B,v] = [ A\otimes B, d^1_{A,B}\circ u\otimes v\circ (d^2_{A,B})^{-1} ], \]
and
\bean [ B,v ] \cdot [A,u] &=& [B\otimes A, d^1_{B,A}\circ v\otimes u\circ (d^2_{B,A})^{-1}] \\
  &=& [B\otimes A, d^1_{B,A}\circ \Sigma_{E_1(A),E_2(B)} \circ u\otimes v\circ
   \Sigma_{E_2(B),E_1(A)}\circ (d^2_{B,A})^{-1}]  \\
  &=& [B\otimes A, d^1_{B,A}\circ E_1(c_{B,A}) \circ u\otimes v\circ
   E_2(c_{B,A})\circ (d^2_{B,A})^{-1}] \\
  &=& [B\otimes A, E_1(c_{A,B}) \circ d^1_{A,B}\circ u\otimes v \circ (d^2_{A,B})^{-1}
   \circ E_2(c_{B,A})] 
\eean
With $X=A\otimes B, Y=B\otimes A, s=c_{A,B}$ and 
$a=d^1_{A,B}\circ u\otimes v\circ (d^2_{A,B})^{-1}\circ E_2(c_{B,A})$
we obtain
\[ [A,u]\cdot[B,v]=[X, a\circ E_2(s)] \]
\[ [B,v]\cdot[A,u]=[Y, E_1(s)\circ a] \]
Thus 
\[ [A,u]\cdot[B,v]-[B,v]\cdot[A,u]=[X, a\circ E_2(s)]-[Y, E_1(s)\circ a] \in I(E_1,E_2), \]
implying $[A_0(E_1,E_2),A_0(E_1,E_2)]\subset I(E_1,E_2)$. Thus
$A(E_1,E_2)=A_0(E_1,E_2)/I(E_1,E_2)$ is commutative.
\qed

\bprop \label{prop-star} \index{$*$-operation}
Let $\2C$ be a TC$^*$ and let $E_1,E_2:\2C\rarr\2H$ be $*$-preserving fiber functors. Then
$A(E_1,E_2)$ has a positive $*$-operation, i.e.\ an antilinear and antimultiplicative involution
such that $a^*a=0$ implies $a=0$.
\eprop

\prf We define a $*$-operation $\star$ on $A_0(E_1,E_2)$. Let $[X,s]\in A_0(E_1,E_2)$. 
Pick a standard conjugate $(\ol{X_i},r_i,\ol{r}_i)$ and define $[X,s]^\star:=[\ol{X},t]$, where 
\[ t= \id_{E_1(\ol{X})} \otimes E_2(\ol{r}^*)\mcirc \id_{E_1(\ol{X})} \otimes
    s^*\otimes\id_{E_2(\ol{X})}  \mcirc E_1(r)\otimes \id_{E_2(\ol{X})} 
   \ \in \ \Hom_\2H(E_2(\ol{X}),E_1(\ol{X})). \]
(Of course, $s^*$ is defined using the inner products on the Hilbert spaces $E_1(X),E_2(X)$.)
If we pick another standard conjugate $(\ol{X}',r',\ol{r}')$ of $X$, we know that there is a unitary
$u:\ol{X}\rarr\ol{X}'$ such that $r'=u\otimes\id_X\mcirc r$ and $\ol{r}'=\id_X\otimes u\mcirc\ol{r}$. Using
$(\ol{X}',r',\ol{r}')$ we obtain $([X,s]^\star)':=[\ol{X}',t']$ with $t'$ defined by replacing
$r,\ol{r}$ by $r',\ol{r}'$. Now, 
\bean [\ol{X},t]-[\ol{X}',t'] &=& [\ol{X},\id_{E_1(\ol{X})} \otimes E_2(\ol{r}^*)\mcirc
    \id_{E_1(\ol{X})} \otimes  s^*\otimes\id_{E_2(\ol{X})} \mcirc E_1(r)\otimes \id_{E_2(\ol{X})}] \\
    && -[\ol{X}',\id_{E_1(\ol{X}')} \otimes E_2({\ol{r}'}^*)\mcirc \id_{E_1(\ol{X}')} \otimes
    s^*\otimes\id_{E_2(\ol{X}')}  \mcirc E_1(r')\otimes \id_{E_2(\ol{X}')} ] \\ 
 &=& [\ol{X}, ( \id_{E_1(\ol{X})} \otimes E_2({\ol{r}'}^*)\mcirc \id_{E_1(\ol{X})} \otimes  
    s^*\otimes\id_{E_2(\ol{X}')} \mcirc E_1(r)\otimes \id_{E_2(\ol{X}')} )\mcirc E_2(u)] \\
    && -[\ol{X}', E_1(u)\mcirc( \id_{E_1(\ol{X})} \otimes E_2({\ol{r}'}^*) \mcirc \id_{E_1(\ol{X})} 
   \otimes s^*\otimes\id_{E_2(\ol{X}')} \mcirc E_1(r)\otimes \id_{E_2(\ol{X}')} )], 
\eean
which is in the ideal $I(E_1,E_2)$ defined in Proposition \ref{prop-A1}. Thus, while $[X,s]^\star$
depends on the chosen conjugate $(\ol{X},r,\ol{r})$ of $X$, its image 
$\gamma([X,s]^\star)\in A(E_1,E_2)$ doesn't. 

In order to be able to define a $*$-operation on $A(E_1,E_2)$ by
$x^*:=\gamma\circ\star\circ\gamma^{-1}(x)$ we must show that the composite map
$\gamma\circ\star: A_0(E_1,E_2)\rarr A(E_1,E_2)$ maps $I(E_1,E_2)$ to zero. To this purpose, let  
$X,Y\in\2C, s:X\rarr Y,\ a\in\Hom(E_2(Y),E_1(X))$ and choose conjugates
$(\ol{X},r_X,\ol{r}_X),(\ol{Y},r_Y,\ol{r}_Y)$. Then
\bean \lefteqn{ [X,a\circ E_2(s)]^\star-[Y,E_1(s)\circ a]^\star } \\
  && = [\ol{X}, \id_{E_1(\ol{X})} \otimes E_2(\ol{r_X}^*)\mcirc \id_{E_1(\ol{X})} \otimes
    (a\circ E_2(s))^*\otimes\id_{E_2(\ol{X})}  \mcirc E_1(r_X)\otimes \id_{E_2(\ol{X})} ]  \\
  && \ \ \ - [\ol{Y},\id_{E_1(\ol{X})} \otimes E_2(\ol{r_Y}^*)\mcirc \id_{E_1(\ol{X})} \otimes
    (E_1(s)\circ a)^*\otimes\id_{E_2(\ol{X})}  \mcirc E_1(r_Y)\otimes \id_{E_2(\ol{X})} ] \\
  && = [\ol{X},\tilde{a}\circ E_2(\tilde{s})]-[\ol{Y},E_1(\tilde{s})\circ\tilde{a}], 
\eean
where
\bean \tilde{a} &=& \id_{E_1(\ol{X})}\otimes E_2(\ol{r}_X^*)\mcirc \id_{E_1(\ol{X})}\otimes a^*\otimes
   \id_{E_2(\ol{Y})} \mcirc E_1(r_X)\otimes \id_{E_2(\ol{Y})} 
  \ \in \ \Hom_\2H(E_2(\ol{Y}),E_1(\ol{X})),         \\
  \tilde{s} &=& \id_{\ol{Y}}\otimes \ol{r}_X^* \mcirc\id_{\ol{Y}}\otimes s^*\otimes\id_{\ol{X}} \mcirc
   r_Y\otimes\id_{\ol{X}} \ \in \ \Hom(\ol{X},\ol{Y}).
\eean
This clearly is in $I(E_1,E_2)$, thus $x^*:=\gamma\circ\star\circ\gamma^{-1}(x)$ defines a
$*$-operation on $A(E_1,E_2)$. 

Now it is obvious that the resulting map $*$ on
$A(E_1,E_2)$ is additive and antilinear. It also is
involutive and antimultiplicative as one verifies by an
appropriate use of the conjugate equations.  We omit
the tedious but straightforward computations. It
remains to show positivity of the
$*$-operation. Consider $[X,s]\in A_0(E_1,E_2)$, pick a
conjugate $(\ol{X},r,\ol{r})$ and compute
$[X,s]^*\cdot[X,s]=[\ol{X}\otimes X,t]$, where
\[ t=     d^1_{\ol{X},X}\mcirc   \left(
\id_{E_1(\ol{X})} \otimes E_2(\ol{r}^*)\mcirc \id_{E_1(\ol{X})} \otimes
    s^*\otimes\id_{E_2(\ol{X})}  \mcirc E_1(r)\otimes \id_{E_2(\ol{X})} 
   \right)\otimes s \mcirc (d^2_{\ol{X},X})^*.
\]
Now, 
\bean [\ol{X}\otimes X,t] &=& [\ol{X}\otimes X,E_1(r^*)\circ E_1(r)\circ t]
   = [\11,E_1(r)\circ t\circ E_2(r^*)] \\
  &=& \left[\11, E_1(r^*)\mcirc\left( \id_{E_1(\ol{X})} \otimes E_2(\ol{r}^*)\mcirc\id_{E_1(\ol{X})} 
  \otimes s^*\otimes\id_{E_2(\ol{X})}
\right. \right. \\
&& \left. \left. \mcirc E_1(r)\otimes \id_{E_2(\ol{X})} \right)\otimes s 
   \mcirc E_2(r) \right] \\
  &=& [\11, E_1(r^*)\mcirc \id\otimes (s\circ s^*)\mcirc E_1(r)] =[\11,u^*u], 
\eean
where we have used the conjugate equations and put $u=\id\otimes s^*\mcirc E_1(r)$.
Thus, $[X,s]^*\cdot[X,s]=[\11,u^*u]$ is zero iff $u^*u$ is
zero. By positivity of the $*$-operation in $\2H$, this holds iff $u=0$. Using once again the
conjugate equations we see that this is equivalent to $s=0$. Thus for elements $a\in A(E_1,E_2)$ of
the form $[X,s]$, the implication $a^*a=0\impl a=0$ holds. For a general $a=\sum_i [X_i,s_i]$ we
pick isometries $v_i:X_i\rarr X$ such that $\sum_i v_i\circ v_i^*=\id_X$ (i.e.\ $X\cong\oplus_iX_i$).
Then $[X_i,s_i]=[X,E_1(v_i)\circ s_i\circ E_2(v_i^*)]$, thus
\[ \sum_i[X_i,s_i]=[X,\sum_iE_1(v_i)\circ s_i\circ E_2(v_i^*)], \]
implying that every element of $A(E_1,E_2)$ can be written as $[X,s]$, and we are done.
\qed

\bprop \label{prop-norm}
Let $\2C$ be a $TC^*$ and let $E_1,E_2:\2C\rarr\2H$ be $*$-preserving fiber functors. Then
\[ \| a\| = {\inf_b}' \sup_{X\in\2C} \| b_X\|_{\End E(X)}, \]
where the infimum is over all representers $b\in A_0(E_1,E_2)$ of $a\in A(E_1,E_2)$, defines a
$C^*$-norm on $A(E_1,E_2)$.  
\eprop

\prf Let $[X,s], [Y,t]\in A_0(E_1,E_2)$. Then $[X,s]\cdot[Y,t]=[X\otimes Y,u]$, where 
$u=d^1_{X,Y}\circ s\otimes t\circ(d^2_{X,Y})^{-1}$. Since $d^1_{X,Y},d^2_{X,Y}$ are unitaries, we
have $\|[X\otimes Y,u]\|=\|u\|\le\|s\|\cdot\|t\|$. Thus $\|b\|=\sup_{X\in\2C} \| b_X\|_{\End E(X)}$
defines a submultiplicative norm on $A_0(E_1,E)$, and the above formula for $\|a\|$ is the usual
definition of a norm on the quotient algebra $A_0(E_1,E_2)/I(E_1,E_2)$. This norm satisfies
$\|[X,s]\|=\|s\|$. Since every $a\in A(E_1,E_2)$ can be written as $[X,s]$, we have $\|a\|=0\impl a=0$.
Finally, the computations in the proof of Proposition \ref{prop-star} imply
\[ \|[X,s]^*[X,s]\|=\|[\11,u^*u]\|=\|u^*u\|=\|u\|^2=\|s\|^2=\|[X,s]\|^2, \]
which is the $C^*$-condition.
\qed

\bdefin Let $\2C$ be a $TC^*$ and let $E_1,E_2:\2C\rarr\2H$ be $*$-preserving fiber functors. Then
$\2A(E_1,E_2)$ denotes the $\|\cdot\|$-completion of $A(E_1,E_2)$. (This is a unital $C^*$-algebra,
which commutative if $\2C,E_1,E_2$ are symmetric.)
\edefin


\subsection{Uniqueness of fiber functors} \label{ss-unique}
\blemma \cite{JS2} \label{lem-iso}
Let $\2C$ be a TC$^*$, $\2D$ a strict tensor category and $E_1,E_2: \2C\rarr\2D$ strict tensor
functors. Then any monoidal natural transformation $\alpha: E_1\rarr E_2$ is a natural isomorphism. 
\elemma

\prf It is sufficient to show that every component $\alpha_X: E_1(X)\rarr E_2(X)$ has a two-sided
inverse $\beta_X: E_2(X)\rarr E_1(X)$. The family $\{\beta_X, X\in\2C\}$ will then automatically be
a natural transformation. If $(\ol{X},r,\ol{r})$ is a conjugate for $X$, monoidality of $\alpha$ implies
\be \label{eq-monnat} E_2(r^*)\mcirc\alpha_{\ol{X}}\otimes\alpha_X=E_2(r^*)\mcirc\alpha_{\ol{X}\otimes X}
   =\alpha_\11\mcirc E_1(r^*)=E_1(r^*). \ee
If we now define
\[ \beta_X=\id_{E_1(X)}\otimes E_2(r^*)\mcirc\id_{E_1(X)}\otimes \alpha_{\ol{X}}\otimes\id_{E_2(X)}
   \mcirc E_1(\ol{r})\otimes\id_{E_2(X)}, \]
we have
\bean \beta_X\circ\alpha_X &=& (\id_{E_1(X)}\otimes E_2(r^*)\mcirc\id_{E_1(X)}\otimes
   \alpha_{\ol{X}}\otimes\id_{E_2(X)} \mcirc E_1(\ol{r})\otimes\id_{E_2(X)})\mcirc\alpha_X \\
  &=& \id_{E_1(X)}\otimes E_2(r^*)\mcirc \id_{E_1(X)}\otimes\alpha_{\ol{X}}\otimes\alpha_X
  \mcirc E_1(\ol{r})\otimes\id_{E_1(X)} \\
  &=& \id_{E_1(X)}\otimes E_1(r^*)\mcirc E_1(\ol{r})\otimes\id_{E_1(X)} = \id_{E_1(X)}.
\eean
The argument for $\alpha_X\circ\beta_X=\id_{E_2(X)}$ is similar. 
\qed

\brem The lemma remains correct if one allows $E_1,E_2$ (or even $\2C,\2D$) to be non-strict. 
To adapt the proof one must replace $E_1(r)$ (which is a morphism $E_1(\11)\rarr E_1(\ol{X}\otimes X)$)
by  $(d^{E_1}_{\ol{X},X})^{-1}\circ E_1(r)\circ e^{E_1}$ (which is a morphism 
$\11_{\Vect}\rarr E_1(\ol{X})\otimes E_1(X)$). Similarly with $E_2(\ol{r})$. 
\erem

\bprop \label{prop-A1} Let $\2C$ be a $TC^*$ and
$E_1,E_2:\2C\rarr\Vect_\7C$ fiber functors.  The
pairing between $A_0(E_1,E_2)$ and the vector space
\bean \Nat(E_1,E_2) &=& \Bigl\{ (\alpha_X)_{X\in\2C} \in
  \prod_{X\in\2C}\Hom(E_1(X),E_2(X)) \ \ | \Bigr. \\
& & \Bigl. \ \ E_2(s)\circ\alpha_X=\alpha_Y\circ E_1(s)\ \forall s:X\rarr Y \Bigr\} \eean
of natural transformations $E_1\rarr E_2$ that is given, for $(\alpha_X)\in\Nat(E_1,E_2)$ and 
$a\in A_0(E_1,E_2)$, by  
\be \label{eq-p} \langle \alpha,a\rangle = \sum_{X\in\2C} \, Tr_{E_1(X)} (a_X\alpha_X) \ee
descends to a pairing between $\Nat(E_1,E_2)$ and the quotient algebra
$A(E_1,E_2)=A_0(E_1,E_2)/I(E_1,E_2)$ such that 
\[ \Nat(E_1,E_2)\ \cong A(E_1,E_2)^*. \]
Under this isomorphism, an element $a\in A(E_1,E_2)^*$ corresponds to an element of
$\Nat_\otimes(E_1,E_2)$, i.e.\ a {\it monoidal} natural transformation (thus isomorphism by Lemma
\ref{lem-iso}), iff it is a character, to wit multiplicative. 
\eprop

\prf The dual vector space of the direct sum $A_0(E_1,E_2)$ is the direct product
$\prod_{X\in\2C}\Hom(E_2(X),E_1(X))^*$, and since the pairing between 
$\Hom(E_2(X),E_1(X))\times\Hom(E_1(X),E_2(X)), \ s\times t\mapsto Tr(s\circ t)$ 
is non-degenerate, we have
\[ A_0(E_1,E_2)^* \cong \prod_{X\in\2C}\Hom(E_1(X),E_2(X)) \]
w.r.t.\ the pairing given in (\ref{eq-p}). Now, $A(E_1,E_2)$ is the quotient of $A_0(E_1,E_2)$ by
the subspace $I(E_1,E_2)$, thus the dual space $A(E_1,E_2)^*$ consists precisely of those elements
of $A_0(E_1,E_2)^*$ that are identically zero on $I(E_1,E_2)$. Assume $(a_X)_{X\in\2C}$ satisfies
$\langle\alpha,a\rangle=0$ for all $a\in I(E_1,E_2)$, equivalently
$\langle \alpha, [X, a\circ E_2(s)]-[Y,E_1(s)\circ a ] \rangle=0$ for all $s:X\rarr Y$ and
$a\in\Hom(E_2(Y),E_1(X))$. By definition (\ref{eq-p}) of the pairing, this is equivalent to 
\[ Tr_{E_1{X}}( a\circ E_2(s)\circ \alpha_X)-Tr_{E_1(Y)}(E_1(s)\circ a \circ \alpha_Y) =0 
  \quad\quad \forall s:X\rarr Y, a\in\Hom(E_2(Y),E_1(X)).
\]
Non-degeneracy of the trace implies that $\alpha=(\alpha_X)_{X\in\2C}$ must satisfy 
$E_2(s)\circ\alpha_X=\alpha_Y\circ E_1(s)$ for all $s:X\rarr Y$, thus $\alpha\in\Nat(E_1,E_2)$,
implying 
\[ A(E_1,E_2)^*\cong\Nat(E_1,E_2). \]

Now we consider the question when the functional $\phi\in A(E_1,E_2)^*$ corresponding to
$\alpha\in\Nat(E_1,E_2)$ is a character, i.e.\ multiplicative. This is the case when 
\[ \langle\alpha, [X,s]\cdot[Y,t]\rangle=\langle\alpha, [X,s]\rangle\langle\alpha, [Y,t]\rangle
  \quad\quad \forall [X,s],[Y,t]\in A(E_1,E_2). \]
(Strictly speaking, $[X,s],[Y,t]$ are representers in $A_0(E_1,E_2)$ for some elements in
  $A(E_1,E_2)$.)
In view of (\ref{eq-p}) and the definition of the product in $A(E_1,E_2)$ this amounts to
\bean Tr_{E_1(X\otimes Y)}( d^1_{X,Y}\circ s\otimes t\circ (d^2_{X,Y})^{-1}\circ\alpha_{X\otimes Y})
   &=& Tr_{E_1(X)}(s\circ\alpha_X)\,Tr_{E_1(Y)}(t\circ\alpha_Y) \\
  &=& Tr_{E_1(X)\otimes E_2(X)}((s\circ\alpha_X)\otimes(t\circ\alpha_Y)) \\
  &=& Tr_{E_1(X)\otimes E_2(X)}(s\otimes t \mcirc \alpha_X\otimes\alpha_Y) 
\eean
In view of the cyclic invariance and non-degeneracy of the trace, this is true for all 
$s: E_2(X)\rarr E_1(X)$ and $t: E_2(Y)\rarr E_1(Y)$, iff
\[ \alpha_{X\otimes Y} = d_{X,Y}^2\circ\alpha_X\otimes\alpha_Y \circ (d^1_{X,Y})^{-1} \quad\quad
   \forall X,Y\in\2C. \]
This is precisely the condition for $\alpha\in\Nat(E_1,E_2)$ to be monoidal, to wit 
$\alpha\in\Nat_\otimes(E_1,E_2)$. 
\qed

\bprop \label{prop-starchar} Let $\2C$ be a $TC^*$ and let $E_1,E_2:\2C\rarr\2H$ be $*$-preserving
fiber functors. Then a monoidal natural transformation $\alpha\in\Nat_\otimes(E_1,E_2)$ is unitary
(i.e.\ each $\alpha_X$ is unitary) iff the corresponding character $\phi\in A(E_1,E_2)$ is a
$*$-homomorphism (i.e.\ $\phi(a^*)=\ol{\phi(a)}$).
\eprop

\prf Let $\alpha\in\Nat_\otimes(E_1,E_2)$ and $[X,s]\in A(E_1,E_2)$. By definition of the pairing of
$A(E_1,E_2)$ and $\Nat(E_1,E_2)$,
\[ \phi([X,s])=\langle\alpha,[X,s]\rangle=Tr_{E_1(X)}(s\circ\alpha_X), \]
and therefore, using $\ol{Tr(AB)}=Tr(A^*B^*)$,
\[ \ol{\phi([X,s])}=Tr_{E_1(X)}(s^*\circ\alpha_X^*). \]
On the other hand,
\bean \phi([X,s]^*) &=& \langle\alpha,[\ol{X},\id_{E_1(\ol{X})} \otimes E_2(\ol{r}^*)\mcirc \id_{E_1(\ol{X})} 
   \otimes s^*\otimes\id_{E_2(\ol{X})}  \mcirc E_1(r)\otimes \id_{E_2(\ol{X})} ]\rangle \\
   &=& Tr_{E_1(\ol{X})}(\id_{E_1(\ol{X})} \otimes E_2(\ol{r}^*)\mcirc \id_{E_1(\ol{X})} \otimes 
   s^*\otimes\id_{E_2(\ol{X})}  \mcirc E_1(r)\otimes \id_{E_2(\ol{X})} \mcirc\alpha_{\ol{X}} )  \\
   &=& E_2(\ol{r}^*)\mcirc s^*\otimes \alpha_{\ol{X}}\mcirc E_1(\ol{r}) \\
   &=& E_2(\ol{r}^*)\mcirc (\alpha_X\circ\alpha_X^{-1}\circ s^*)\otimes \alpha_{\ol{X}}\mcirc E_1(\ol{r}) \\
   &=& E_1(\ol{r}^*)\mcirc (\alpha_X^{-1}\circ s^*)\otimes\id_{E_2(\ol{X})}\mcirc E_1(\ol{r}) \\
   &=& Tr_{E_1(X)}(\alpha_X^{-1}\circ s^*).
\eean
(In the fourth step we have used the invertibility of $\alpha$ (Lemma \ref{lem-iso}) and in the
fifth equality we have used (\ref{eq-monnat}) with $X$ and $\ol{X}$ interchanged and $r$ 
replaced by $\ol{r}$.). Now non-degeneracy of the trace implies that
$\ol{\phi([X,s])}=\phi([X,s]^*)$ holds for all $[X,s]\in (E_1,E_2)$ iff $\alpha_X^*=\alpha_X^{-1}$
for all $X\in\2C$, as claimed.  
\qed

Now we are in a position to prove the first of our outstanding claims: \\

\noindent{\it Proof of Theorem \ref {theor-uniq1}.}  \index{fiber functor!uniqueness}
By the preceding constructions, the $\|\cdot\|$-closure $\2A(E_1,E_2)$ of $A(E_1,E_2)$ is a
commutative unital $C^*$-algebra. As such it has (lots of) characters, i.e.\ unital
$*$-homomorphisms into $\7C$. (Cf.\ e.g.\ Theorem \ref{theor-cstar} below.) Such a character
restricts to $A(E_1,E_2)$ and corresponds, by Propositions \ref{prop-A1} and \ref{prop-starchar}, to
a unitary monoidal natural transformation $\alpha\in\Nat(E_1,E_2)$.  
\qed

\brem 1. The discussion of the algebra $A(E_1,E_2)$ is inspired by the one in the preprint
\cite{bichon1} that didn't make it into the published version \cite{bichon}. The above proof of 
Theorem \ref {theor-uniq1} first appeared in \cite{bichon2}.

2. Lemma \ref{lem-iso} implies that the category consisting of fiber functors and monoidal natural 
transformations is a groupoid, i.e.\ every morphism is invertible. Theorem \ref{theor-uniq1} then
means that the category consisting of symmetric $*$-preserving fiber functors and unitary monoidal
natural transformations is a transitive groupoid, i.e.\ all objects are isomorphic. That this
groupoid is non-trivial is the statement of Theorem \ref{theor-C}, whose proof will occupy the bulk
of this section, beginning in Subsection \ref{ss-starpres}. 
\erem


\subsection{The concrete Tannaka theorem. Part II} \label{ss-concrete2}
In order to prove Proposition \ref{prop-dense} we need the formalism of the preceding subsections. 
 We write $\2A(E)$ for the commutative unital $C^*$-algebra $\2A(E,E)$ defined earlier. In order to
study this algebra we need some results concerning commutative unital $C^*$-algebras that can be
gathered, e.g., from \cite{ped}.

\btheor \label{theor-cstar}
Let $\2A$ be a commutative unital $C^*$-algebra. Let $\2A^*$ be its Banach space dual and let
\bean P(\2A) &=& \{\phi\in\2A^*\ | \ \phi(1)= 1,\ \|\phi\|\le 1\}, \\
   X(\2A) &=& \{ \phi\in\2A^* \ | \ \phi(1)=1,\ \phi(ab)=\phi(a)\phi(b), \ \phi(a^*)=\ol{\phi(a)}
  \ \ \forall a,b\in\2A\}. 
\eean
$P(\2A)$ and $X(\2A)$ are equipped with the $w^*$-topology on $\2A$ according to which
$\phi_\iota\rightarrow\phi$ iff $\phi_\iota(a)\rightarrow\phi(a)$ for all $a\in\2A$. Then:
\begin{itemize}
\item[(i)] $X(\2A)\subset P(\2A)$  (thus $*$-characters have norm $\le 1$.)
\item[(ii)] $X(\2A)$ is compact w.r.t.\ the $w^*$-topology on $P(\2A)$.
\item[(iii)] The map $\2A\rarr C(X(\2A))$ given by $a\mapsto (\phi\mapsto\phi(a))$ is an isomorphism of
  $C^*$-algebras.
\item[(iv)] The convex hull
\[ \left\{ \sum_{i=1}^N c_i \phi_i \ , \ N\in\7N, \ c_i\in\7R_+, \ \sum_ic_i=1, \ \phi_i\in X(\2A) \right\} \]
of $X(\2A)$ is $w^*$-dense in $P(\2A)$.
\end{itemize}
\etheor

\prf (i) Any unital $*$-homomorphism $\alpha$ of Banach algebras satisfies $\|\alpha(a)\|\le\|a\|$.

(ii) By Alaoglu's theorem \cite[Theorem 2.5.2]{ped}, the unit ball of $\2A^*$ is compact w.r.t.\ the
$w^*$-topology, and so are the closed subsets $X(\2A)\subset P(\2A)\subset\2A^*$. 

(iii) This is Gelfand's theorem, cf.\ \cite[Theorem 4.3.13]{ped}.

(iv) This is the Krein-Milman theorem, cf.\ Theorem 2.5.4 together with Proposition 2.5.7 in
\cite{ped}. 
\qed

Theorem \ref{theor-cstar}, (ii) implies that the set $X\equiv X(\2A(E))$ of $*$-characters of
$\2A(E)$ is a compact Hausdorff space w.r.t.\ the $w^*$-topology. By (iii) and Proposition
\ref{prop-starchar}, the elements of $X$ are in bijective correspondence with the set $G_E$ of
unitary monoidal transformations of the functor $E$.    

\blemma
The bijection $X\cong G_E$ is a homeomorphism w.r.t.\ the topologies defined above.
\elemma

\prf By definition of the product topology on $\prod_{X\in\2C}\2U(E(X))$, a net $(g_\iota)$ in $G_E$
converges iff the net $(g_{\iota,X})$ in $\2U(E(X))$ converges for every $X\in\2C$. On the other
hand, a net $(\phi_\iota)$ in $X$ converges iff $(\phi_\iota(a))$ converges in $\7C$ for every
$a\in\2A(E)$. In view of the form of the  correspondence $\phi\leftrightarrow g$ established in
Proposition \ref{prop-A1}, these two notions of convergence coincide.
\qed

The homeomorphism $X\cong G_E$ allows to transfer the topological group structure that $G_E$
automatically has to the compact space $X$. Now we are in a position to complete the proof of our
second outstanding claim. \\

\noindent{\it Proof of Proposition \ref{prop-dense}.}
Since $\2C$ is semisimple and essentially small, there exist a set $I$ and a family 
$\{X_i, i\in I\}$ of irreducible objects such that every object is (isomorphic to) a finite direct
sum of objects from this set. If $\Nat(E)\equiv\Nat(E,E)$ is the space of natural transformations
from $E$ to itself, with every $\alpha\in\Nat(E)$ we can associate the family
$(\alpha_i=\alpha_{X_i})_{i\in I}$, which is an element of $\prod_{i\in  I}\End\,E(X_i)$. 
Semisimplicity of $\2C$ and naturality of $\alpha$ imply that every such element arises from exactly
one natural transformation of $E$. (In case it is not obvious, a proof can be found in
\cite[Proposition 5.4]{MRT}.) In this way we obtain an isomorphism  
\[ \gamma: \Nat(E)\rarr\prod_{i\in I}\End\,E(X_i), \quad \alpha\mapsto(\alpha_{X_i})_{i\in I} \]
of vector spaces. Now consider the linear map
\[ \delta:\ \bigoplus_{i\in I} \End\,E(X_i)\rarr A(E), \quad (a_i)\mapsto \sum_i [X_i,a_i]. \]
Since every $a\in A(E)$ can be written as $[X,s]$ (proof of Proposition \ref{prop-star}) and every
$[X,s]$ is a sum of elements $[X_i,s_i]$ with $X_i$ irreducible,
$\delta$ is surjective. When understood as a map to $A_0(E)$,
$\delta$ obviously is injective. As a consequence of $\Hom(X_i,X_j)=\{0\}$ for $i\ne j$ ,
the image in $A_0(E)$ of of $\delta$ has trivial intersection with the ideal $I(E)$, which
is the kernel of the quotient map $A_0(E)\rarr A(E)$, thus $\delta$ is injective and
therefore an isomorphism (of vector spaces, not algebras). 
If the $C^*$-norm on $A(E)$ is pulled back via $\delta$ we obtain the norm
\[ \| (a_i)_{i\in I} \| = \sup_{i\in I} \|a_i\|_{\End\,E(X_i)} \]
on $\bigoplus_{i\in I} \End\,E(X_i)$. Thus we have an isomorphism 
$\ol{\delta}:\ol{\bigoplus_{i\in I}\End\,E(X_i)}^{\|\cdot\|}\rarr\2A(E)$ of the norm closures.
W.r.t.\ the isomorphisms $\gamma,\delta$,
the pairing $\langle\cdot,\cdot\rangle: \Nat(E)\times A(E)\rarr\7C$ of Proposition \ref{prop-A1}
becomes
\[ \langle\cdot,\cdot\rangle^\sim:\ \prod_{i\in I}\End\,E(X_i) \ \times\ \bigoplus_{i\in I}
   \End\,E(X_i)\rarr\7C, \quad\quad 
   (\alpha_{X_i})\times(a_i)\mapsto \sum_{i\in I} Tr_{E(X_i)}(\alpha_ia_i). \]
(More precisely: $\langle \cdot,\delta(\cdot)\rangle=\langle\gamma(\cdot),\cdot\rangle^\sim$ as maps
$\Nat(E)\times\bigoplus_{i\in I} \End\,E(X_i)\rarr\7C$.)
Thus if $\alpha\in\Nat(E)$ is such that $\gamma(\alpha)\in\prod_{i\in I}\End\,E(X_i)$ has only
finitely many non-zero components (i.e.\ $\gamma(\alpha)\in\oplus_{i\in I}\End\,E(X_i)$), then
$\langle\alpha,\cdot\rangle\in A(E)^*$  extends to an element of $\2A(E)^*$. 

Now (iv) of Theorem \ref{theor-cstar} implies that every $\phi\in\2A(E)^*$ is the $w^*$-limit of a
net $(\phi_\iota)$ in the $\7C$-span of the $*$-characters $X(\2A(E))$ of $\2A(E)$.  Thus for every 
$(\alpha_i)\in\bigoplus_{i\in I}\End\,E(X_i)$ there is a such a net $(\phi_\iota)$ for which
\[ w^*-\lim \phi_\iota =
\langle\gamma^{-1}((\alpha_i)),\cdot\rangle\in\2A(E)^*. \] Restricting
the $\phi_\iota$ to $A(E)$ and using the isomorphism $\Nat\,E\cong
A(E)^*$, we obtain a net in $\Nat\,E$ that converges to
$\gamma^{-1}((\alpha_i))$. By Propositions \ref{prop-A1},
\ref{prop-starchar}, the isomorphism $A(E)^*\rarr\Nat\,E$ maps the
elements of $X(\2A(E))$ to the unitary natural monoidal
transformations of $E$, i.e.\ to elements of $G_E$.  Thus, in
particular for every finite $S\subset I$ we have

\[ \overline{\mathrm{span}} _{\7C} \{
  \underbrace{ \pi_{s_1}(g)\oplus\cdots\oplus\pi_{s_{|S|}}(g)}_{\mathrm{all}\
    s\in S},\ g\in G_E \}  = \bigoplus_{s\in S}\End\,E(X_s), \]
which clearly is a good deal more than claimed in Proposition
\ref{prop-dense}.  \qed

This concludes the proof of all ingredients that went into the proof
of Theorem \ref{theor-T2}. From the proof it is obvious that the
commutative $C^*$-algebra $\2A(E)$ is just the algebra of continuous
functions on the compact group $G_E$, whereas $A(E)$ is the linear
span of the matrix elements of the finite dimensional representations
of $G_E$.


\subsection{Making a symmetric fiber functor $*$-preserving} \label{ss-starpres}
The aim of his subsection is to prove the following result, which seems to be new:

\btheor \label{theor-F} An even $STC^*\ \2C$ that admits a symmetric
fiber functor $\2C\rarr\Vect_\7C$ also admits a symmetric
$*$-preserving fiber functor $\2C\rarr\2H$.  \etheor

\blemma Let $\2C$ be an $STC^*$ and $E:\2C\rarr\Vect_\2C$ a symmetric
fiber functor. Choose arbitrary positive definite inner products
$\langle\cdot,\cdot\rangle^0_X$ (i.e.\ Hilbert space structures) on
all of the spaces $E(X), X\in\2C$. Then the maps $X\mapsto E(X)$ and
$s\mapsto E(s^*)^\dagger$, where $E(s^*)^\dagger$ is the adjoint of
$E(s^*)$ w.r.t.\ the inner products $\langle\cdot,\cdot\rangle^0_X$,
define a faithful functor $\widetilde{E}:\7C\rarr\Vect_\7C$. With
$d^{\widetilde{E}}_{X,Y}=((d^E_{X,Y})^\dagger)^{-1}$ and
$e^{\widetilde{E}}=((e^E)^\dagger)^{-1}$, this is a symmetric fiber
functor.  \elemma

\prf First note that $s\mapsto\widetilde{E}(s)$ is $\7C$-linear and really defines a functor, since 
$\widetilde{E}(\id_X)=\id_{\widetilde{E}(X)}$ and 
\[ \widetilde{E}(s\circ t)=E((s\circ t)^*)^\dagger=E(t^*\circ
s^*)^\dagger=(E(t^*)\circ E(s^*))^\dagger =E(s^*)^\dagger\circ
E(t^*)^\dagger=\widetilde{E}(s)\circ\widetilde{E}(t). \] Faithfulness
of $E$ clearly implies faithfulness of $\widetilde{E}$.  With
$d^{\widetilde{E}}_{X,Y}=((d^E_{X,Y})^\dagger)^{-1}$ and
$e^{\widetilde{E}}=((e^E)^\dagger)^{-1}$, commutativity of the
diagrams (\ref{eq-A1}) and (\ref{eq-A2}) is obvious. Since $E$ is a
tensor functor, we have
\[ E(s\otimes t)\mcirc d^E_{X,Y}=d^E_{X',Y'}\mcirc E(s)\otimes E(t) \]
for all $s:X\rarr X',\ t:Y\rarr Y'$, which is equivalent to
\[ (E(s\otimes t))^\dagger\mcirc((d^E_{X',Y'})^{-1})^\dagger 
   =((d^E_{X,Y})^{-1})^\dagger\mcirc(E(s)\otimes E(t))^\dagger. \]
Since this holds for all $s,t$, we have proven naturality of the family
$(d^{\widetilde{E}}_{X,Y})$, thus $\widetilde{E}$ is a tensor functor. The computation
\[
\widetilde{E}(c_{X,Y})=E(c_{X,Y}^*)^\dagger=E(c_{Y,X})^\dagger=\Sigma_{E(Y),E(X)}^\dagger
=\Sigma_{E(X),E(Y)}, \] where we have used
$\Sigma_{H,H'}^\dagger=\Sigma_{H',H}$, shows that $\widetilde{E}$ is
also symmetric. Thus $\widetilde{E}$ is a symmetric fiber functor.
\qed

Now the discussion of Subsection \ref{ss-algebras} applies and
provides us with a commutative unital $\7C$-algebra
$A(E,\widetilde{E})$. However, we cannot appeal to Proposition
\ref{prop-star} to conclude that $A(E,\widetilde{E})$ is a
$*$-algebra, since $E,\widetilde{E}$ are not $*$-preserving.  In fact,
for arbitrary symmetric fiber functors $E_1,E_2$ there is no reason
for the existence of a positive $*$-operation on $A(E_1,E_2)$, but in
the present case, where the two functors are related by
$E_2(s)=E_1(s^*)^\dagger$, this is true:

\bprop Let $\2C$ be an $STC^*$, $E:\2C\rarr\Vect_\7C$ a symmetric fiber functor and $\widetilde{E}$
as defined above. Then 
\[ [X,s]^\star=[X,s^\dagger] \] is well defined and is a positive
$*$-operation on $A(E,\widetilde{E})$. With respect to this
$*$-operation, the norm $\|\cdot\|$ from Proposition \ref{prop-norm}
is a $C^*$-norm, i.e.\ $\|a^\star a\|=\|a\|^2$ for all $a\in
A(E,\widetilde{E})$.  \eprop

\prf For $[X,s]\in A_0(E,\widetilde{E})$ we define
$[X,s]^\star=[X,s^\dagger]$, where $s^\dagger$ is the
adjoint of $s\in\End E(X)$ w.r.t.\ the inner product on
$E(X)$. Clearly, $\star$ is involutive and
antilinear. Now, if $s:X\rarr Y,\
a\in\Hom(E_2(Y),E_1(X))$, then \bean \lefteqn{ \left(
    [X, a\circ E_2(s)]-[Y,E_1(s)\circ a ] \right)^\star
  =[X, a\circ E(s^*)^\dagger]^\star-[Y,E(s)\circ a ]^\star } \\
&& =[X, E(s^*)\circ a^\dagger]-[Y,a^\dagger\circ
E(s)^\dagger] =[X, E_1(s^*)\circ
a^\dagger]-[Y,a^\dagger\circ E_2(s^*)], \eean Since
$s^*\in\Hom(Y,X)$ and $a^\dagger\in\Hom(E(X),E(Y))$,
the right hand side of this expression is again in
$I(E,\widetilde{E})$. Thus $I(E,\widetilde{E})$ is
stable under $\star$, and $\star$ descends to an
antilinear involution on $A(E,\widetilde{E})$. In
$A_0(E,\widetilde{E})$ we have 
\bean
 ([X,s]\cdot[Y,t])^\star &=& [X\otimes Y, d^{\widetilde{E}}_{X,Y}\circ s\otimes
t\circ(d^E_{X,Y})^{-1}]^\star \\
&=& [X\otimes Y, ({d^E_{X,Y}}^\dagger)^{-1}\circ s\otimes t\circ(d^E_{X,Y})^{-1}]^\star \\
&=&[X\otimes Y, ({d^E_{X,Y}}^\dagger)^{-1}\circ
s^\dagger\otimes t^\dagger\circ({d^E_{X,Y}})^{-1}] \\
&=& [X\otimes Y, d^{\widetilde{E}}_{X,Y}\circ s^\dagger\otimes t^\dagger\circ({d^E_{X,Y}})^{-1}] \\
&=& [X,s]^\star\cdot[Y,t]^\star.  \eean Together with
commutativity of $A(E,\widetilde{E})$ this implies that
$\star$ is antimultiplicative.  Recall that there is an
isomorphism $\delta:\bigoplus_{i\in I}\End\,E(X_i)\rarr
A(E,\widetilde{E})$ such that $\|\delta((a_i)_{i\in
  I})\|=\sup_i \|a_i\|$, where $\|\cdot\|$ is the norm
defined in Subsection \ref{ss-algebras}. By definition
of $\star$ we have
$\delta((a_i))^\star=\delta((a_i^\dagger))$, implying
$\|a^\star a\|=\|a\|^2$. Thus
$(A(E,\widetilde{E}),\star,\|\cdot\|)$ is a
pre-$C^*$-algebra.  \qed

(Note that the involution $\star$ has nothing at all to do with the
one defined in Subsection \ref{ss-algebras}!)

\bprop 
Let $\2C$ be an $STC^*$ and $E:\2C\rarr\Vect_\7C$ a symmetric fiber functor. With $\widetilde{E}$ 
as defined above, there exists a natural monoidal isomorphism $\alpha:E\rarr\widetilde{E}$, whose
components $\alpha_X$ are positive, i.e.\ $\langle u,\alpha_Xu\rangle_X^0>0$ for all nonzero $u\in E(X)$.
\eprop

\prf As in Subsection \ref{ss-unique}, the norm-completion
$\2A(E,\widetilde{E})$ of $A(E,\widetilde{E})$ is a commutative unital
$C^*$-algebra and therefore admits a $*$-character
$\phi:\2A(E,\widetilde{E})\rarr\7C$. Restricting to
$A(E,\widetilde{E})$, Proposition \ref{prop-A1} provides a monoidal
natural isomorphism $\alpha:E\rarr\widetilde{E}$. But we know more:
The character $\phi$ is positive, i.e.\ $\phi(a^\star a)>0$ for all
$a\ne 0$. With $a=[X,s]$ and taking (\ref{eq-p}) into account, we have
\[ \begin{aligned} \phi(a^\star a) &=\phi([X,s^\dagger s])=Tr_{E(X)}(s^\dagger s\alpha_X)=Tr_{E(X)}(s\alpha_Xs^\dagger)
  \\
&= \sum_i \langle e_i, s\alpha_X s^\dagger e_i\rangle_X^0 =\sum_i \langle s^\dagger e_i, \alpha_X
  s^\dagger e_i\rangle_X^0, \end{aligned} \]
where $\{e_i\}$ is any basis of $E(X)$ that is orthonormal w.r.t.\ $\langle\cdot,\cdot\rangle_X^0$.
This is positive for all $a=[X,s]\in\2A(E,\widetilde{E})$ iff $\langle u,\alpha_Xu\rangle_X^0>0$ for
all nonzero $u\in E(X)$. 
\qed

Now we are in a position to prove the main result of this subsection, which is a more specific
version of Theorem \ref{theor-F}.

\btheor \label{theor-starfunc} Let $\2C$ be an even $STC^*$ and
$E:\2C\rarr\Vect_\7C$ a symmetric fiber functor. Then there exist
Hilbert space structures (i.e.\ positive definite inner products
$\langle\cdot,\cdot\rangle_X$) on the spaces $E(X),X\in\2C$ such that
$X\mapsto (E(X),\langle\cdot,\cdot\rangle_X)$ is a $*$-preserving
symmetric fiber functor $\2C\rarr\2H$.  \etheor

\prf Pick non-degenerate inner products $\langle\cdot,\cdot\rangle_X^0$ on the spaces $E(X),X\in\2C$. 
Since $E(\11)$ is one-dimensional and
spanned by $e^E1$, where $1\in\7C=\11_{\Vect_\7C}$, we can define $\langle\cdot,\cdot\rangle_\11^0$ by
$\langle ae^E1,b e^E1\rangle_\11^0=\ol{a}b$, as will be assumed in the sequel. Let $\widetilde{E}$
and $\alpha\in\Nat_\otimes(E,\widetilde{E})$ as above. Defining new inner products
$\langle\cdot,\cdot\rangle_X$ on the spaces $E(X)$ by  
\[ \langle v,u\rangle_X=\langle v,\alpha_X u\rangle^0_X, \] the
naturality
\[ \alpha_Y\circ E(s) = \widetilde{E}(s)\circ\alpha_X=E(s^*)^\dagger\circ\alpha_X
   \quad\forall s:X\rarr Y, \]
of $(\alpha_X)$ implies
\[ \langle v,E(s)u\rangle_Y=\langle v,\alpha_Y E(s)u\rangle^0_Y=\langle v,E(s^*)^\dagger\alpha_Xu\rangle^0_Y 
   =\langle E(s^*)v,\alpha_Xu\rangle^0_X  =\langle E(s^*)v, u\rangle_X \]
for all $s:X\rarr Y,\ u\in E(X),\ v\in E(Y)$. This is the same as $E(s^*)=E(s)^*$, where now
$E(s)^*$ denotes the adjoint of $E(s)$ w.r.t.\ the inner products $\langle\cdot,\cdot\rangle$. Thus
the functor $X\mapsto(E(X),\langle\cdot,\cdot\rangle_X)$ is $*$-preserving. The new inner products 
$\langle\cdot,\cdot\rangle_X$ are non-degenerate since the $\alpha_X$ are invertible,
and the positivity property $\langle u,\alpha_Xu\rangle_X^0> 0$ for $u\ne 0$ implies that
$(E(X),\langle\cdot,\cdot\rangle_X)$ is a Hilbert space. The monoidality
\[ \alpha_{X\otimes Y}\mcirc d^E_{X,Y}=d^{\widetilde{E}}_{X,Y}\mcirc
\alpha_X\otimes\alpha_Y =((d^E_{X,Y})^\dagger)^{-1}\mcirc
\alpha_X\otimes\alpha_Y \quad\forall X,Y \] of the natural isomorphism
$\alpha:E\rarr\widetilde{E}$ is equivalent to \be
\alpha_X\otimes\alpha_Y=(d^E_{X,Y})^\dagger\mcirc\alpha_{X\otimes
  Y}\mcirc d^E_{X,Y}. \label{eq-alpha}\ee Using this we have \bean
\lefteqn{ \langle d^E_{X,Y}(u'\otimes v'),d^E_{X,Y}(u\otimes v)
  \rangle_{X\otimes Y}
  = \langle d^E_{X,Y}(u'\otimes v'),\alpha_{X\otimes Y}\circ d^E_{X,Y}(u\otimes v)\rangle^0_{X\otimes Y} } \\
&& = \langle (u'\otimes v'),(d^E_{X,Y})^\dagger\mcirc\alpha_{X\otimes
  Y}\mcirc d^E_{X,Y}(u\otimes v)\rangle^0_{X\otimes Y}
= \langle (u'\otimes v'), (\alpha_X\otimes\alpha_Y)(u\otimes v)\rangle^0_{X\otimes Y}  \\
&& = \langle u',\alpha_X u\rangle^0_X \langle v',\alpha_Y v\rangle^0_Y
= \langle u',u\rangle_X \langle v', v\rangle_Y, \eean thus the
isomorphisms $d^E_{X,Y}:E(X)\otimes E(Y)\rarr E(X\otimes Y)$ are
unitary w.r.t.\ the inner products $\langle\cdot,\cdot\rangle$.

Now, the compatibility (\ref{eq-A2}) of $d^E$ and $e^E$ implies that
$d^E_{\11,\11}\circ e^E1\otimes e^E1=e^E1$ and therefore, using our
choice of the inner product $\langle\cdot,\cdot\rangle_\11^0$,
\begin{eqnarray*} \langle d^E_{\11,\11}(ae^E1\otimes be^E1),d^E_{\11,\11}(ce^E1\otimes de^E1)\rangle^0_{\11\otimes\11}
   = \langle abe^E1,cde^E1\rangle^0_\11 \\
 =\ol{ab}cd 
  = \langle ae^E1,ce^E1\rangle^0_\11\langle
  be^E1,de^E1\rangle^0_\11. \end{eqnarray*}
This means that $d^E_{\11,\11}:E(\11)\otimes E(\11)\rarr E(\11)$ is unitary w.r.t.\ the inner
product $\langle\cdot,\cdot\rangle_\11^0$. Taking $X=Y=\11$ in (\ref{eq-alpha}) and using
$\alpha_\11=\lambda\id_{E(\11)}$, we get $\lambda^2=\lambda$. Since $\alpha_\11$ is invertible, we have
$\lambda=1$, thus $\alpha_\11=\id_{E(\11)}$ and therefore
$\langle\cdot,\cdot\rangle_\11=\langle\cdot,\cdot\rangle_\11^0$. Now,
\[ \langle e^E 1,e^E1\rangle_\11=\langle e^E
1,\alpha_\11e^Eu\rangle^0_\11 =\langle e^E
1,e^E1\rangle^0_\11=1=\langle1,1\rangle_\7C, \] thus
$(e^E)^*e^E=\id_\7C$. By one-dimensionality of the spaces involved, we
also have $e^E(e^E)^*=\id_{E(\11)}$, thus $e^E:\11\rarr E(\11)$ is
unitary w.r.t.\ the inner new products $\langle\cdot,\cdot\rangle$.
\qed


\subsection{Reduction to finitely generated categories}
\label{ss-fingen} \bdefin \index{tensor!category, finitely
  generated}\index{finitely generated!tensor category} An additive
tensor category $\2C$ is finitely generated if there exists an object
$Z\in\2C$ such that every object $X\in\2C$ is a direct summand of some
tensor power $Z^{\otimes n}=\underbrace{Z\otimes\cdots\otimes
  Z}_{n\,\mathrm{factors}},\ n\in\7N$, of $Z$.  \edefin

\blemma \label{lem-ind}
Let $\2C$ be a $TC^*$. Then the finitely generated tensor subcategories of $\2C$ form a directed
system, and $\2C$ is the inductive limit of the latter:
\[ \2C\cong\lim_{\stackrel{\longrightarrow}{\iota\in I}} \2C_i. \]
\elemma

\prf Consider all full tensor subcategories of $\2C$. Since $\2C$ is
essentially small, the equivalence classes of such subcategories form
a set, partially ordered by inclusion.  If $\2C_1,\2C_2\subset\2C$ are
finitely generated, say by the objects $X_1,X_2$, then then the
smallest tensor subcategory containing $\2C_1$ and $\2C_2$ is
generated by $X_1\oplus X_2$, thus we have a directed system. Clearly
there is a full and faithful tensor functor
$\lim_{\stackrel{\longrightarrow}{\iota\in I}}\2C_i\rarr\2C$. Since
every object $X$ is contained in a finitely generated tensor
subcategory (e.g., the one generated by $X$), this functor is
essentially surjective and thus an equivalence of categories, cf.\
\cite{cwm}, in fact of tensor categories, cf.\ \cite{SR}.  \qed

\brem \label{rem-fingen} 1. The reason for considering finitely
generated categories is that the existence problem of fiber functors
for such categories can be approached using powerful purely algebraic
methods. The general case can then be reduced to the finitely
generated one using Lemma \ref{lem-ind}.

2. Note that we don't require the generator $Z$ to be irreducible. Thus if
we a priori only know that $\2C$ is generated by a finite set $Z_1,\ldots,Z_r$ of objects, the
direct sum  $Z=\oplus_i Z_i$ will be a (reducible) generator of $\2C$. This is why only a single
generating object appears in the definition.

3. If $G$ is a compact group, the category $\Rep_fG$ is finitely generated iff $G$ is a Lie group.
(Proof: $\Leftarrow$ is a consequence of the well known representation theory of compact Lie
groups. $\Rightarrow$: It is well known that the finite dimensional representations of $G$
separate the elements of $G$.  Therefore, if $(H,\pi)$ is a generator of $\Rep_fG$, it is clear that
$\pi$ must be faithful. Thus $G$ is isomorphic to a closed subgroup of the compact Lie group
$\2U(H)$, and as such it is a Lie group.)

4. The index set $I$ in Lemma \ref{lem-ind} can be taken countable iff
$\2C$ has countably many isomorphism classes of irreducible objects.
The category $\Rep_fG$, where $G$ is a compact group, has this
property iff $G$ is second countable, equivalently metrizable.  \erem

In Subsections \ref{ss-monoid1}-\ref{ss-monoid2} we will prove the following result, which we take
for granted for the moment:

\btheor \label{theor-D} \index{fiber functor!existence}
A finitely generated even $STC^*$ admits a symmetric  fiber functor $E: \2C\rarr\Vect_\7C$.  
\etheor

\noindent {\it Proof of Theorem \ref{theor-C}}: By Lemma
\ref{lem-ind}, we can represent $\2C$ as an inductive limit
$\lim_{\stackrel{\longrightarrow}{\iota\in I}}\2C_i$ of finitely
generated categories.  Now Theorem \ref{theor-D} provides us with
symmetric fiber functors $E_i:\2C_i\rarr\Vect_\7C,\ i\in I$, and
Theorem \ref{theor-starfunc} turns the latter into $*$-preserving
symmetric fiber functors $E_i:\2C_i\rarr\2H$. By Theorem
\ref{theor-T2}, we obtain compact groups $G_i=\Nat_\otimes E_i$ (in
fact compact Lie groups by Remark \ref{rem-fingen}.3) with
representations $\pi_{i,X}$ on the spaces $E_i(X),X\in\2C_i$ such that
the functors $F_i: \2C_i\rarr\Rep_fG_i,\ X\mapsto(E_i(X),\pi_{i,X}$
are equivalences. Let now $i\le j$, implying that $\2C_i$ is a full
subcategory of $\2C_j$. Then $E_j\restr\2C_i$ is a fiber functor for
$\2C_i$ and thus Theorem \ref{theor-uniq1} implies the existence of a
unitary natural isomorphism $\alpha^{i,j}: F_1\rarr F_2\restr\2C_i$.
(Note that $\alpha^{i,j}$ is not unique!) Now, by definition every
$g\in G_2$ is a family of unitaries $(g_X\in\2U(E_2(X)))_{X\in\2C_2}$
defining a monoidal natural automorphism of $E_2$. Defining, for every
$X\in\2C_1$, $h_X:=\alpha^{i,j}_X\circ g_X\circ(\alpha^{i,j}_X)^*$ we
see that the family $(h_X\in\2U(E_1(X)))_{X\in\2C_1}$ is a unitary
monoidal natural automorphism of $E_1$, to wit an element of $G_1$. In
this way we obtain a map $\beta^{i,j}: G_j\rarr G_i$ that clearly is a
group homomorphism and continuous. By Schur's lemma, the unitary
$\alpha^{i,j}_X$ is unique up to a phase for irreducible $X$. Thus for
such $X$, $\beta^{i,j}_X$ is independent of the chosen $\alpha^{i,j}$,
and thus $\beta^{i,j}$ is uniquely determined. It is also surjective
in view of the Galois correspondence between the full tensor
subcategories of $\Rep_fG$ and the quotients $G/N$, where $N\subset G$
is a closed normal subgroup. Now the inverse limit
\[ G=\lim_{\stackrel{\longleftarrow}{i\in I}} G_i = \{ (g_i\in
G_i)_{i\in I}\ | \ \beta^{i,j}(g_j)=g_i\ \mathrm{whenever}\ i\le j \}
\] is a compact group with obvious surjective homomorphisms
$\gamma_i:G\rarr G_i$ for all $i\in I$.  Now we define a functor $E:
\2C\rarr\Rep_fG$ as follows: For every $X\in\2C$ pick an $i\in I$ such
that $X\in\2C_i$ and define $F(X)=(E_i(X),\pi_i(X)\circ\gamma_i)$.
Clearly this is an object in $\Rep_fG$, and its isomorphism class is
independent of the chosen $i\in I$. In this way we obtain a functor
from $\2C=\lim_\rightarrow\2C_i$ to
$\Rep_fG\cong\lim_\rightarrow\Rep_fG_i$ that restricts to equivalences
$\2C_i\rarr\Rep_fG_i$. Thus $E$ is full and faithful. Finally, $E$ is
essentially surjective since every finite dimensional representation
of $G=\lim_\leftarrow G_i$ factors through one of the groups $G_i$.
\qed

\brem In view of Remark \ref{rem-fingen}.3, the preceding proof also
shows that every compact group is an inverse limit of compact Lie
groups.  \erem


\subsection{Fiber functors from monoids} \label{ss-monoid1}
\index{Deligne} Our strategy to proving Theorem \ref{theor-D} will be
essentially the one of Deligne \cite{del}, replacing however the
algebraic geometry in a symmetric abelian category by fairly
elementary commutative categorical algebra. There are already several
expositions of this proof \cite{bichon,rosenb,phh}, of which we find
\cite{bichon} the most useful, see also \cite{bichon1}. However, we
will give more details than any of these references, and we provide
some further simplifications.

The following result clearly shows the relevance of the notions introduced in Subsection
\ref{ss-comm-alg} to our aim of proving Theorem \ref{theor-D}:

\bprop \label{prop-embed} \index{fiber functor!existence} \index{monoid!absorbing}
Let $\2C$ be a $TC^*$ and $\widehat{\2C}$ be a $\7C$-linear strict tensor category containing $\2C$
as a full tensor subcategory. Let $(Q,m,\eta)$ be a monoid in $\widehat{\2C}$ satisfying
\begin{itemize}
\item[(i)] $\dim\Hom_{\widehat{\2C}}(\11,Q)=1$. (I.e., $\Hom_{\widehat{\2C}}(\11,Q)=\7C\eta$.)
\item[(ii)] For every $X\in\2C$, there is $n(X)\in\7Z_+$ such that $n(X)\ne 0$ whenever 
$X\not\cong 0$ and an isomorphism $\alpha_X:(Q\otimes X,m\otimes\id_X)\rarr n(X)\cdot(Q,m)$ of
$Q$-modules. 
\end{itemize}
Then the functor $E: \ \2C\rarr \Vect_\7C$ defined by 
\[ E: \ \2C\rarr\2H, \quad X\mapsto \Hom_{\widehat{\2C}}(\11,Q\otimes
X), \] together with
\begin{equation} \label{e-Emor}
E(s)\phi=\id_Q\otimes s\mcirc  \phi, \quad s:X\rarr Y,\ \phi\in\Hom(\11,Q\otimes X) 
\end{equation}
is a faithful (strong) tensor functor and satisfies $\dim_\7C E(X)=n(X)$.

If $\widehat{\2C}$ has a symmetry $c$ w.r.t.\ which $(Q,m,\eta)$ is
commutative then $E$ is symmetric monoidal w.r.t.\ the symmetry
$\Sigma$ of $\Vect_\7C$, i.e.\ $E(c_{X,Y})=\Sigma_{E(X),E(Y)}$.
\eprop

\prf We have $E(X) = \Hom(\11,Q\otimes X)\cong\Hom(\11,n(X)Q)\cong d(X)\Hom(\11,Q)\cong \7C^{n(X)}$,
thus $E(X)$ is a vector space of dimension $n(X)$. Since $E(X)\ne 0$ for every non-zero
$X\in\2C$, the functor $E$ is faithful. 

To see that $E$ is monoidal first observe that by (ii) we have
$E(\11)=\Hom(\11,Q)=\7C\eta$. Thus there is a canonical isomorphism
$e: \7C=\11_{\Vect_\7C} \rarr E(\11 )=\Hom(\11,Q)$ defined by
$c\mapsto c\eta$. Next we define morphisms
\[ d^E_{X,Y}: E(X)\otimes E(Y)\rarr E(X\otimes Y), \quad \phi\otimes \psi\mapsto
   m\otimes\id_{X\otimes Y}\mcirc\id_Q\otimes\phi\otimes\id_Y\mcirc\psi. \]
By definition (\ref{e-Emor}) of the map $E(s): E(X)\rarr E(Y)$ it is obvious that the family
$(d^E_{X,Y})$ is natural w.r.t.\ both arguments. The equation
\[ d^E_{X_1\otimes X_2,X_3}\mcirc  d^E_{X_1,X_2}\otimes\id_{E(X_3)}=
   d^E_{X_1, X_2\otimes X_3}\mcirc \id_{E_1}\otimes d^E_{X_2,X_3}  \quad\forall X_1,X_2,X_3\in\2C
\]  
required from a tensor functor is a straightforward consequence of the associativity of $m$. The
verification is left as an exercise.

That $(E, (d_{X,Y}),e)$ satisfies the unit axioms is almost obvious. The first condition follows by 
\[ d_{X,\11}(\id_{E(X)}\otimes
e)\phi=d_{X,\11}(\phi\otimes\eta)=m\otimes\id_X\mcirc\id_Q\otimes\phi\mcirc\eta=\phi,
\] and the second is shown analogously.

So far, we have shown that $E$ is a weak tensor functor for which $e:
\11_\2H\rarr E(\11_\2C)$ is an isomorphism. In order to conclude that
$E$ is a (strong) tensor functor it remains to show that the morphisms
$d^E_{X,Y}$ are isomorphisms. Let $X,Y\in\2C$. We consider the
bilinear map \bean \gamma_{X,Y}: && \Hom_Q(Q,Q\otimes X)\boxtimes
\Hom_Q(Q,Q\otimes Y)\rarr
\Hom_Q(Q,Q\otimes X\otimes Y), \\
&& s\boxtimes t\mapsto s\otimes\id_Y\mcirc t.  \eean (We write
$\boxtimes$ rather than $\otimes_\7C$ for the tensor product of
$\Vect_\7C$ in order to avoid confusion with the tensor product in
$Q-\Mod$.) By 2., we have Q-module morphisms $s_i: Q\rarr Q\otimes X,
s_i': Q\otimes X\rarr Q$ for $i=1,\ldots,n(X)$ satisfying $s_i'\circ
s_j=\delta_{ij}\id_Q$, and $\sum_i s_i\circ s_i'=\id_{Q\otimes X}$,
and similar morphisms $t_i, t_i', \ i=1,\ldots,n(Y)$ for $X$ replaced
by $Y$. Then the $\gamma_{ij}=\gamma_{X,Y}(s_i\otimes t_j)$ are
linearly independent, since they satisfy
$\gamma'_{i'j'}\circ\gamma_{ij}=\delta_{i'i}\delta_{j'j}\id_Q$ with
$\gamma'_{i'j'}=t'_j \mcirc s'_i\otimes\id_Y$. Bijectivity of
$\gamma_{X,Y}$ follows now from the fact that both domain and codomain
of $\gamma_{X,Y}$ have dimension $n(X)n(Y)$.  Appealing to the
isomorphisms $\delta_X: \Hom_Q(Q,Q\otimes X)\mapsto\Hom(\11,Q\otimes
X)$ one easily shows
\[ d^E_{X,Y}= \delta_{X\otimes Y}\mcirc \gamma_{X,Y} \mcirc \delta_X^{-1} \boxtimes \delta_Y^{-1}, \]
which implies that $d^E_{X,Y}$ is an isomorphism for every $X,Y\in\2C$.

We now assume that $\widehat{\2C}$ has a symmetry $c$ and that $(Q,m,\eta)$ is commutative. In order
to show that $E$ is a symmetric tensor functor we must show that 
\[ E(c_{X,Y})\mcirc d^E_{X,Y}=\Sigma_{E(X),E(Y)}\mcirc d^E_{Y,X} \]
for all $X,Y\in\2C$. Let $\phi\in E(X), \psi\in E(Y)$. 

By definition of $E$ we have
\[ (E(c_{X,Y})\mcirc d^E_{X,Y})(\phi\otimes\psi)= \id_Q\otimes c_{X,Y}\mcirc m\otimes\id_{X\otimes Y}
   \mcirc \id_Q\otimes\phi\otimes\id_Y\mcirc \psi \]
\[ = \quad\quad
\begin{tangle}
  \hstep\object{Q}\step[1.5]\object{Y}\step\object{X}\\
  \hstep\id\step[1.5]\hXX\\
  \hh\step[-1]\obj{m}\step\hcd\step\id\step\id\\
  \hh\id\step\id\step\id\step\id\\
  \hh\id\step\frabox{\phi}\step\id\\
  \d\step\dd\\
  \hh\step\frabox{\psi}
\end{tangle}
\quad\quad = \quad\quad
\begin{tangle}
\hstep\object{Q}\step[2.5]\object{Y}\step\object{X}\\
\step[-1]\obj{m}\step\hcd\Step\id\step\id\\
\id\step\X\step\id\\
\frabox{\psi}\Step\frabox{\phi}
\end{tangle}
\quad = \quad
\begin{tangle}
\hstep\object{Q}\step[1.5]\object{Y}\step\object{X}\\
\hh\step[-1]\obj{m}\step\hcd\step\id\step\id\\
\hXX\step\id\step\id\\
\hh\id\step\id\step\id\step\id\\
\hh\id\step\frabox{\psi}\step\id\\
\d\step\dd\\
\hh\step\frabox{\phi}
\end{tangle}
\]
On the other hand,
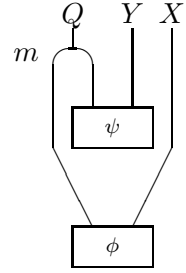
\[ (d^E_{Y,X}\mcirc c_{E(X),E(Y)})(\phi\otimes\psi)=(d^E_{Y,X}\mcirc
  \Sigma_{E(X),E(Y)})(\phi\otimes\psi) 
   =d^E_{Y,X}(\psi\otimes\phi)
=\quad\quad
\begin{tangle}
\hstep\object{Q}\step[1.5]\object{Y}\step\object{X}\\
\hh\step[-1]\obj{m}\step\hcd\step\id\step\id\\
\hh\id\step\id\step\id\step\id\\
\hh\id\step\frabox{\psi}\step\id\\
\d\step\dd\\
\hh\step\frabox{\phi}
\end{tangle}
\]   
If $m$ is commutative, i.e.\ $m=m\circ c_{Q,Q}$, these two expressions coincide, and we are done. 
\qed

\brem 1. The property (ii) in the proposition is called the `absorbing property'.

2. The conditions in Proposition \ref{prop-embed} are in fact necessary for the existence of a fiber
functor! Assume that a tensor $*$-category $\2C$ admits a $*$-preserving fiber functor
$E:\2C\rarr\2H$. By \cite{MRT}, which reviews and extends work of Woronowicz, Yamagami and others,
there is a discrete algebraic quantum group $(A,\Delta)$ such that $\2C\simeq\Rep_f(A,\Delta)$. In
\cite{MT} it is shown that taking 
$\widehat{\2C}\simeq\Rep(A,\Delta)$ (i.e.\ representations of any dimension) and $Q=\pi_l$, there is
a monoid $(Q,m,\eta)$ satisfying the conditions of Proposition \ref{prop-embed}. Namely, one can
take $Q=\pi_l$, the left regular representation. In \cite{MT} it shown that (i)
$\dim\Hom(\pi_0,\pi_l)=1$, i.e.\ there exists a non-zero morphism $\eta:\pi_0\rarr\pi_l$, unique  up
to normalization; (ii) $\pi_l$  has the required absorbing property; (iii) there exists a morphism
$m:\pi_l\otimes\pi_l\rarr\pi_l$ such  that $(Q=\pi_l,m,\eta)$ is a monoid.

3. In the previous situation, the left regular representation $\pi_l$ lives in $\Rep_f(A,\Delta)$
iff $A$ is finite dimensional. This already suggests that the category $\2C$ in general is too small
to contain a monoid of the desired properties. In fact, assume we can take $\widehat{\2C}=\2C$. Then
for every irreducible $X\in\2C$ we 
have $\dim\Hom(X,Q)=\dim\Hom(\11,Q\otimes\ol{X})=n(\ol{X})>0$. Thus $Q$ contains all irreducible
objects as direct summands. Since every object in $\2C$ is a finite direct sum of simple objects,
$\widehat{\2C}=\2C$ is possible only if $\2C$ has only finitely many isomorphism classes of simple
objects. In fact, even in this case, our construction of $(Q,m,\eta)$ will require the use of a
bigger category $\widehat{\2C}$. It is here that the category $\mathrm{Ind}\,\2C$ of Subsection
\ref{ss-ind} comes into play.
\erem

Since we have already reduced the problem of constructing a fiber functor to the case of finitely
generated tensor categories, we want a version of the preceding result adapted to that situation:

\bcoro \label{coro-monoid}
Let $\2C$ be a $TC^*$ with monoidal generator $Z\in\2C$ and let $\widehat{\2C}$ be a $\7C$-linear
strict tensor category containing $\2C$ as a full tensor subcategory. If $(Q,m,\eta)$ is a monoid in
$\widehat{\2C}$ satisfying
\begin{itemize}
\item[(i)] $\dim\Hom_{\widehat{\2C}}(\11,Q)=1$.
\item[(ii)] There is $d\in\7N$ and an isomorphism 
$\alpha_Z:(Q\otimes Z,m\otimes\id_Z)\rarr d\cdot(Q,m)$ of $Q$-modules.
\end{itemize}
Then the hypothesis (ii) in Proposition \ref{prop-embed} follows. Thus 
$E: X\mapsto\Hom_{\widehat{\2C}}(\11,Q\otimes X)$ is a fiber functor.
\ecoro

\prf If $X\in\2C$, there exists $n\in\7N$ such that $X\prec Z^{\otimes n}$. Concretely, there are
morphisms $u:X\rarr Z^{\otimes n}$ and $v:Z^{\otimes n}\rarr X$ such that $v\circ u=\id_X$.
Then the morphisms $\tilde{u}=\id_Q\otimes u: Q\otimes X\rarr Q\otimes Z^{\otimes n}$ and
$\tilde{v}=\id_Q\otimes v:Q\otimes Z^{\otimes n}\rarr Q\otimes X$ are morphisms of $Q$-modules.
Thus the $Q$-module $(Q\otimes X,m\otimes\id_X)$ is a direct summand of 
$(Q\otimes Z^{\otimes n},m\otimes\id_{Z^{\otimes n}})$. By assumption, the latter is isomorphic to a
direct sum of $d^n$ copies of $(Q,m)$. By Lemma \ref{lem-end} and assumption (i),
$\End_Q((Q,m))\cong\7C$, thus $(Q,m)\in Q-\Mod$ is irreducible. Thus the direct summand 
$(Q\otimes X,m\otimes\id_X)$ of $d^n\cdot(Q,m)$ is a direct sum of $r$ copies of $(Q,m)$ with 
$r\le d^m$ and $r\ne 0$ whenever $X\ne 0$. Thus hypothesis (ii) in Proposition \ref{prop-embed} holds.
\qed

In view of Corollary \ref{coro-monoid}, proving Theorem \ref{theor-D} amounts to finding a symmetric
tensor category $\widehat{\2C}$ containing $\2C$ as a full subcategory and a commutative monoid
$(Q,m,\eta)$ in $\widehat{\2C}$ such that $\dim\Hom(\11,Q)=1$ and $Q\otimes Z\cong d\otimes Q$ as
$Q$-modules for a suitable monoidal generator $Z$ of $\2C$. This will be achieved in Subsection
\ref{ss-monoid2}, based on thorough analysis of the permutation symmetry of the category $\2C$.


\subsection{Symmetric group action, determinants and
  integrality of dimensions} \label{saggy}
We now turn to a discussion of certain representations of the symmetric groups $P_n, n\in\7N$,
present in tensor $*$-categories with a unitary symmetry. It is well known that the symmetric group
$P_n$ on $n$ labels has the presentation 
\[ \begin{aligned} P_n=(\sigma_1,\ldots,\sigma_{n-1} \
  | & \ |i-j|\ge 2\impl\sigma_i\sigma_j =
  \sigma_j\sigma_i, \\ & \ \ 
   \sigma_i\sigma_{i+1}\sigma_i=\sigma_{i+1}\sigma_i\sigma_{i+ 1} \ \forall i\in\{1,\ldots,n-1\},
   \ \ \sigma_i^2=1 \ \forall i). \end{aligned} \]

   Since $\2C$ is strict we may define the tensor powers $X^{\otimes
     n},\ n\in\7N$, in the obvious way for any $X\in\2C$. We posit
   $X^{\otimes 0}=\11$ for every $X\in\2C$.

\blemma Let $\2C$ be an $STC^*$. Let $X\in\2C$ and $n\in\7N$. Then 
\[ \Pi^X_n: \ \sigma_i \ \mapsto \ \id_{X^{\otimes i-1}}\otimes c_{X,X}\otimes\id_{X^{\otimes
  n-i-1}} \] 
uniquely determines a homomorphism $\Pi^X_n$ from the group $P_n$ into the unitary group of
$\End\,X^{\otimes n}$. 
\elemma
\prf It is clear that $\Pi^X_n(\sigma_i)$ and $\Pi^X_n(\sigma_j)$ commute if $|i-j|\ge 2$. 
That $\Pi^X_n(\sigma_i)^2=\id_{X^{\otimes n}}$ is equally obvious. Finally, 
\[ \Pi^X_n(\sigma_i)\mcirc \Pi^X_n(\sigma_{i+ 1})\mcirc \Pi^X_n(\sigma_i)=
  \Pi^X_n(\sigma_{i+ 1})\mcirc \Pi^X_n(\sigma_i)\mcirc \Pi^X_n(\sigma_{i+ 1})  \]
follows from the Yang-Baxter equation satisfied by the symmetry $c$.
\qed

\brem 
Dropping the relations $\sigma_i^2=1$ the same formulae as above define homomorphisms of the Artin
braid groups $B_n$ into $\End\,X^{\otimes n}$. However, none of the following considerations has
known analogues in the braided case. 
\erem

Recall that there is a homomorphism $\mathrm{sgn}: P_n\rarr\{1,-1\}$, the signature map.

\blemma \label{l-proj} 
Let $\2C$ be an $STC^*$. For any $X\in\2C$ we define orthogonal projections in 
$\End\,X^{\otimes 0}=\End\,\11$ by $S_0^X=A_0^X=\id_\11$. For any $n\in\7N$, the morphisms 
\begin{eqnarray*} 
S_n^X &=& \frac{1}{n!} \sum_{\sigma\in P_n} \Pi_n^X(\sigma), \\
A_n^X &=& \frac{1}{n!} \sum_{\sigma\in P_n} \mathrm{sgn}(\sigma) \,\Pi_n^X(\sigma)
\end{eqnarray*}
satisfy 
\[ \Pi_n^X(\sigma) \mcirc  S_n^X = S_n^X\mcirc  \Pi_n^X(\sigma) = S_n^X, \]
\[ \Pi_n^X(\sigma) \mcirc  A_n^X = A_n^X\mcirc  \Pi_n^X(\sigma) =  \mathrm{sgn}(\sigma)\,A_n^X \]
for all $\sigma\in P_n$ and are thus orthogonal projections in the $*$-algebra $\End\,X^{\otimes n}$. 
\elemma

\prf Straightforward computations. \qed

\bdefin 
The subobjects (defined up to isomorphism) of $X^{\otimes n}$ corresponding to the idempotents
$S_n^X$ and $A_n^X$ are denoted by $S_n(X)$ and $A_n(X)$, respectively.
\edefin

The following was proven both in \cite{DR} and \cite{del}:

\bprop \label{p-trace-an} 
Let $\2C$ be an even $STC^*$. For any $X\in\2C$ we have
\begin{equation} \label{e-traxn}
Tr_{X^{\otimes n}}\,A_n^X = \frac{d(X)(d(X)-1)(d(X)-2)\cdots (d(X)-n+1)}{n!} \quad \forall n\in\7N.
\end{equation}
\eprop

\prf {\it (Sketch)} Making crucial use of the fact that $\2C$ is even,
i.e.\ $\Theta(X)=\id_X$ for all $X\in\2C$, one can prove
\[   Tr_{X^{\otimes n}}\,\Pi_n^X(\sigma)= d(X)^{\#\sigma} \quad\forall X\in\2C, \sigma\in P_n, \]
where $\#\sigma$ is the number of cycles into which the permutation $\sigma$ decomposes. (The reader
familiar with tangle diagrams will find this formula almost obvious: Triviality of the twist
$\Theta(X)$ implies invariance under the first Reidemeister move. Thus the closure of the
permutation $\sigma$ is equivalent to $\#\sigma$ circles, each of which contributes a factor $d(X)$.)
Now the result follows at once from the definition of $A_n^X$ and the formula
\[  \sum_{\sigma\in P_n} \mathrm{sgn}(\sigma)\,z^{\#\sigma}=z(z-1)(z-2)\cdots (z-n+1), \]
which holds for all $n\in\7N$ and $z\in\7C$, as one can prove by induction over $n$.
\qed

\bcoro \label{c-int} \index{dimension!integrality of}
In an $STC^*$ we have $d(X)\in\7N$ for every non-zero $X\in\2C$.
\ecoro 

\prf Assume first that $\2C$ is even, and let $X\in\2C$. Since $\2C$ has subobjects there exist an
object $A_n(X)\in\2C$ and a morphism $s: A_n(X)\rarr X^{\otimes n}$ such that 
$s^*\mcirc  s=\id_{A_n(X)}$ and $s\mcirc  s^*=A^X_n$. Then by part 1 and 2 in Proposition
\ref{prop-trace}, we get 
\[ Tr_{X^{\otimes n}}\,A_n^X = Tr_{X^{\otimes n}}(s\mcirc s^*) =
Tr_{A_n(X)}(s^*\mcirc s) =Tr_{A_n(X)}\,\id_{A_n(X)} = d({A_n(X)}). \]
Since the dimension of any object in a $*$-category is non-negative we
thus conclude that $Tr_{X^{\otimes n}}\,A_n^X\ge 0$ for all $n\in\7N$.
From the right-hand side in the formula (\ref{e-traxn}) for
$Tr_{X^{\otimes n}}\,A_n^X$ we see that $Tr_{X^{\otimes n}}\,A_n^X$
will become negative for some $n\in\7N$ unless $d(X)\in\7N$.

If $\2C$ is odd, the above argument gives integrality of the dimensions in the bosonized category
$\widetilde{\2C}$. Since the categorical dimension is independent of the braiding, we have
$d_\2C(X)=d_{\widetilde{\2C}}(X)$ and are done.
\qed


Let $\2C$ be an $STC^*$ and $X\in\2C$ non-zero and set $d=d(X)\in\7N$. Consider the subobject 
$A_d(X)$ of $X^{\otimes d}$, introduced in the proof of Corollary \ref{c-int}, which corresponds to
the orthogonal projection $A_d^X\in\End\,X^{\otimes d}$ defined in Lemma \ref{l-proj}. Then
\[ d(A_d (X))=Tr_{X^{\otimes d}}\,A_d^X=\frac{d!}{d!}=1 ,\]
we see that $A_d (X)$ is an irreducible and invertible object of $\2C$ (with inverse $\overline{A_d (X)}$).

\bdefin \index{determinant}
The isomorphism class of $A^{d(X)}(X)$ is called the {\it determinant $\det(X)$ of $X$}. 
\edefin

\blemma \label{l-dets}
Let $\2C$ be an $STC^*$ and $X,Y\in\2C$. Then 
\begin{itemize}
\item[(i)] $\det(\ol{X})\cong \ol{\det(X)}$.
\item[(ii)] $\det(X\oplus Y)\cong \det(X)\otimes \det(Y)$.
\item[(iii)] $\det(X\oplus \ol{X})\cong\11$.
\end{itemize}
\elemma

\prf (i) Let $(\ol{X},r,\ol{r})$ be a standard left inverse of $X$. By inductive use of Lemma
\ref{lem-mult} one obtains standard left inverses $(\ol{X}^{\otimes n},r_n,\ol{r}_n)$ of
$X^{\otimes n}$ for any $n\in\7N$. If now $\sigma=\sigma_{i_1}\cdots\sigma_{i_r}\in P_n$, one can
verify that
\[ \Pi_n^{\ol{X}}(\sigma')=r_n^*\otimes\id_{\ol{X}^{\otimes n}}
  \mcirc\id_{\ol{X}^{\otimes n}}\otimes\Pi_n^X(\sigma)\otimes\id_{\ol{X}^{\otimes n}}
  \mcirc\id_{\ol{X}^{\otimes n}}\otimes\ol{r}_n, \]
where $\sigma'=\sigma_{n-i_r}^{-1}\cdots\sigma_{n-i_1}^{-1}$. In particular,
$\mathrm{sgn}\,\sigma'=\mathrm{sgn}\,\sigma$, implying
\[ A_n^{\ol{X}}=r_n^*\otimes\id_{\ol{X}^{\otimes n}}
  \mcirc\id_{\ol{X}^{\otimes n}}\otimes A_n^X\otimes\id_{\ol{X}^{\otimes n}}
  \mcirc\id_{\ol{X}^{\otimes n}}\otimes\ol{r}_n, \]
for any $n\in\7N$. Now the claim follows from Lemma \ref{lem-conj}.

(ii) For any $X\in\2C$ we abbreviate $d_X=d(X)$ and $A^X=A_{d_X}^X\in\End\,X^{\otimes d_X}$. Let 
$u:X\rarr Z, v:Y\rarr Z$ be isometries implementing $Z\cong X\oplus Y$. Then $X^{\otimes d_X}$ is a
subobject of $Z^{\otimes d_X}$, and similarly for $Y^{\otimes d_Y}$. By definition, $\det(Z)$ is
the subobject of $Z^{\otimes d_Z}$ corresponding to the projector 
$A^Z\in\End\,Z^{\otimes d_Z}$. On the other hand, $\det(X)\otimes \det(Y)$ is the subobject of
$X^{\otimes d_X}\otimes Y^{\otimes d_Y}$ corresponding to the projector $A^X\otimes A^Y$, and
therefore it is isomorphic to the subobject of  $Z^{\otimes d_Z}$ corresponding to the projector
\[ u\otimes\cdots\otimes u\otimes v\otimes\cdots\otimes v\mcirc A^X\otimes A^Y\mcirc
u^*\otimes\cdots\otimes u^*\otimes v^*\otimes\cdots\otimes v^*\ \in\End\,Z^{\otimes d_Z}, \] 
where there are $d_X$ factors $u$ and $u^*$ and $d_Y$ factors $v$ and $v^*$. This equals
\[ \frac{1}{d_X!d_Y!}\sum_{\sigma\in P_{d_X}\atop\sigma'\in P_{d_Y}} \mathrm{sgn}(\sigma)\mathrm{sgn}(\sigma')\,
   u\otimes\cdots\otimes u\otimes v\otimes\cdots\otimes v\mcirc \Pi_{d_X}^X(\sigma)\otimes \Pi_{d_Y}^Y(\sigma')
  \mcirc u^*\otimes\cdots\otimes u^*\otimes v^*\otimes\cdots\otimes v^*
\] 
By naturality of the braiding, this equals
\[ \frac{1}{d_X!d_Y!}\sum_{\sigma\in P_{d_X}\atop\sigma'\in P_{d_Y}}
   \mathrm{sgn}(\sigma)\mathrm{sgn}(\sigma')\,    \Pi_{d_X}^Z(\sigma)\otimes \Pi_{d_Y}^Z(\sigma') 
  \mcirc p_X\otimes\cdots\otimes p_X\otimes p_Y\otimes\cdots\otimes p_Y, \]
where $p_X=u\circ u^*, p_Y=v\circ v^*$. With the juxtaposition 
$\sigma\times\sigma'\in P_{d_X+d_Y}=P_{d_Z}$  of $\sigma$ and $\sigma'$ this becomes 
\be \label{eq-antisymm}
 \frac{1}{d_X!d_Y!}\sum_{\sigma\in P_{d_X}\atop\sigma'\in P_{d_Y}} \mathrm{sgn}(\sigma\times\sigma')\,
   \Pi_{d_Z}^Z(\sigma\times\sigma') \mcirc p_X\otimes\cdots\otimes p_X\otimes p_Y\otimes\cdots\otimes p_Y, \ee
On the other hand,
\[ A^Z=\frac{1}{d_Z!}\sum_{\sigma\in
  P_{d_Z}}\mathrm{sgn}(\sigma)\,\Pi_{d_Z}^Z(\sigma)
=\left(\sum_{\sigma\in
    P_{d_Z}}\mathrm{sgn}(\sigma)\,\Pi_{d_Z}^Z(\sigma)\right) \mcirc
(p_X+p_Y)\otimes\cdots\otimes(p_X+p_Y). \] Of the $2^{d_Z}$ terms into
which this can be decomposed, only those with $d_X$ factors $p_X$ and
$d_Y$ factors $p_Y$ are nonzero since $A^X_n=0$ for $n>d_X$ and
$A^Y_n=0$ for $n>d_Y$. We are thus left with a sum of $d_Z!/d_X!d_Y!$
terms, and working out the signs we see that they all equal to
$d_X!d_Y!/d_Z!$ times (\ref{eq-antisymm}), thus the sum equals
(\ref{eq-antisymm}). This proves the isomorphism
$\det(Z)\cong\det(X)\otimes\det(Y)$.

Finally, (iii) follows from
\[ \det(X\oplus \ol{X})\cong\det X\otimes\det\ol{X}\cong\det X\otimes\ol{\det X}\cong\det
  X\otimes(\det X)^{-1}\cong\11, \]
where we have used (i) and (ii) of this lemma, $d(\det\,X)=1$ and (iii) of Lemma \ref{lem-dim}.
\qed

For later use we state a computational result:

\blemma \label{l-conj}
Let $X$ satisfy $\det X\cong\11$ and write $d=d(X)$. If $s: \11\rarr X^{\otimes d}$ is an isometry
for which $s\circ  s^*=A_d^X$ then 
\begin{equation} \label{e-cc}
  s^*\otimes\id_X\mcirc\id_X\otimes s=(-1)^{d-1}d^{-1}\,\id_X. 
\end{equation}
\elemma
\prf We abbreviate $x=s^*\otimes\id_X\mcirc\id_X\otimes s$ and observe that by non-degeneracy of
the trace it is sufficient to show that
$Tr_X(ax)=(-1)^{d-1}d^{-1}Tr_X(a)$ for all $a\in\End\,X$. In order to show this, let
$(\ol{X},r,\ol{r})$ be a standard solution of the conjugate equations and compute
\begin{eqnarray*} \lefteqn{ Tr_X(ax)= } \\
&& 
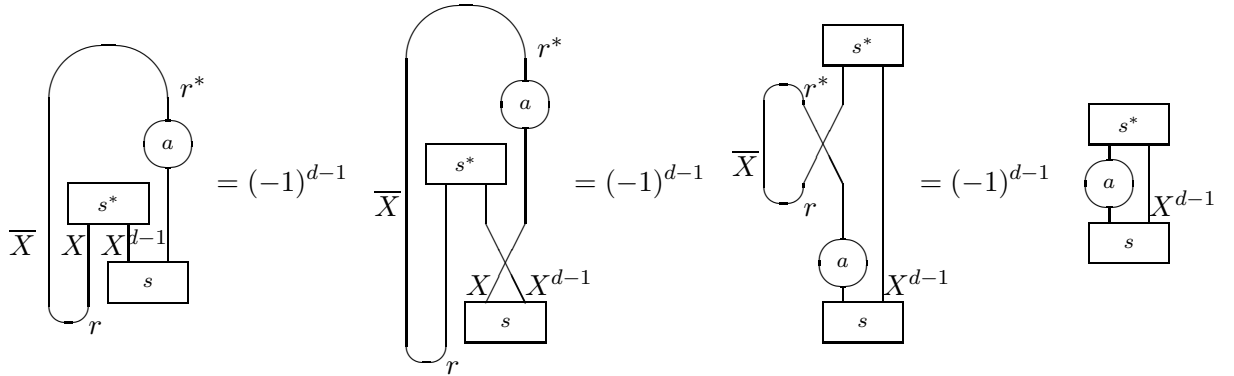
\begin{tangle}
\mcoev\step[.3]\obj{r^*}\\
\id\step[3]\O a\\
\hh\id\step\frabox{s^*}\step\id\\
\step[-1]\obj{\ol{X}}\step\hh\id\step[.3]\obj{X}\step[.7]\id\step[.3]\obj{X^{d-1}}\step[.7]\id\step\id\\
\hh\id\step\id\step\frabox{s}\\
\hh\ev\obj{r}
\end{tangle}
\quad = (-1)^{d-1}\quad\quad
\begin{tangle}
\mcoev\step[.3]\obj{r^*}\\
\id\step[3]\O a\\
\hh\id\step\frabox{s^*}\step\id\\
\step[-.8]\obj{\ol{X}}\step[.8]\hh\id\step\id\step\id\step\id\\
\id\step\id\hstep\obj{X}\hstep\hXX\obj{X^{d-1}}\\
\hh\id\step\id\step\frabox{s}\\
\hh\ev\obj{r}
\end{tangle}
\quad = (-1)^{d-1}\quad\quad
\begin{tangle}
\hh\Step\frabox{s^*}\\
\hh\hcoev\obj{r^*}\step\id\step\id\\
\step[-.8]\obj{\ol{X}}\step[.8]\id\step\hXX\step\id\\
\hh\hev\obj{r}\step\id\step\id\\
\Step\O a\step\id\obj{X^{d-1}}\\
\Step\hh\frabox{s}
\end{tangle}
\quad = (-1)^{d-1}\quad\quad
\begin{tangle}
\hh\frabox{s^*}\\
\O a\step\id\obj{X^{d-1}}\\
\hh\frabox{s}
\end{tangle}
\end{eqnarray*}
We have in turn used the total antisymmetry of $s$ (Lemma \ref{l-proj}), the naturality properties 
of the braiding and the triviality of the twist $\Theta_X$. Now, 
\begin{eqnarray*} \lefteqn{ s^*\circ a\otimes\id_{X^{\otimes d-1}}\circ s
  = Tr_\11(s^*\circ a\otimes\id_{X^{\otimes d-1}}\circ
  s) } && \\
  &=& Tr_{X^{\otimes d}}(a\otimes\id_{X^{\otimes
      d-1}}\circ s\circ s^*)   
  = Tr_{X^{\otimes d}}(a\otimes\id_{X^{\otimes d-1}}\circ A^X_d).
\end{eqnarray*}
In order to complete the proof we need to show that this equals $d^{-1}Tr_Xa$, which is done by
suitably modifying the proof of Proposition \ref{p-trace-an}. By the same argument as given there,
it suffices to prove 
$Tr_{X^{\otimes d}}(a\otimes\id_{X^{\otimes d-1}}\mcirc \Pi^X_d(\sigma))=d^{\#\sigma -1}Tr_Xa$.
Again, the permutation $\sigma$ decomposes into a set of cyclic permutations, of which now precisely
one involves the index $1$. It is therefore sufficient to prove
$Tr_{X^{\otimes n}}(a\otimes\id_{X^{\otimes n-1}}\mcirc \Pi^X_n(\sigma))=Tr_Xa$ for every cyclic
permutation $\sigma$ of all $n$ indices. Inserting $a$ at the appropriate place, the calculation
essentially proceeds as before. The only difference is that instead of $Tr_X\id_X=d(X)$ one is left
with $Tr_Xa$, giving rise to the desired result.
\qed

\brem Objects with determinant $\11$ were called special in \cite{DR}, where also all results of
this subsection can be found. 
\erem


This concludes our discussion of antisymmetrization  and determinants, and we turn to
symmetrization and the symmetric algebra. It is here that we need the Ind-category that was
introduced in Subsection \ref{ss-ind}.


\subsection{The symmetric algebra} \label{ss-symm}
In ``ordinary'' algebra one defines the symmetric algebra $S(V)$ over a vector space $V$. Unless
$V=\{0\}$, this is an infinite direct sum of non-trivial vector spaces. We will need a
generalization of this construction to symmetric tensor categories other than $\Vect$. While
infinite direct sums of objects make sense in the setting of $C^*$-tensor categories (Definition
\ref{def-cstar}), a more convenient setting for the following considerations is given by the theory
of abelian categories. 

\blemma \label{lem-mij}
Let $\2C$ be an $STC^*$ and $X\in\2C$. For every $n\in\7N$ choose an object $S_n(X)$ and an isometry  
$u_n:S_n(X)\rarr X^{\otimes n}$ such that $u_n\circ u_n^*=S_n^X$. Also, let $u_0=\id_\11$,
interpreted as a morphism from $S_0(X)=\11$ to $X^0=\11$. The the morphisms 
$m_{i,j}: S_i(X)\otimes S_j(X)\rarr S_{i+j}(X)$ defined by 
\[ \begin{diagram} 
  m_{i,j}: S_i(X)\otimes S_j(X) & \rTo^{u_i\otimes u_j}& X^{\otimes i}\otimes X^{\otimes j} \equiv
     X^{\otimes(i+j)} &\rTo^{u_{i+j}^*} & S_{i+j}(X) \end{diagram} \]
satisfy
\[ m_{i+j,k}\circ m_{i,j}\otimes\id_{S_k(X)}=m_{i,j+k}\circ \id_{S_i(X)}\otimes m_{j,k} \]
for all $i,j,k\in\7Z_+$. Furthermore, 
\[ m_{i,j}=m_{j,i}\circ c_{S_i(X),S_j(X)} \quad \forall i,j \]
and $m_{i,0}=m_{0,i}=\id_{S_i(X)}$.
\elemma

\prf As a consequence of $S_n^X\circ\Pi_n^X(\sigma)=S_n^X(\sigma)$ for all $\sigma\in P_n$, cf.\
Lemma \ref{l-proj}, we have
\[ S_{i+j+k}^X\mcirc S_{i+j}^X\otimes \id_{X^{\otimes k}}\mcirc S_i^X\otimes S_j^X\otimes\id_{X^{\otimes k}}
   =S_{i+j+k}^X\mcirc S_{i+j}^X\otimes \id_{X^{\otimes k}}= S_{i+j+k}^X, \]
\[ S_{i+j+k}^X\mcirc\id_{X^{\otimes k}}\otimes S_{j+k}^X \mcirc \id_{X^{\otimes k}}\otimes S_j^X\otimes S_k^X
  = S_{i+j+k}^X\mcirc\id_{X^{\otimes i}}\otimes S_{j+k}^X = S_{i+j+k}^X. \]
Multiplying all this with $u_{i+j+k}^*$ on the left and with $u_i\otimes u_j\otimes u_k$ on the
right and using $u_i^*\circ S_i^X=u_n^*$ and $S_i^X\circ u_i=u_i$ this implies
\[ u_{i+j+k}^*\mcirc S_{i+j}^X\otimes \id_{X^{\otimes k}}\mcirc u_i\otimes u_j\otimes u_k
 = u_{i+j+k}^*\mcirc u_i\otimes u_j\otimes u_k 
 = u_{i+j+k}^*\mcirc\id_{X^{\otimes k}}\otimes S_{j+k}^X \mcirc u_i\otimes u_j\otimes u_k \]
Using again that $S_{i+j}^X=u_{i+j}\circ u_{i+j}^*$, we have the first identity we wanted to prove.
Furthermore,
\bean \lefteqn{ m_{j,i}\mcirc c_{S_i(X),S_j(X)} = u_{i+j}^*\circ u_j\otimes u_i\mcirc c_{S_i(X),S_j(X)}
  = u_{i+j}^*\mcirc c_{X^{\otimes i},X^{\otimes j}}\mcirc u_i\otimes u_j } \\ 
  &&= u_{i+j}^*\mcirc \Pi_{i+j}^X(\sigma)\mcirc u_i\otimes u_j 
   =u_{i+j}^*\mcirc S_{i+j}^X\mcirc\Pi_{i+j}^X(\sigma)\mcirc u_i\otimes u_j 
   =u_{i+j}^*\mcirc S_{i+j}^X\mcirc u_i\otimes u_j \\
  &&= u_{i+j}^*\mcirc u_i\otimes u_j =m_{i,j},
\eean
where $\sigma\in P_{i+j}$ is the permutation exchanging the first $i$ with the remaining $j$
strands. The last claim is obvious in view of $S_0(X)=\11$.
\qed

In view of Lemma \ref{lem-abel}, $\2C$ (with a zero object thrown in) is an abelian category, thus
there exists an abelian $\7C$-linear strict symmetric tensor category $\mathrm{Ind}\,\2C$ containing
$\2C$ as a full subcategory and complete w.r.t.\ filtered inductive limits. Therefore, for any
object $X$ in the $STC^*\ \2C$, there exists an object 
\[ S(X)=\lim_{\stackrel{\longrightarrow}{n\rarr\infty}} \bigoplus_{i=0}^n S_n(X) \]
together with monomorphisms $v_n: S_n(X)\rarr S(X)$.

\bprop \index{symmetric algebra}
Let $\2C$ be an $STC^*$ and $X\in\2C$. Then there exists a morphism
$m_{S(X)}: S(X)\otimes S(X)\rarr S(X)$ such that 
\[ m_{S(X)}\circ v_i\otimes v_j = v_{i+j}\circ m_{i,j}: \ S_i(X)\otimes S_j(X)\rarr S(X) \]
and $(S(X),m_{S(X)},\eta_{S(X)}\equiv v_0)$ is a commutative monoid in $\mathrm{Ind}\,\2C$.
\eprop

\prf
This amounts to using
\[ \Hom_{\mathrm{Ind}\,\2C} (S(X)\otimes S(X),S(X))
   =\lim_{\longleftarrow\atop m}\lim_{\longrightarrow\atop n}
   \Hom_\2C\left(\bigoplus_{i,j=0}^m S_i(X)\otimes S_j(X),\bigoplus_{k=0}^n S_k(X)\right) \]
to assemble the morphisms $m_{i,j}: S_i(X)\otimes S_j(X)\rarr S_{i+j}(X)$ into one big 
morphism $S(X)\otimes S(X)\rarr S(X)$. 
We omit the tedious but straightforward details. Associativity
($m_{S(X)}\circ m_{S(X)}\otimes\id_{S(X)}=m_{S(X)}\circ\id_{S(X)}\otimes m_{S(X)}$) and
commutativity ($m_{S(X)}=m_{S(X)}\circ c_{S(X),S(X)}$) then follow  from the respective properties
of the $m_{i,j}$ established in Lemma \ref{lem-mij}. The unit property 
$m_{S(X)}\circ\id_{S(X)}\otimes v_0=\id_{S(X)}\otimes v_0=\id_{S(X)}$ follows from
$m_{i,0}=m_{0,i}=\id_{S_i(X)}$. 
\qed


We now study the interaction between the operations of symmetrization and antisymmetrization, i.e.\
between determinants and symmetric algebras, that lies at the core of the embedding theorem. We begin 
by noting that given two commutative monoids $(Q_i,m_i,\eta_i),\ i=1,2$ in a strict symmetric tensor
category, the triple $(Q_1\otimes Q_2,m_{Q_1\otimes   Q_2}, \eta_{Q_1\otimes Q_2})$, where 
$\eta_{Q_1\otimes Q_2}=\eta_1\otimes\eta_2$ and
\[ m_{Q_1\otimes Q_2}=m_1\otimes m_2\mcirc \id_{Q_1}\otimes c_{Q_2,Q_1}\otimes\id_{Q_2}, \]
defines a commutative monoid, the direct product $(Q_1,m_1,\eta_1)\times(Q_2,m_2,\eta_2)$.
The direct product $\times$ is strictly associative, thus multiple direct products are unambiguously 
defined by induction.

\blemma \label{lem-det}
Let $\2C$ be a STC and assume $Z\in\2C$ satisfies $\det Z\cong\11$. We write $d=d(Z)$ and pick 
$s: \11\rarr Z^{\otimes d},\ s': Z^{\otimes d}\rarr\11$ such that $s'\circ s=\id_\11$ and
$s\circ s'=A_d^Z$. Let $S(Z)$ be the symmetric tensor algebra over $Z$ with the canonical 
embeddings $v_0:\11\rarr S(Z), v_1: Z\rarr S(Z)$.   
Consider the commutative monoid structure on $Q=S(Z)^{\otimes d}$ given by 
\[ (Q,m_Q,\eta_Q) = (S(Z),m_{S(Z)},\eta_{S(Z)})^{\times d}. \]
Define morphisms $f: \11\rarr Q$ and 
$u_i: Z\rarr Q,\ \ t_i: Z^{\otimes (d-1)}\rarr Q,\ \ i=1,\ldots,d$ by
\[ f= \underbrace{v_1\otimes \ldots\otimes v_1}_{d\ {\scriptstyle\mbox{factors}}}\ \circ\,s, \]
\[ u_i=\underbrace{v_0\otimes\ldots\otimes v_0}_{i-1\ {\scriptstyle \mbox{factors}}}\,\otimes\, v_1
   \otimes\,\underbrace{v_0\otimes\ldots\otimes v_0}_{d-i\ {\scriptstyle \mbox{factors}}}\ ,
\]
\[ t_i= (-1)^{d-i}\, \underbrace{v_1\otimes\ldots\otimes v_1}_{i-1\ {\scriptstyle \mbox{factors}}}  
  \,\otimes\, v_0 \otimes\,\underbrace{v_1\otimes\ldots\otimes v_1}_{d-i\ 
   {\scriptstyle \mbox{factors}}}. 
\]
Then $s,f,u_i,t_j$ satisfy
\begin{equation} \label{e-tu}
   m_Q\mcirc t_j\otimes u_i\mcirc s=\delta_{ij}\,f \quad\quad\forall i,j\in\{1,\ldots,d\}.
\end{equation}
\elemma

\prf First note that $s:\11\rarr Z^{\otimes d}$ as required exists since $\det Z\cong\11$ and that 
$f$ is a composition of monics, thus non-zero. We compute
\begin{eqnarray*} m_Q\mcirc t_i\otimes u_i\mcirc s &=&
   (-1)^{d-i}\ \id_{S(Z)^{(i-1)}} \otimes c_{S(Z)^{\otimes(d-i)},S(Z)} \ \circ\ 
     v_1\otimes v_1\otimes\cdots\otimes v_1\ \circ\ s  \\
 &=& (-1)^{d-i}\  v_1\otimes v_1\otimes\cdots\otimes v_1\ \circ\
    \id_{Z^{\otimes(i-1)}}\otimes c_{Z^{\otimes(d-i)},Z}\ \circ\ s\\ 
 &=& v_1\otimes v_1\otimes\cdots\otimes v_1\ \circ\ s\\
 &=& f.
\end{eqnarray*}
In the first equality we used the definition of $(Q,m_Q,\eta_Q)$ as $d$-fold direct product of
$(S(Z),m_{S(Z)},\eta_{S(Z)})$ and the fact that $v_0=\eta_{S(Z)}$ is the unit, naturality of the
braiding in the second and Lemma \ref{l-proj} in the third. To see that 
$m_Q\mcirc t_j\otimes u_i\mcirc s=0$ if $i\ne j$ consider $j=d-1, i=d$. Then 
$m_Q\mcirc t_j\otimes u_i\mcirc s$ is the composite
\[\begin{diagram}
\11 & \rTo^{s} & Z^{\otimes d} & \rTo^{ \overbrace{v_1\otimes\cdots\otimes v_1}^{d-2\ {\scriptstyle
  \mbox{factors}}} \otimes v_0\otimes v_1\otimes v_1} & S(Z)^{\otimes (d+1)} &  
 \rTo^{\id_{S(Z)^{\otimes(d-1)}}\otimes m_{S(Z)}} & S(Z)^{\otimes d} \equiv Q.
\end{diagram}\]
Now,
\begin{eqnarray*} \lefteqn{
   \id_{S(Z)^{\otimes(d-1)}}\otimes m_{S(Z)}\ \circ\ v_1\otimes\cdots\otimes v_1\otimes v_0\otimes 
   v_1\otimes v_1 \ \circ\ s } \\
  && = \id_{S(Z)^{\otimes(d-1)}}\otimes (m_{S(Z)}\circ c_{S(Z),S(Z)})\ \circ\ v_1\otimes\cdots\otimes
    v_1\otimes v_0\otimes v_1\otimes v_1 \ \circ\ s \\
  && = \id_{S(Z)^{\otimes(d-1)}}\otimes m_{S(Z)}\ \circ\ \id_{S(Z)^{\otimes(d-1)}}\otimes c_{S(Z),S(Z)}
   \ \circ\ v_1\otimes\cdots\otimes v_1\otimes v_0\otimes v_1\otimes v_1 \ \circ\ s \\
  && = \id_{S(Z)^{\otimes(d-1)}}\otimes m_{S(Z)} \ \circ\ v_1\otimes\cdots\otimes v_1\otimes
     v_0\otimes v_1\otimes v_1\ \circ\ \id_{Z^{\otimes(d-2)}}\otimes c_{Z,Z} \ \circ\ s \\ 
  &&= -\    \id_{S(Z)^{\otimes(d-1)}}\otimes m_{S(Z)}\ \circ\ v_1\otimes\cdots\otimes v_1\otimes
   v_0\otimes v_1\otimes v_1 \ \circ\ s,
\end{eqnarray*}
where we used the commutativity of $m_{S(Z)}$ in the first step and the total antisymmetry of $s$ in
the last. Thus $m_Q\mcirc u_d\otimes t_{d-1}\mcirc s=-m_Q\mcirc u_d\otimes t_{d-1}\mcirc s=0$. For
general $i\ne j$ the argument is exactly the same, but becomes rather tedious to write up in detail.
\qed

\brem Lemma \ref{lem-det} and Proposition \ref{prop-absorb} below, both taken from \cite{bichon},
are the crucial ingredients in our approach to the reconstruction theorem.
\erem


\subsection{Construction of an absorbing commutative monoid} \label{ss-monoid2}
Throughout this subsection, let $\2C$ be an even $STC^*$ with monoidal generator $Z$. Consider the
commutative monoid $(Q,m,\eta)=(S(Z),m_{S(Z)},\eta_{S(Z)})^{\times d(Z)}$ in $\mathrm{Ind}\,\2C$ and
the morphisms $s,s',f,u_i,t_j$ as defined in Lemma \ref{lem-det}. Then $m_0\in\End\,Q$ defined by 
\[  m_0 = m_Q\mcirc\id_Q\otimes (f-\eta_Q)=m_Q\mcirc\id_Q\otimes f \,-\, \id_Q \]
is a $Q$-module map, thus $m_0\in\End_Q((Q,m_Q))$. Then its image 
$j=\im\,m_0:(J,\mu_J)\rarr(Q,m_Q)$ (in the abelian category $Q-\Mod$) defines 
an ideal $j:(J,\mu_J)\rarr(Q,m)$ in $(Q,m,\eta)$. This ideal is proper iff $j$ is not an
isomorphism iff $m_0$ is not an isomorphism. Postponing this issue for a minute, we have:

\bprop \label{prop-absorb} \index{monoid!absorbing}
Let $\2C$ be an even symmetric $STC^*$ and let $Z\in\2C$ be such that $\det Z\cong\11$. Let
$(Q,m,\eta)$ and $s,s',f,u_i,t_j$ be as defined in Lemma \ref{lem-det} and $m_0$ as above. Let
$j':(J',\mu')\rarr(Q,m)$ be any proper ideal in $(Q,m,\eta)$ containing the ideal
$j:(J,\mu)\rarr(Q,m)$, where $j=\im\,m_0$. Let $(B,m_B,\eta_B)$ be the quotient monoid. Then
there is an  isomorphism 
\[ (B\otimes Z,m\otimes\id_Z)\cong d(Z)\cdot(B,m_B) \]
of $B$-modules. 
\eprop

\prf Since the ideal is proper, the quotient $(B,m_B,\eta_B)$ is nontrivial and we have an epi 
$p: Q\rarr B$ satisfying
\bea   p\circ m_Q &=& m_B\mcirc p\otimes p, \label{p1} \\
  p\mcirc f &=& p\mcirc\eta_Q \ = \ \eta_B. \label{p2}
\eea
In order prove the claimed isomorphism $B\otimes Z\cong d\,B$ of $B$-modules we define
morphisms $\tilde{q}_i\in\Hom(\11,B\otimes Z),\ \tilde{p}_i\in\Hom(Z,B),\ i=1,\ldots,d$ as the
following compositions:
\bean \tilde{q}_i: &&
\begin{diagram}  \11 & \rTo^{s} & Z^{\otimes d} & \rCongruent &  Z^{\otimes
   (d-1)}\otimes Z & \rTo^{t_i\otimes \id_Z} & Q\otimes Z  & \rTo^{p\otimes\id_Z} & B\otimes Z,
\end{diagram} \\
  \tilde{p}_i: &&
\begin{diagram}  Z &  \rTo^{u_i} & Q & \rTo^{p} & B. \end{diagram} 
\eean
Using, first (\ref{p1}), then (\ref{e-tu}) and (\ref{p2}) we compute
\begin{equation} \label{e-orth}
\begin{tangle}
\hstep\object{B}\\
\hh\hcd\obj{m_B} \\
\hh\id\step\id\\
\hh\id\step\frabox{\tilde{p}_i} \\
\hh\id\step\id\obj{Z}\\
\hh\frabox{\tilde{q}_j} 
\end{tangle}
\quad=\quad
\begin{tangle}
\hstep\object{B}\\
\hh\hcd\obj{m_B} \\
\id\step\O p \\
\id\step\O{u_i} \\
\O p\step\id\\
\O{t_j}\step\id\\
\hh\id\step\id\\
\hh\frabox{s}
\end{tangle}
\quad\quad=\quad\quad
\begin{tangle}
\step\object{B}\\
\step\O p\\
\hstep\obj{m_Q}\step[-.5]\cd\\
\O{t_j}\Step\O{u_i}\\
\hh\step[-1.5]\obj{Z^{d-1}}\step[1.5]\d\step\dd\obj{Z}\\
\hstep\frabox{s}
\end{tangle}
\quad=\ \ 
\delta_{ij}\, p\circ f\ =\ \delta_{ij}\,\eta_B.
\end{equation}
Defining, for $i=1,\ldots,d$,
\[ q_i=\quad\quad
\begin{tangle}
\hstep\object{B}\step[1.5]\object{Z}\\
\hh\step[-1.3]\obj{m_B}\step[1.3]\hcd\step\id \\
\hh\id\step\id\step\id\\
\hh\id\step\frabox{\tilde{q}_i} \\
\hh\id\\
\object{B}
\end{tangle}
\quad\quad\quad\quad
p_i=\quad\quad
\begin{tangle}
\hstep\object{B}\\
\hh\step[-1.3]\obj{m_B}\step[1.3]\hcd \\
\hh\id\step\id\\
\hh\id\step\frabox{\tilde{p}_i} \\
\hh\id\step\id\\
\object{B}\step\object{Z}
\end{tangle}
\]
we find
\[ p_i\mcirc q_j= \quad\quad\quad
\begin{tangle}
\step\object{B}\\
\hh\hstep\step[-1.3]\obj{m_B}\step[1.3]\hcd \\
\hh\hstep\id\step\id\\
\hh\hstep\id\step\frabox{\tilde{p}_i} \\
\hh\step[-1.3]\obj{m_B}\step[1.3]\hcd\step\id \\
\hh\id\step\id\step\id\obj{Z}\\
\hh\id\step\frabox{\tilde{q}_j} \\
\hh\id\\
\object{B}
\end{tangle}
\quad = \quad
\begin{tangle}
\step\object{B}\\
\hstep\obj{m_B}\step[-.5]\cd\\
\hh\id\step[1.5]\hcd\obj{m_B} \\
\hh\id\step[1.5]\id\step\id\\
\hh\id\step[1.5]\id\step\frabox{\tilde{p}_i} \\
\hh\id\step[1.5]\id\step\id\obj{Z}\\
\hh\id\step[1.5]\frabox{\tilde{q}_j} \\
\hh\id\\
\object{B}
\end{tangle}
\quad = \ \ \delta_{ij}\quad\quad
\begin{tangle}
\hstep\object{B}\\
\step[-1.3]\obj{m_B}\step[1.3]\hcd\\
\id\step\counit\obj{\eta_B}\\
\object{B}
\end{tangle}
\quad = \ \delta_{ij}\, \id_B,
\]
where in the next to last step we used (\ref{e-orth}). It is obvious from their definitions
that $p_i, q_i$ are morphisms of $B$-modules. We have thus shown that the $B$-module 
$(B\otimes Z, m_B\otimes\id_Z)$ has $d$ direct summands $(B,m_B)$, and therefore 
\[ (B\otimes Z, m_B\otimes\id_Z)\ \cong \
\underbrace{(B,m_B)\oplus\ldots\oplus(B,m_B)}_{d\
  {\scriptstyle \mbox{summands}}} \ \oplus\
(N,\mu_N). \] It remains to be shown that $N=0$ or,
equivalently, $\sum_{i=1}^d q_i\circ p_i=\id_{B\otimes
  Z}$.  A short argument to this effect is given in
\cite{del,bichon}, but since it is somewhat abstract we
give a pedestrian computational proof. We calculate
\begin{eqnarray*} \sum_{i=1}^d q_i\circ p_i &=& \sum_{i=1}^d \quad
\begin{tangle}
\step\object{B}\step[1.5]\object{Z}\\
\hh\hstep\hcd\step\id\\
\hh\hstep\id\step\id\step\id\\
\hh\hstep\id\step\frabox{\tilde{q}_i}\\
\hh\hcd\\
\id\step\O{\tilde{p}_i}\\
\hh\id\step\id\\
\object{B}\step\object{Z}
\end{tangle}
\quad=\sum_{i=1}^d\quad
\begin{tangle}
\step\object{B}\step[2.5]\object{Z}\\
\hstep\obj{m_B}\step[-.5]\cd\step[1.5]\id\\
\hh\id\hstep\obj{m_B}\step\hcd\step\id\\
\hh\id\step[1.5]\id\step\frabox{\tilde{q}_i}\\
\id\step[1.5]\O{\tilde{p}_i}\\
\hh\id\step[1.5]\id\\
\object{B}\step[1.5]\object{Z}
\end{tangle} \\
&=& \sum_{i=1}^d \quad\quad
\begin{tangle}
\step\object{B}\step[3]\object{Z}\\
\hstep\obj{m_B}\step[-.5]\cd\Step\id\\
\id\step\hstep\obj{m_B}\step[-.5]\cd\step\id\\
\id\step\id\Step\O p\step\id\\
\id\step\id\Step\O{t_i}\step\id\\
\hh\id\step\id\Step\frabox{s}\\
\id\step\O p\\
\id\step\O{u_i}\\
\object{B}\step\object{Z}
\end{tangle}
\quad=\sum_{i=1}^d \quad\quad
\begin{tangle}
\step\object{B}\step[3]\object{Z}\\
\hstep\obj{m_B}\step[-.5]\cd\Step\id\\
\id\step\step\O p\Step\id\\
\id\step[1.5]\obj{m_Q}\step[-.5]\cd\step\id\\
\id\step\O{u_i}\Step\O{t_i}\step\id\\
\hh\id\step\id\hstep\obj{Z^{d-1}}\step[1.5]\id\step\id\\
\hh\id\step\id\Step\frabox{s}\\
\object{B}\step\object{Z}
\end{tangle}
\end{eqnarray*}
Composition with $\eta_B\otimes\id_Z$ shows that this equals $\id_{B\otimes Z}$ iff
\begin{equation} \label{e-iff} 
\sum_{i=1}^d \quad\quad
\begin{tangle}
\step\object{B}\Step\object{Z}\\
\step\O p\Step\id\\
\hstep\obj{m_Q}\step[-.5]\cd\step\id\\
\O{u_i}\Step\O{t_i}\step\id\\
\hh\id\hstep\obj{Z^{d-1}}\step[1.5]\id\step\id\\
\hh\id\Step\frabox{s}\\
\object{Z}
\end{tangle}
\quad=\quad\quad
\begin{tangle}
\object{B}\step\object{Z}\\
\step[-1]\obj{\eta_B}\step\counit\step\id\\
\step\object{Z}
\end{tangle}
\end{equation}
In view of the definition of $(Q,m_Q,\eta_Q)$, the left hand side of (\ref{e-iff}) equals
\begin{eqnarray} \label{e-bla} \lefteqn{ 
   \sum_{i=1}^d (-1)^{d-i}\ \left( p \mcirc c_{S(Z),S(Z)^{\otimes(i-1)}}\otimes\id_{S(Z)^{\otimes(d-i)}}
   \mcirc v_1\otimes \cdots\otimes v_1 \right) \otimes\id_Z \ \circ\ \id_Z\otimes s } \\
 &&= (p \mcirc v_1\otimes \cdots\otimes v_1) \otimes\id_Z \ \circ\
   \left(\sum_{i=1}^d (-1)^{d-i}\ c_{Z,Z^{\otimes(i-1)}}\otimes\id_{Z^{\otimes(d-i)}} \otimes\id_Z\
   \circ\ \id_Z\otimes s \right). \nonumber
\end{eqnarray}
Writing $K_i=c_{Z,Z^{\otimes(i-1)}}\otimes\id_{Z^{\otimes(d-i+1)}}\ \circ\ \id_Z\otimes s$, where
$i\in\{1,\ldots,d\}$, one easily verifies
\[ \Pi^Z_{d+1}(\sigma_j)\mcirc K_i= \left\{ 
\begin{array} {r@{\quad:\quad}l}
  K_{i-1} & j=i-1 \\ K_{i+1} & j=i \\ -K_i & \mbox{otherwise}
\end{array}\right. \]
for all $j\in\{1,\ldots,i-1\}$. This implies that the morphism $Z\rarr Z^{\otimes (d+1)}$ in the
large brackets of (\ref{e-bla}) is totally antisymmetric w.r.t.\ the first $d$ legs, i.e.\ changes
its sign upon multiplication with $\Pi_{d+1}^Z(\sigma_j),\ j=1,\ldots,d-1$ from the left. 
We can thus insert $A_d^Z=s\circ s'$ at the appropriate place and see that (\ref{e-bla}) equals
\begin{eqnarray*} &=& (p \mcirc v_1\otimes \cdots\otimes v_1) \otimes\id_Z \ \circ\ 
  (s\circ s')\otimes\id_Z \\
&& \circ\ \left(\sum_{i=1}^d (-1)^{d-i}\ c_{Z,Z^{\otimes(i-1)}}\otimes
     \id_{Z^{\otimes(d-i)}} \otimes\id_Z\ \circ\ \id_Z\otimes s \right) \\
 &=& (p\mcirc v_1\otimes \cdots\otimes v_1\mcirc
 s)\otimes\id_Z \\
&& \circ\ \left(\sum_{i=1}^d (-1)^{d-i}\
   s'\otimes\id_Z\ \circ\ c_{Z,Z^{\otimes(i-1)}}\otimes\id_{Z^{\otimes(d-i)}} \otimes\id_Z\ \circ\
  \id_Z\otimes s \right) 
\end{eqnarray*}
Now, $p\mcirc v_1\otimes \cdots\otimes v_1\mcirc s=p\mcirc f=\eta_B$.
On the other hand, by the total antisymmetry of $s$ we have 
$s'\mcirc c_{Z,Z^{\otimes(i-1)}}\otimes\id_{Z^{\otimes(d-i)}}=(-1)^{i-1}s'$ and thus
\begin{eqnarray*} \lefteqn{ \sum_{i=1}^d (-1)^{d-i}\ s'\otimes\id_Z\ \circ\
   c_{Z,Z^{\otimes(i-1)}}\otimes\id_{Z^{\otimes(d-i)}} \otimes\id_Z\ \circ\ \id_Z\otimes s } \\
   && = \sum_{i=1}^d (-1)^{d-i}(-1)^{i-1}\ s'\otimes\id_Z\ \circ\ \id_Z\otimes s
   \ =\ d(-1)^{d-1}\,s'\otimes\id_Z\ \circ\ \id_Z\otimes s\ =\ \id_Z, 
\end{eqnarray*}
where the last equality is provided by Lemma \ref{l-conj}. Thus (\ref{e-iff}) is true, implying
$\sum_{i=1}^d q_i\circ p_i=\id_{B\otimes Z}$ and therefore the claimed isomorphism
$B\otimes Z\cong d(Z)\,B$ of $B$-modules.
\qed

\blemma  \label{lem-gamma}
Let $\2C, Z$ and the monoid $(Q,m,\eta)$ be as in Lemma \ref{lem-det}. Then the
commutative algebra $\Gamma_Q=\Hom(\11,Q)$ is $\7Z_+$-graded and has at most countable dimension.
\elemma

\prf By construction of $Q$ we have
\[ \Gamma_Q=\Hom(\11,Q)=\lim_{\stackrel{\longrightarrow}{n}}\bigoplus_{i=0}^n\Hom(\11,S_i(Z))
   =\bigoplus_{i\ge 0}\Hom(\11,S_i(Z)). \] 
Each of the direct summands on the right hand side lives in $\2C$ and thus has finite dimension. It
follows that $\Gamma_Q$ has at most countable dimension. That $\Gamma_Q$ is a $\7Z_+$-graded algebra
is evident from the definition of $m_Q$ in terms of the morphisms 
$m_{i,j}: S_i(X)\otimes S_j(X)\rarr S_{i+j}(X)$ of Lemma \ref{lem-mij}.
\qed

\btheor \label{theor-absorb}
Let $Z\in\2C$ be such that $\det Z\cong\11$. Then there exists a commutative monoid $(B,m_B,\eta_B)$ 
in $\mathrm{Ind}\,\2C$ such that $\dim\Hom_{\mathrm{Ind}\,\2C}(\11,B)=1$ and there is an isomorphism 
$B\otimes Z\cong d(Z)B$ of $B$-modules.
\etheor

\prf Let $(Q,m,\eta)$ and the ideal $j=\mathrm{im}\,m_0:(J,\mu)\rarr(Q,m)$ as before. Assume that
$j$ is an isomorphism, thus epi. Then $m_0$ is epi and thus an isomorphism by Lemma
\ref{l-epiiso}. In particular, the map $\Gamma_Q\rarr\Gamma_Q$ given by $s\mapsto s\bullet(f-\eta)$ 
is an isomorphism, thus $f-\eta\in\Gamma_Q$ is invertible. This, however, is impossible since
$\Gamma_Q$ is $\7Z_+$-graded and $f-\eta\in\Gamma_Q$ is not in the degree-zero part. Thus the ideal
$j$ is proper. By Lemma \ref{l-maxid} there exists a maximal ideal
$\widetilde{j}:(\widetilde{J},\widetilde{\mu})\rarr(Q,m)$ 
containing $j:(J,\mu)\rarr(Q,m)$. If the monoid $(B,m_B,\eta_B)$ is the quotient of $(Q,m,\eta_Q)$
by $j:(\widetilde{J},\widetilde{\mu})\rarr(Q,m)$, Proposition \ref{prop-absorb} implies the
isomorphism $B\otimes Z\cong d(Z)\cdot B$ of $B$-modules. By Lemma \ref{lem-corr}, the quotient 
module $(B,m_B,\eta_B)$ has no proper non-zero ideals, thus by Lemma \ref{l-field}, the commutative
$\7C$-algebra $\End_B((B,m_B))$ is a field extending $k$. By Lemma \ref{lem-end}, 
$\End_B((B,m))\cong\Hom(\11,B)=:\Gamma_B$ as a $\7C$-algebra. By Lemma \ref{lem-proj}, the unit
$\11\in\mathrm{Ind}\,\2C$ is projective, thus Lemma \ref{l-quot} implies that $\Gamma_B$ is a
quotient of $\Gamma_Q$, and by Lemma \ref{lem-gamma} it has at most countable dimension. Now Lemma
\ref{lem-extC} below applies and gives $\Gamma_B=\7C$ and therefore $\dim\Hom(\11,B)=1$ as desired. 
\qed

\blemma \label{lem-extC} \index{finitely generated!algebra}
Let $K\supset\7C$ a field extension of $\7C$. If $[K:\7C]\equiv\dim_\7CK$ is at most countable then
$K=\7C$.
\elemma

\prf Assume that $x\in K$ is transcendental over $\7C$. We claim that the set 
$\{ \frac{1}{x+a}\ | \ a\in\7C\}\subset K$ is linearly independent over $\7C$: Assume that
$\sum_{i=1}^N \frac{b_i}{x+a_i}=0$, where  the $a_i$ are pairwise different and $b_i\in\7C$.  
Multiplying with $\prod_i(x+a_i)$ (which is non-zero in $K$) we obtain the polynomial equation
$\sum_{i=1}^N b_i\prod_{j\ne i}(x+a_j)=0=\sum_{k=0}^{N-1}c_kx^k$ for $x$. Since $x$ is
transcendental, we have $c_k=0$ for all $k=0,\ldots,N-1$. This gives us $N$ linear equations
$\sum_{i=1}^N M_{ki}b_i=0,\ k=1,\ldots,N$, where
$M_{ki}= \sum_{S\subset\{1,\ldots,N\}-\{i\} \atop \# S={k-1}} \prod_{s\in S}a_s$. This matrix can be
transformed into the matrix $(V_{ki}=a_i^{k-1})$ by elementary row transformations. By Vandermonde's
formula, $\det\,V=\prod_{i<j}(a_j-a_i)\ne 0$, thus the only solution of $M{\bf b}=0$ is
$b_1=\cdots=b_N=0$, proving linear independence. Since $\7C$ is uncountable this contradicts the
assumption that $K$ has countable dimension over $\7C$. Thus $K/\7C$ is algebraic and therefore
$K=\7C$ since $\7C$ is algebraically closed.  
\qed

Finally we have: \\

\noindent{\it Proof of Theorem \ref{theor-D}.} If $\2C$ is an even $STC^*$ with monoidal generator
$Z$, Lemma \ref{l-dets} allows us to assume $\det Z\cong\11$ (replacing $Z$ by $Z\oplus\ol{Z}$). Now
Theorem \ref{theor-absorb} provides a monoid $(B,m,\eta)$ in $\mathrm{Ind}\,\2C$ satisfying the
assumptions of Corollary \ref{coro-monoid}, which gives rise to a symmetric fiber functor
$E:\2C\rarr\Vect_\7C$. 
\qed

\brem \label{rem-new} \index{Deligne}
It seems instructive to point out the main difference of our proof of Theorem \ref{theor-D} w.r.t.\
the approaches of \cite{del,bichon}. In \cite{del}, a commutative monoid $(Q,m,\eta)$ for which
there is an isomorphism $Q\otimes Z\cong d(Z)Q$ of $Q$-modules is constructed by a somewhat
complicated inductive procedure. The explicit construction of the monoid that we gave is due to
\cite{bichon}. Deligne proceeds by observing that, for every $X\in\2C$, the $k$-vector space
$\Hom(\11,Q\otimes X)$ is a module over the commutative ring
$\Gamma_Q:=\End_Q((Q,m))\cong\Hom(\11,Q)$, and the functor $\tilde{E}: X\mapsto\Hom(\11,Q\otimes X)$
is monoidal w.r.t.\ the tensor product of $\Gamma_Q-\Mod$ (rather than that of $\Vect_\7C$). 
Now, a quotienting procedure w.r.t.\ a maximal ideal $J$ in $\Gamma_Q$ is used to obtain a tensor
functor $E: \2C\rarr K-\Vect$, where $K=\Gamma_Q/J$ is a field extension of the ground field $k$. If
$\Hom(\11,Q)$ is of at most countable dimension then $[K:k]\le\aleph_0$, and if $k$ is uncountable
and algebraically closed it follows that $K=k$.  

Our approach differs in two respects. Less importantly, our insistence on $\det\,Z\cong\11$ makes the
construction of the monoid $(Q,m,\eta)$ slightly more transparent than in \cite{bichon}. More
importantly, we perform the quotienting by a maximal ideal inside the category of $Q$-modules 
in $\mathrm{Ind}\,\2C$ rather than in the category of $\Gamma_Q$-modules, yielding a monoid
$(Q',m',\eta')$ in $\mathrm{Ind}\,\2C$ with $\Gamma_{Q'}=\7C$. Besides giving rise to a symmetric
fiber functor $E:\2C\rarr\Vect_\7C$ in a more direct fashion, this has the added benefit, as we will
show in the final subsection, of allowing to recover the group $\Nat_\otimes E$ without any
reference to the fiber functor and its natural transformations! The ultimate reason for 
this is that, due to uniqueness of the embedding functor, the monoid $(Q',m',\eta')$ in
$\mathrm{Ind}\,\2C$ is nothing but the monoid $(\pi_l,\tilde{m},\tilde{\eta})$ in $\Rep\,G$ that
arises from the left regular representation of $G$, cf.\ \cite{MT}. 
\erem


\subsection{Addendum} \label{ss-add}
In the previous subsection we have concluded the proof of the existence of a fiber functor and, by
the concrete Tannaka theorem, of the equivalence $\2C\simeq\Rep_f(G,k)$, where $(G,k)$ is a compact
supergroup. However, we would like to show how the group $\mathrm{Nat}_\otimes E$, and in some cases
also $G$, can be read off directly from the monoid $(Q,m,\eta)$, bypassing fiber functors, natural
transformations etc. 

\bdefin 
The automorphism group of a monoid $(Q,m,\eta)$ in a strict tensor category $\2C$ is 
\[ \Aut(Q,m,\eta)=\{ g\in\Aut\,Q\ | \ g\circ m=m\circ g\otimes g, \ g\circ\eta=\eta\}. \]
\edefin

\bprop 
Let $\2C$ be an $STC^*$ and $(Q,m,\eta)$ a monoid in $\mathrm{Ind}\,\2C$ satisfying
\begin{itemize}
\item[(i)] $\dim\Hom_{\mathrm{Ind}\,\2C }(\11,Q)=1$.
\item[(ii)] For every $X\in\2C$, there is $n(X)\in\7Z_+$ such that $n(X)\ne 0$ whenever 
$X\not\cong 0$ and an isomorphism 
$\alpha_X:(Q\otimes X,m\otimes\id_X)\rarr n(X)\cdot(Q,m)$ of $Q$-modules.
\end{itemize}
Then the group $\Nat_\otimes E$ of monoidal natural automorphisms of the functor constructed in
Proposition \ref{prop-embed} is canonically isomorphic to the group $\Aut(Q,m,\eta)$.
\eprop

\prf Let $g\in\Aut(Q,m,\eta)$. For every $X\in\2C$ define $g_X\in\End\,E(X)$ by 
\[ g_X\,\psi=g\otimes\id_X\circ\psi \quad \forall \psi\in E(X)=\Hom(\11,Q\otimes X). \]
From the definition of $(g_X)_{X\in\2C}$ and of the functor $E$ it is immediate that
$(g_X)_{X\in\2C}$ is a natural transformation from $E$ to itself. We must show this natural
transformation is monoidal, i.e.\ 
\[ \begin{diagram} E(X)\otimes E(Y) & \rTo^{d_{X,Y}} & E(X\otimes Y) \\
  \dTo^{g_X\otimes g_Y} && \dTo_{g_{X\otimes Y}} \\
   E(X)\otimes E(Y) & \rTo^{d_{X,Y}} & E(X\otimes Y)
\end{diagram}\]
commutes. To this end consider $\phi\in E(X)=\Hom(\11,Q\otimes X),\psi\in E(X)=\Hom(\11,Q\otimes Y)$
and $g\in\Aut(Q,m,\eta)$ with $(g_X)_{X\in\2C}$ as just defined. Then the image of 
$\phi\boxtimes\psi\in E(X)\otimes E(Y)$ under $g_{X\otimes Y}\circ d_{X,Y}$ is
\[  g\otimes\id_{X\otimes Y}\mcirc m\otimes\id_{X\otimes Y}\mcirc\id_Q\otimes\phi\otimes\id_Y\mcirc\psi, \]
whereas its image under $d_{X,Y}\circ g_X\otimes g_Y$ is
\[ m\otimes\id_{X\otimes Y}\mcirc g\otimes g\otimes\id_{X\otimes
   Y}\mcirc\id_Q\otimes\phi\otimes\id_Y\mcirc\psi. \] 
In view of $g\circ m=m\circ g\otimes g$, these two expressions coincide, thus 
$(g_X)\in\Nat_\otimes E$. It is very easy to see that the map
$\sigma:\Aut(Q,m,\eta)\rarr\Nat_\otimes E$ thus obtained is a group homomorphism.

We claim that $\sigma$ is an isomorphism. Here it is important that we work in $\mathrm{Ind}\,\2C$
rather than any category $\widehat{\2C}$, since this implies that $Q$ is an inductive limit of
objects in $\2C$. The assumptions (i),(ii) then give 
$\Hom(X,Q)\cong\Hom(\11,Q\otimes\ol{X})\cong \7C^{n(\ol{X})}$ for all $X\in\2C$ and thus (using
$n(X)=n(\ol{X})=\dim E(X)$) 
\be \label{eq-Q} Q\cong\lim_{\longrightarrow\atop S\subset I}\bigoplus_{i\in S}n(X_i)X_i \quad\quad
  \mbox{and}\quad\quad
   \End\,Q\cong\prod_{i\in I}\End\,E(X_i), \ee
where $S$ runs though the finite subsets of $I$. Assume now that $\sigma(g)$ is the identity natural
transformation, i.e.\ $g\otimes\id_X\mcirc\phi=\phi$ for all $X\in\2C$ and 
$\phi\in\Hom(\11,Q\otimes X)$. Be the existence of conjugates in $\2C$, this is equivalent to
$g\circ s=s$ for all $Y\in\2C$ and $s\in\Hom(Y,Q)$. Since $Q$ is an inductive limit of objects in
$\2C$, this implies $g=\id_Q$. 

If now $\alpha\in\Nat_\otimes E$, we first observe that $\alpha$ is a
natural isomorphism by \ref{lem-iso}. By the isomorphisms
$\Nat\,E\cong\prod_{i\in I}\End\,E(X_i)$ (cf.\ the proof of
Proposition \ref{prop-dense}) and (\ref{eq-Q}), we have a map
$\Nat_\otimes E\rarr\Aut\,Q$.  Reversing the preceding computations
shows that every $\alpha\in\Nat_\otimes E$ gives rise to an element of
$\Aut(Q,m,\eta)$.  \qed

\brem This result shows that the group $\mathrm{Nat}_\otimes E$ can be
recovered directly from the absorbing monoid $(Q,m,\eta)$ in
$\mathrm{Ind}\,\2C$. In general the compact group $G$ as defined in
Subsection \ref{ss-concrete1} is a true subgroup of
$\mathrm{Nat}_\otimes E$, the latter being the pro-algebraic envelope
of $G$. (In the cases of $G=U(1), SU(2), U(2)$, e.g., that would be
$\7C^\times, SL(2,\7C), GL(2,\7C)$, respectively.) But if $\2C$ is
finite (i.e.\ has finitely many isomorphism classes of simple objects)
then $\Nat_\otimes E$ is finite and $G=\Nat_\otimes E$.
Interestingly, even in the case of finite $\2C$, where the monoid
$(Q,m,\eta)$ actually lives in $\2C$, there seems to be no way to
recover $G$ without using $\mathrm{Ind}\,\2C$ at an intermediate
stage.  \erem



\end{document}